\documentclass[12pt,a4paper]{book}
\usepackage[utf8]{inputenc}
\usepackage{graphicx}
\usepackage{lettrine}
\usepackage{scalefnt}
\usepackage{amsmath}
\usepackage{subfigure}
\usepackage[colorlinks, citecolor = magenta]{hyperref}
\usepackage{color}
\usepackage{xcolor}
\usepackage{dsfont}
\usepackage{mathrsfs}
\usepackage{mathtools}
\usepackage{setspace}
\usepackage[T1]{fontenc}
\usepackage{txfonts}
\graphicspath{ {figures/} }
\usepackage{braket}
\usepackage[margin=3.0cm]{geometry}
\usepackage{tikzpagenodes}
\usepackage{lipsum}
\usepackage{tikz} 
\usepackage{caption}
\usepackage[toc,page]{appendix}

\usetikzlibrary{shapes.geometric, arrows}
\tikzstyle{startstop} = [rectangle, rounded corners, minimum width=3cm, minimum height=1cm,text centered, text width=4cm, draw=black, fill=red!30]
\tikzstyle{io} = [trapezium, trapezium left angle=70, trapezium right angle=110, minimum width=3cm, minimum height=1cm, text centered, text width=4cm, draw=black, fill=blue!30]
\tikzstyle{process} = [rectangle, minimum width=3cm, minimum height=1cm, text centered, text width=4cm, draw=black, fill=orange!30]
\tikzstyle{decision} = [diamond, minimum width=3cm, minimum height=1cm, text centered, text width=3.4cm, draw=black, fill=green!30]
\tikzstyle{ov} = [ellipse, minimum width=3cm, minimum height=2cm,text centered, text width=4cm, draw=black, fill=green!30]
\tikzstyle{arrow} = [thick,->,>=stealth]
\usepackage{braket}
\begin{document}

\begin{center}
{\Large {\bf Quantum Simulation of Neutrino Oscillation
  and Dirac Particle Dynamics in Curved Space-time }}
\vskip 0.70cm
{\bf {\em By}} 
\vskip -0.2cm
{\bf {\large Arindam Mallick}}
\vskip 0.0cm
{\bf {\large PHYS10201205004}}
\vskip 0.5cm
{\bf {\large The Institute of Mathematical Sciences, Chennai}}
\vskip 2.6cm
{\bf {\em {\large A thesis submitted to the
\vskip 0.05cm
Board of Studies in Physical Sciences
\vskip 0.05cm
In partial fulfillment of requirements
\vskip 0.05cm
For the Degree of 
}}}
\vskip 0.05cm
{\bf {\large DOCTOR OF PHILOSOPHY}}
\vskip 0.1cm
{\bf {\em of}}
\vskip 0.1cm
{\bf {\large HOMI BHABHA NATIONAL INSTITUTE}}

\vfill

\includegraphics[height=3.5cm, width=3.5cm]{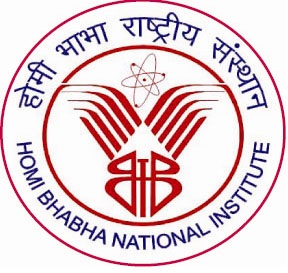}

\vfill

{\bf {\large June, 2018}}

\vfill
\end{center}

\newpage
\centerline{{\bf {\large STATEMENT BY AUTHOR}}}
\vskip 1.00cm
\doublespacing
This dissertation has been submitted in partial fulfillment of
requirements for an advanced degree at Homi Bhabha National Institute
(HBNI) and is deposited in the Library to be made available to borrowers
under rules of the HBNI.
\vskip 0.6cm
Brief quotations from this dissertation are allowable without special
permission, provided that accurate acknowledgment of source is made.
Requests for permission for extended quotation from or reproduction of
this manuscript in whole or in part may be granted by the Competent
Authority of HBNI when in his or her judgment the proposed use of the
material is in the interests of scholarship. In all other instances,
however, permission must be obtained from the author.

\vskip 1.8cm


$~$\hspace{10.2cm} Arindam Mallick 
\newpage

{\bf \centerline{\Large{List of Publications (included in the Thesis)}}}

\vspace{1cm}

{\bf {\large{Journal}}}
\begin{enumerate}
\item\label{one} \textbf{Dirac cellular automaton from split-step quantum walk}.\\
Arindam Mallick, \& C. M. Chandrashekar.\\
\href{https://www.nature.com/articles/srep25779}{{\it Scientific Reports}, {\bf 6}, 25779 (2016)}.

\item\label{two} \textbf{Neutrino oscillations in discrete-time quantum walk framework}.\\
Arindam Mallick, Sanjoy Mandal, \& C. M. Chandrashekar.\\
\href{https://doi.org/10.1140/epjc/s10052-017-4636-9}{{\it Eur. Phys. J. C}  {\bf 77}, 85, (2017)}.

\end{enumerate}

 \vspace{0.2cm}

{\bf {\large{Communicated}}}

\begin{enumerate}
\item\label{three} \textbf{Simulating Dirac Hamiltonian in curved space-time by split-step quantum walk}.\\
Arindam Mallick, Sanjoy Mandal, Anirban Karan, \& C. M. Chandrashekar.\\
\href{https://arxiv.org/abs/1712.03911v2}{arXiv:1712.03911v2 [quant-ph] (2018)}.

\end{enumerate}

\newpage
\centerline{{\bf {\Large ~~~}}}
\vskip 8cm
 \centerline{\Huge {\it Dedicated }}
 \vskip 1cm
 
 \centerline{\Huge {\it To }}
 \vskip 1cm
 
 \centerline{\Huge {\it My Grandparents}}

\newpage
~

\centerline{{\bf{\large ACKNOWLEDGMENTS}}}

\vskip 0.5cm

 This thesis would not have been possible without the support of various people. I want to show my special gratitude to my supervisor 
 Dr.~C.~M.~Chandrashekar for helping me to learn the subject and guiding me through my research works. During my PhD life, I interacted a lot with
 Prof.~Sibasish Ghosh and I always had something new to learn from him. I am very grateful for his affection and supports. 
 
 \vspace{0.6cm}

 I thank Prof.~Sitabhra Sinha and Dr.~S Sridhar for guiding me during my Master degree project, and helping to expand my knowledge. 
 I am thankful to my senior researchers importantly Dr.~Manik Banik, Dr.~Aravinda S and Dr.~George Thomas who were postdoctoral fellows at our group during my PhD.       
 It was my great fortune that I got collaborators and friends like Sanjoy Mandal, Sagnik Chakraborty, Dipanjan Mandal, Anirban Karan who always kindly participate 
 in academic discussion and help me a lot in my research works. I would also like to thank all the IMSc faculties who taught me during my Master degree course works.
 I am also thankful to all of my doctoral committee members who help me to pursue my PhD by giving their valuable feedbacks.

 \vspace{0.6cm}

I am also showing my gratitude towards all of my teachers specially Mr.~Pranab Chakraborty, Mr.~Jayanta Sinha, and my friends. 
Above all, the love and affection of my grandparents, parents and other family members go beyond my acknowledgment.  

\newpage
~

\centerline{{\bf{\large ABSTRACT}}}

\vskip 0.5cm
Quantum simulation become a necessary step to learn physics about a system when theoretical analysis, direct experimental observation, and numerical investigations in classical computers are difficult.
The Dirac particles in a general situation is one such system. Discrete quantum walk (DQW)---a U(2) coin operations followed by a
coin state dependent positional shift operations is a powerful quantum simulation scheme, and implementable in well controllable table-top set-ups.
In the thesis, we first identify that the conventional DQW can't exactly simulate Dirac Cellular Automaton (DCA) which is a discretized theory 
of free Dirac Hamiltonian (DH). We found some particular choice of coin parameters of the split-step (SS) DQW---a generalization of DQW can fully
simulate single-particle DCA. Next we question whether the same SS-DQW can simulate dynamics of free Dirac particle with extra degrees of freedom like colors, 
flavors besides the spin or chirality. One such example is Neutrino oscillation. 
By moving from the U(2) coined SS-DQW to the U(6) coined SS-DQW we have simulated the exact probability profile of Neutrino flavor transitions.  
We further probe towards simulating single particle massive DH in presence of background potentials and space-time curvature. By using a SS-DQW with position-time
dependent coin parameters, and we realize that it will give us an unbounded effective Hamiltonian, at the continuum limit of position-time. So we have introduced a modified version of
inhomogeneous SS-DQW which will produce a bounded effective Hamiltonian. This modified SS-DQW with U(2) coin operations produces single-particle massive DH in presence of abelian
gauge potentials and space-time curvature. Introducing higher dimensional---U(N) coin operations in the modified SS-DQW we can include
non-abelian potentials in the same DH. In order to simulate two-particle DH in presence of curved space-time and external potentials, we have used two particle modified SS-DQW, 
where the shift operations act separately on each particle, the coin operations which act simultaneously on both particles and contain all kinds of interactions.
 
\newpage

\tableofcontents
 \pagenumbering{arabic}

 \newpage
\vspace{-2cm}
 
 \chapter*{Synopsis}
\addcontentsline{toc}{chapter}{Synopsis} 

\vspace{-1cm}

  \section*{Introduction and motivation}
\vspace{-.5cm}  
  
For many decades now, mimicking operations, behaviors and functions of one system by another has played an important role in understanding the real system that are difficult to be engaged in direct 
study due to practical constraints. This has also led to identify many unobserved behaviors and for new discoveries. When it comes to systems in nature with prominent quantum behavior,
mimicking them using classical systems or computers is limited due to insufficient memory of the available states (classical bits) in classical system. This was one of the main motivations 
for the development of the field, quantum computation and quantum information\,\cite {feyn}.  One of the important steps in the direction of mimicking quantum systems is to set the criteria 
of the system that can be qualified as an efficient mimicker. It turns out that not all experimentally accessible quantum systems are efficient mimickers of other quantum systems. 
Quantum simulation is such a subject where we scientifically investigate these criteria and develop efficient algorithmic schemes and models for mimicking selected quantum systems.

Classical random walk (CRW) has played an important role in simulation of various dynamics in classical systems.  So, it appears natural to explore the use of its quantum analog 
in discrete space-time, discrete quantum walk (DQW)\,\cite{aharonov2} as a basic tool to model dynamics and develop algorithms for quantum simulations. It is based on unitary 
evolution and is implementable in table-top experimental quantum imitators like cold atoms in optical lattice, superconductor qubits, NMR, photonic and other controllable 
quantum devices and systems. Compared to quantum simulators based on the Lagrangian or Hamiltonian formalism whose operations are global,
DQW provides an approach where operations can be local and applied in discrete space and time. 

 {\bf DQW:} As single-step CRW evolution is described using a coin toss operation followed by a outcome, head or tail dependent positional movement operation. In a similar way,
 DQW evolution operator is described using a quantum coin operation $C$ which is a general unitary operator followed by a spatial shift operation $S$ that depends on the coin state
 of the particle (walker), $U_{DQW} = S \cdot C$. In contrast to the CRW case, it is always unitary, and superposition of different coin states  as well as different position states
 are allowed. This enables the spreading behavior of DQW in position space to grow quadratically faster than the classical one.  $U_{DQW}$ is defined on a composite Hilbert 
 space $\mathcal{H}_c \otimes \mathcal{H}_x$ where $\mathcal{H}_c = \text{span}\big\{\ket{\uparrow} = (1~~0)^T, \ket{\downarrow} = (0~~1)^T \big\}$ represents coin Hilbert 
 space and $\mathcal{H}_x = \text{span}\big\{x: x \in a~\mathbb{Z}~\text{or}~a~\mathbb{Z}_\mathcal{N}\big\}$ represents discrete lattice Hilbert space (position Hilbert space). 
 The coin operator $C = e^{- i \sum\limits_{q=0}^3 \theta^q(\delta t)~ \sigma_q} \otimes \sum\limits_x \ket{x}\bra{x}$ acts as a general $2 \times 2$ unitary matrix
 on $\mathcal{H}_c$ where $\{\sigma_0, \sigma_1, \sigma_2, \sigma_3\}$ are conventional Pauli matrices, and identity on $\mathcal{H}_x$; the shift
 operator $S =  \sum\limits_{x}  \ket{\uparrow}\bra{\uparrow} \otimes \ket{x + a}\bra{x} + \ket{\downarrow}\bra{\downarrow} \otimes \ket{x-a}\bra{x}$
 defined as projection operation on $\mathcal{H}_c$ and lattice translation operation on $\mathcal{H}_x$. In the shift operator $S$, $a$ is the
 lattice step-size and it will be unitary when the lattice is either periodic ($\mathbb{Z}_\mathcal{N}$ case) or infinite ($\mathbb{Z}$ case).
 Operator $U_{DQW}$ takes a system state $\ket{\psi(t)}$ at time $t$ to a state at time $t + \delta t$, $U_{DQW} \ket{\psi(t)} = \ket{\psi( t + \delta t)}$ where
 $\delta t$ is the time step-size, and, any time $t \in \delta t \times \big(\{0\} \bigcup \mathbb{N}\big)$. 

Description of Dirac particles, one of the fundamental constituents of our nature using Dirac equation has played an important role in understanding various 
phenomenon in both, low energy and high energy regime.  Some previous studies have shown that with proper choice of coin operation, DQW reproduces the form of a
Dirac Hamiltonian in the short length scale limit (which we will call as continuum limit)  of it's background discrete space and time-step: $a \to 0, \delta t \to 0$. 
Therefore, DQW can be thought as a discretization and simulation scheme for Dirac dynamics. However, the existing connections 
between DQW and Dirac equations still leaves some gaps to be explored. 
In this thesis using a generalized version of DQW---the spit-step discrete quantum walk (SS-DQW) we address the following questions :

 1.~ The form of the conventional DQW doesn't capture all the properties of the Dirac cellular automaton (DCA) which is a discretization of free Dirac Hamiltonian (FDH). 
 What modification are needed for DQW to capture all relevant features of the Dirac dynamics ?
 
 2.~ Is the conventional form of the DQW able to simulate phenomena related to Dirac particles with additional degrees of freedom, such as the neutrino flavor oscillation ?
 If not, what are the modifications needed for initial state preparation and DQW evolution operator to mimic the neutrino flavor oscillation probability ?
 
 3.~ Introducing position and time dependency in the coin parameters $\{\theta^{q}\}_{q=0}^3$ of DQW doesn't 
 easily capture space-time curvature, abelian, nonabelian gauge potential effects on a massive Dirac particle in a single Hamiltonian framework. 
 So, what are the modification required for DQW to capture all these and simulate Dirac Hamiltonian in curved space-time and effect of potentials?
 
 Below we will discuss them section wise. 

\begin{figure}[h]\label{depiction}
\centering
\subfigure[][]{
\includegraphics[width=0.45\textwidth]{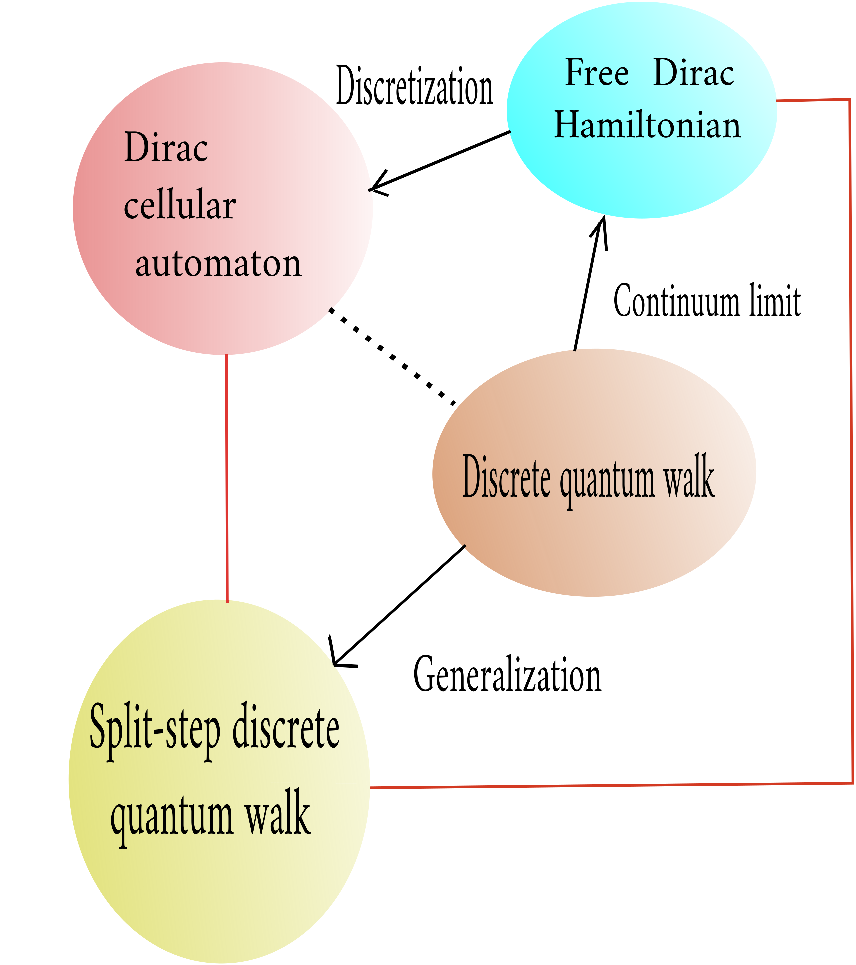}
\label{fig:sub1}}~~~
\subfigure[][]{\vspace{1pt}
 \includegraphics[width=0.45\textwidth]{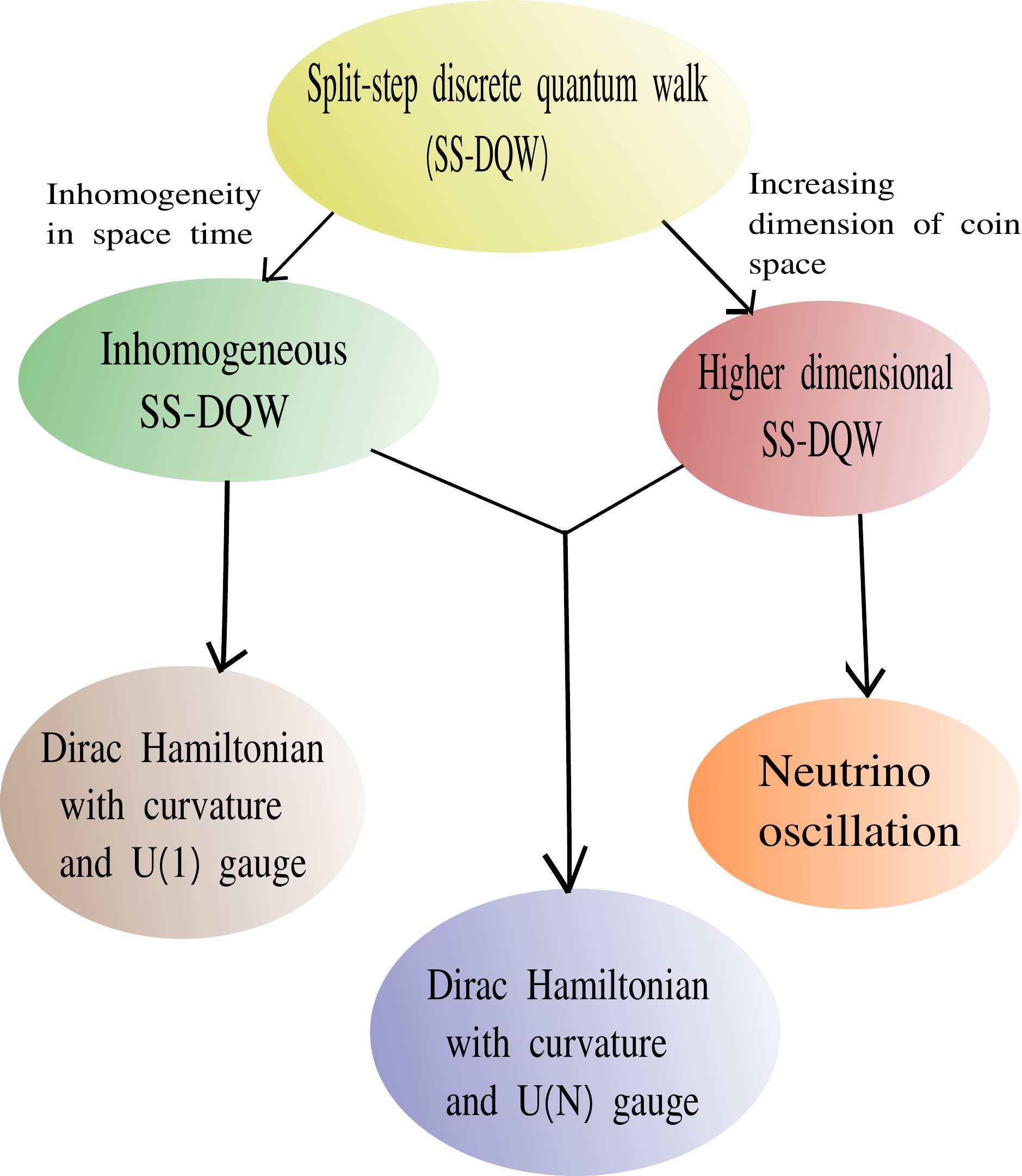}
\label{fig:sub2}}
 \caption[Pictorial description of the content of the thesis]{Pictorial description of the content of the thesis.
 Fig.~\ref{fig:sub1} answers the {\it first} issue, where we identify the missing link of DCA-DQW-FDH 
 by using SS-DQW. Fig.~\ref{fig:sub2} is about the {\it second} and {\it third} issues. 
 By properly increasing the coin space dimension in SS-DQW, we can simulate the three-flavor neutrino oscillation.
 Introducing inhomogeneity in coin operator we reproduce massive Dirac Hamiltonian in $(1+1)$ 
 dimensional curved space-time in presence of $U(1)$ gauge potential and shows how the combination of inhomogeneity 
 and higher dimensional coin operator can capture more general $U(N)$ nonabelian potential effect.}
\end{figure} 
\vspace{-1cm}

\section*{Connecting DCA with DQW}
\vspace{-.5cm}

One of the discretization scheme of free Dirac particle dynamics is DCA defined on $\mathcal{H}_c \otimes \mathcal{H}_x$ and 
based on the four basic assumptions: (1) the evolution operator $U_{DCA}$ is local unitary, (2) $U_{DCA}$  is invariant
under spatial translation (discrete sense), (3)$U_{DCA}$ is covariant under parity and time reversal transformation (discrete sense),
(4) $U_{DCA}$ contains a minimum two controller or internal degrees of freedom. The local unitary means, the dynamics at one site is only 
influenced by it's nearest neighbors. The form of the evolution operator is  
\begin{align}
U_{DCA} =  \sum_{x} \eta_1 \big[~ \ket{\uparrow}\bra{\uparrow} \otimes \ket{x + a}\bra{x} 
+ \ket{\downarrow}\bra{\downarrow} \otimes \ket{x-a}\bra{x} ~\big] - i~\eta_2~ \sigma_1 \otimes \ket{x}\bra{x}~.
\end{align} 
The $\eta_1$ is the hopping strength in lattice and $\eta_2$ corresponds to the mass term. 
This evolution operator produces free Dirac particle dynamics at the continuum limit of
position, time-step for smaller values of $\eta_2$ and larger values of $\eta_1$. But DCA is not in a ready-to-use implementable form (operational form). 
On the other hand, implementation and control over the dynamics of DQW has already been demonstrated in various state-of-art experimental set-ups like cold-atom, ions, NMR and photonic devices. 
And DQW is shown to reproduce the FDH at the continuum limit of position and time steps.
In one spatial dimension the single step of standard DQW evolution operator takes the from:
\begin{align}
 U_{DQW} = S \cdot C = \sum_{x}  \ket{\uparrow}\bra{\uparrow}e^{- i \sum_{q=0}^3 \theta^q(\delta t)~ \sigma_q} 
\otimes \ket{x + a}\bra{x} + \ket{\downarrow}\bra{\downarrow}e^{- i \sum_{q=0}^3 \theta^q(\delta t)~ \sigma_q} \otimes \ket{x-a}\bra{x} \nonumber\\
 = e^{- i~ \theta^0(\delta t)} \Big[ \big(\ket{\uparrow}\bra{\uparrow} F + \ket{\uparrow}\bra{\downarrow} G \big) \otimes \ket{x + a}\bra{x}
 + \big( -\ket{\downarrow}\bra{\uparrow}G^*  + \ket{\downarrow}\bra{\downarrow}F^* \big) \otimes \ket{x-a}\bra{x}\Big],
\end{align}
\vspace{-1cm}

where $F$, $G$ are complex functions of coin parameters $\{\theta^q\}_{q=1}^3$. 
There doesn't exist any choice of $\theta^q$ for which, $U_{DQW}$ exactly matches with $U_{DCA}$. The noticeable difference appears in the 
positional probability distribution in the form of fine oscillation in case of DCA, and, it's absence in DQW scenario \cite{perez}. 

\begin{figure}[h]
\centering
 \includegraphics[height = 6.5 cm]{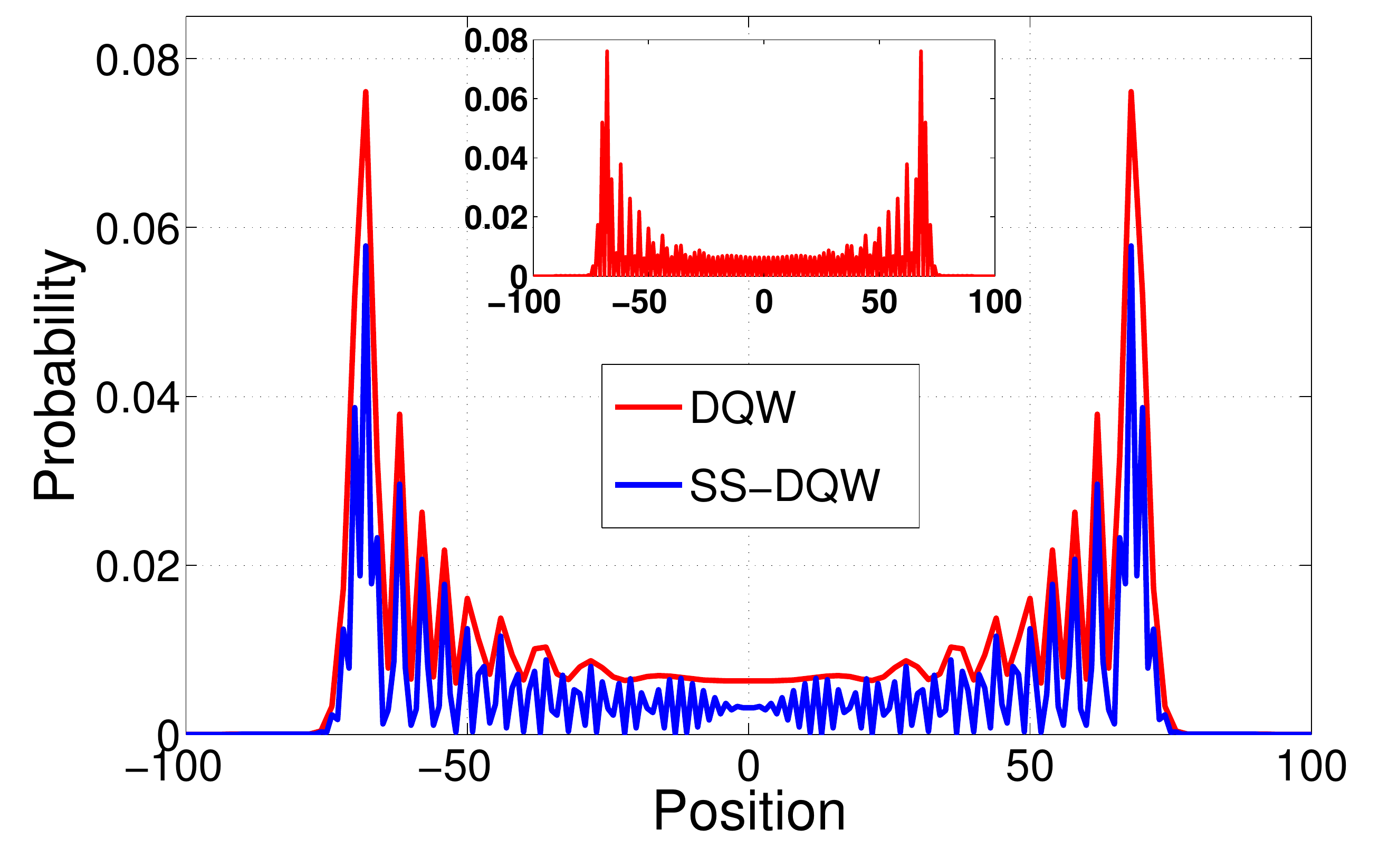}
 \caption[Positional probabilities for DQW and SS-DQW]{The probability of finding the particle in one-dimensional position space after 100 time-steps of DQW and SS-DQW.
The initial state of the particle is $\frac{1}{\sqrt{2}} ( \ket{\uparrow} + \ket{\downarrow} )\otimes \ket{x=0}$ and 
the coin operators are $C_j = e^{- i~\theta^1_j \sigma_1} \otimes \sum_x \ket{x}\bra{x}$. 
 Blue distribution is for SS-DQW with the choice: $\theta^1_1 = 0$, $\theta^1_2 = \frac{\pi}{4}$, 
 which is identical to DCA when $\eta_1 = \eta_2 = \frac{1}{\sqrt{2}}$ and the red line is for the DQW with $\theta^1 = \frac{\pi}{4}$; the
points with zero probability is removed from the main plot whereas, it is retained in the inset.}\label{dcafig}
\end{figure}

To reproduce those fine oscillations missing in DQW,  we use the SS-DQW \cite{topo} which is a generalized version of DQW. In this case, the single-step
evolution operator is given as \begin{align}
 U_{SQW} = S_+ \cdot C_2 \cdot S_- \cdot C_1 
\end{align}
\vspace{-1.5cm}

acting on $\mathcal{H}_c \otimes \mathcal{H}_x$ 
where 
\begin{align}
 C_j = e^{- i \sum_{q=0}^3 \theta_j^q(\delta t)~\sigma_q} \otimes \sum_x \ket{x}\bra{x}~\text{for}~~j=1,2;\\
 S_+ =  \sum_{x}  \ket{\uparrow}\bra{\uparrow} \otimes \ket{x + a}\bra{x} + \ket{\downarrow}\bra{\downarrow} \otimes \ket{x}\bra{x};\\
S_- =  \sum_{x}  \ket{\uparrow}\bra{\uparrow} \otimes \ket{x}\bra{x} + \ket{\downarrow}\bra{\downarrow} \otimes \ket{x-a}\bra{x}. 
\end{align}

For the choice of operators $\theta_1^q(\delta t)= 0$ for all $q$ and for $\sum_{q=0}^3 \theta_2^q(\delta t)~\sigma_q = \theta(\delta t)~\sigma_1$,
\begin{align}U_{SQW} =
\sum_{x} \cos \theta \big[~ \ket{\uparrow}\bra{\uparrow} \otimes \ket{x + a}\bra{x} 
+ \ket{\downarrow}\bra{\downarrow} \otimes \ket{x-a}\bra{x} ~\big] - i \sin \theta ~ \sigma_1 \otimes \ket{x}\bra{x}. \label{ssev}
\end{align}
This exactly matches with the $U_{DCA}$ and capture the fine oscillation in the positional probability distribution 
which was missing for the DQW case as shown in fig.~\ref{dcafig}. By using the definition $U_{SQW} = e^{- \frac{i \delta t}{\hbar} H}$,
where $H$ is the Hamiltonian, one can show $H$ matches with FDH at the continuum limit $\delta t \to 0, a \to 0$. 
In this way the missing connection: FDH-DQW-DCA has been established \cite{dca}.

In this work we also discussed the entanglement between position and spins, as a function of 
time-steps and initial states for various choice of coin parameters. We have shown that the entanglement at long time limits, 
for SS-DQW configuration which results in recovery of DCA is more sensitive to the initial states than for
the case of regular DQW. 
\vspace{-1cm}

\section*{Simulating neutrino oscillation} 
\vspace{-.5cm}

When we have a discrete space-time quantum simulation scheme for FDH, one of the natural question would be to explore simulation of other phenomena related to Dirac particles. 
Simulation of phenomena related to Dirac particle while it carries internal degrees of freedom other than spin is one such question we asked. One such phenomenon is 
the three flavor neutrino oscillation.

Considering the case that neutrino is a massive Dirac particle, it can be described by SS-DQW as evident from the previous section. 
Neutrino carries extra three flavor degrees of freedom described by the states: $\ket{\nu_e}$, $\ket{\nu_\mu}$, $\ket{\nu_\tau}$ 
or three possible mass eigenstates: $\ket{\nu_{m_1}}$, $\ket{\nu_{m_2}}$, $\ket{\nu_{m_3}}$; 
so instead of 2 dimensional coin space, here  
we need at least 6 dimensional coin space in $(1+1)$ dimensional space-time.
So, in this case, we define the evolution operator 

\vspace{-2cm}

\begin{align}
 U_\nu = U(\theta_{m_1}) \oplus U(\theta_{m_2}) \oplus U(\theta_{m_3})~~\text{where}~~
U(\theta_{m_j}) \coloneqq U_{SQW}(\theta_{m_j}).
\end{align}
\vspace{-1.5cm}

Neutrinos in nature is usually not detected as mass eigenstates but as the flavor states which are not the eigenstates of their governing Dirac Hamiltonian, so during propagation 
the neutrinos remain as a superposition of three flavors states, the corresponding probability 
amplitudes are determined by PMNS matrix elements ($U_{\alpha j}$) and the free Dirac Hamiltonians with three different masses.
After evolution for time-step $t$ an initial flavor state of neutrino: $\ket{\nu_\alpha} = \sum_j U^*_{\alpha j} \ket{\nu_{m_j}}$ for $\alpha \in \{e, \mu, \tau\}, j \in \{1, 2, 3\}$, 
becomes 
\begin{align}
 \ket{\psi(t)} = [U_\nu ]^{\frac{t}{\delta t}} \ket{\nu_\alpha} = \sum_j U^*_{\alpha j}  [U_\nu ]^{\frac{t}{\delta t}} \ket{\nu_{m_j}} 
= \sum_j U^*_{\alpha j}  e^{- i \frac{E_j t}{\hbar} } \ket{\nu_{m_j}}
\end{align}
 where $E_j$ is the energy eigenvalue of the state $\ket{\nu_{m_j}}$. 
The oscillation probability, i.e., the transition probability from flavor $\alpha$  to flavor $\beta$ is determined by the 
quantity
\vspace{-1.5cm}

\begin{align}
       P_{\beta\alpha}(t) = |\braket{\nu_\beta| \psi(t)}|^2.   
\end{align}
\vspace{-1.5cm}

For given PMNS matrix, the mass difference is responsible for the oscillation \cite{ponte} and the masses can be controlled by 
the coin parameters $\theta_{m_j}$. By choosing proper coin parameters we are 
able to reproduce the exact oscillation profile \cite{neu} as same as the current real experimental observation \cite{theoryneu}. 
These coin parameter values can be used to investigate other neutrino physical phenomena.
We also discussed the entanglement between spins and position space. During neutrino propagation this entanglement can quantify
the wave function delocalization around instantaneous average position of the neutrino.
We provide a simulation scheme by a three-qubit and a qubit-qutrit system which can be used in place of a 6 dimensional single quantum state. 

This shows a way to construct DQW algorithm for high energy particle phenomena where particles carry other internal degrees of freedom besides that of the spins.  
\vspace{-1cm}

  \section*{Simulating Dirac particle dynamics in presence of general external potential and curved space-time}
  \vspace{-.5cm}
  
In some previous literatures it was shown a single-step DQW is not sufficient to reproduce the Hamiltonian in curved space-time. In order to capture the curvature effect on the Dirac dynamics,
the coin parameters are made position ($x$) and time-step ($t$) dependent: 
$C = \sum_x e^{- i \sum_{q=0}^3 \theta^q(x,t, \delta t)~ \sigma_q} \otimes \ket{x}\bra{x}$ as well as the two-step DQW evolution has been treated as 
a single-step evolution operator \cite{molfetta}. But these are unable to capture curvature effects in addition to the external gauge potential effects on a massive Dirac
particle in a single framework.

Here we started from a single-step SS-DQW with position, time-step dependent coin operators:
$C_j = \sum_x e^{- i \sum_{q=0}^3 \theta_j^q(x,t, \delta t)~\sigma_q} \otimes \ket{x}\bra{x}$. But in this case 
$U_{SQW}$ doesn't approach identity operator on $\mathcal{H}_c \otimes \mathcal{H}_x$ at continuum limit of space time---which causes ill-defined Hamiltonian. 
To circumvent this, we define our evolution operator as 
\vspace{-1cm}

\begin{align}
\mathscr{U}(\delta t) = \bigg( \lim_{\delta t \to 0,~ a \to 0} U_{SQW} \bigg)^\dagger \cdot U_{SQW}
= e^{- \frac{i \delta t}{\hbar} \mathscr{H}}
\end{align}
\vspace{-1cm}

and now $\mathscr{H}$ will be treated as the effective Hamiltonian. 
By properly choosing the coin parameters in $\mathscr{U}(\delta t)$, 
at the continuum limit of space-time, we derive the massive Dirac Hamiltonian $\mathscr{H}$
in $(1+1)$D curved space-time in presence of $U(1)$ abelian gauge potential such as electromagnetic potential \cite{curve}. For a choice like
\begin{align}
\sum_{q=0}^3\theta_j^q(x,t, \delta t)~ \sigma_q = \theta_j^0(x,t, \delta t)~ \sigma_0 + \theta_j^1(x,t, \delta t)~ \sigma_1 \nonumber\\
\text{such that}~~\theta_j^1(x,t, \delta t) = \theta_j^1(x,t,0) + (\delta t) \vartheta_j^1(x,t) + \mathcal{O}(\delta t^2), 
\end{align}
 the parameters $\theta_j^1(x,t,0)$ control the curvature effects, $\vartheta_j^1(x,t)$ give rise to mass term
and $\theta_j^0(x,t, \delta t)$ carry the details about the $U(1)$ potentials. The same Hamiltonian $\mathscr{H}$ can reproduce 
curved $(2+1)$ D massive Dirac Hamiltonian if one of the spatial momenta remains fix. 

\begin{figure}[htbp]
\centering
\subfigure[][]{
\includegraphics[width=0.7\textwidth]{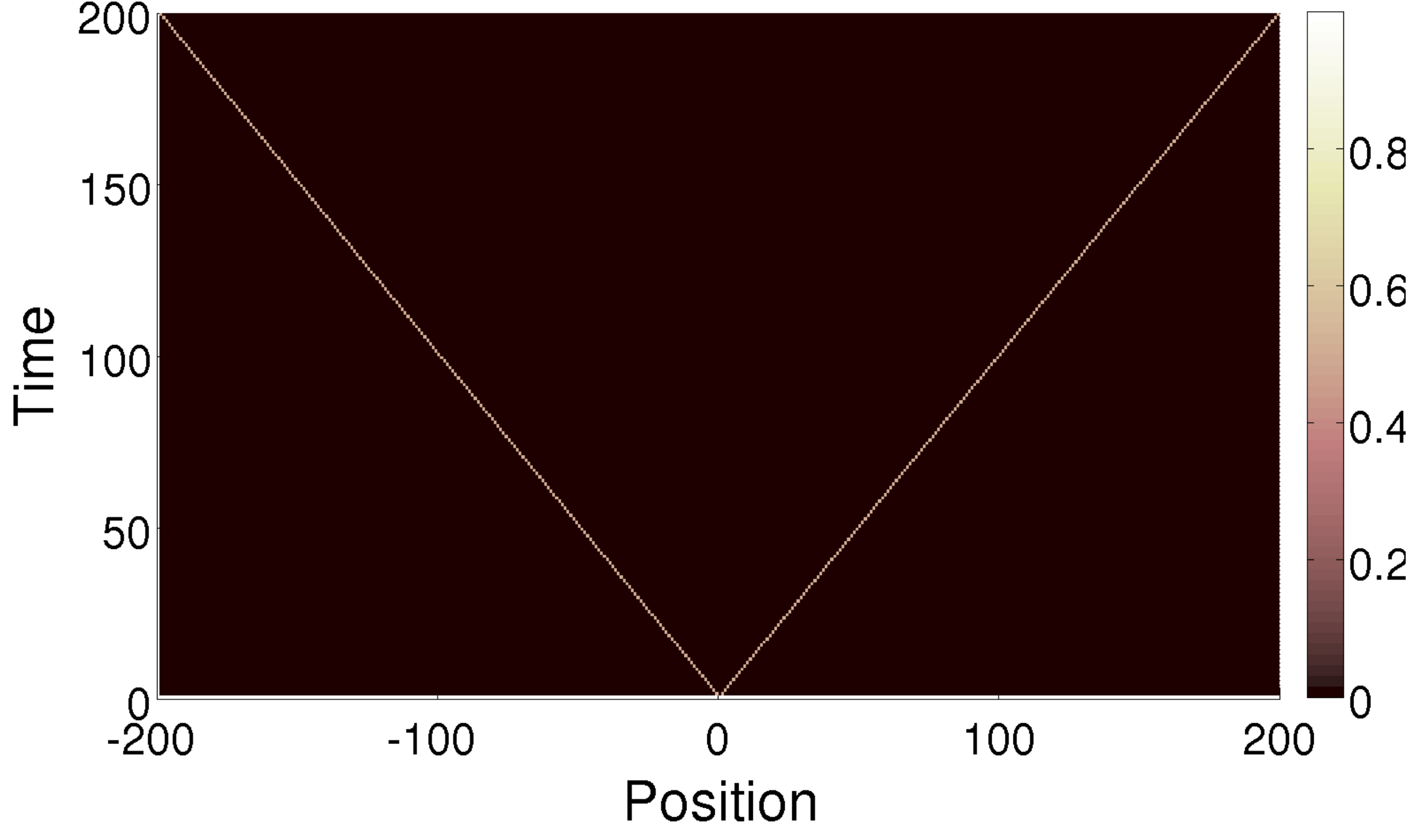}
\label{fig21}}
\subfigure[][]{\vspace{2pt}
 \includegraphics[width=0.7\textwidth]{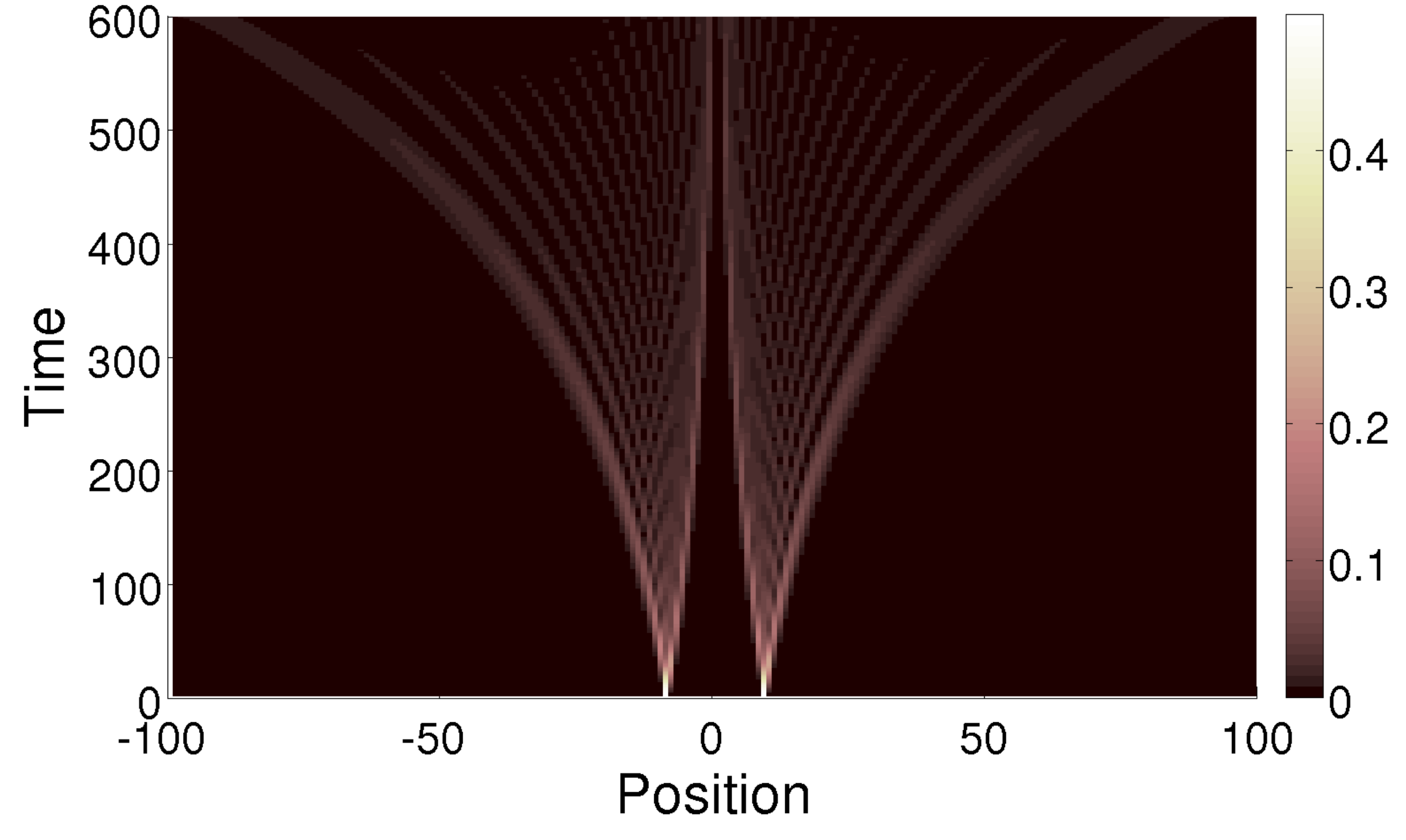}
\label{fig22}}
 \caption[Probability as a function of position and time-step in one-dimensional (a) Minkowski space-time, (b) curved space-time]
 {Probability as a function of position and time-step in one-dimensional flat lattice (position space) during 
 SS-DQW evolution. The fig.~\ref{fig21} is for the initial state $\frac{1}{\sqrt{2}}(\ket{\uparrow} + i \ket{\downarrow}) \otimes \ket{x = 0}$,  
 Minkowski space-time and the fig.~\ref{fig22} is for the initial state  $\frac{1}{2}(\ket{\uparrow}
 + i \ket{\downarrow}) \otimes \big( \ket{x = -9 a} + \ket{x = 9 a} \big)$, curved space-time described by the metric: $g^{00} = 1$, $g^{01} = g^{10} = 0$, $g^{11} = - x^2$. The mass of the particle is 0.04 unit.}
\end{figure} 
In order to include the effect of a general nonabelian $U(N)$ gauge potential such as potentials due to the weak force, strong force;
we need $2N$ dimensional coin operator instead of the 2 dimensional one. The shift operators has to be defined like, 
\begin{align}
 S_+ =  \sum_{x}  \ket{\uparrow}\bra{\uparrow} \otimes \mathds{1}_N \otimes \ket{x + a}\bra{x}
+ \ket{\downarrow}\bra{\downarrow} \otimes \mathds{1}_N \otimes \ket{x}\bra{x}, \\
 S_- =  \sum_{x}  \ket{\uparrow}\bra{\uparrow} \otimes \mathds{1}_N \otimes \ket{x}\bra{x} + \ket{\downarrow}\bra{\downarrow}
\otimes \mathds{1}_N \otimes \ket{x-a}\bra{x},
\end{align}
\vspace{-1.5cm}

 where $\mathds{1}_N$ is $N \times N$ identity operator on coin space. 
The coin operators are defined as
\begin{align}
 C_j = \sum_x \Big[ e^{- i \sum_{q=0}^3 \theta_j^q(x, t, \delta t)~\sigma_q} \otimes \mathds{1}_N \Big] \nonumber\\
\cdot \Big[ \ket{\uparrow}\bra{\uparrow} \otimes e^{- i \delta t \sum_{q=0}^{N^2 - 1} \omega_j^q(x,t) \Lambda_q}
+  \ket{\downarrow}\bra{\downarrow} \otimes  e^{- i \delta t \sum_{q=0}^{N^2 - 1} \Omega_j^q(x,t) \Lambda_q} \Big] \otimes \ket{x}\bra{x},
\end{align}
 where $\Lambda_q$ are the generators of $U(N)$ group and 
$\omega^q_j(x,t)$, $\Omega_j^q(x,t)$ are the corresponding coefficients which carry the effect of the nonabelian potential functions on the Dirac particle. 

We also extend this study for the two-particle case where the coin dependent shift operators are in separable form and the interactions among
the particles are incorporated in the global coin operators.  
The two-particle evolution operator =
\begin{align}
 \mathscr{U}^\text{two}(\delta t) = \bigg(\lim\limits_{\delta t \to 0, a \to 0} U^\text{two}_{SQW} \bigg)^\dagger \cdot U^\text{two}_{SQW}~\text{where} \nonumber\\ 
 U^\text{two}_{SQW} = (S^\text{first}_+ \otimes S^\text{second}_+) \cdot C_2 \cdot (S^\text{first}_- \otimes S^\text{second}_-) \cdot C_1.
\end{align}
The coin operators $C_1, C_2$ will now act on both the first and second 
particles. This helps us to study two-body problem in curved space-time.

\vspace{-.8cm}

\subsection*{Future directions}
\vspace{-.3cm}

We can think to extend our scheme of neutrino oscillation for other particle physics phenomena by constructing proper DQW algorithms. 
The scheme for the two-particle SS-DQW can be extended to study 
physics for distinguishable as well as indistinguishable particles for the case of constant
number of particles and also where particle annihilation and creation are possible.   

 \addcontentsline{toc}{chapter}{\listfigurename}
\listoffigures


\chapter{Introduction}\label{first}

Most of the progress we, the human race have made comes from our pursuit in
understanding of the natural world. Our understanding has helped the race to
progress by engineering things around us beyond what natural world has
provided. Our insight about nature and specific physical problems has most
of the time come from  mimicking one system by another controllable system
according to the prescribed theory, and analyze the outcomes obtained from
the mimicker system. This path of simulation becomes the only way when neither the direct experimental observations are available due to energy constraints or less controllability, nor 
analytical and numerical investigation are possible by existing computational devices. 
If the mimicker is well understood, it can help to understand the dynamics of the actual system, helps to develop new theory of unexplained but already observed 
phenomena. This also show ways to design experimental apparatuses for exploring the actual system. Analytical and numerical
investigation by existing computation devices has helped a lot in this process of studies using simulations.

The basic laws of our nature are well described by quantum theory and the
unusual behaviors of quantum phenomena put questions on the applicability of classical laws in the further
development of science and technology \cite{qlimitmoore}. The first requirement to deal with quantum objects, is to understand the nature of nonclassicality in them.
Because of the unusual behavior of quantum mechanics, quantum objects are less controllable and often simulation becomes the only option to study it.
The requirement become more prominent in case of dynamics related to the fundamental particles or complex quantum systems \cite{schaetz}.
But in such cases, not all systems specially classical information  
carriers (bits) are efficient mimickers.
We need to work with the quantum information carriers---qubits or more generally qudits.
This is one of the motivations behind the development of the branch of physics --- ``Quantum Information and Computation'' \cite{feyn2}.

The reasons behind the insufficiency of bits compared to qubits are following.

{\bf (1)} Quantum objects inherently possess superposition of orthogonal states and are described by probability amplitude which is in general complex number. 
So in case of simulation of quantum phenomena, to run algorithm and store the results, the required memory
scales exponentially with the degrees of freedom of the inputs. This is hard to realize with the classical objects for which this superposition 
is not possible and outcomes are analyzed in terms of probability, not by the probability amplitude \cite{galvao}. 

{\bf (2)} The quantum nonlocality and sharp collapse of the system wavevector to an eigenstate of the measurement operator cannot easily be realized by any 
classical means \cite{bharat, massar}.  The quantum nonlocality becomes even more
important in case of implementation of communication, teleportation \cite{bennett} between distant parties.

Also not all controllable systems that show nonclassical nature, are efficient quantum mimickers.
It is important to know the kinds of systems that are qualified to be the efficient mimickers. 
In general the actual system and the mimicker are different in nature, and because of the dissimilarities between them,
the mimicker has to follow some algorithms which are either encoded version of the actual theories or toy-models of the actual theories.
The ``Quantum Simulation'' \cite{georgescu, buluta, johnson} is a specialized quantum computation, primitive to universal quantum computations,
where we investigate about the proper quantum mimickers and develop efficient algorithms for simulation. 
Simulation involves efficient execution of three steps: (1) state preparation, (2) evolution (3) extraction of information about desired observables. 
Here efficient execution means the required resources for simulation scales polynomially with the size or number,
i.e., the number of input particles, time of evolutions etc.~of the simulated system.
While developing quantum simulation algorithms and simulators, we focus on the efficiency in this sense.

In this direction many quantum algorithms have been developed. Often these
algorithms
are developed by keeping in mind, the properties of particular simulators.
In these cases algorithms are efficiently implementable in the
corresponding simulator while it appears poor for some other kind of
simulators. But different mimickers have their own limitations in different
environmental conditions. So, to build a universal quantum simulators, it
is reasonable to think of a particular class of algorithms that can be
implementable on a large number of different mimickers.

The quantum walk which comes out as a quantum analog of classical random walk (CRW) and has played an important role in development of quantum walk based
algorithms for quantum computational tasks.  
It has recently emerged as a
powerful algorithms for quantum simulations and also be experimentally
implemented in a large class of table-top quantum set-ups. Existing
literatures discuss two main kind of quantum walk: continuous time quantum
walk (CQW) and discrete quantum walk
(DQW). In CQW which was first introduced in ref.~\cite{farhi}, the evolution operator acts continuously in time, measurement of the system state can be done at any time; while in DQW case 
evolution operation need finite time to execute, measurement 
on states can be done only after the end of a finite time-span. That is why sometimes DQW approach is identified with the piecewise continuous process which is continuous only within the finite time-span. 
The underlying position space of both CQW and DQW is usually defined on some graph structure, and the nodes are identified with the discrete position.
In CQW the evolution is defined in terms of a Hamiltonian and corresponding Schrodinger equation, while in DQW case the unitary evolution operator is more basic than Hamiltonian operator.   
For the DQW case both the time and position are discrete. In DQW the presence of another degrees of freedom (coin) makes the system state space larger than that for CQW in which 
the coin degrees of freedom is absent. Later we will see the form of the DQW is such that it naturally arises as an analog simulator for the Dirac Hamiltonian. 
All the works in this thesis are based on DQW where the position space will be described by one dimensional lattice.

\section{Discrete quantum walk}
In case of classical algorithms, probabilistic algorithm based on CRW often appear faster
than the existing deterministic algorithms \cite{karp}.
So investigation of the algorithm power of its quantum version, the DQW, would be an interesting topic.
DQW is initially introduced as a quantum analog of classical random walk (CRW) in discrete space-time \cite{aharonov}.
The DQW, we are going to present here is named ``Coined Quantum Walk'' in the recent literatures.
We will describe DQW mathematically after providing a brief description of CRW.  

\subsection{Classical random walk}
In one dimensional position space, a single step CRW evolution is a coin toss operation followed by a head or tail dependent positional movement operation.
We will use the notation that head state = $\ket{H}$ and tail state = $\ket{T}$; where classicality implies orthogonality of the coin states: $\braket{H|T} = 0$.
Here the coin Hilbert space = $\mathcal{H}_c \coloneqq \text{span}\big\{ \ket{H}, \ket{T} \big\} \cong \mathbb{C}^2$. 

Classically there are four possible forms of coin operation. After coin toss the probability of the coin to remain in the head state = $p_H$ if the initial state is head 
and if the initial coin state is tail the probability of the coin to remain in the tail state = $p_T$.
But usually it is assumed that $p_H = 1 - p_T$. Hence, irrespective of the coin state before tossing, the probability to get head state = $p_H$ and that of tail = $1 - p_H$.  
The Hilbert space associated with discrete positions of the particle can be defined as
$\mathcal{H}_x = \text{span}\big\{\ket{x} : x \in a \mathbb{Z}~\text{or}~x \in a \mathbb{Z}_\mathcal{N}\big\}$, $a$ is the lattice-step size that takes same value throughout the lattice. 
For CRW all the operations either coin tossing or spatial shift act at the density matrix $\in \mathcal{D}(\mathcal{H}_c \otimes \mathcal{H}_x)$ level.    \\
The action of the coin operation $C$ on the basis states are as follows.
\begin{align}\label{classcoin}
C  \big(\ket{H}\bra{H} \otimes \ket{x}\bra{x} \big) C^\dagger  = p_H  \ket{H}\bra{H} \otimes \ket{x}\bra{x} + (1 - p_H) \ket{T}\bra{T} \otimes \ket{x}\bra{x},\nonumber\\
C  \big(\ket{T}\bra{T} \otimes \ket{x}\bra{x} \big) C^\dagger = p_H  \ket{H}\bra{H} \otimes \ket{x}\bra{x} + (1 - p_H) \ket{T}\bra{T} \otimes \ket{x}\bra{x}.
\end{align} Here the coin operator acts as identity on the position space.  The action of the shift operator at the density matrix level can be 
expressed as:
\begin{align}
 S  \big(\ket{H}\bra{H} \otimes \ket{x}\bra{x} \big) S^\dagger  =  \ket{H}\bra{H} \otimes \ket{x + a}\bra{x+a}, \nonumber\\
 S \big(\ket{T}\bra{T} \otimes \ket{x}\bra{x} \big) S^\dagger =  \ket{T}\bra{T} \otimes \ket{x-a}\bra{x-a}.
\end{align}  This shift-operator shifts the particle one-step forward in $x$-axis if the particle is in the $\ket{H}$ state and 
one-step backward in $x$-axis if it is in $\ket{T}$ state, but does not change the coin state. 

\begin{figure}[htbp]
\centering
\subfigure[][]{
\includegraphics[width=0.475\textwidth]{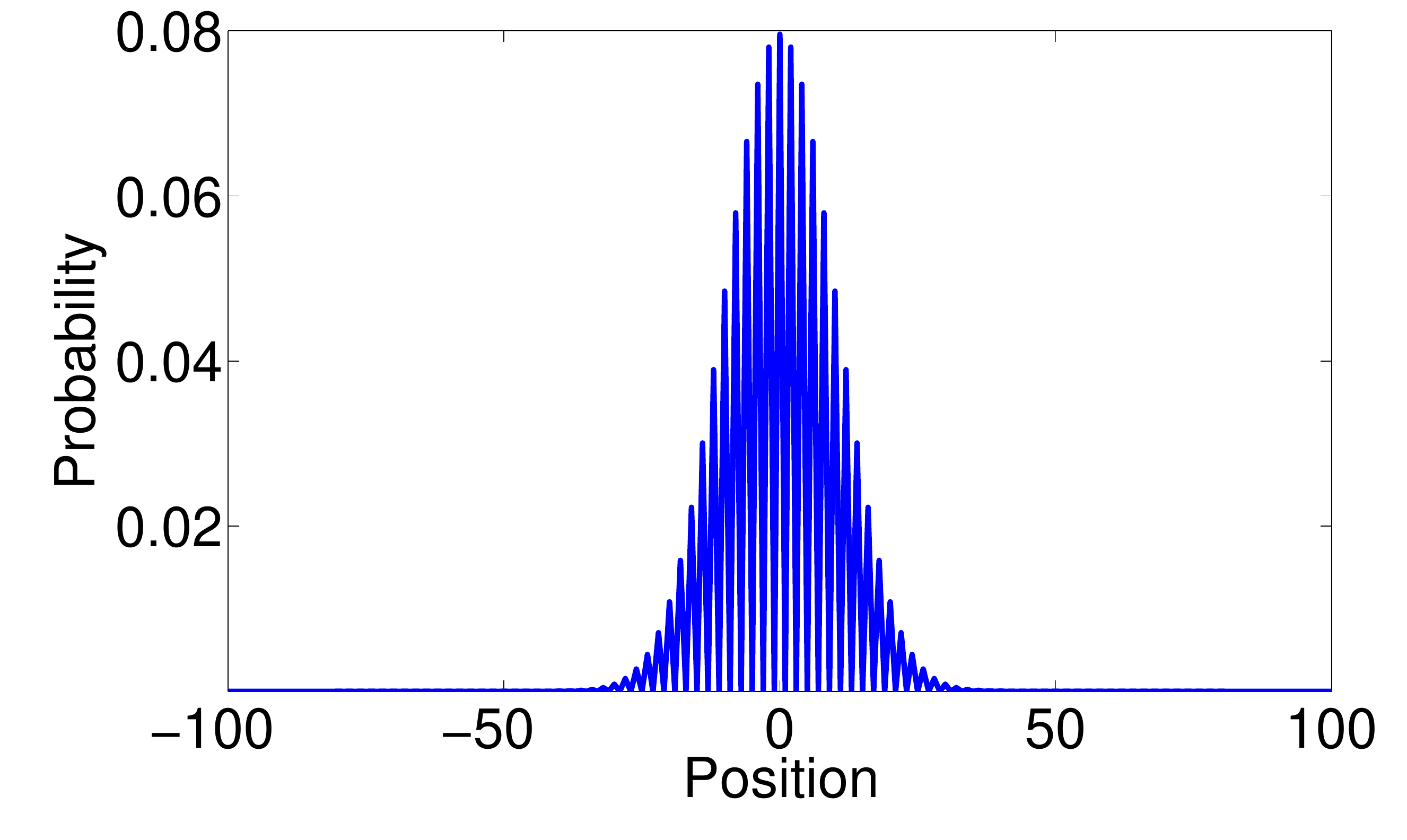}
\label{cwfig}}
\subfigure[][]{\vspace{2pt}
 \includegraphics[width=0.475\textwidth]{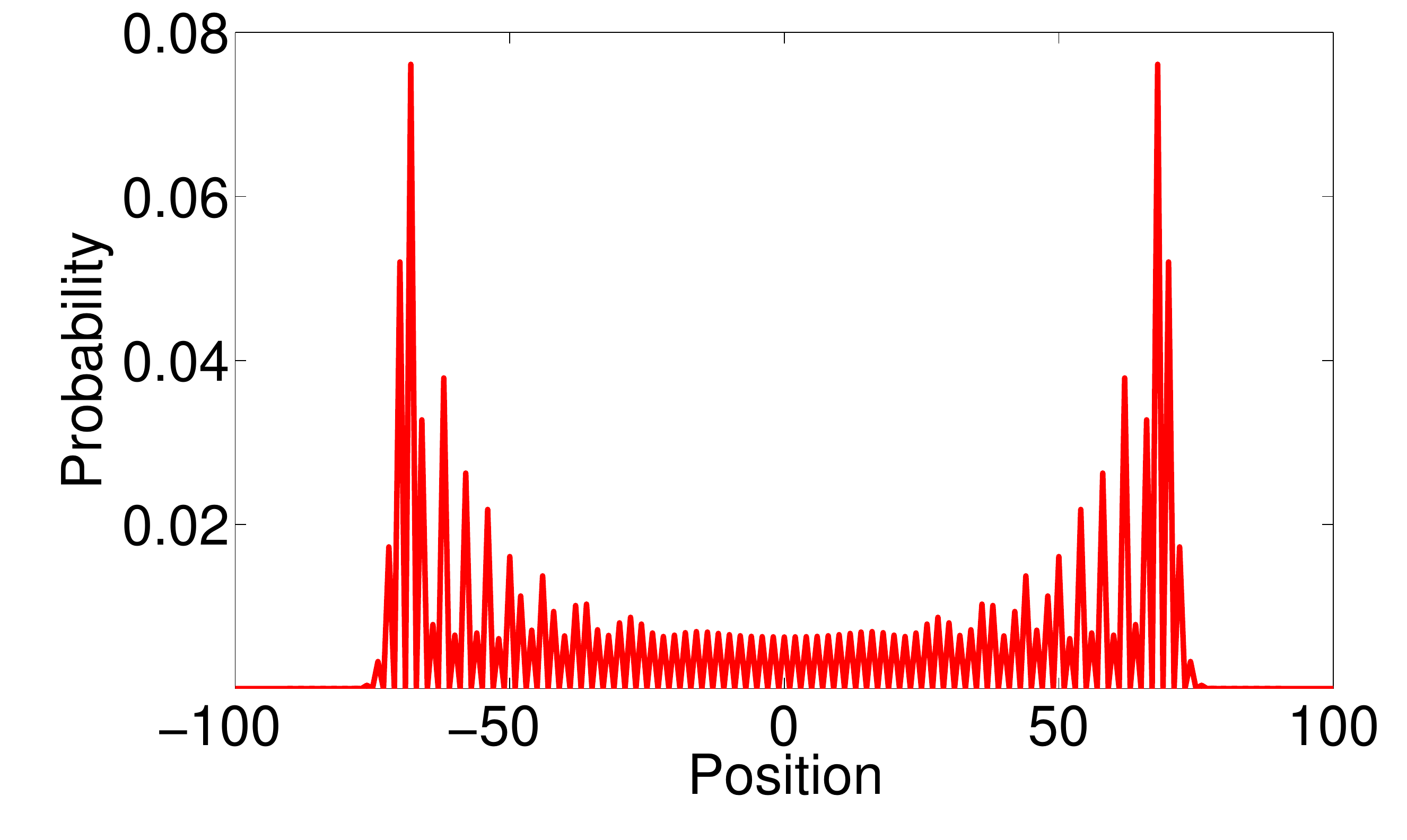}
\label{qwfig}}
 \caption[Positional probability for (a) CRW, (b) DQW]{(a) Probability as a function of position after 100 steps 
 of CRW evolution for unbiased coin: $p_H = \frac{1}{2}$, initial state: $\ket{H}\bra{H} \otimes \ket{x=0}\bra{x=0}$. 
 (b) Probability as a function of position after 100 steps 
 of DQW evolution for $\theta^0 = 0$, $\theta^1 = \frac{\pi}{4}$, $\theta^2 = 0$, $\theta^3 = 0$,
 initial state: $\ket{H}\bra{H} \otimes \ket{x=0}\bra{x=0}$.}
\end{figure}

This is evident that CRW evolution is not in general unitary as it can map a pure state to a mixed state. 
For this case starting from a initial state $\ket{H}\bra{H}\otimes \ket{x = 0}\bra{x = 0}$, after 100 steps of CRW 
the system state will be \begin{align}
\rho(100 \delta t) =  \big(S \cdot C \big)^{100}
\Big(\ket{H}\bra{H}\otimes \ket{x = 0}\bra{x = 0}\Big)\big(C^\dagger \cdot S^\dagger\big)^{100},
\end{align} where $\delta t$ is the time-step or the required time to execute a single step CRW and 
any time-step $t = \delta t \times \big( \{0\} \bigcup \mathbb{N} \big)$.
In this case the positional probability profile takes usually Gaussian like structure. In fig.~\ref{cwfig},
a positional probability profile has been shown for a unbiased coin operator, where in the figure the particle can be either in head or tail state.

One can also include the possibility of walker staying in same place with certain probability after single-step evolution.  Mathematically
that can be done by changing the form of the coin-state dependent position shift operation.   

\subsection{Discrete quantum walk operator form}
In case of DQW superposition of coin states as well as position states are possible, which implies coherence can play important role in DQW. 
Hence, it is necessary to know the action of the coin and shift operators at the wave-vector level.
Further, these operators have been defined such that they respect quantum superposition of states. For detailed introduction 
about DQW one can look at the refs.~\cite{kempe, chandru2}. In this case the coin operation analogous to that in eq.~(\ref{classcoin}) 
is \begin{align}
    C = \left( \begin{array}{cc}
                p_H & p_H \\
                1 - p_H & 1 - p_H \\
               \end{array} \right) \otimes \mathds{1}_x
   \end{align} where we have used the notations $\ket{H} = (1~~0)^T$, $\ket{T} = (0~~1)^T$, $\mathds{1}_x = \sum\limits_x \ket{x}\bra{x}$.
   This coin operator $C$ has the ability to transform a head or tail state into a superposition of head and tail states.  
 Here $C$ is a special $2 \times 2$ operator for $\mathcal{H}_c$. But in quantum mechanics a close system allows general but only unitary 
 operations. So a general coin operator is defined as  
\begin{align}\label{dqwcoinp}
 C = e^{- i \sum_{q=0}^3 \sigma_q \theta^q} \otimes \mathds{1}_x = e^{- i \theta^0} \left(\begin{array}{cc}
                                                                                           F & G \\
                                                                                           -G^* & F^*\\
                                                                                          \end{array} \right) \otimes \mathds{1}_x 
\end{align}
where $e^{- i \theta^0} \left(\begin{array}{cc}
                                                                                           F & G \\
                                                                                           -G^* & F^*\\
                                                                                          \end{array} \right)$
is in general a unitary matrix $\in U(2)$ group where $\sigma_q \in \{ \sigma_0 \coloneqq \mathds{1}_{2 \times 2}, \sigma_1, \sigma_2, \sigma_3 \}$
are conventional Pauli matrices; $F, G$ are complex functions of the real parameters $\{\theta^q\}_{q = 1}^3$, satisfying $|F|^2 + |G|^2 = 1$.
The form of the coin operator in eq.~(\ref{dqwcoinp}) indicates that after quantum coin toss the probability 
of staying at the same classical state: either head or tail = $|F|^2$ and flipping probability = $|G|^2$. Explicitly 
\begin{align}\label{FG}
F = \cos\Big(|\vec{\theta}|\Big) - i \frac{\theta^3}{|\vec{\theta}|} \sin\Big(|\vec{\theta}|\Big),~~
G = - i \frac{\theta^1}{|\vec{\theta}|} \sin\Big(|\vec{\theta}|\Big) - \frac{\theta^2}{|\vec{\theta}|} \sin\Big(|\vec{\theta}|\Big), \nonumber\\  
\text{where}~~|\vec{\theta}| = \sqrt{(\theta^1)^2 + (\theta^2)^2 + (\theta^3)^2}~\text{and}~\vec{\theta} = (\theta^1, \theta^2, \theta^3).
\end{align}

The positional shift operator can be defined as\begin{align}
        S = \ket{H}\bra{H} \otimes \sum_x \ket{x+a}\bra{x} + \ket{T}\bra{T}\otimes \sum_x \ket{x-a}\bra{x}.
       \end{align} This shift operator is unitary if the lattice is either periodic with $\mathcal{N}$ number of lattice sites (the $\mathbb{Z}_\mathcal{N}$ case)
       or contains infinite lattice sites (the $\mathbb{Z}$ case). This shift operator has the ability to make a walker superposed in many position eigenstates $\ket{x}$.
       Hereafter in this thesis we alternatively use up-spin state: $\ket{\uparrow}$ and down-spin state: $\ket{\downarrow}$ in place of $\ket{H}$ and $\ket{T}$, respectively.  
       
       The single-step DQW evolution operator is defined as
\begin{align}\label{dqwop}
 U_{DQW} = S \cdot C = e^{- i~\theta^0} \left( \begin{array}{cc}
                               F \sum_x \ket{x+a}\bra{x} & G \sum_x \ket{x+a}\bra{x} \\
                               -G^* \sum_x \ket{x-a}\bra{x} & F^* \sum_x \ket{x-a}\bra{x} \\
                              \end{array}\right).
\end{align} This unitary operator acts on the Hilbert space $\mathcal{H}_c \otimes \mathcal{H}_x$. $U_{DQW}$ takes a state $\ket{\psi(t)}$ at time $t$ to a state 
$\ket{\psi(t + \delta t)}$ at time $t + \delta t$, so starting from a state $\ket{\psi(0)}$ after $n$-steps of DQW evolution the system state will be  
$U^n_{DQW} \ket{\psi(0)} = \ket{\psi(n \times \delta t)}$.
Note that, the presented DQW in the ref.~\cite{aharonov2} is different from the DQW described by eq.~(\ref{dqwop}). 
In that ref.~\cite{aharonov2}, every step of DQW evolution is followed by a coin state measurement and 
the coin state is changed to the initial coin state and this is repeated after every single step evolution operation. 
During DQW evolution the superpositions in coin and position space allow interference which has ability to make its spreading behavior in position space 
quadratically faster than the classical one \cite{nayak}. Starting from the same state $\rho(0)$ as in the CRW case, after 100 time-steps DQW we 
get an inverted bell-shaped probability profile as shown in fig.~\ref{qwfig}, which has much more spreading in position space 
compared to the case of CRW. 

The above defined coin operation is independent of both space and time steps: $(x, t)$. In our analysis we 
have to consider the continuum limits of space-time: $\delta t \to 0$, $a \to 0$. So, in general we 
will consider the coin parameters $\{\theta^q\}_{q = 0}^3$ as functions of $\delta t$, and the $a$ dependence will be  
taken care by the relation $a = c \times \delta t$, where $c$ is a constant.

\subsection{Physical implementation of DQW in quantum devices}
DQW has already been realized in many state-of-art table-top experimental setups. Below I am going to describe them briefly. 

\begin{enumerate}
 \item {\bf NMR system} \cite{ryan, lu}: Here nuclear spins are treated as the qubits, some of them can be used to 
represent the particle or walker's internal states and others as the position space states. The spin-Z axis can be identified with the 
direction of the applied external magnetic field. The single-qubit gates, i.e., single qubit coin operations are implemented by electromagnetic pulses, 
such as radio frequency (RF) pulses, by moving to the rotating frame of references \cite{laflamme}.
The shift operators can be expressed as combination of controlled-NOT gate and single-qubit gates. 
The two-qubit gate C-NOT can be implementable by tuning spin coupling among two adjacent nuclei and RF pulse.

\item {\bf Ion trap} \cite{schmitz} : Here the hyperfine states of a ion are defined as the coin states and RF pulse is used to implement
the coin operation. The ion is trapped by external electric potential which causes motion like simple harmonic oscillation. The energy quantized states 
due to this oscillation, can be treated as positions. Sometimes, the coherent states formed by these energy fock states are used as position Hilbert space. 
These fock state occupation can be controlled by cooling the system. The shift operator is implemented by hyperfine-state dependent optical dipole force.

\item {\bf Photonic devices}: In ref.~\cite{zhang}, polarizations and orbital angular momenta (OAM) of photon form coin space and position space, respectively.
 Coin operation is performed using polarization (quarter, half) wave-plate, shift operator is performed using the combination of q-plate and polarization wave-plates.
 The q-plate changes the 
 OAM and polarization depending on the polarization state of the input photon. In ref.~\cite{broome}, instead of OAM, 
 the longitudinal spatial modes of photons are treated as the positions.
 Shift operation is performed by birefringent calcium beam displacer. In ref.~\cite{schreiber}, 
 the arrival time of photons at the detector are treated as position space instead of OAM. The 
 shift operation is implemented by optical network loop that displaces the photon spatially and temporally depending on the polarization degrees of freedom.  
 In ref.~\cite{do} DQW is implemented by the optical quantum quincunx set-up which is a combination of polarizing beam splitters (PBS) and half-wave plates.
 The each of the PBS port is considered as position,
 polarization of the photon along the direction of entering to PBS is treated as the coin state.

\item {\bf Cavity QED}: In ref.~\cite{di}, electronic levels of atom are used as the coin states and the photon number states are used as the positions. 
The resonant interaction of the atomic level with the classical field can do the job of a coin operation, the quantized field changes the photon numbers as well as the atomic levels.   
So proper combination of them can be used to implement DQW operation.  
 
\item {\bf Cold atom}: The ref.~\cite{alberti} describes the cooling of atom via Bose-Einstein condensation in a static optical lattice. 
The quasi momenta states have been used as the position states and the hyperfine levels of the atom used as the coin states. The coin operation 
is implemented by external electromagnetic pulses which induce Rabi-oscillation dynamics. An internal state dependent force is applied which causes 
displacement in momentum space. In refs.~\cite{dur, karski} instead of momentum space the displacement is engineered in optical lattice potential traps.

\item {\bf Quantum dots}: The ref.~\cite{manou} implemented DQW in array of quantum dots. Each dot represents as position state 
and electronic energy levels at each dot used as coin states. External laser pulses are used to perform coin and shift operations. 
 
\end{enumerate}

\subsection{On the possibility of DQW implementation by classical wave and nonclassicality in single particle DQW}
The basic difference that pointed by people is the ballistic spreading behavior of DQW compared to the diffusive behavior of the CRW in position space.
But some works \cite{knight, francisco, jeong} discuss about the possibility of DQW simulation using the property of classical waves, which 
can show the ballistic spreading. But there is a difference that become prominent when the question of measurement comes, and that makes the 
classical wave implementation of a general quantum system doubtful. 
The quantum mechanical system shows wave-particle duality \cite{shadbolt} while the classical system shows either wave property 
or particle property. Recently it is shown experimentally in ref.~\cite{rab} that quantum object can be in a superposition of particle and wave states.
This nature becomes evident when wave-vector collapse during measurement plays major role, like particle the system moves to stay at 
 one of the possible eigenstate of the measurement operator. In the case, the measurement operator is not projective on the system, 
 one can include environment to make a general POVM projective on the system + environment.
 But for classical wave case, during measurement the system simultaneously 
can stay in many possible orthogonal eigenstates. The entanglement between coin and position degrees of 
freedom has measurement contextual consequences \cite{basu, markiewicz}, while 
classical behavior is always non-contextual.
The noticeable difference appears in the case of two or many distant particles, 
where the classical wave is unable to reproduce quantum nonlocality \cite{lee}. The indistinguishability is another 
quantum aspect that is difficult to simulate by classical means \cite{ortiz}. 

\section{Importance of simulating Dirac particle dynamics}
Dirac equation was introduced by Paul Dirac \cite{dirac} to describe electrons.
Now Dirac particles are qualified to be one of the fundamental constituent of our nature. The  
matter is made of $\frac{1}{2}$-spinors which follows the properties of Dirac particle. 
Examples include leptons, quarks, etc.~and their antiparticles. So studying Dirac particle phenomena is 
very crucial if one wants to explore all physics behind any materialistic phenomena.  

In first quantized version the proposed Dirac Hamiltonian is
\begin{align}
 H  = c \kappa^1  p_1  + c \kappa^2  p_2  + c \kappa^3  p_3  + m c^2 \varkappa
\end{align}
in flat $(3 + 1)$ dimensional space-time. Here each of the matrices $\kappa^1, \kappa^2, \kappa^3, \varkappa$ has to anticommute with any other,
and square of each matrix has to be equal to the identity operator, in order to obey the relation: $H^2 = \sum_{j= 1}^3 p_j^2 c^2 + m^2 c^4$. 
The $p_j$ are momentum operators, $c$ is the velocity of light in free-space, $m$ is the 
mass of the Dirac particle. To be precise, the spin and spatial degrees of freedom belong to different Hilbert spaces, so 
they should be in tensor product form in the Hamiltonian, and hence the proper Hamiltonian should be written as 
\begin{align}
 H  = c \kappa^1 \otimes  p_1  + c \kappa^2 \otimes p_2  + c \kappa^3 \otimes p_3  + m c^2 \varkappa \otimes \mathds{1}_x~.
\end{align} where $\mathds{1}_x$ is the identity operator on position space. 
In Schrödinger formalism the evolution of any wavefunction $\psi = \braket{x_1, x_2, x_3 |\psi}$ is written as 
\begin{align}
i \hbar \partial_t \psi = \braket{x_1, x_2, x_3 | H |\psi} = - i \hbar c \sum_{j=1}^3 \kappa^j \partial_{j} \psi + m c^2  \varkappa~\psi. 
\end{align}In covariant form this equation can be written as 
\begin{align}\label{flatdirham}
\big(i \hbar \gamma^{(a)} \partial_{(a)}- m c^2 \big)~\psi = 0,  
\end{align} where we have used Einstein's summation rule and $\partial_{(a)} \in \{\partial_t, c \partial_1,  c \partial_2,  c \partial_3 \}$. \\
Here $\big\{ \gamma^{(a)}, \gamma^{(b)}\big\} = 2 \eta^{(a)(b)} \times$ identity matrix,
where $\eta^{(a)(b)}$ is the flat space-time metric, and we have considered the sign-convention: $\eta^{(0)(0)} = 1$, $\eta^{(j)(j)} = -1$ for all $j \in \{1, 2, 3\}$.
In this thesis we will confine mainly to the $(1+1)$ dimensional case, by setting other momentum components to zero.  
For proper investigation we should move to the second quantized version of the Dirac particle dynamics as usually done in fermionic field theory,
but to understand from the basic we are starting from the first quantized version.  

Both theoretical and experimental study related to the Dirac particle dynamics has been done. For free Dirac Hamiltonian it is easy to get analytical solution, 
but it appears difficult in presence of complicated background potential or when interaction among many particles become important. For these cases we have to rely on the 
approximate theoretical analysis mainly in the perturbative coupling regions; and also numerical analysis by classical computer appears unsuitable. Below we will 
discuss the single particle Dirac equation in presence of general external potentials. In analogy to free Hamiltonian in the flat space-time (while special theory of relativity 
works) in eq.~(\ref{flatdirham}) one can write the Dirac Hamiltonian in general relativistic curved space-time as the following.

\begin{align}\label{curvedirham}
\Big(i \hbar e^\mu_{(a)} \gamma^{(a)} \nabla_\mu - m c^2 \Big)~\psi = 0~. 
\end{align}

In this case we will use latin indices within first brakets to denote local inertial coordinate and greek indices for the global general coordinate, where the tetrad $e^\mu_{(a)}$
transforms the local coordinate to the global one or vice versa. In the absence of gravity or space-time curvature the vielbeins $e^\mu_{(a)} = \delta^\mu_{(a)}$ the Kronecker delta function.
The $\nabla_\mu = \partial_\mu + \Gamma_\mu$ such that $\Gamma_\mu = \frac{1}{8} \big[\gamma_{(c)}$, $\gamma_{(d)}\big] e^{(c) \nu} \Big( \partial_\mu e^{(d)}_\nu - \Gamma^\lambda_{\mu \nu} e^{(d)}_\lambda \Big)$
where $\Gamma^\lambda_{\mu \nu} = \frac{1}{2} g^{\alpha \lambda}\big(\partial_\mu g_{\nu \alpha} + \partial_\nu g_{\mu \alpha} - \partial_\alpha g_{\nu \mu}\big)$. 
The metric in global space-time coordinate is 
$g_{\mu \nu}  = e^{(a)}_\mu e^{(b)}_\nu \eta_{(a)(b)}$.

In presence of the external gauge potential the term $\nabla_\mu$ in eq.~(\ref{curvedirham}) has to be replaced by $\nabla_\mu - i A_\mu^q \Lambda_q$
where $A_\mu^q$ acts as the external gauge potential function and $\Lambda_q$ is a generator of the corresponding gauge group. 

 There are many phenomena of interests
 where it is theoretically predicted that Dirac particle plays crucial roles, but still 
require proper investigation; examples include fermion confinement, near Planck scale physics \cite{hossenfelder, faraoni}, unruh effect with massive particles \cite{kialka}. 
Sometimes these are hard to realize in real experiment or by direct observation and
classical numerical analysis can not capture many properties of them. Quantum simulation for Dirac particle dynamics becomes necessary to understand these kinds 
of phenomena at the current stage of time. 
\vspace{-.2cm}

\section{DQW as a simulation tool}
As today's many classical algorithms are based on the CRW, DQW also appear as a basic to construct algorithms for search problem \cite{shenvi, ambainis1},
 state transfer \cite{stefanak, chandru4, kurzynski, yang}. Moreover, DQW appear as a study tool for thermodynamics: localization-thermalization 
 \cite{romanelli1, romanelli2, machida2, chandru5, ambainis2, konno, vakulchyk}, realization of various topological phases \cite{kitagawa, kitagawa2}.  
 It is shown that DQW applied in simple graphs while the coin operation is restricted to only Grover coins, can simulate all the features
of the universal quantum gates. And, hence it can serve as a 
universal computing algorithm \cite{lovett}. But the interesting algorithmic application of DQW is that it captures many properties of relativistic quantum mechanics \cite{chandru3}. 

{\bf DQW as a simulation tool of free Dirac Hamiltonian :}~
It is well known that for particular choice of coin parameters DQW produces free Dirac Hamiltonian (FDH) at the 
continuum limit \cite{strauch, bracken,sato, chandru1, arrighi}.
One simple way to derive the FDH from the DQW evolution operator $U_{DQW}$ is by 
using the definitions 
\vspace{-1cm}

\begin{align}\label{defi1}
U_{DQW} = \exp\bigg(- i \frac{\delta t}{\hbar} H \bigg) \end{align}

and moving to the discrete Fourier space from the position space. The operator $H$ is the effective Hamiltonian. Using Fourier transformation we get the following forms of 
the translation operators: 
\begin{align}
= \sum_x \ket{x \pm a}\bra{x} = \exp\bigg(\mp i \frac{a}{\hbar} p\bigg) = \sum_{k} \exp\bigg(\mp i \frac{a}{\hbar} k\bigg) \ket{k}\bra{k}
\end{align} where $k$ is the eigenvalue of the momentum operator: $p$ corresponding to the eigenstate $\ket{k}$.  In this case 
 \begin{align}\label{fouruni}
  U_{DQW} = e^{- i~\theta^0} \sum_k \left( \begin{array}{cc}
                               F e^{- i  \frac{a}{\hbar} k} &  G e^{- i  \frac{a}{\hbar} k}  \\
                               -G^* e^{ i  \frac{a}{\hbar} k} & F^* e^{ i  \frac{a}{\hbar} k} \\
                              \end{array}\right) \otimes \ket{k}\bra{k}
 \end{align} is already diagonal in $\{\ket{k}\}$ basis, so after diagonalization in coin basis we 
 can derive the effective Hamiltonian: $H \coloneqq \sum\limits_k H_k \otimes \ket{k}\bra{k}$. So we get
 \begin{align}\label{hamdqw}
  H_k = - \frac{\hbar}{\delta t} \frac{\cos^{-1}\big[\Re(F e^{- i  \frac{a}{\hbar} k})\big]}{\sqrt{1 - \big[\Re(F e^{- i  \frac{a}{\hbar} k})\big]^2}}
  \bigg[ \Im(F e^{- i  \frac{a}{\hbar} k}) \sigma_3 + \Re(G e^{- i  \frac{a}{\hbar} k}) \sigma_2 + \Im(G e^{- i  \frac{a}{\hbar} k}) \sigma_1 \bigg]
  + \frac{\hbar}{\delta t} \theta^0 \sigma_0.
 \end{align}
 Here we have used the property: if matrix $V$ diagonalizes $U_{DQW}$ and the diagonal form of it is $U_{diag}$ then 
 
\begin{align}
  U_{DQW} = V \cdot U_{diag} \cdot V^\dagger  \Rightarrow - i \frac{\delta t}{\hbar} H = \ln (V \cdot U_{diag} \cdot V^\dagger) = V \cdot \ln (U_{diag}) \cdot V^\dagger.
\end{align}
The general forms of the functions $F$, $G$ are given in eq.~(\ref{FG}).
A choice like $\theta^0 = 0$, $F = \cos[\theta^1(\delta t)]$, $G = - i \sin[\theta^1(\delta t)]$ i.e., when the coin operation
$C = e^{- i \theta^1(\delta t) \sigma_1}$, 
will make the Hamiltonian in (\ref{hamdqw}) to 
\begin{align}\label{dqwham}
 H_k  =  \frac{\hbar}{\delta t} \frac{\bar{E}(k, \theta^1)}{|\sin \bar{E}(k, \theta^1)|}
  \bigg[\cos \theta^1 \sin \bigg(\frac{a}{\hbar} k \bigg)~ \sigma_3 + \sin \theta^1 \sin \bigg(\frac{a}{\hbar} k \bigg)~ \sigma_2
  + \sin \theta^1 \cos \bigg(\frac{a}{\hbar} k \bigg)~ \sigma_1 \bigg],
\end{align}
where the energy eigenvalues are \begin{align} \pm E(k, \theta^1)~~\text{such that}~~
E(k, \theta^1) = \frac{\hbar}{\delta t} \cos^{-1}\bigg[\cos \theta^1 \cos \bigg(\frac{a}{\hbar} k \bigg)\bigg].\end{align}
Because this energy is a periodic function of both momentum and angle $\theta^1$, energy value is defined as 
the principle value of it = $E(k, \theta^1)$ modulus $2 \pi$. Sometimes it is called quasienergy in contrast to 
the realistic case where energies can take continuous real values. But this problem will go when we consider the continuum
limit for which the periodicity itself takes infinitely large value. One important point to be noticed is that, for a given $\theta^1$
energy eigenvalue $E(k, \theta^1)$ is a monotonic function for the domains $k  \in \Big[0, \frac{\hbar \pi}{a}\Big]$,  $\Big[-\frac{\hbar \pi}{a}, 0\Big]$.
So, in this case the fermion doubling problem, i.e., the existence of low-energy excitation for both lower and higher values of momentum can not arise 
here. 

At the continuum limit $\delta t \to 0, a \to 0$ if the following limits exist
\begin{align}
 \lim_{\delta t \to 0} \frac{\theta^1}{\delta t} = \frac{m c^2}{\hbar}, ~  \lim_{\delta t \to 0} \frac{a}{\delta t} = c, 
\end{align} the expression (\ref{dqwham}) takes the form 
\begin{align}\label{dqwhamdir}
 H_k  = c k~\sigma_3 + m c^2~\sigma_1~ \Rightarrow H = c \sigma_3 \otimes p + m c^2 \sigma_1 \otimes \mathds{1}_x
\end{align}which is the Dirac Hamiltonian in $(1 +1)$ dimensional flat space-time. The constant $c$ has to be identified with the 
velocity of light in free space. The coin parameter $\theta^1$ determines the mass of the Dirac particle.
If the limit: $\lim\limits_{\delta t \to 0} \frac{\theta^0}{\delta t}$ exist and gives nonzero value, 
the term $\frac{\hbar}{\delta t}\theta^0$ in eq.~(\ref{hamdqw}) corresponds to a position-time independent potential operator, so that it makes just an additional energy shift.

Here we have discussed the derivation of free Dirac Hamiltonian from position, time-steps independent DQW evolution operators. 
The Dirac Hamiltonian in presence of general potentials and the curved space-time can also be derived if we start from the 
coin operators whose parameters depend on the position and time-steps, as discussed in refs.~\cite{Molfetta, molfettacurve}.
We will discuss them latter in chapter \ref{3} and point out the gaps that are filled in this thesis.

One can question about the importance of simulating the first quantized Dirac Hamiltonian by DQW, because of the incompleteness associated with 
the relativistic quantum mechanics. For proper study we should deal with quantum field theory. To develop the DQW simulation scheme of  
quantum field theory it is better first to understand the maps between the DQW parameters and parameters that control relativistic quantum mechanical dynamics.
This will also help to apply well known results of relativistic quantum mechanics in different situations where DQW algorithm is applicable.
Another point to be made,  almost all of the results of quantum information theory are in the nonrelativistic quantum domain, in order to 
apply it to quantum field theory, one can go via relativistic quantum mechanics.

One interesting point is that the maximum finite hopping velocity $c$ in DQW framework, 
keeps wavefunction spreading within the light-cone and hence obey locality principle,
in contrast to the conventional quantum mechanical approach (follows from the calculation of Hamiltonian eigenstates in 
continuous space-time) where wave-function can spread beyond light cone. This spreading behavior beyond light-cone was a problem that also remain in relativistic quantum mechanics 
and solved by quantum field theory formulation.

\subsection{Comparison with other existing simulation schemes}
 \begin{itemize}
 
 \item The DQW is based on the discrete space-time, so it can be directly applied to test the theories that are defined on discrete space-time background. 
 One can further explore developing low energy table-top simulation schemes for Planck energy phenomena using DQW tools. These are not easily doable for the 
 simulation schemes where position space or time or both are treated as continuous.

 \item  In this case unitary operator is more basic than the Hamiltonian or Lagrangian, the effective Hamiltonian is derived from the unitary evolution operation. 
 In general quantum theory the state evolution is calculated by the form of the evolution operator and if Hamiltonian is given one need to exponentiate it for accurate calculation.
 So the problem of exponentiation of Hamiltonian operator using Trotter Product formula and the precision problem will not arise in DQW case.  
 
 \item Many existing simulation algorithms of Dirac particle phenomena are specifically made for some kind of simulators or to capture some particular 
 kind of dynamics. On the otherhand DQW has already huge application in construction of data search algorithms and state transfer, and in realization of topological phases.
 Compared to other simulation schemes of Dirac particle phenomena, it can serve as universal basic alternative to the universal quantum logic gates.  
 Moreover, DQW is simulable by a large class of quantum simulators. So, if we can connect DQW tools to existing predictions of relativistic quantum 
 phenomena, it helps to grow our intuition in the other algorithmic application of DQW.  
 \end{itemize}
 
\section{The main questions and our contributions}

As discussed in previous sections, the DQW can be thought as a discretization and simulation scheme for Dirac particle dynamics. 
However, existing connection between DQW and Dirac equation still had some gaps when I started my research work. They have been addressed by asking the following questions.

\begin{enumerate}
 \item The form of the conventional DQW in eq.~(\ref{dqwop}) does not capture all the properties of the Dirac cellular automaton (DCA) 
 which is a discretization of free Dirac Hamiltonian (FDH). 
 What modifications are needed for DQW to capture all relevant features of the Dirac particle dynamics ?
 
\vspace{.2cm}  
 
~~~$\bullet$ We have exactly reproduced the DCA by a particular parameter choice of---the split-step discrete quantum walk (SS-DQW)---a generalized version of DQW.
This is reported in our paper ``\href{https://www.nature.com/articles/srep25779}{Dirac Quantum Cellular Automaton from Split-step Quantum Walk} \cite{dca1}'' and 
discussed in chapter \ref{firstq}.
 
\vspace{.8cm} 
 
 \item Is the conventional form of the DQW able to simulate phenomena related to Dirac particles with
 additional degrees of freedom like color, flavor and various charges besides the spin?
 One such example we picked is the neutrino flavor oscillation. If the conventional DQW is not enough,
 what are the modifications needed for initial state preparation and DQW evolution operator to mimic the neutrino flavor oscillation probability?
 
\vspace{.2cm}  
 
~~~$\bullet$ As the DCA is exactly reproducible by the SS-DQW, we have taken SS-DQW as our basic tool to answer this Dirac particle related problem.   
The conventional two dimensional coin space in SS-DQW in not enough. We can reproduce the exact neutrino oscillation profile by moving to the higher (six in this case) 
dimensional coin space and particular choice of the parameter values. This is reported in our paper 
``\href{https://doi.org/10.1140/epjc/s10052-017-4636-9}{Neutrino oscillations in discrete-time quantum walk framework} \cite{neu1}'' and 
discussed in chapter \ref{secondq}.
 
\vspace{.8cm}  
 
 \item Introducing position and time-step dependency in the coin parameters $\{\theta^{q}\}_{q=0}^3$ of DQW does not 
 easily capture space-time curvature, abelian, and nonabelian gauge potential effects on a massive Dirac particle in a single Hamiltonian framework. 
 So, what are the modifications of DQW required to solve this and simulate massive Dirac Hamiltonian in curved space-time and the effect of the other background potentials?
 
\vspace{.2cm}  
 
~~~$\bullet$ In this case position and time-step dependent coin parameters of SS-DQW are necessary. But this questions the existence of well-defined
 operator values at the continuum space-time limit. We modified the SS-DQW evolution operator form and got rid of the continuum problem. The general form of this 
 modified SS-DQW operator with two-dimensional coin operations is able to capture the abelian gauge potential and the curved space-time effects on the 
 single massive Dirac particle. To capture the nonabelian gauge potential effects, we need to work with higher dimensional coin operators whose parameters also 
 depend on the position and time-steps.  
 This is reported in our paper ``\href{https://arxiv.org/abs/1712.03911v2}{Simulating Dirac Hamiltonian in Curved Space-time by Split-step Quantum Walk} \cite{curve1}''
 and discussed in chapter \ref{3}.

 \end{enumerate}  


This approach is more than just simulation, as we are trying to formulate all kind of fundamental particle dynamics in terms of two basic operations: coin operation, coin state dependent positional shift operation.
In this sense it is a kind of algorithmic unification.


\chapter{Connecting Dirac cellular automaton with discrete quantum walk}\label{firstq}

\section{Dirac cellular automaton}
Cellular automaton (CA) is developed as a generalized tool for computation. This is defined on discrete cellular structure and discrete time.
In this case the state evolution rule of the system state is local i.e., the state at cell $x$ and time-step $t$ 
depends only on the states of nearest neighboring cells of $x$
including $x$ itself, at the previous time-step $(t - \delta t)$. The state update rule acts simultaneously
at every possible cell. The word cell here is a general term, in basic physical analysis we can treat it as a lattice point 
where the unit building block of the lattice are all identical and the associated 
graph is regular, i.e., each vertex carries equal number of edges \cite{neumann, wolfram}.

\begin{figure}[h]
\centering 
\includegraphics[width = 0.7\textwidth]{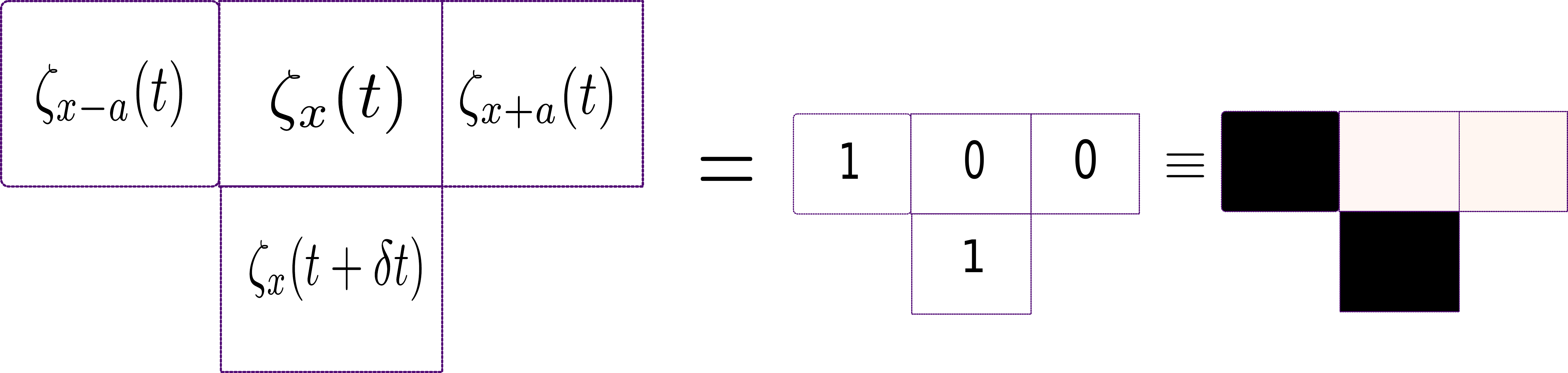}
\caption[Notation used for cellular automata rules]{The notation used for the cellular automata rules, depicted for one possible state
when $\zeta_{x-a}(t) = 1$, $\zeta_x(t) = 0$, $\zeta_{x+a}(t) = 0$ and $\zeta_x(t + \delta t) = 1$. 
Each black box or ``1'' denotes the presence of the particle and white box or ``0'' denotes absence of that.} \label{depi}
\end{figure}

In our thesis I will focus the one-dimensional CA which are of two kinds: deterministic and stochastic (probabilistic) CA.
Let us first concentrate on the deterministic CA, where I denote the existence of a particle at a site as ``1'' and absence as ``0''. 
Local rule of CA implies in the three-neighborhood scheme that the state of three sequential cells 
$(x-a, x, x+a)$: $\zeta_{x-a}(t)$ $\zeta_x(t)$ $\zeta_{x+a}(t)$ at time-step $t$  
determines the state of the cell $x$: $\zeta_x(t+\delta t)$ at time-step $t + \delta t$, 
where we have denoted $\zeta_{x}(t) \in \{0, 1\}$ $\forall$  $x$, $t$, $\delta t$. I will denote the presence of the 
particle by black box and the absence of it by light-yellow box. For depiction of an example please see the fig.~\ref{depi}. 

\begin{figure}[h]
\centering
\includegraphics[width = 0.65\textwidth]{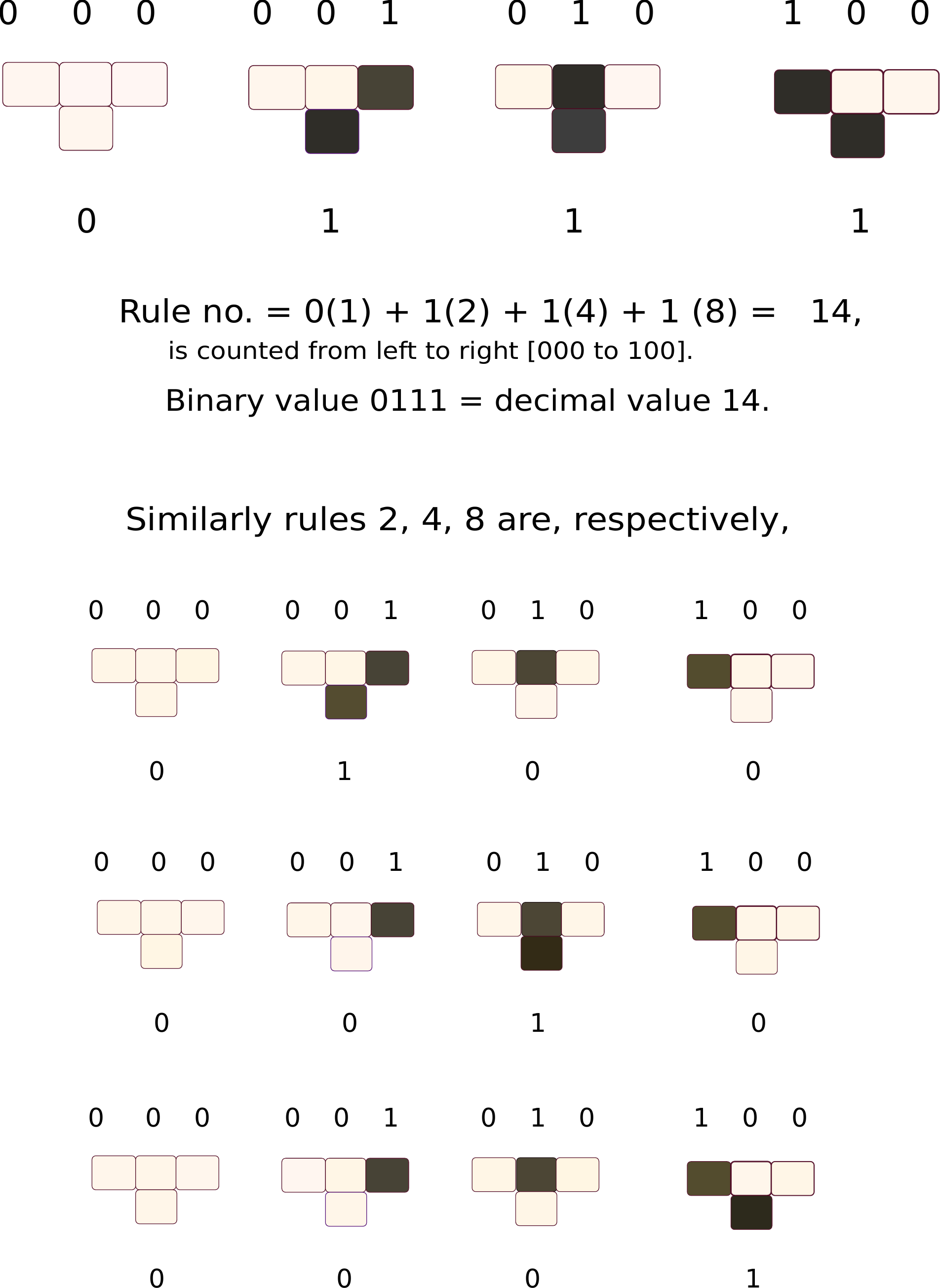}
\caption[Deterministic cellular automaton for single-particle case]{One example
of classical deterministic cellular automaton for the single particle case. 
The above three cells in all the four configuration are at the time-step $t$, and 
the single cell attached with them is at the time $t + \delta t$.}
\label{ddca}\end{figure}

 Here I will work with the single or no particle case. So the possible state of the cells at time-step $t$: 
 $\zeta_{x-a}(t)$ $\zeta_x(t)$ $\zeta_{x+a}(t)$ $\in$ $\{000, 100, 010, 001\}$. For particle number conservation 
 $\zeta_x(t+\delta t) = 1$ when $\zeta_{x-a}(t)$ $\zeta_x(t)$ $\zeta_{x+a}(t)$ $\in$ $\{100, 010, 001\}$, and 
 $\zeta_x(t+ \delta t) = 0$ when $\zeta_{x-a}(t)$ $\zeta_x(t)$ $\zeta_{x+a}(t) = 000$.  In case of extended lattice sites we will 
 see that there are restrictions even among these possibilities. The binary values of the states of cells at time-step $t + \delta t$ determines 
 the deterministic CA rules. As shown in the top picture of the fig.~\ref{ddca}, we have shown rule ``14'' which in binary value = 1110, counted from the left to right. 
 $\zeta_{x-a}(t)$ $\zeta_x(t)$ $\zeta_{x+a}(t) = 000$, if $\zeta_x(t + \delta t) = 0$ the decimal assignment = $0 \times 2^0 = 0$;  
  $\zeta_{x-a}(t)$ $\zeta_x(t)$ $\zeta_{x+a}(t) = 001$, if $\zeta_x(t + \delta t) = 1$ the decimal assignment = $1 \times 2^1 = 2$;  
 $\zeta_{x-a}(t)$ $\zeta_x(t)$ $\zeta_{x+a}(t) = 010$, if $\zeta_x(t + \delta t) = 1$ the decimal assignment = $1 \times 2^2 = 4$;  
 $\zeta_{x-a}(t)$ $\zeta_x(t)$ $\zeta_{x+a}(t) = 100$, and if $\zeta_x(t + \delta t) = 1$ the decimal assignment = $1 \times 2^3 = 8$. 
 In this sense, the rule number described at the top of the fig.~\ref{ddca} is $0 + 2 + 4 + 8$ = $14$. 
 But if we consider conservation of particle number we find only three valid possible rules 2, 4, 8.
The reason is that, if one particle moves to the right-cell it can not either stay at the 
 same cell or move to the left-cell---the rule 8 which implies that the particle can only move to right cell at the next time-step, 
if one particle moves to the left-cell it can not either stay at the same cell 
or move to the right-cell---the rule 2 which implies that the particle can only move to left cell at the next time-step,
if one particle remains at the same cell it can not move either right or
left-cell---the rule 4 which implies that the particle will remain at the same cell at the next time-step.

\begin{figure}[h]
\centering
\includegraphics[width = 0.5\textwidth]{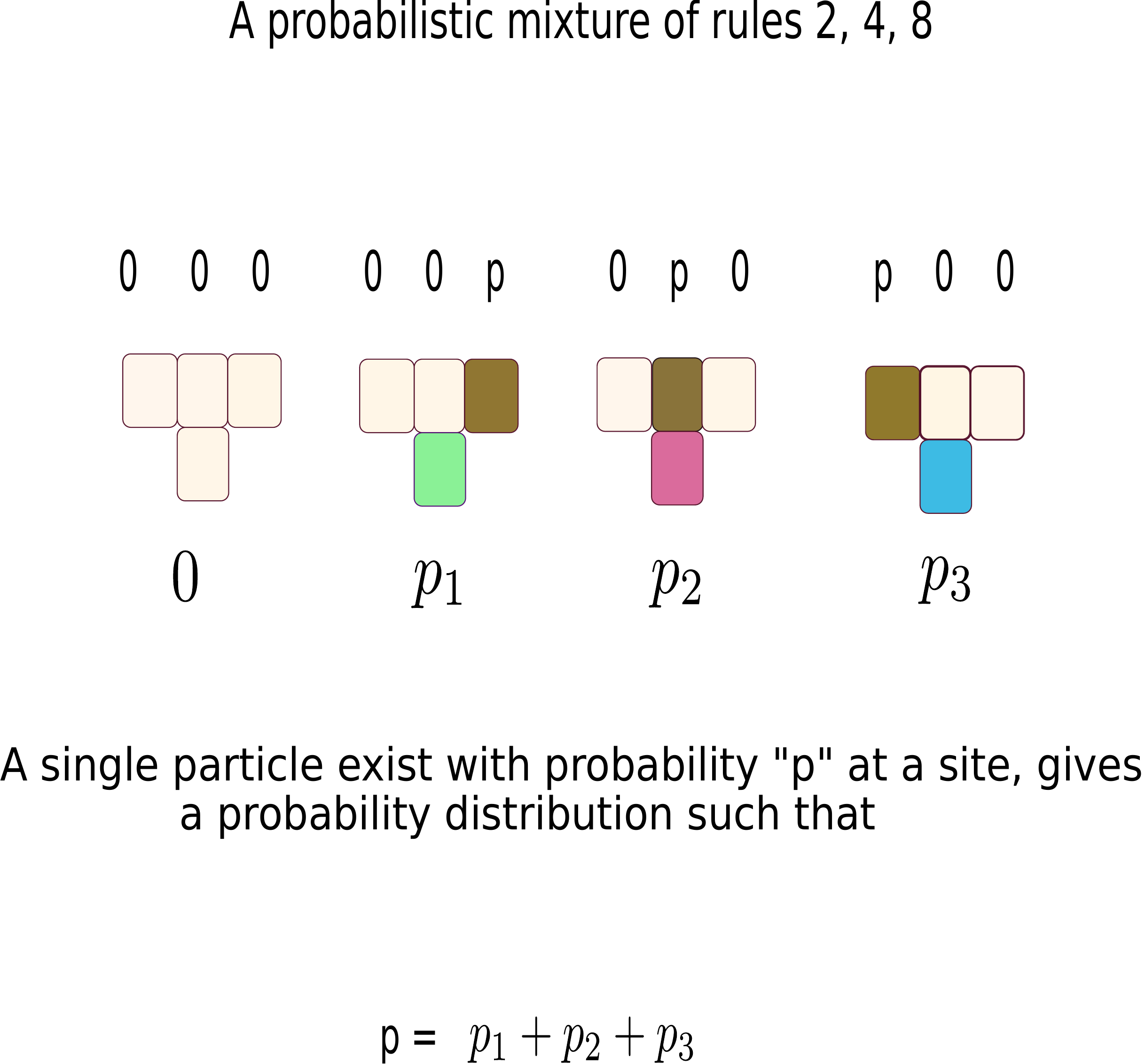}
\caption[Probabilistic cellular automaton for single-particle case]{One example of classical probabilistic cellular automaton for the single particle case exists with certain probability p at a position or site.
The different color indicates different value of probability. Particle number conservation imposed the condition: p $= p_1 + p_2 + p_3$.}\label{ddproca}
\end{figure}
The probabilistic CA can be obtained by considering the statistical mixture of the deterministic CA as shown in fig.~\ref{ddproca}. 
In this case existence of the particle has probabilistic nature so denoted by p, $p_1, p_2, p_3 \in [0,1]$ at each cell. If the existence probability 
has to be conserved we must have p = $p_1 + p_2 + p_3$, in analogy to the particle conservation in the deterministic CA case. Here the moving 
probability to the left cell is $p_1/$p, the probability to stay at the same cell is $p_2/$p and the moving probability to the right cell is $p_3/$p. 
Therefore rules 2, 4 and 8 occur with probabilities $p_1/$p, $p_2/$p and $p_3/$p, respectively.

The cellular automaton is called quantum, when the state evolution rules are quantum
mechanical \cite{wiesner}. The term quantum cellular automata (QCA) 
was first introduced in ref.~\cite{grossing} where the existence of particle denoted by ``0'' or ``1'' are replaced by the 
complex probability amplitudes $\psi(x)$ such that $0 \leq |\psi(x)| \leq 1$, and rules of the evolution are defined in such a way that the coherent superposition as well as 
statistical mixture are possible. One can think of general QCA as a completely positive process---as the most general realizable quantum mechanical process is completely positive.
So, we can define QCA as a map $M_{QCA}: \mathcal{H}_\text{in} \otimes \mathcal{H}_\text{cell} \to \mathcal{H}_\text{in} \otimes \mathcal{H}_\text{cell}$. 
Here $\mathcal{H}_\text{in}$ is the Hilbert space associated with the internal degrees of freedom of the system,
and usually it is assumed that $\text{dim}(\mathcal{H}_\text{in})$ is finite. $\mathcal{H}_\text{cell}$ is the Hilbert space associated with 
the cellular structure of the underlying graph.
If we restrict ourselves only to the unitary QCA, it is deterministic at the level of evolution without the measurement,
and probabilistic at the level of measurement outcome. 
But this determinism is different from the determinism used in classical CA,
here probability amplitudes are deterministic after every steps of evolution, but the sure existence like ``0'' or ``1'' are not determined. 
In the ref.~\cite{meyer}, it is shown that a nontrivial homogeneous unitary and scalar QCA can not exist in one spatial dimension---the QCA is either 
identity in position or equivalent to lattice translation operator.
The scalar implies that there is no intrinsic degrees of freedom of the particle, i.e., $\text{dim}(\mathcal{H}_\text{in}) = 1$.
It is shown that breaking translation symmetry and demanding the unitary evolution operator remained invariant under two times positional translation operation,
scalar QCA can show non-trivial signatures, which is equivalent to a homogeneous unitary and spinor 
(two-component) QCA, i.e., $\text{dim}(\mathcal{H}_\text{in}) = 2$ \cite{meyer, birula}.

Dirac cellular automaton (DCA) is one of the form of QCA which produces Dirac particle dynamics at the continuum limit of underlying cell structures \cite{birula, bisio}.
For details one can look at the ref.~\cite{mosco}. It can be thought as one of the discretization scheme of Dirac particle 
dynamics defined in discrete space-time. One can identify the cells as the positions of the particle and internal degrees as the chiralities or spins,    
for that case: $\mathcal{H}_\text{in} \equiv \mathbb{C}^d$, $\mathcal{H}_\text{cell} = \mathcal{H}_x$, 
where we have considered the spin-space dimension = $d$ and the position-space is regular one-dimensional lattice. In DCA the state of each cell is actually the 
Dirac field operator, but in case of single particle we will analyze this for wavevectors. 

One can derive the form of the DCA evolution operator: $U_{DCA} \coloneqq M_{QCA}$ starting from four basic assumptions \cite{bisio}:
\begin{enumerate}
\item {\bf  $U_{DCA}$ is local unitary.}
 The locality implies the state of a cell at present time-step depends only on the states at the nearest 
 neighbor cells including the cell itself at the previous time-step --- a Markovian process. 
 
\item {\bf $U_{DCA}$  is invariant under translation in position space.} 
So in the lattice the operator will be invariant under the transformation $x \to x + n a$ 
 where $n \in \mathbb{Z}$ or $\mathbb{Z}_\mathcal{N}$.

\item {\bf $U_{DCA}$ is covariant under parity and time reversal transformation.}
Parity transformation is imposed by the change $x \to -x$.
Time reversal transformation is imposed by the change $t \to -t$ which acts as antiunitary operator.
The form of the $U_{DCA}$ remains unchanged under these transformations. 

\item {\bf $U_{DCA}$ contains a minimum two controller or internal degrees of freedom.} 
This is necessary if $U_{DCA}$ has to obey the first three conditions, otherwise this acts as simply 
identity or translation operator in position space. 
\end{enumerate}
Then the derived DCA evolution operator: 
\begin{align}\label{dcaform}
U_{DCA} =  \sum_{x} \eta_1 \big[~ \ket{\uparrow}\bra{\uparrow} \otimes \ket{x + a}\bra{x} 
+ \ket{\downarrow}\bra{\downarrow} \otimes \ket{x-a}\bra{x} ~\big] - i~\eta_2~ \sigma_1 \otimes \ket{x}\bra{x}~\nonumber\\
= \left(\begin{array}{cc}
         \eta_1 \sum_x \ket{x + a}\bra{x} & - i~\eta_2 \sum_x \ket{x}\bra{x} \\
          - i~\eta_2 \sum_x \ket{x}\bra{x} & \eta_1 \sum_x \ket{x - a}\bra{x} \\
        \end{array}\right)~~\text{in spin basis},
\end{align} which acts on the wavevectors $\in \mathcal{H}_c \otimes \mathcal{H}_x$. 
Here $\eta_1$ is the lattice hopping strength, $\eta_2$ acts as the mass term of the particle, subject to the 
condition: $|\eta_1|^2 + |\eta_2|^2 = 1$. It is shown that for smaller values of $\eta_2$,
at the continuum limit of position and time-steps the derived Hamiltonian from this operator
obeys free Dirac particle Hamiltonian form \cite{bisio}. This appears as a special case of the general single-particle QCA described in section 5 of ref.~\cite{meyer}. 
Therefore DCA is more than the discretization of free DH, as it is derived from the basic principles of symmetry, is a not a formulation developed from the form of free DH.  
Using the same formulation as in eqs.~(\ref{defi1})-(\ref{hamdqw}) we can derive the effective Hamiltonian:
\begin{align}\label{dcahamde}
 H_{DCA} = \frac{\hbar}{\delta t} \frac{\cos^{-1}\Big[\eta_1 \cos\Big(\frac{p a}{\hbar}\Big)\Big]}{\sqrt{1 - \eta^2_1 \cos^2\Big(\frac{p a}{\hbar}\Big)}}
 \Bigg[ \eta_1 \sigma_3 \otimes \sin\bigg(\frac{p a}{\hbar}\bigg) + \eta_2 \sigma_1 \otimes \mathds{1}_x \Bigg]
\end{align} which at the continuum limit, takes the form of the 
Dirac Hamiltonian in (1+1) flat space-time:
\begin{align}\label{dcahamdir}
 H_{DCA} =  \sigma_3 \otimes p c + m c^2 \sigma_1 \otimes \mathds{1}_x,
\end{align}
 if we demand the existence of the relations: $\lim\limits_{\delta t \to 0} \frac{a}{\delta t} = c$,
$\lim\limits_{\delta t \to 0} \frac{\eta_2}{\delta t} =  \frac{m c^2}{\hbar}$. The last limiting condition implies $\lim\limits_{\delta t \to 0} \eta_1 = 1$.

It is conventionally thought that single particle DCA is nothing but the DQW of the Dirac particle---because of similarity of their
continuum behaviors---compare Hamiltonians given in eq.~(\ref{dcahamdir}) and eq.~(\ref{dqwhamdir}).
In this case the dimension $d = 2$ and $\mathcal{H}_\text{in} = \mathcal{H}_c$ the coin Hilbert space. 
But there is a noticeable difference: because of the presence of the term $ i~\eta_2~ \sigma_1 \otimes \ket{x}\bra{x}$ there does not exist any choice of coin parameters in 
DQW, for which the DQW evolution operator as in eq.~(\ref{dqwop}) exactly matches with $U_{DCA}$.
For massless case it is possible for the choice: $\theta^0 = 0, F = F^*, G = 0$, and $\eta_2 = 0$, but for a general massive case it is not.  
The presence of this term shows fine oscillation in the positional probability profile while it is absent in the DQW \cite{perez2}, and 
it becomes important when the wavelengths of the system are comparable to the few lattice step-lengths. 

Next, I am going to describe the SS-DQW which is a generalization of the DQW, and the way to get rid of the dissimilarities between DCA and DQW.

\section{Split-step discrete quantum walk}  
 The SS-DQW is first introduced in the ref.~\cite{kitagawa} to develop simulation scheme for various topological phases
 and already has implementation schemes in state-of-art simulators: photonic devices \cite{kitagawa2, zhang2}, neutral atoms in optical lattices \cite{groh},
 IBM-Q five-qubit quantum computer \cite{balu}, superconducting circuits \cite{flurin}.   
Here the single-step evolution operator is defined as 
\begin{align}
 U_{SQW} = S_+ \cdot C_2 \cdot S_- \cdot C_1
\end{align} acting on $\mathcal{H}_c \otimes \mathcal{H}_x$ where coin and shift operators are, respectively,
\begin{align}
C_j = e^{- i \sum_{q=0}^3 \theta_j^q(\delta t)~\sigma_q} \otimes \sum_x \ket{x}\bra{x} ~\text{for}~ j=1,2;  \nonumber\\
S_+ =  \sum_{x}  \ket{\uparrow}\bra{\uparrow} \otimes \ket{x + a}\bra{x} + \ket{\downarrow}\bra{\downarrow} \otimes \ket{x}\bra{x}, \nonumber\\
S_- =  \sum_{x}  \ket{\uparrow}\bra{\uparrow} \otimes \ket{x}\bra{x} + \ket{\downarrow}\bra{\downarrow} \otimes \ket{x-a}\bra{x}.
\end{align} This is called split-step because the whole evolution is now split into two substeps each of which is a coin operation followed by a shift operator. 
But the shift operators used here are different from the shift operator defined in DQW. 
The coin operators $C_j$ act the same way as for the DQW case. The shift operator $S_+$ shifts a particle one-step forward along positive $x$-direction 
if the particle is in up-spin state and does nothing if the particle is in down-spin state. The shift operator $S_-$ does nothing if the particle is in 
up-spin state and shifts it by one-step backward in $x$-axis if the particle is in down-spin state. 
So, in general, single-step  SS-DQW is not equivalent to two-step DQW evolution with two different set of coin parameters. 

In matrix representation, 
\vspace{-2cm}

\begin{align}\label{soperator}
 U_{SQW} = ~~~~\hspace{8cm} \nonumber\\
 e^{- i~\big[\theta_2^0 + \theta_1^0\big]}\sum_x \left( \begin{array}{cc}
                                      F_2  \ket{x+a}\bra{x} & G_2  \ket{x+a}\bra{x} \\
                                     - G^*_2  \ket{x}\bra{x} & F^*_2 \ket{x}\bra{x} \\
                                      \end{array} \right)
                                      \cdot \sum_x \left( \begin{array}{cc}
                                      F_1  \ket{x}\bra{x} & G_1  \ket{x}\bra{x} \\
                                     - G^*_1 \ket{x-a}\bra{x} & F^*_1 \ket{x-a}\bra{x} \\
                                      \end{array} \right) \nonumber\\
       =  e^{- i~\big[\theta_2^0 + \theta_1^0\big]} \sum_x \left(\begin{array}{cc}
                                                           F_2 F_1 \ket{x+a}\bra{x} - G_2 G_1^* \ket{x}\bra{x} & F_2 G_1 \ket{x+a}\bra{x} + G_2 F_1^* \ket{x}\bra{x} \\
                                                           - G_2^* F_1 \ket{x}\bra{x} - F_2^* G_1^* \ket{x-a}\bra{x}  & -G_2^* G_1 \ket{x}\bra{x} + F_2^* F_1^* \ket{x-a}\bra{x}\\
                                                           \end{array} \right)                              
\end{align}

where $F_j, G_j$ are the similar functions of the coin parameters as in eq.~(\ref{FG}) with $\theta^q$ replaced by $\theta^q_j$ for $j = 1, 2$. Using the 
same process as in the DQW case we can derive the effective Hamiltonian. The Hamiltonian is $H = \sum_k H_k \otimes \ket{k}\bra{k}$ where 
\begin{align}
 H_k =  - \frac{\hbar}{\delta t} \frac{\cos^{-1}\big[\Re(F_2 F_1 e^{- i  \frac{a}{\hbar} k} - G_2 G_1^*)\big]}
 {\sqrt{1 - \big[\Re(F_2 F_1 e^{- i  \frac{a}{\hbar} k} - G_2 G_1^*)\big]^2}}
  \bigg[ \Im(F_2 F_1 e^{- i  \frac{a}{\hbar} k} - G_2 G_1^*) \sigma_3 \nonumber\\
  + \Re(F_2 G_1 e^{- i  \frac{a}{\hbar} k} + G_2 F_1^*) \sigma_2 
  + \Im(F_2 G_1 e^{- i  \frac{a}{\hbar} k} + G_2 F_1^*) \sigma_1 \bigg] + \frac{\hbar}{\delta t} \big[ \theta_1^0 + \theta_2^0 \big] \sigma_0. 
\end{align}
In this case eigenvalues of the Hamiltonian are $\pm E(k)$ where \begin{align}
                                                 E(k) = \frac{\hbar}{\delta t} 
                                                 \cos^{-1}\big[\Re(F_2 F_1 e^{- i  \frac{a}{\hbar} k} - G_2 G_1^*)\big]~~~\text{for all}~k .
                                                \end{align} 
We will use the notation: $\tilde{E}(k) = \frac{\delta t}{\hbar} E(k)$. 
The corresponding eigenvectors, respectively, are \begin{align}
                                \ket{\phi^\pm(k)} \otimes \ket{k} = \frac{ i \big[F_2 G_1 e^{- i  \frac{a}{\hbar} k} + G_2 F_1^*\big]  \ket{\uparrow} 
                                +  \Big( \Im \big[ F_2 F_1 e^{- i  \frac{a}{\hbar} k} - G_2 G_1^*\big] \pm |\sin\tilde{E}(k)| \Big)
                                 \ket{\downarrow} }{\sqrt{2|\sin\tilde{E}(k)|\Big(|\sin\tilde{E}(k)|
                                \pm \Im \big[F_2 F_1 e^{- i  \frac{a}{\hbar} k} - G_2 G_1^* \big]  \Big)}} \otimes \ket{k}~.
                               \end{align}
\begin{figure}[h]\centering
 \includegraphics[width = 0.95\textwidth]{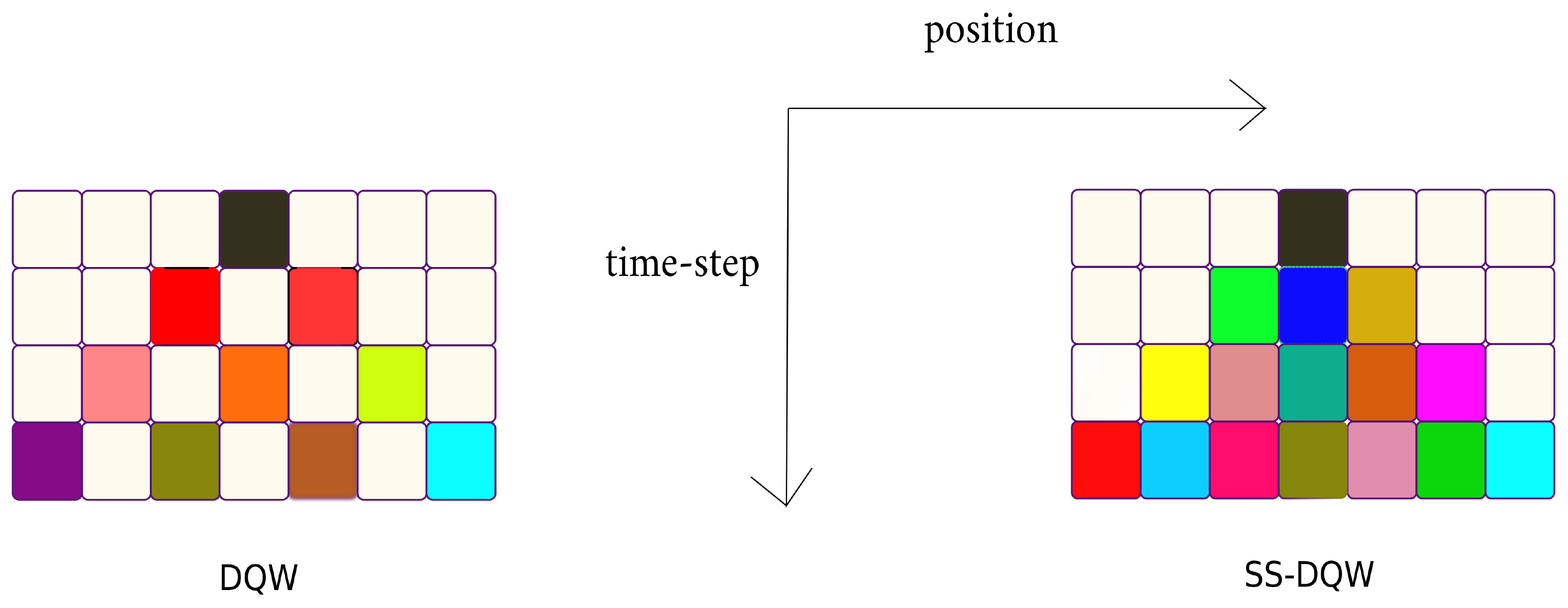}
 \caption[Quantum cellular automata for single initially localized particle]{Figure describes evolution of a initially localized particle for four time-steps of quantum walk. 
 In terms of elementary CA, DQW is coherent superposition of rules 2 and 4, where SS-DQW is the superposition of rules 2, 4 and 8.}\label{qrule}
 \end{figure}
                          
We can see the difference between the conventional DQW and SS-DQW in terms of the elementary CA rules.
The DQW can be thought as a quantum superposition of the rules ``2'' and ``8'', but 
SS-DQW is a quantum superposition of ``2'', ``4'' and ``8''. This implies SS-DQW possesses more richer structure than the DQW.  
In the fig.~\ref{qrule} we have depicted this issue for seven lattice sites and four time steps, where the different colors denote 
different probability amplitude for the existence of a quantum particle.

\section{Connection of DCA and SS-DQW} 

For the choice: $\theta_1^0 + \theta_2^0 = 0,~ F_1 = 1,~G_1 = 0$
                we have  \begin{align} U_{SQW} = 
                \sum_x \left(\begin{array}{cc}
                 F_2 \ket{x+a}\bra{x} &  G_2 \ket{x}\bra{x} \\
               - G_2^* \ket{x}\bra{x}  &  F_2^* \ket{x-a}\bra{x}\\
                                                           \end{array} \right).
                                                           \end{align}
                
       Further choice: $F_2 = \cos \theta^1_2,~G_2 = - i \sin \theta_2^1$ will give rise to                 
     \begin{align}\label{ssqwope}
U_{SQW} =  \sum_{x} \cos \theta_2^1 \big[~ \ket{\uparrow}\bra{\uparrow} \otimes \ket{x + a}\bra{x} 
+ \ket{\downarrow}\bra{\downarrow} \otimes \ket{x-a}\bra{x} ~\big] - i~\sin \theta_2^1~ \sigma_1 \otimes \ket{x}\bra{x}~.
\end{align}      
 Now identifying $\cos \theta_2^1$ with $\eta_1$ and $\sin \theta^1_2$ with $\eta_2$ we can show this exactly matches with the form of the 
 DCA evolution operator in eq.~(\ref{dcaform}). Hence, it will capture all the properties of the DCA. 
 \begin{figure}[h]
\centering
 \includegraphics[height = 7.5cm]{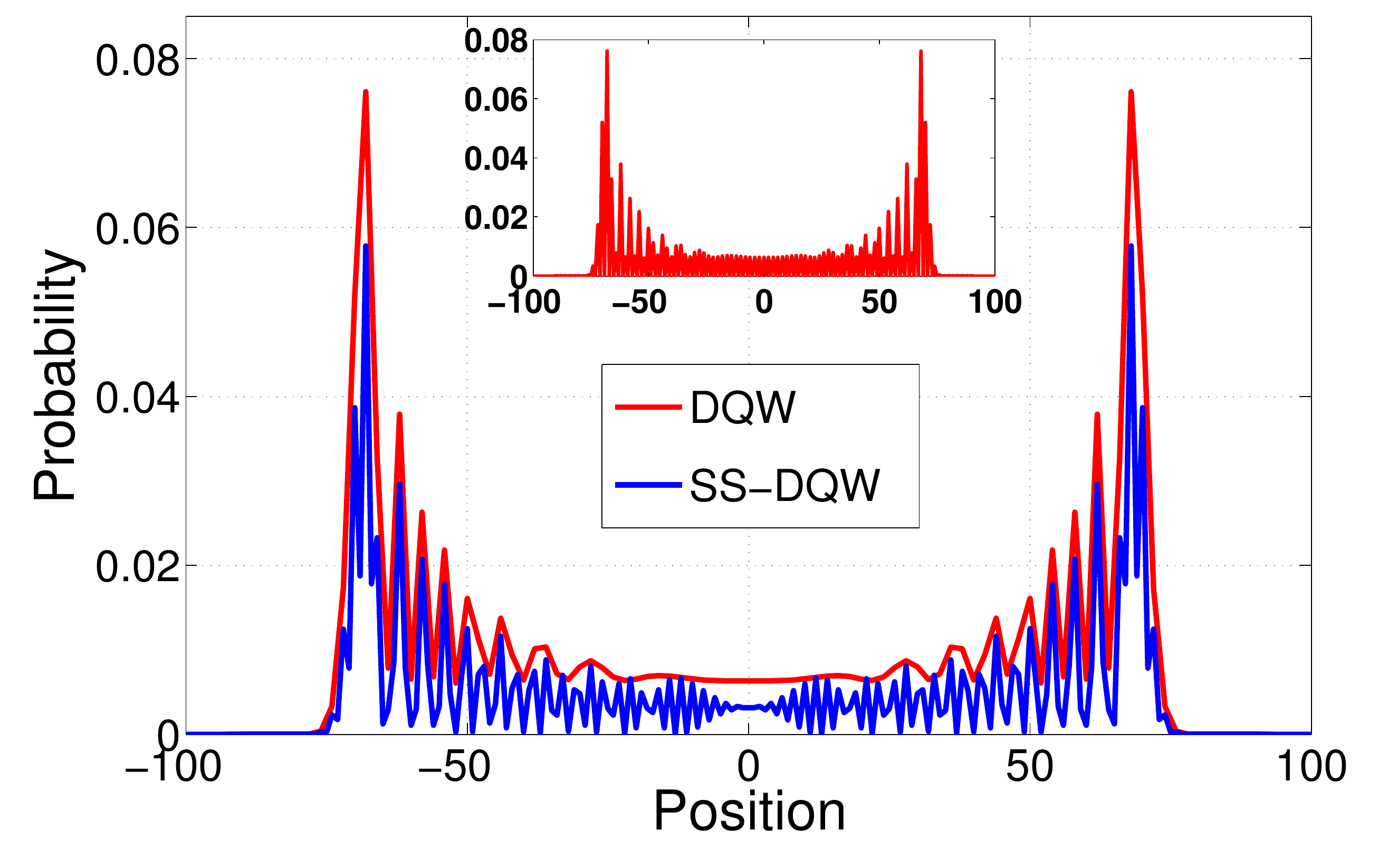}
 \caption[Positional probabilities for DQW and SS-DQW]{The probability of finding the particle in one-dimensional position space after 100 steps of DQW and SS-DQW.
The initial state of the particle is $\frac{1}{\sqrt{2}} ( \ket{\uparrow} + \ket{\downarrow} )\otimes \ket{x=0}$ and 
the coin operators are $C_j = e^{- i~\theta^1_j \sigma_1} \otimes \sum_x \ket{x}\bra{x}$. 
 Blue distribution is for SS-DQW with the choice: $\theta^1_1 = 0$, $\theta^1_2 = \frac{\pi}{4}$, 
 which is identical to DCA when $\eta_1 = \eta_2 = \frac{1}{\sqrt{2}}$ and the red line is for the DQW with $\theta^1 = \frac{\pi}{4}$; the
points with zero probability is removed from the main plot whereas, it is retained in the inset.}\label{dcafig2}
\end{figure}

 Using the same procedure as in the DQW case we can derive the effective Hamiltonian $H$ which has the following form.          
  \begin{align}\label{ssham}
    H = \frac{\hbar}{\delta t} \sum_k \frac{\bar{E}(k, \theta_2^1)}
 {|\sin \bar{E}(k, \theta_2^1)|}
  \bigg[ \cos \theta_2   \sin \bigg( \frac{a}{\hbar} k\bigg)~\sigma_3 
  + \sin \theta_2^1~ \sigma_1 \bigg] \otimes \ket{k}\bra{k}. 
  \end{align}
For this case, the eigenvalues and eigenvectors of the Hamiltonian are, respectively, \begin{align}\label{eigenval}
  E(k) = \frac{\hbar}{\delta t} 
 \cos^{-1}\bigg[\cos \theta^1_2 \cos\bigg(\frac{k a}{\hbar}\bigg) \bigg],~~~\text{and}~\hspace{5cm} \nonumber\\ \nonumber\\
                                \ket{\phi^\pm(k)} = \frac{ \sin \theta^1_2   \ket{\uparrow} 
                                +  \bigg(- \cos \theta^1_2 \sin \bigg(\frac{ka}{\hbar}\bigg) \pm |\sin\tilde{E}(k)| \bigg)
                                 \ket{\downarrow} }{\sqrt{2|\sin\tilde{E}(k)|\bigg(|\sin\tilde{E}(k)|
                                \mp \cos \theta^1_2 \sin \bigg(\frac{ka}{\hbar}\bigg)  \bigg)}}~~~\text{for all}~k .
                               \end{align}

  The Hamiltonian in eq.~(\ref{ssham}) will give the free Dirac Hamiltonian 
  \begin{align}
    H = c \sigma_3 \otimes p  + m c^2 \sigma_1 \otimes \mathds{1}_x ~,~
  \text{if the limits:}~ \lim_{\delta t \to 0} \frac{a}{\delta t} = c,~ \lim_{\delta t \to 0} \frac{\theta_2^1}{\delta t} = \frac{m c^2}{\hbar}~ \text{exist}. 
  \end{align}

\subsection{Comparison of position-coin entanglement between DQW and SS-DQW cases}
 The presence of coin state dependent shift operation in quantum walk evolution, is responsible for the interaction among coin and positions of the particle. 
 Entanglement between position and spin (coin), can be thought as a quantity of interaction between them. 
 Presence of this between two parties implies that, one party carries some information about the 
 other party. As this entanglement is a correlation between two different degrees of freedom of the same Dirac particle, 
 its usefulness in information processing where nonlocality is an important resource, is under question.  
 But it is obviously useful in another perspective.
 Considering the coin as the system and the position lattice as the bath, ref.~\cite{romanelli1} describes how one can study thermodynamics by this 
 kind of coin-position entanglement dynamics, in DQW framework.  
  As it is shown that DCA is more near to the particular choice of SS-DQW  
 than the DQW case, this kind of study helps to understand thermodynamical perspective in Dirac particle dynamics.   
 
\begin{figure}[htbp]
\centering
\subfigure[][]{
\includegraphics[width=0.47\textwidth]{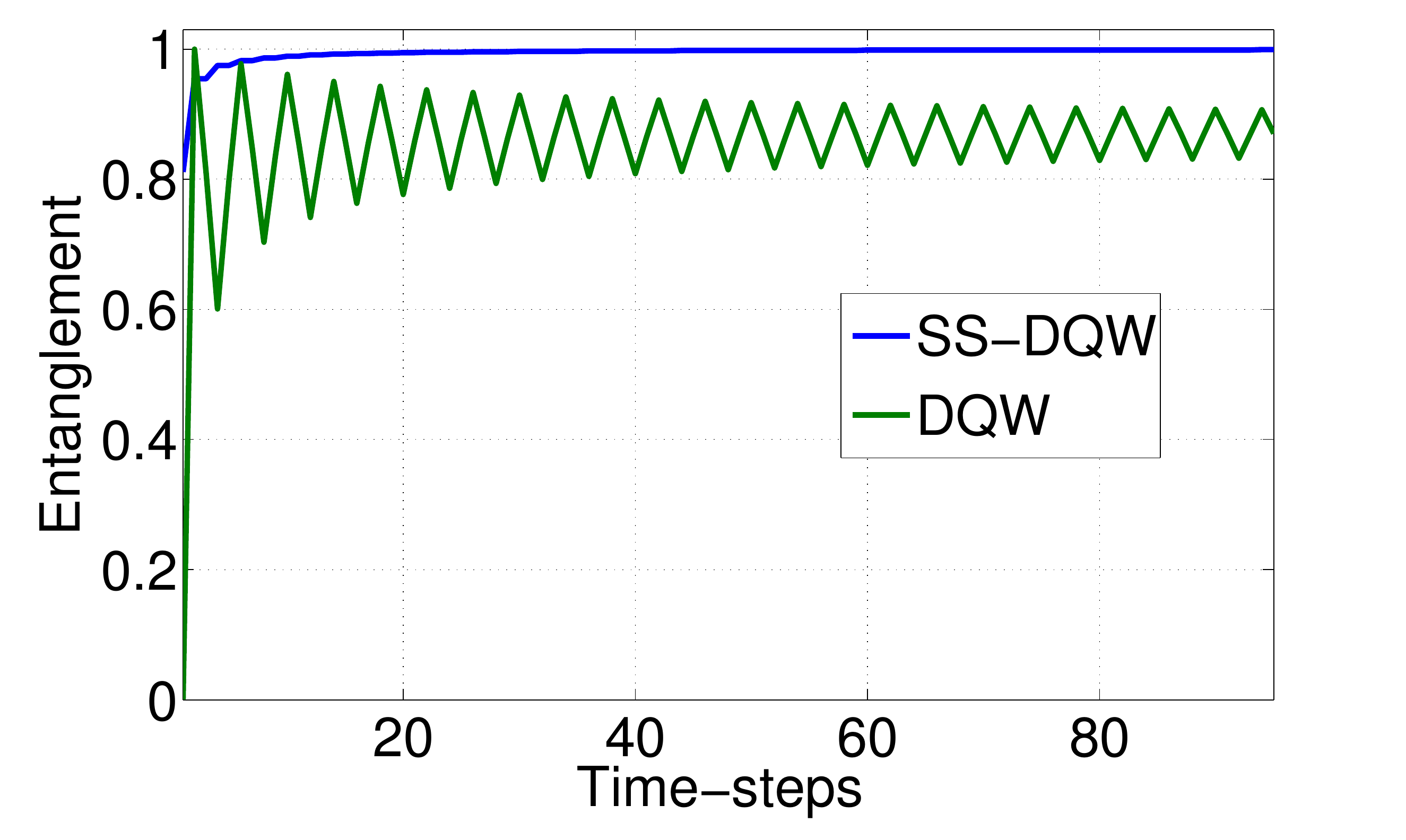}
\label{}}
\subfigure[][]{
 \includegraphics[width=0.47\textwidth]{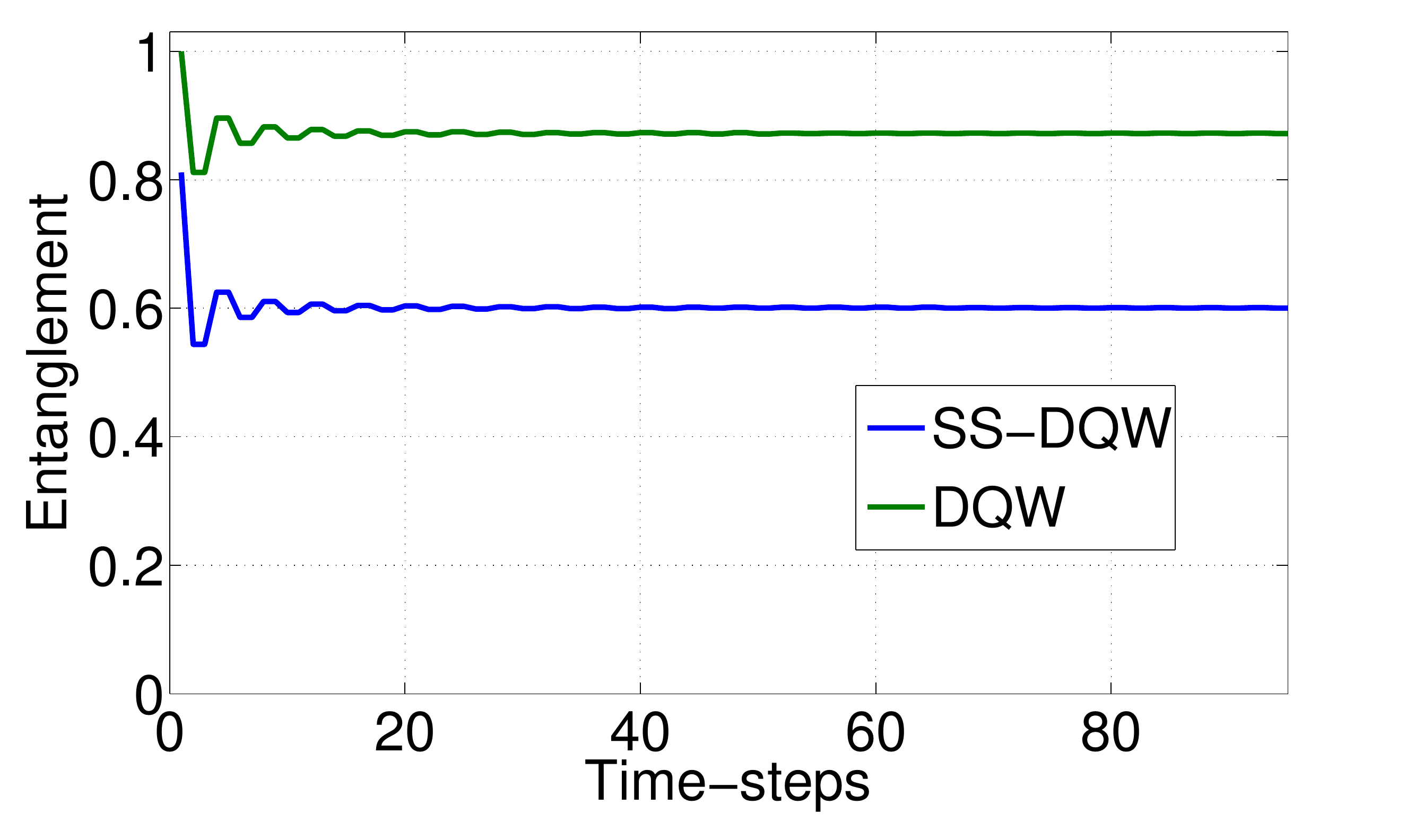}
\label{}}
\subfigure[][]{
 \includegraphics[width=0.47\textwidth]{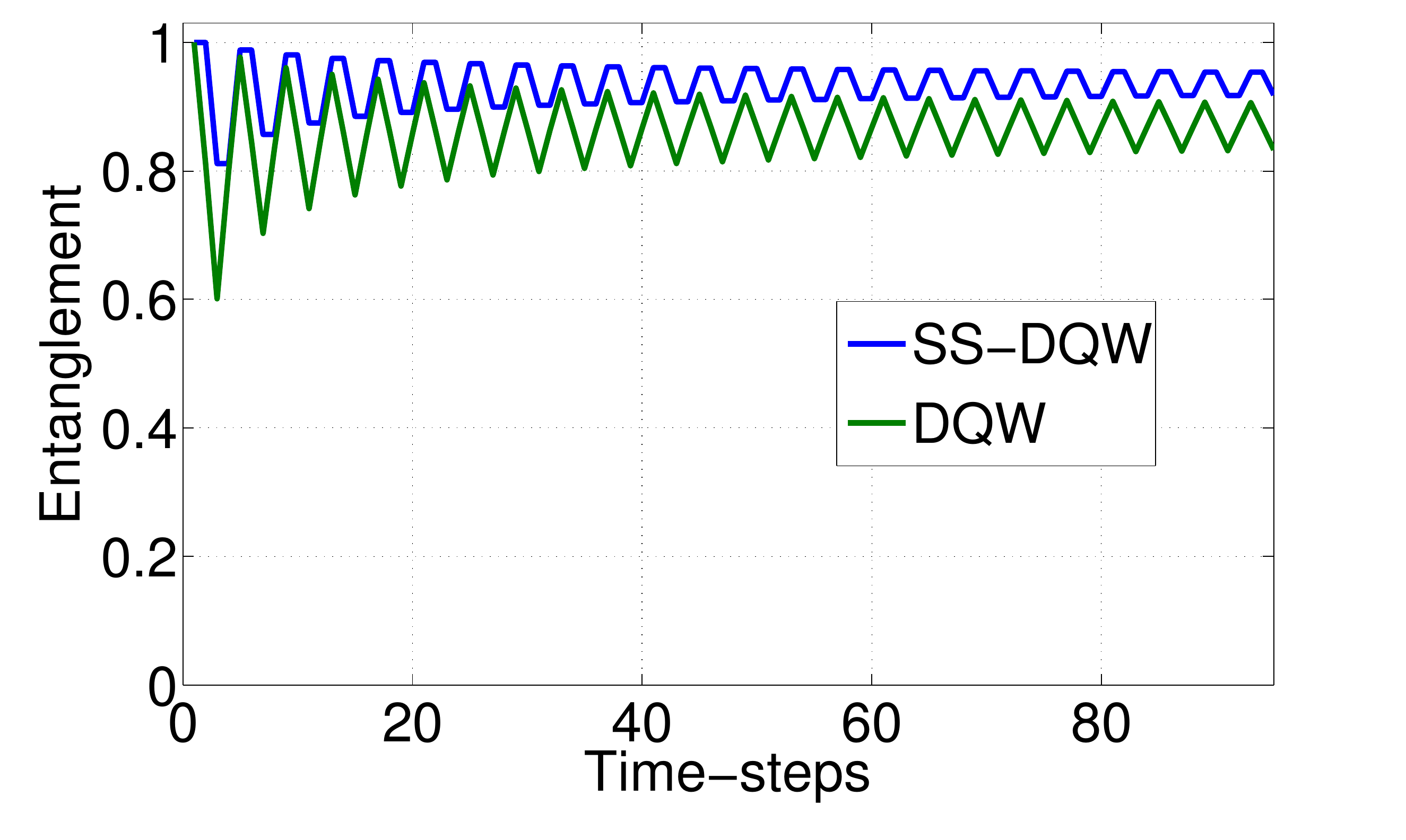}
\label{}}
 \caption[Coin-position entanglement as a function of time with different initial states]
 {Coin-position entanglement as a function of time with different initial states. For conventional DQW coin parameter
$\theta = \frac{\pi}{4}$ and for SS-DQW $\theta^q_1 = 0, \theta^1_2 = \frac{\pi}{4}$. The initial states in 
(a) $\frac{1}{\sqrt{2}} (\ket{\uparrow} + i \ket{\downarrow}) \otimes \ket{x=0} $ 
(b) $\frac{1}{\sqrt{2}} (\ket{\uparrow} +  \ket{\downarrow}) \otimes \ket{x=0} $ and 
(c) $ \ket{\uparrow} \otimes \ket{x=0} $. Dependency of entanglement value on the initial state is higher for split-step QW compared to the conventional DQW.}\label{entan1}
\end{figure}

 Here we will use partial entropy as measure of entanglement. All of our evolution operators are unitary 
 and hence, it preserves the purity of a state. Thus if we start from a pure state $ \ket{\psi(0)} \in \mathcal{H}_c \otimes \mathcal{H}_x$, 
 this partial entropy of the state can be treated as a proper measure of entanglement. Mathematically:
\begin{align}
 \rho(t) = \ket{\psi(t)}\bra{\psi(t)}~~\Rightarrow~\rho_c(t) = \text{Tr}_x [\rho(t)] = \sum_x \braket{x| \rho(t) |x},\nonumber\\
 \text{partial entropy}~ = - \text{Tr}_c [ \rho_c(t) \ln \rho_c(t)],  
\end{align} where $\rho(t)$ is the density matrix of the system at time-step $t$ and $\rho_c(t)$ is the partial state of the system defined on the coin 
Hilbert space $\mathcal{H}_c$. The ref.~\cite{romanelli2} discusses the relation between the asymptotic value of the coin-position entanglement 
and the initial coin state in DQW set-up. In contrary to the classical Markov process where the asymptotic entanglement is independent of the initial state, here 
 it is sensitive to the initial states. 
 In this section we will compare the entanglement arises in SS-DQW with that in DQW.
  In the fig.~\ref{entan1} we have shown the entanglement evolution as a function of time-steps in case of SS-DQW and DQW 
for particular choice of parameters and three different localized initial states.  
Note that in every case entanglement moves near to a saturated value starting from the initial zero value.
The evolution operator is diagonalized in momentum space, not in the position space.  So, for a state in momentum eigenstate 
the entanglement always remains zero. From that we can conclude, 
for nonlocalized initial states the entanglement can have lower value compared to the localized case, 
depending on how much the state is localized in momentum space. 
\begin{figure}[htbp]
\centering
\includegraphics[width=0.7\textwidth]{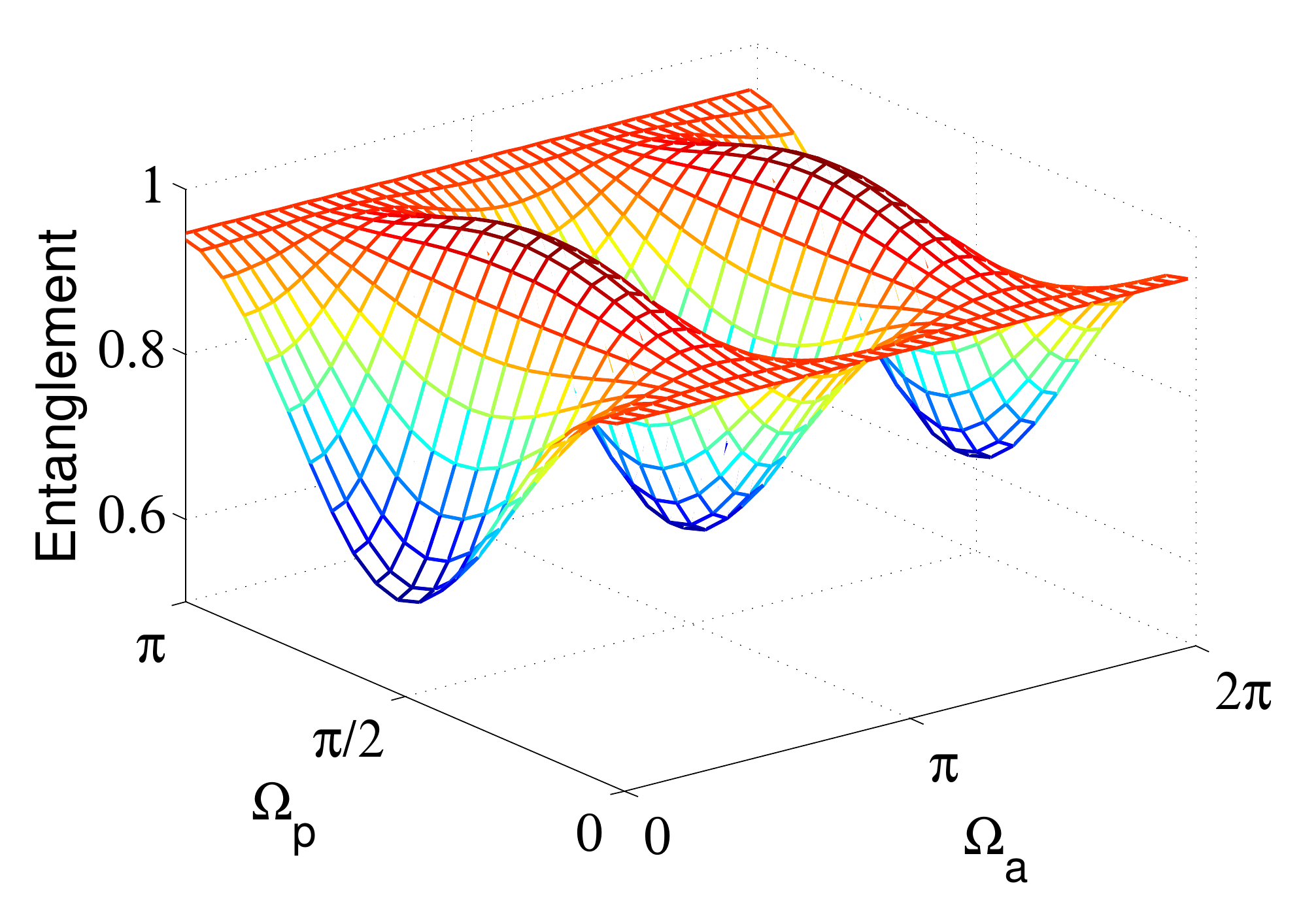}
 \caption[Coin-position entanglement as a function of localized initial state and coin parameters for SS-DQW]
 {Entanglement (averaged over the entanglement values for last few steps of walk, near 100 time-steps)
 between space and internal degree of freedom as a function of localized initial state and coin parameters. 
The entanglement as a function of initial state parameters for SS-DQW with $\theta^q_1 = 0$ 
for all $q \in \{0, 1, 2, 3\}$, $\theta^1_2 = \frac{\pi}{4}$, $\theta^r_2 = 0$ for all $r \in \{0, 2, 3\}$ is shown---according to eq.~(\ref{soperator}).}\label{entan2}
\end{figure} 
 \begin{figure}[h]\centering
 \includegraphics[width=0.7\textwidth]{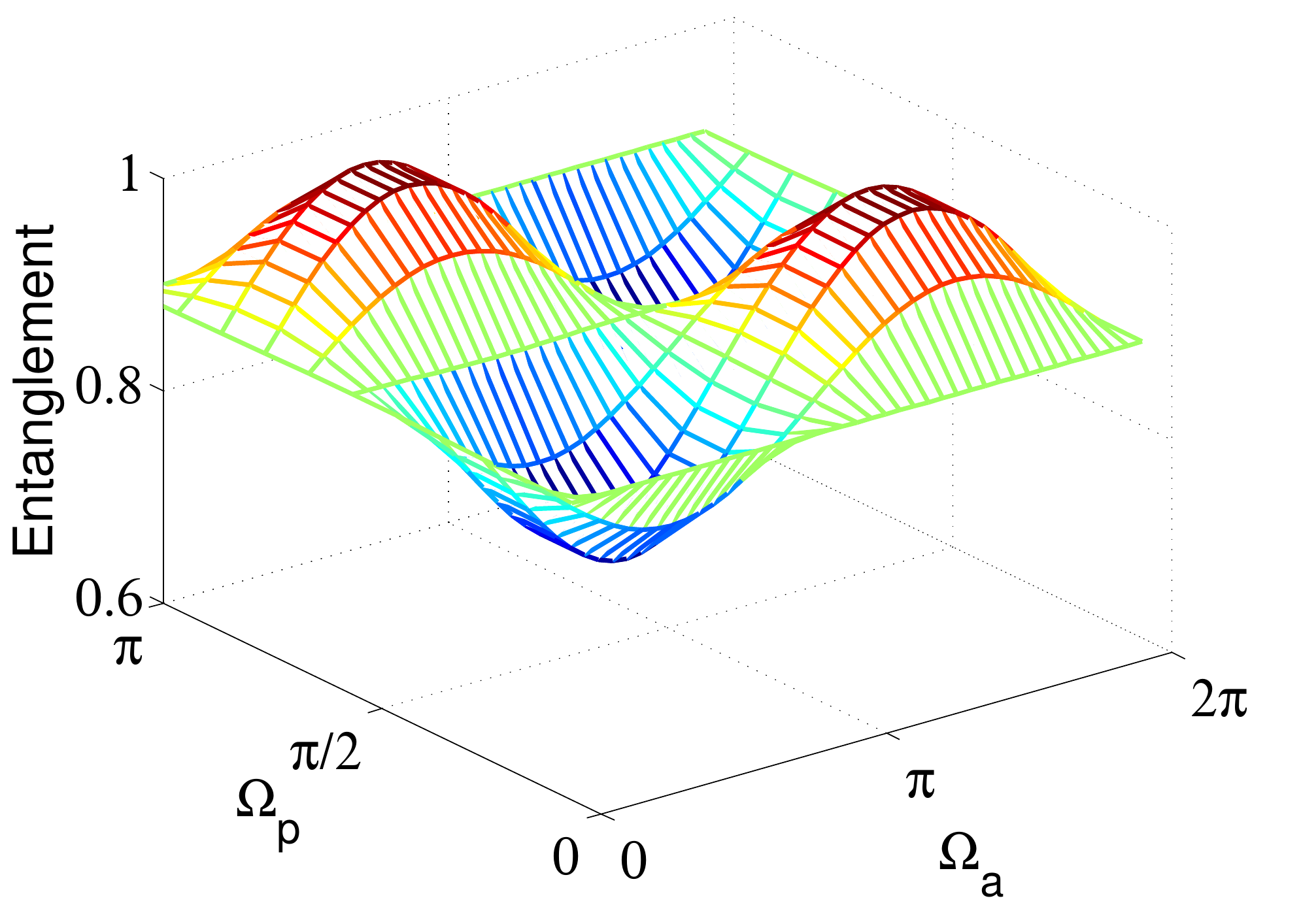}
 \caption[Coin-position entanglement as a function of localized initial state and coin parameters for DQW]
 {Entanglement (averaged over the entanglement values for last few steps of walk, near 100 time-steps)
 between space and internal degree of freedom as a function of localized initial state and coin parameters. 
The entanglement as a function of initial state parameters for the conventional DQW with 
$\theta^1 = \frac{\pi}{4}$, $\theta^q = 0$ for all $q \in \{0, 2, 3\}$ is shown---according to eq.~(\ref{dqwop}).}\label{entan3}
 \end{figure}

For figs. \ref{entan2} and \ref{entan3} the azimuthal angle: $\Omega_a$ and polar angle: $\Omega_p$ 
correspond to the spherical coordinate angles of Bloch sphere associated with the internal degree (coin space).
It is evident that in figs.~\ref{entan1}, \ref{entan2} and \ref{entan3} the 
entanglement is more sensitive to the initial coin state in SS-DQW case for which DCA is recovered, compared to the case of DQW.

The above entanglement analysis is done for pure initial states, for mixed initial states partial entropy is not a good measure of the entangled.
Because quantum walk system can be treated as a qubit-qudit system where qubit represents coin state and the qudit represents position space with lattice sites $d$, 
we can use negativity of the partial transposed state of the system as a measure of entanglement in case of mixed state.

\section{Zitterbewegung oscillation}
This is a property of vibrating motion during the evolution of any quantum mechanical observable $\mathscr{A}$ which does not commute with the Hamiltonian operator,
i.e., $[\mathscr{A}, H] \neq 0$. For our case the noncommutivity results in mixing of positive and negative energy eigenvalue 
solutions during the evolution. This mixing is responsible for oscillation of the expectation value of the observable 
and is known as Zitterbewegung oscillation~\cite{zit}. 
Zitterbewegung oscillations is a very common phenomenon that describes the jittering motion of free Dirac particles. 
Here we will look into this phenomenon as a function of SS-DQW parameter for which we get the equivalence with DCA. 

\begin{figure}[h]
\centering
\includegraphics[width = 0.7\textwidth]{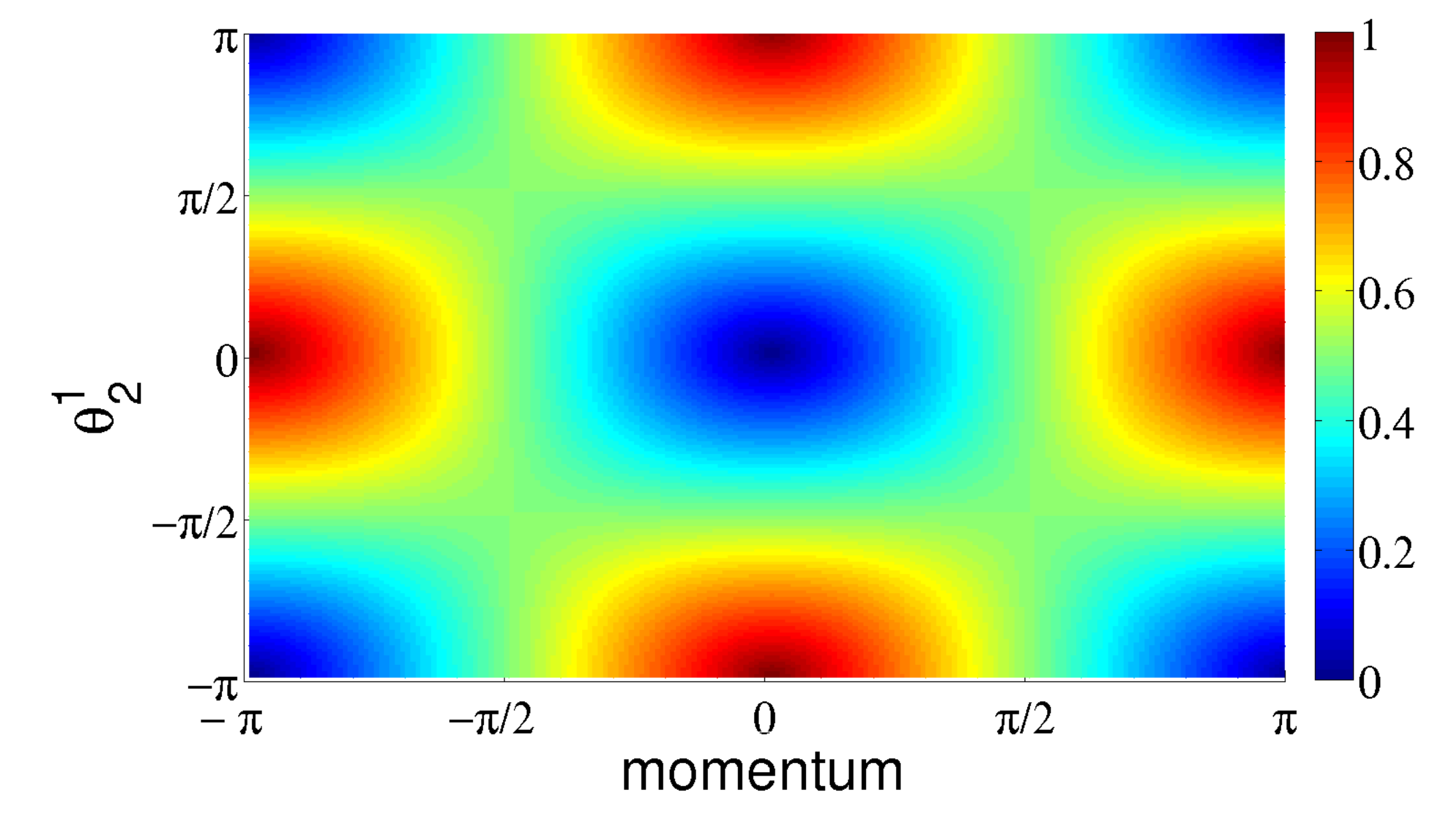}
\caption[Zitterbewegung frequency as a function of coin parameter and momentum]{Zitterbewegung frequency as a function of $\theta^1_2$
(other parameters are set to the values for which SS-DQW is equivalent to DCA) and momentum = $k$ with $\frac{a}{\hbar} = 1, \delta t = 1$.}\label{zitt}
\end{figure}

For the case of SS-DQW, the state $\ket{\psi_s(t)} \in \mathcal{H}_c$ of a particle moving with momentum $k$,
 can be expressed as a linear superposition of the eigenvectors of $U_{SQW}$: $\ket{\phi^\pm_k}$ (normalized) with the same momentum $k$,
 so  \begin{align}   \ket{\psi_s(t=0)} = c_1 \ket{\phi^+(k)} + c_2 \ket{\phi^-(k)} 
                        \Rightarrow \ket{\psi_s(t)} = c_1~(U_{SQW})^t \ket{\phi^+(k)}  +  c_2~(U_{SQW})^t \ket{\phi^-(k)} \nonumber\\
                       \Rightarrow \ket{\psi_s(t)} = c_1 e^{-\frac{i}{\hbar} E(k) t } \ket{\phi^+(k)} 
                       + c_2 e^{-\frac{i }{\hbar}E(k) t} \ket{\phi^-(k)} \nonumber\\
                        \Rightarrow  \braket{\mathscr{A}}_t =  \braket{\psi_s(t)|\mathscr{A}|\psi_s(t)}             
                        = |c_1|^2 \bra{\phi^+(k)} \mathscr{A}  \ket{\phi^+(k)} + |c_2|^2 \bra{\phi^-(k)} \mathscr{A} \ket{\phi^-(k)} \nonumber\\
                        +   c^*_1 c_2 e^{\frac{2 i }{\hbar}E(k) t} \bra{\phi^+(k)} \mathscr{A}  \ket{\phi^-(k)}
        +  c_1 c^*_2 e^{ - \frac{2 i }{\hbar}E(k) t} \bra{\phi^-(k)}\mathscr{A}  \ket{\phi^+(k)}.
\label{eqpecval}                                                         
                                                                          \end{align}
From the equation~(\ref{eqpecval}) we can see that the time dependent part is
\begin{align}
  c^*_1 c_2 e^{ \frac{2 i}{\hbar} E(k) t} \bra{\phi^+(k)} \mathscr{A}  \ket{\phi^-(k)} +  c_1 c^*_2 e^{ - \frac{2 i E(k) t}{\hbar}}
  \bra{\phi^-(k)} \mathscr{A} \ket{\phi^+(k)}.
  \end{align} This time-dependent part contains the frequency = $\frac{2 E(k)}{2 \pi \hbar} = \frac{E(k)}{\pi \hbar}$ which is identified as 
   the Zitterbewegung frequency:  
            \begin{align}
           Z_{SQW} \coloneqq  \frac{ 1 }{\pi \times \delta t} \cos^{-1}
           \big[\Re(F_2 F_1 e^{- i  \frac{a}{\hbar} k} - G_2 G_1^*)\big] = \frac{ 1 }{\pi \times \delta t} \cos^{-1}
           \bigg[ \cos \theta^1_2 \cos \bigg(\frac{k a}{\hbar} \bigg) \bigg].
            \end{align} In fig.~\ref{zitt} we have plotted this Zitterbewegung frequency as a function of coin parameter and momentum. 


\chapter{Simulating neutrino oscillation}\label{secondq}

After identifying the SS-DQW as a simulation scheme of free Dirac particle dynamics, 
one can question whether the same SS-DQW scheme is sufficient to simulate Dirac particle related phenomena where the particle
contains other internal degrees of freedom besides of the spin. 
In this chapter I will show by slight modification, i.e., increasing the coin space dimension we can capture some of the phenomena. 
One important phenomenon in this direction is neutrino oscillation.

In order to get rid of the energy, angular momentum and momentum conservation anomaly during the nuclear beta decay Pauli first
proposed the possible existence of neutrino \cite{pauli}.
This is an electric chargeless particle, participates only in the  
weak and  gravitational interactions. The oscillation theory is first proposed in 1957 by ref.~\cite{pontecorvo}. 
Neutrino appears in three flavor degrees of freedom, the corresponding flavor states are described by electron neutrino: $\ket{\nu_e}$, 
muon neutrino: $\ket{\nu_\mu}$, and tauon neutrino: $\ket{\nu_\tau}$. Each of the flavor states can be written as linear superposition of 
three mass eigenstates: $\{ \ket{\nu_{m_1}}, \ket{\nu_{m_2}}, \ket{\nu_{m_3}}\}$, where the coefficients of the superposition are described 
by the PMNS matrix element $U_{\alpha m_j}$ for $\alpha \in \{e, \mu, \tau\}$, $j \in \{1, 2, 3\}$, 
\begin{align}
\text{and the PMNS \cite{maki2} matrix}~=  U = \left( \begin{array}{ccc}
             U_{e m_1} & U_{e m_2} & U_{e m_3} \\
             U_{\mu m_1} & U_{\mu m_2} & U_{\mu m_3} \\
             U_{\tau m_1} & U_{\tau m_2} & U_{\tau m_3} \\ 
            \end{array}\right) = \nonumber\\
 \left( \begin{array}{ccc}
  1 & 0 & 0\\
  0 & c_{23} & s_{23}\\
  0 & -s_{23} & c_{23}
\end{array}\right)
 \left( \begin{array}{ccc}
  c_{13} & 0 & s_{13}e^{-i\delta}\\
  0 & 1 & 0\\
  -s_{13}e^{i\delta} & 0 & c_{13}
\end{array}\right)
\left( \begin{array}{ccc}
  c_{12} & s_{12} & 0\\
  -s_{12} & c_{12} & 0\\
  0 & 0 & 1
\end{array}\right)
\left( \begin{array}{ccc}
 e^{i\alpha_{1}/2} & 0 & 0\\
 0 & e^{i\alpha_{2}/2} & 0\\
 0 & 0 & 1
\end{array}\right),
\end{align} which is a unitary matrix.
Here $c_{ij} \coloneqq \text{cos}\,\theta_{ij}$ and $s_{ij} \coloneqq \text{sin}\,\theta_{ij}$ with  $\theta_{ij}$ being the mixing angle, 
and $\alpha_{1}$, $\alpha_{2}$, $\delta$ are CP-violating phases. This implies neutrino does not have definite mass when it is in a particular 
flavor state.
These mass eigenstates are actually the eigenstates of the free Dirac Hamiltonian;
but of course we are avoiding here the controversy that whether neutrino is a Dirac or Majorana particle,
and we will simply consider it as a Dirac particle. Majorana is such particle which is indistinguishable with its anti-particle.
Neutrinos are usually detected as the flavors, not as mass eigenstates. So, starting from a initial flavor state, a neutrino changes its flavor during the evolution, as the evolution is 
governed by the Dirac Hamiltonian. This change happens at the probability level and shows oscillating behavior with time. This phenomenon is known as the neutrino oscillation.
Neutrino with a fixed momentum eigenstate can show this kind of oscillation, 
and hence the three-flavor oscillation can survive if the masses: $m_j$ are different for all $j \in \{1, 2, 3\}$.
Neutrino oscillation is a beyond standard model (SM) phenomenon because SM consider it as a massless particle, and hence SM is unable to explain this oscillation. 
Mathematically, if the initial neutrino state = \begin{align}\label{neutheo}
                  \ket{\psi_\nu(0)} = \ket{\nu_\alpha} =  \sum_{j=1}^3 U^*_{\alpha j} \ket{\nu_{m_j}} 
                  \Rightarrow~\text{the state at time}~t~\text{is}~\ket{\psi_\nu(t)} =  U^n_\nu \ket{\psi_\nu(0)} \nonumber\\
                  \Rightarrow~ \ket{\psi_\nu(t)} = \sum_{j=1}^3 U^*_{\alpha j} e^{- \frac{i E_j t}{\hbar}} \ket{\nu_{m_j}} 
                  \Rightarrow \braket{\nu_\beta | \psi_\nu(t)} =  \sum_{j=1}^3 U^*_{\alpha j} e^{- \frac{i E_j t}{\hbar}} \braket{\nu_\beta |\nu_{m_j}}
                 \end{align} where we have considered the unitary evolution operator for the neutrino as $U_\nu =  e^{- \frac{i H_D \delta t}{\hbar}}$,
                 with $t = n \times \delta t$ and $H_D$ as the corresponding evolution Hamiltonian.                
So, if the initial neutrino state is at flavor $\alpha$, the transition probability from flavor $\alpha$ to flavor $\beta$ at time $t$ is 
 \begin{align}\label{probaneu}
  P_{\alpha \to \beta}(t) = |\braket{\nu_\beta | \psi_\nu(t)}|^2 = \Big|\sum_{j=1}^3 U^*_{\alpha j} e^{- \frac{i E_j t}{\hbar}} U_{\beta j} \Big|^2 \nonumber\\
  =  \sum_{j=1}^3 | U^*_{\alpha j} U_{\beta j}|^2
  +  \sum_{j \neq l; j,l=1}^3 U^*_{\alpha j} U_{\alpha l} e^{- \frac{i ( E_j -E_l) t}{\hbar}} U_{\beta j} U^*_{\beta l}.
 \end{align}
Presence of the terms $e^{- \frac{i (E_j -E_l) t}{\hbar}}$, is responsible for the oscillation in the flavor transition probability. 
Neutrino can be treated as ultra-relativistic particle, i.e., momentum of the $j$-th mass eigenstate: $k_j >> m_j c$. Hence, in this case 
the energy eigenvalue 
\begin{align} E_j = \sqrt{k_j^2 c^2 + m_j^2 c^4} \approx k_j c + \frac{m_j^2 c^3}{2 k_j}\end{align} upto first order approximation in $\frac{m_j^2 c^2}{k^2_j}$.
Because of this ultra-relativistic nature, the velocity is set to the $c$, so if the traveling distance of it is $L$ for time $t$, then $L \approx c t$.    
In the above eqs.~(\ref{neutheo}), it is considered that the initial neutrino is in a momentum eigenstate, 
i.e., \begin{align}\ket{\nu_\alpha} = \ket{\nu_\alpha(k)} \otimes \ket{k}~~\text{and also}~~\ket{\nu_{m_j}} = \ket{\nu_{m_j}(k)} \otimes \ket{k}.\end{align}   
So, we have to deal with the case when $k_1 = k_2 = k_3 = k$. So, obeying the condition $k >> m_j c$, $k_j c \approx E_j \approx E$ for all $j$.
\begin{align}
 \Rightarrow E_j \approx E + \frac{m_j^2 c^4}{2 E} \Rightarrow  (E_j - E_l) t \approx \frac{c^4}{2 E}(m_j^2 - m_l^2) t =  \frac{L c^3}{2 E}(m_j^2 - m_l^2).
\end{align}
Below we will show the oscillation probability profile as a function of propagation length-energy ratio,
according to a recent experimental data \cite{garcia}. This is done for $\ket{\psi_\nu(0)} = \ket{\nu_e}$ 
assuming a normal ordered neutrino mass spectrum $(m_3 > m_2 > m_1 )$.
  \begin{align}\label{actualpara}
 m^2_2 -  m^2_1 = 7.50 \times 10^{-5}~\text{eV}^2,~~ m^2_3 -  m^2_1 = 2.457 \times 10^{-3}~\text{eV}^2,~\nonumber\\
 m^2_3 -  m^2_2 = 2.382 \times 10^{-3}~\text{eV}^2,~~ E = 1~~\text{GeV}~~.~~~~~~~~~~~~~ 
\end{align}
As $\delta$ has not been determined by experiments, 
it can take a value anywhere between 0 to 2$\pi$ and for simplicity we have taken $\delta$=0 for our oscillation plots. 
Here we have considered neutrino as Dirac particle, so we can choose $\alpha_1 = \alpha_2 = 0$ which imply all the elements 
of the PMNS matrix are real. Then the oscillation probability in eq.~(\ref{probaneu}):
\begin{align}
 P_{\alpha \to \beta}(t) 
  =  \sum_{j=1}^3 | U_{\alpha j} U_{\beta j}|^2
  +  2 \sum_{j > l;~ j,l = 1}^3 U_{\alpha j} U_{\alpha l}U_{\beta j} U_{\beta l} \cos \frac{( E_j -E_l) t}{\hbar}.
\end{align}

\begin{itemize}
 \item  Neutrino oscillation probabilities for an initial electron neutrino (obtained using the real experimental data). 
Here, we show oscillation probability $\nu_{e}(0)\rightarrow\nu_{e}(t)$ (blue),
$\nu_{e}(0)\rightarrow\nu_{\mu}(t)$ (green), $\nu_{e}(0)\rightarrow\nu_{\tau}(t)$ (red).
\begin{figure}[htbp]\centering
\includegraphics[width = 0.7\textwidth]{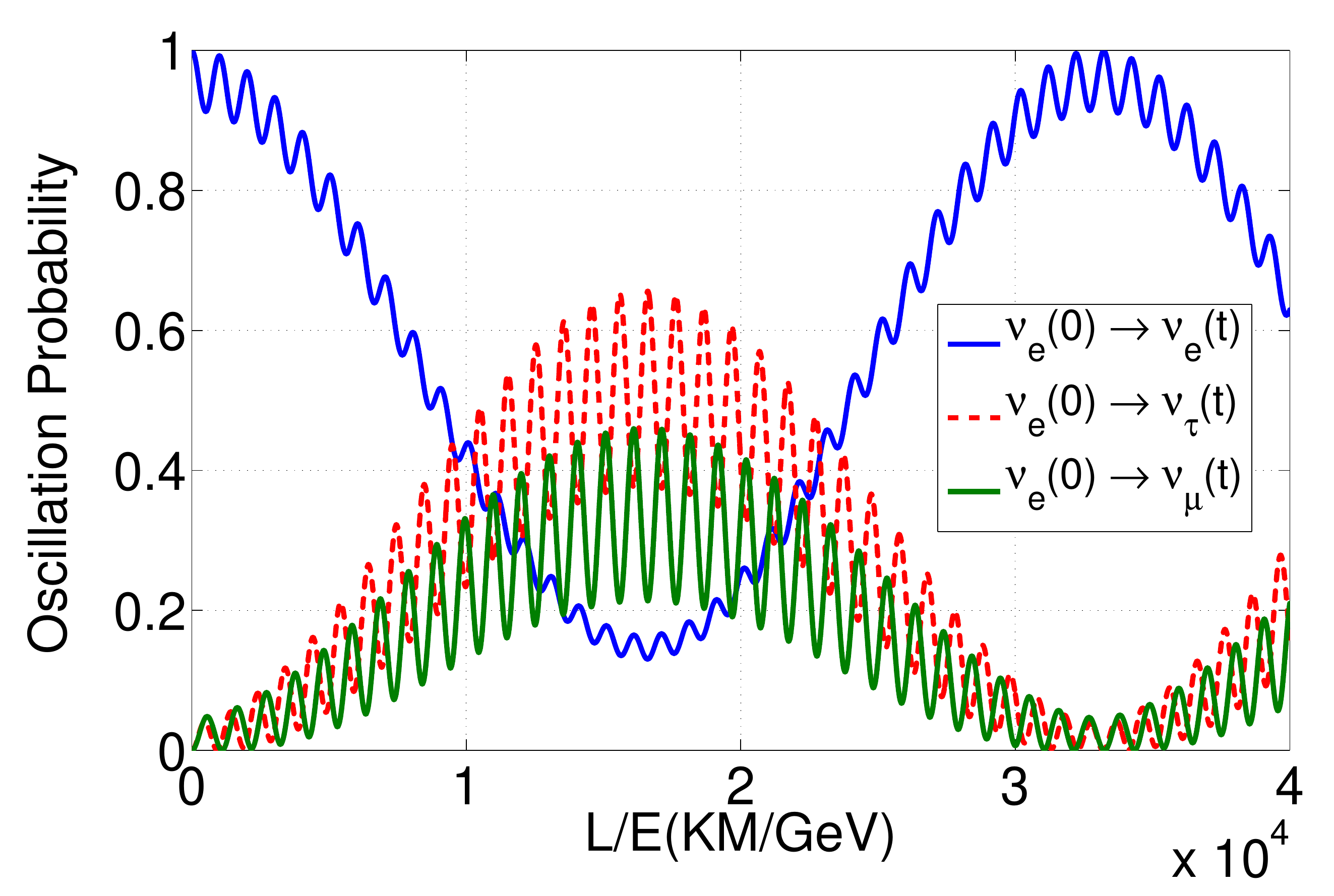} 
\caption[Theoretical long range neutrino oscillation probabilities for an initial electron neutrino]
{Theoretical long range neutrino oscillation probabilities for an initial electron neutrino}\label{theoryl}
\end{figure}
\begin{figure}[htbp]\centering
\includegraphics[width = 0.7\textwidth]{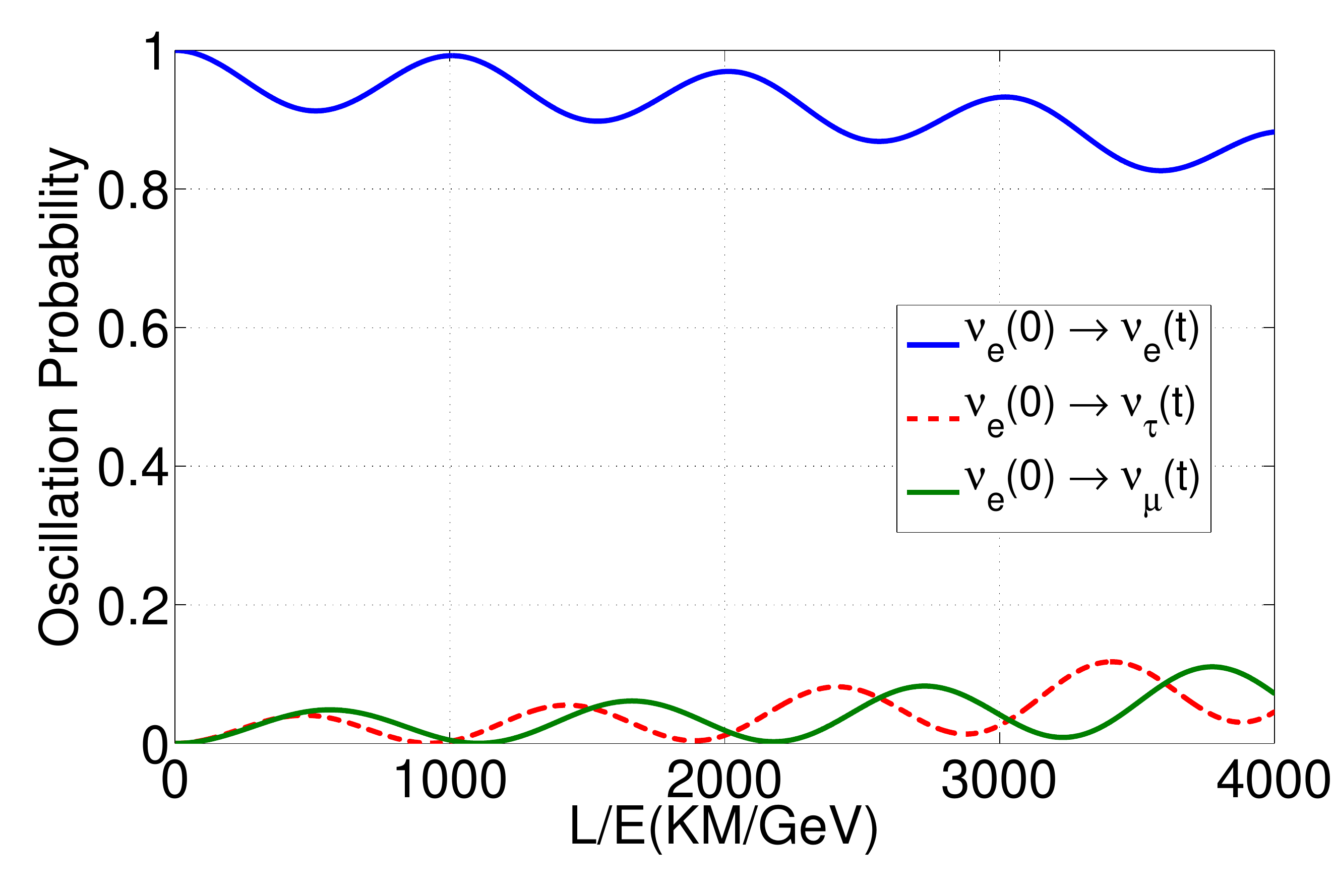} 
\caption[Theoretical short range neutrino oscillation probabilities for an initial electron neutrino]
{Theoretical short range neutrino oscillation probabilities for an initial electron neutrino}\label{theorys}
\end{figure}
\end{itemize}

\section{Problem in conventional SS-DQW and solution}

Our aim is to see whether this same oscillation profile as plotted in figs.~\ref{theoryl} and \ref{theorys} can be reproduced by the SS-DQW scheme which captures the properties of DCA. 
In the previous chapter it is shown that SS-DQW serves as a simulation scheme of the free Dirac Hamiltonian.
In (1+1) dimension SS-DQW carries only two spin degrees of freedom. But here we have three flavors and each flavor state is orthogonal to the other flavor state, 
and hence flavor states irrespective of the spin state can be described by three dimensional Hilbert space. 
So, if each flavor state carries two spin degrees of freedom, spin of one flavor state is independent of the spin of the other one.
In our SS-DQW formalism we can write $\ket{\nu_{m_j}} \in \mathcal{H}_c \otimes \mathcal{H}_x$ and $\braket{\nu_{m_l}| \nu_{m_j}} = \delta_{jl}$ for all $j$, $l$. 
Thus two internal degrees of the quantum walker are not enough, we need at least $2 \times 3$ = $6$ internal degrees of freedom in order to describe it in (1+1) dimension.
One possible way to incorporate this thing is to increase the coin space dimension from two to six.
Instead of the coin Hilbert space = $\text{span}\{(1~~0)^T, (0~~1)^T\}$ as used in the previous chapter, now the whole coin Hilbert space has to be defined as   
 \begin{align} 
\mathcal{H}_{c6} = \text{span}\big\{ \ket{\zeta_r} : r \in \{1, 2, 3, 4, 5, 6\} \big\}.
\end{align} The basis coin vectors are defined as:
\begin{align}
 \ket{\zeta_1} = \big[ \ket{1,\uparrow} + 0  \ket{1,\downarrow} \big] \oplus 
 \big[ 0  \ket{2,\uparrow} + 0  \ket{2,\downarrow} \big] \oplus
 \big[ 0  \ket{3,\uparrow} + 0  \ket{3,\downarrow} \big] = (1~0~0~0~0~0)^T, \nonumber\\
 \ket{\zeta_2} = \big[ 0 \ket{1,\uparrow} + \ket{1,\downarrow} \big] \oplus 
 \big[ 0  \ket{2,\uparrow} + 0  \ket{2,\downarrow} \big] \oplus
 \big[ 0  \ket{3,\uparrow} + 0  \ket{3,\downarrow} \big] = (0~1~0~0~0~0)^T, \nonumber\\
 \ket{\zeta_3} = \big[ 0\ket{1,\uparrow} + 0 \ket{1,\downarrow} \big] \oplus 
 \big[ \ket{2,\uparrow} + 0  \ket{2,\downarrow} \big] \oplus
 \big[ 0  \ket{3,\uparrow} + 0  \ket{3,\downarrow} \big] = (0~0~1~0~0~0)^T, \nonumber\\
 \ket{\zeta_4} = \big[0 \ket{1,\uparrow} + 0  \ket{1,\downarrow} \big] \oplus 
 \big[ 0  \ket{2,\uparrow} +  \ket{2,\downarrow} \big] \oplus
 \big[ 0  \ket{3,\uparrow} + 0  \ket{3,\downarrow} \big] = (0~0~0~1~0~0)^T, \nonumber\\
 \ket{\zeta_5} = \big[ 0\ket{1,\uparrow} + 0  \ket{1,\downarrow} \big] \oplus 
 \big[ 0  \ket{2,\uparrow} + 0  \ket{2,\downarrow} \big] \oplus
 \big[ \ket{3,\uparrow} + 0  \ket{3,\downarrow} \big] = (0~0~0~0~1~0)^T, \nonumber\\
 \ket{\zeta_6} = \big[ 0\ket{1,\uparrow} + 0 \ket{1,\downarrow} \big] \oplus 
 \big[ 0  \ket{2,\uparrow} + 0  \ket{2,\downarrow} \big] \oplus
 \big[ 0  \ket{3,\uparrow} + \ket{3,\downarrow} \big] = (0~0~0~0~0~1)^T, \nonumber\\
\text{where we have used the vector representation equivalence},~~~~~~~~~~~~~~~~ \nonumber\\
\ket{j,\uparrow} = (1 ~0)^T ~~;~~  \ket{j, \downarrow} = (0 ~1)^T~~ \text{for all}~ j = 1, 2, 3~.~~~~~~~~~~~~~~~~~~
\end{align}
The mass eigenstates are expressed as 
\begin{align}
\ket{\nu_{m_1}} = f(k, \theta^1_2(m_1)) \ket{\zeta_1} \otimes \ket{k} + g(k, \theta_2^1(m_1)) \ket{\zeta_2}  \otimes \ket{k}, ~\nonumber\\
\ket{\nu_{m_2}} = f(k, \theta^1_2(m_2)) \ket{\zeta_3} \otimes \ket{k} + g(k, \theta_2^1(m_2)) \ket{\zeta_4} \otimes \ket{k}, ~ \nonumber\\
\ket{\nu_{m_3}} = f(k, \theta^1_2(m_3)) \ket{\zeta_5} \otimes \ket{k} + g(k, \theta_2^1(m_3)) \ket{\zeta_6} \otimes \ket{k},~
\end{align}
where the coefficients are taken from the eigenvectors defined in eq.~(\ref{eigenval})
by 
\begin{align}\ket{\phi^+(k)} = f(k, \theta^1_2) \ket{\uparrow} + g(k, \theta_2^1) \ket{\downarrow}.\end{align}
As the mass is controlled by the parameter $\theta_2^1$, the three different masses will be 
considered in our simulation scheme by the proper choice of $\theta^1_2(m_j)$ for all $j$. 
The whole evolution operator $U_\nu$ should be defined in such a way that it must satisfy
the condition: 
\begin{align} U^n_\nu \ket{\nu_{m_j}} = e^{- \frac{i E_j(k) t}{\hbar}} \ket{\nu_{m_j}}.\end{align} 
So we define it as 
\begin{align}
 U_{\nu} = U_{SQW}(\theta^1_2(m_1)) \oplus U_{SQW}(\theta^1_2(m_2)) \oplus  U_{SQW}(\theta^1_2(m_3)),
\end{align}
where the form of $U_{SQW}$ is given in eq.~(\ref{ssqwope}). $U_{SQW}(\theta^1_2(m_j))$ acts only on the $j$-th sector. 
The `$\oplus$' is taken over the coin Hilbert spaces only, not in the position space, so position Hilbert space remains the same for all the sectors. 
The whole coin operators are $C_j = \tilde{C}_j \otimes \sum_x \ket{x}\bra{x}$ for $j \in \{1, 2\}$ where 
\begin{align}
 \tilde{C}_1 = \Big( \sigma_0 \oplus \sigma_0 \oplus \sigma_0 \Big)   =  \left( \begin{array}{cccccc}
           1 & 0 & 0 & 0 & 0 & 0\\
           0 & 1 & 0 & 0 & 0 & 0\\
           0 & 0 & 1 & 0 & 0 & 0\\
           0 & 0 & 0 & 1 & 0 & 0 \\
           0 & 0 & 0 & 0 & 1 & 0 \\
           0 & 0 & 0 & 0 & 0 & 1 \\
          \end{array}\right) 
          = \sum_{j=1}^6 \ket{\zeta_j}\bra{\zeta_j} \end{align} and, 
          \begin{align}
 \tilde{C}_2 = \bigg( e^{- i \theta^1_2(m_1) \sigma_1} \oplus e^{- i \theta^1_2(m_2) \sigma_1} \oplus e^{- i \theta^1_2(m_3) \sigma_1} \bigg) 
  = \hspace{4cm} \nonumber\\
    \left( \begin{array}{cccccc}
           \cos \theta^1_2(m_1) & - i \sin \theta^1_2(m_1) & 0 & 0 & 0 & 0\\
           - i \sin \theta^1_2(m_1) & \cos \theta^1_2(m_1) & 0 & 0 & 0 & 0\\
           0 & 0 & \cos \theta^1_2(m_2) & - i \sin \theta^1_2(m_2) & 0 & 0\\
           0 & 0 & - i \sin \theta^1_2(m_2) & \cos \theta^1_2(m_2) & 0 & 0 \\
           0 & 0 & 0 & 0 & \cos \theta^1_2(m_3) & - i \sin \theta^1_2(m_3) \\
           0 & 0 & 0 & 0 & - i \sin \theta^1_2(m_3) & \cos \theta^1_2(m_3) \\
          \end{array}\right) \nonumber\\ 
    = \sum_{j = 1}^3 \cos \theta^1_2(m_j) \Big( \ket{\zeta_{2j-1}}\bra{\zeta_{2j-1}}
    +  \ket{\zeta_{2j}}\bra{\zeta_{2j}}\Big)
    - i  \sin \theta^1_2(m_j) \Big( \ket{\zeta_{2j-1}}\bra{\zeta_{2j}} 
    +  \ket{\zeta_{2j}}\bra{\zeta_{2j-1}}\Big)~.
\end{align}

The shift operators are now
\begin{align}
 S_+ = \sum_x\left( \begin{array}{cccccc}
           \ket{x+a}\bra{x} & 0 & 0 & 0 & 0 & 0\\
           0 & \ket{x}\bra{x} & 0 & 0 & 0 & 0\\
           0 & 0 & \ket{x+a}\bra{x} & 0 & 0 & 0\\
           0 & 0 & 0 & \ket{x}\bra{x} & 0 & 0 \\
           0 & 0 & 0 & 0 & \ket{x+a}\bra{x} & 0 \\
           0 & 0 & 0 & 0 & 0 & \ket{x}\bra{x} \\
          \end{array}\right) \nonumber\\
          =  \sum_{j=1}^3 \ket{\zeta_{2j-1}}\bra{\zeta_{2j-1}} \otimes \sum_x \ket{x+a}\bra{x}
          + \sum_{j=1}^3  \ket{\zeta_{2j}}\bra{\zeta_{2j}} \otimes \sum_x \ket{x}\bra{x}
 \end{align}
and 
\begin{align}
 S_- = \sum_x\left( \begin{array}{cccccc}
           \ket{x}\bra{x} & 0 & 0 & 0 & 0 & 0\\
           0 & \ket{x-a}\bra{x} & 0 & 0 & 0 & 0\\
           0 & 0 & \ket{x}\bra{x} & 0 & 0 & 0\\
           0 & 0 & 0 & \ket{x-a}\bra{x} & 0 & 0 \\
           0 & 0 & 0 & 0 & \ket{x}\bra{x} & 0 \\
           0 & 0 & 0 & 0 & 0 & \ket{x-a}\bra{x} \\
          \end{array}\right) \nonumber\\
          =  \sum_{j=1}^3 \ket{\zeta_{2j-1}}\bra{\zeta_{2j-1}} \otimes \sum_x \ket{x}\bra{x}
          + \sum_{j=1}^3  \ket{\zeta_{2j}}\bra{\zeta_{2j}} \otimes \sum_x \ket{x-a}\bra{x}~.
\end{align}

\section{Implementation by lower dimensional system}

In our SS-DQW scheme a six-dimensional quantum particle can fully simulate the neutrino oscillation profile as in figs.~\ref{theoryl} and \ref{theorys}.
 But experimentally it is difficult to find and control a single six-dimensional system. So here we show some alternative ways of implementing  
 the same scheme by lower dimensional quantum systems which are easily available and well controllable.
 We use the simple rules: $6 < 8$ = $2 \times 2 \times 2$, $6 = 2 \times 3$; that is to say,
 we can use three-qubit system while neglecting its two extra degrees of freedom, or a qubit-qutrit system.  
 Below we describe them as follows. As, $C_1$ is nothing but the identity operator, its implementation is not required. So, we will not discuss this here.
 
\subsection{Three-qubit system}
 Qubit has two degrees of freedom denoted by,  $\ket{0} = (1~~0)^T$, $\ket{1} = (0~~1)^T$.   
A three-qubit system formed by tensor product of three vector spaces associated with each qubit. This is equivalent eight dimensional Hilbert space $\equiv \mathbb{C}^8$.
But for simulating three flavor neutrino oscillation we need six dimensions. So we will confine ourselves only on the vector space described by
 \vspace{-1.5cm}

\begin{align}
 \text{span} \Big\{ \ket{000} \equiv \ket{\zeta_1}, \ket{001} \equiv \ket{\zeta_2}, \ket{010} \equiv \ket{\zeta_3}, \ket{011} \equiv \ket{\zeta_4},
 \ket{100} \equiv \ket{\zeta_5}, \ket{101} \equiv \ket{\zeta_6}  \Big\}.\end{align}
  \vspace{-1.5cm}
 
In this case, $C_2 = \tilde{C}_2 \otimes \sum_x \ket{x}\bra{x}$ where 
 \vspace{-1.5cm}

\begin{align}
  \tilde{C}_2 = \cos \theta^1_2(m_1) \ket{000}\bra{000} - i \sin \theta^1_2(m_1) \ket{000}\bra{001} - i\sin \theta^1_2(m_1) \ket{001}\bra{000}\nonumber\\
  + \cos \theta^1_2(m_1) \ket{001}\bra{001} + \cos \theta^1_2(m_2) \ket{010}\bra{010} - i \sin \theta^1_2(m_2) \ket{010}\bra{011} \nonumber\\
  - \sin \theta^1_2(m_2) \ket{011}\bra{010} + \cos \theta^1_2(m_2) \ket{011}\bra{011}  + \cos \theta^1_2(m_3) \ket{100}\bra{100} \nonumber\\
  - i \sin \theta^1_2(m_3) \ket{100}\bra{101} - i \sin \theta^1_2(m_3) \ket{101}\bra{100} + \cos \theta^1_2(m_3) \ket{101}\bra{101}.\end{align}
 The shift operators are 
\begin{align}
 S_+ = \Big(\ket{000} \bra{000} +  \ket{010} \bra{010} +   \ket{100} \bra{100}\Big) \otimes \sum_x \ket{x+a}\bra{x} \nonumber\\
 +  \Big(\ket{001} \bra{001} +  \ket{011} \bra{011} +   \ket{101} \bra{101}\Big) \otimes \sum_x \ket{x}\bra{x}, 
\end{align}
\begin{align}
 S_- = \Big(\ket{000} \bra{000} +  \ket{010} \bra{010} +   \ket{100} \bra{100}\Big) \otimes \sum_x \ket{x}\bra{x} \nonumber\\
 +  \Big(\ket{001} \bra{001} +  \ket{011} \bra{011} +   \ket{101} \bra{101}\Big) \otimes \sum_x \ket{x-a}\bra{x}, 
\end{align}
Here the coin operation $C_2$ and shift operators $S_+$, $S_-$ 
that act on the vector-space described by $\text{span}\{ \ket{110}, \ket{111}\}$ are set to be zero operators. 
Thus from the complete $8$ dimensional coin-space we will be using only six dimensions. 

Therefore, the states that are equivalent to mass eigenstates of neutrino, can be written as, 
 \begin{align}
               \ket{\nu_1} =& \big( f(k, \theta_2^1(m_1)) \ket{000} + g(k, \theta_2^1(m_1))  \ket{001} \big) \otimes \ket{k}, \nonumber\\
               \ket{\nu_2} =  &\big( f(k, \theta_2^1(m_2)) \ket{010} + g(k, \theta_2^1(m_2))  \ket{011} \big) \otimes \ket{k}, \nonumber\\
                \ket{\nu_3} =  &\big( f(k, \theta_2^1(m_3)) \ket{100} + g(k, \theta_2^1(m_3)) \ket{101} \big) \otimes \ket{k}.     \end{align} 
             
 \subsection{Qubit-qutrit system}

    Similarly, we can simulate the same dynamics by a qubit-qutrit system. The coin space is the tensor product of vector spaces associated with qubit and qutrit.   
   Qubit has two degrees of freedom, $\ket{0} = (1~~0)^T$, $\ket{1} = (0~~1)^T$, and 
   qutrit has three degrees of freedom,  $ \ket{0} = (1~~0~~0)^T$,  $\ket{1} = (0~~1~~0)^T$,
   $\ket{2} = (0~~0~~1)^T$ and together they form a six-dimensional Hilbert space $\equiv \mathbb{C}^6$.
  
  In this case, 
   \vspace{-1.5cm}
   
  \begin{align}\ket{00} \equiv \ket{\zeta_1}, \ket{01} \equiv \ket{\zeta_2}, \ket{02} \equiv \ket{\zeta_3},
 \ket{10} \equiv \ket{\zeta_4}, \ket{11} \equiv \ket{\zeta_5}, \ket{12} \equiv \ket{\zeta_6}. \end{align}
 The coin operator $C_2 = \tilde{C}_2 \otimes \sum_x \ket{x}\bra{x}$ where  
 \vspace{-1.5cm}
 
 \begin{align}
 \tilde{C}_2 = \cos \theta^1_2(m_1) \ket{00}\bra{00} - i \sin \theta^1_2(m_1) \ket{00}\bra{01} - i \sin \theta^1_2(m_1) \ket{01}\bra{00} \nonumber\\
  + \cos \theta^1_2(m_1) \ket{01}\bra{01}  + \cos \theta^1_2(m_2) \ket{02}\bra{02} - i \sin \theta^1_2(m_2) \ket{02}\bra{10} \nonumber\\
  - i \sin \theta^1_2(m_2) \ket{10}\bra{02} + \cos \theta^1_2(m_2) \ket{10}\bra{10} + \cos \theta^1_2(m_3) \ket{11}\bra{11} \nonumber\\
  + \sin \theta^1_2(m_3) \ket{11}\bra{12}  - i \sin \theta^1_2(m_3) \ket{12}\bra{11} + \cos \theta^1_2(m_3) \ket{12}\bra{12},
\end{align} The shift operators are
\begin{align}
 S_+ = \Big(  \ket{00} \bra{00} +  \ket{02} \bra{02} +   \ket{11} \bra{11}  \Big) \otimes \sum_x \ket{x+a}\bra{x} \nonumber\\
 + \Big(  \ket{01} \bra{01} +  \ket{10} \bra{10} +   \ket{12} \bra{12} \Big) \otimes  \sum_x \ket{x}\bra{x}, 
\end{align}
\begin{align}
 S_- = \Big(  \ket{00} \bra{00} +  \ket{02} \bra{02} +   \ket{11} \bra{11}  \Big) \otimes \sum_x \ket{x}\bra{x} \nonumber\\
 + \Big(  \ket{01} \bra{01} +  \ket{10} \bra{10} +   \ket{12} \bra{12} \Big) \otimes  \sum_x \ket{x-a}\bra{x}.
\end{align}
The mass eigenstates of neutrino can be written as follows, 
  \begin{align}
               \ket{\nu_1}& = \big( f(k, \theta_2^1(m_1)) \ket{00} + g(k, \theta_2^1(m_1))  \ket{01} \big) \otimes \ket{k}, \nonumber\\
               \ket{\nu_2} &=  \big( f(k, \theta_2^1(m_2)) \ket{02} + g(k, \theta_2^1(m_2))  \ket{10} \big) \otimes \ket{k}, \nonumber\\
                \ket{\nu_3} &=  \big( f(k, \theta_2^1(m_3)) \ket{11} + g(k, \theta_2^1(m_3)) \ket{12} \big) \otimes \ket{k}.
                     \end{align} 
                   
 For physical implementation one can only work with the coin operations, if the neutrino is in a particular momentum eigenstate. 
 As the shift operators are diagonalizable in momenta eigenbasis, so can be absorbed into the coin operation.   
                   
 \section{Numerical simulation}\label{numeric}
 
Two following main points have to be taken care when we map from the actual experimental parameters described in eq.~(\ref{actualpara}) to our SS-DQW parameters.

 (1) The SS-DQW produces free Dirac Hamiltonian at the small time-step size $\delta t$,  if $\frac{ka}{\hbar}$ and $\theta_2^1$ also take small values. 

 (2) As neutrino is ultra-relativistic particle, momentum $k >> m_j c$ must be obeyed, for all $j \in \{1,2,3\}$.

 For simulation purpose we will use the relation \begin{align}
                                                m_j c^2 = \frac{\hbar \theta^1_2(m_j)}{\delta t}
\Rightarrow m_j c = \frac{\hbar \theta^1_2(m_j)}{a}.  
                                                 \end{align} Thus the condition (2) implies 
$\frac{k a}{\hbar} >> \theta^1_2(m_j)$ for all $j$. 
Using this conditions for SS-DQW case we get: \begin{align}
                \frac{\delta t}{\hbar}(E_j - E_l) = \cos^{-1}\bigg[\cos \frac{k a}{\hbar} \cos \theta^1_2(m_j) \bigg]
                  - \cos^{-1}\bigg[\cos \frac{k a}{\hbar} \cos \theta^1_2(m_l) \bigg] \nonumber\\
                  \approx \sqrt{\bigg(\frac{k a}{\hbar}\bigg)^2 + \Big( \theta^1_2(m_j) \Big)^2}
                  - \sqrt{\bigg(\frac{k a}{\hbar}\bigg)^2 + \Big( \theta^1_2(m_l) \Big)^2}
                  = \frac{\hbar}{2ka}\Big[\Big(\theta^1_2(m_j) \Big)^2 - \Big( \theta^1_2(m_l) \Big)^2\Big] \nonumber\\
                 \Rightarrow \frac{t}{\hbar}(E_j - E_l)  = \frac{\hbar}{2ka}\Big[\Big(\theta^1_2(m_j) \Big)^2 - \Big( \theta^1_2(m_l) \Big)^2\Big] \frac{t}{\delta t}~.
                 \end{align}

\begin{itemize}
 \item {\bf Problem in the actual formula:} For the case of neutrino energy 1 GeV, $k c = \mathcal{O}( 10^9 $eV) 
 $\Rightarrow  \frac{k a}{\hbar} = \mathcal{O}(10^{24} s^{-1}) \times \delta t$.
Then, to have small $\frac{k a}{\hbar}$, $\delta t$ should be utmost = $\mathcal{O}(10^{-26} s).$
Hence, to reproduce the same oscillation profile, we have to consider
$[\theta_2^1(m_3)]^2 - [\theta_2^1(m_2)]^2  \approx \Big( \frac{ \delta t}{\hbar} \Big)^2 \approx \mathcal{O} (10^{-25}) $,
$[\theta_2^1(m_2)]^2 - [\theta_2^1(m_1)]^2 \approx \Big( \frac{ \delta t}{\hbar} \Big)^2 \approx \mathcal{O} (10^{-27}) $.
Hence required number of walk steps to produce short range and long range oscillation are $ \mathcal{O}( 10^{25} )$, $ \mathcal{O}(10^{26})$ respectively. 
These kinds of order of $\delta t$, $[\theta_2^1(m_j)]^2 - [\theta_2^1(m_l)]^2 $, number of walk steps are very difficult to achieve in table-top lattice experiments up to date.

\item{\bf At the Planck scale limit:} If we consider, walk time-step size $\delta t = \mathcal{O}( t_p),$ lattice space step-size $ a = \mathcal{O}( l_p),$
where, Planck time = $ t_p = 5.3912 \times 10^{-44} $ s, Planck length = $l_p = 1.6162 \times 10^{-35}$ m,
then $\frac{ka}{\hbar} = \mathcal{O}(10^{-19})$, $[\theta_2^1(m_3)]^2 - [\theta_2^1(m_2)]^2 = \mathcal{O}(10^{-59})$,
$[\theta_2^1(m_2)]^2 - [\theta_2^1(m_1)]^2=\mathcal{O}(10^{-61})$ and required number of walk steps for
short and long range oscillation are  $ \mathcal{O}( 10^{42} )$ and $\mathcal{O}(10^{43})$, respectively. So in principle it is possible to satisfy the two 
conditions (1) and (2) mentioned at the beginning of this section \ref{numeric} and simulate neutrino oscillation exactly by SS-DQW, 
but it is hard to realize in table-top lattice experiments.

\item{\bf Zooming in the frequency:} The oscillation profile is determined by the quantity = frequency of the oscillation $\times$ time of evolution = $f_\nu t$
where $f_\nu(j,l) = \frac{E_{j}-E_{l}}{\hbar}$. The only condition to simulate neutrino oscillation
is that $f_\nu  t$ will be the same in simulation system to the value in the real experiments, for the given PMNS matrix.
It implies that if we increase the frequency $f_\nu$, then we can decrease the number of walk steps which can be realizable in a table-top set-up.
Thus in order to successfully simulate, we have to increase the value of the quantity = 
$\bigg[ \sqrt{\tilde{k}^2 + [\theta^1_2(m_j)]^2 } - \sqrt{\tilde{k}^2 + [\theta^1_2(m_l)]^2 } \bigg]$ such that 
the same oscillation profile can be obtained with lesser no of walk steps $\frac{t}{\delta t}$.
That is to say, we are zooming in the frequency and zooming out the number of SS-DQW evolution steps. 
The Dirac dynamics is only produced at the continuum limit in SS-DQW evolution, when $\theta_2^1(m_j)$ and $ \frac{ka}{\hbar}$ both are small. 
 Respecting this condition, the number of walk-steps we have chosen $450$ and $4500$ for short and long range oscillation profiles, respectively.
We have chosen the parameter values: $\tilde{k} = 0.01$ rad, $\theta_1 = 0.001$ rad, $\theta_2 = 0.00615654$ rad, $\theta_3 = 0.0664688$ rad. 

\end{itemize}

\subsection{Simulated neutrino oscillation profiles}

\begin{figure}[htbp]\centering
 \includegraphics[width =0.7\textwidth]{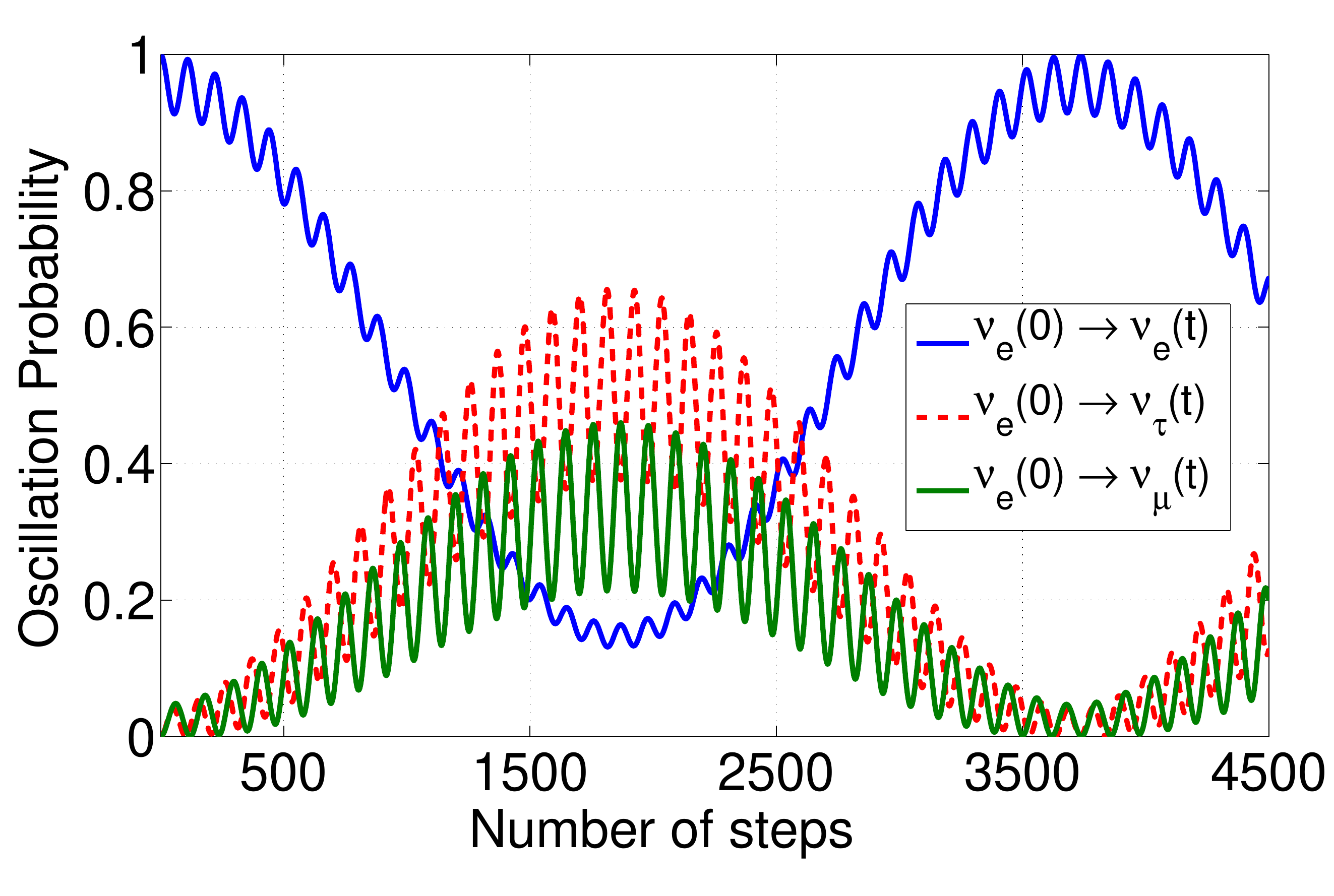}
 \caption[Neutrino long range oscillation probabilities for an initial electron neutrino, using our SS-DQW scheme]
 {Long range neutrino oscillation obtained for 4500 time-steps of SS-DQW.}\label{simul}
 \end{figure}

\begin{figure}[htbp]\centering
 \includegraphics[width =0.7\textwidth]{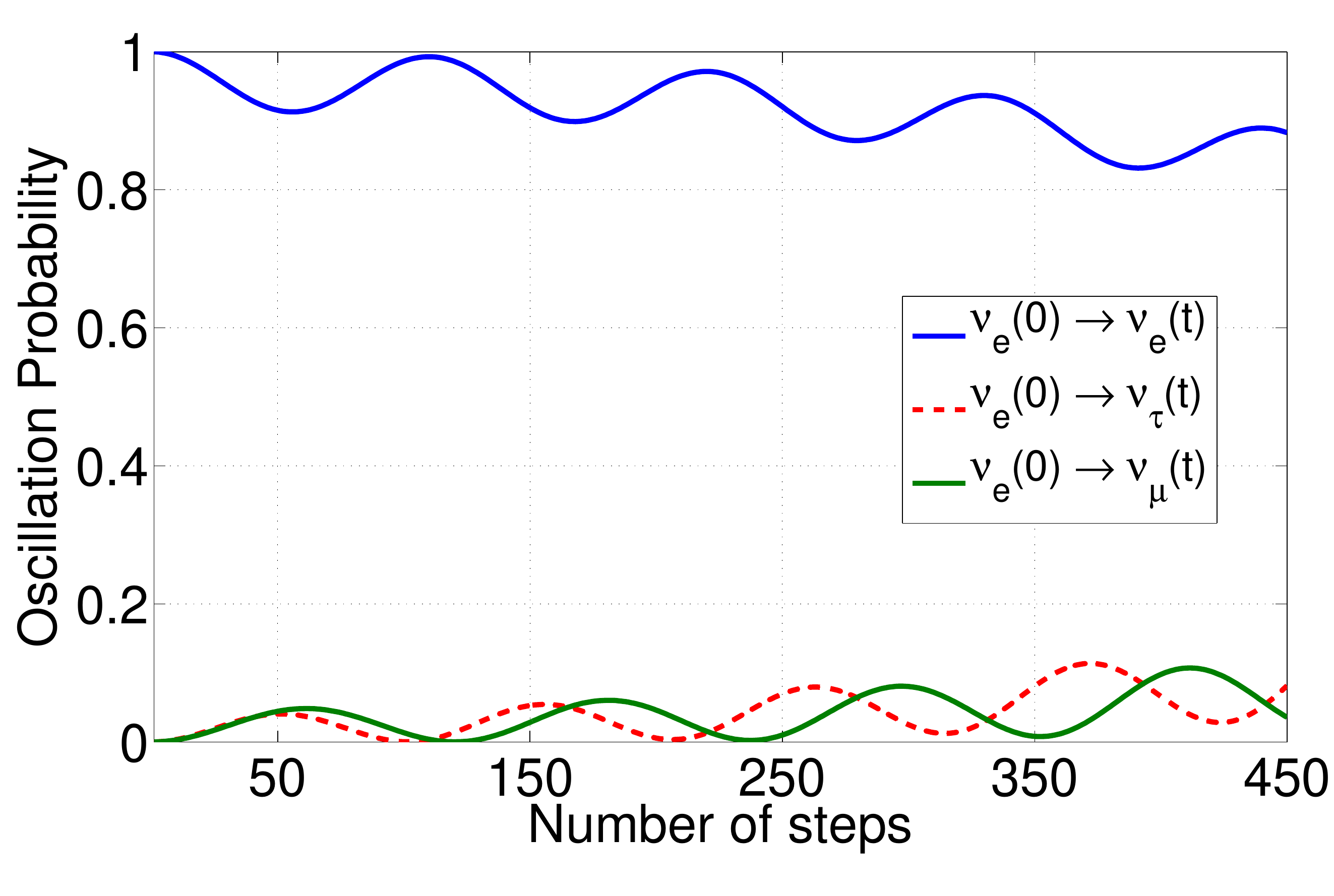}
 \caption[Neutrino short range oscillation probabilities for an initial electron neutrino, using our SS-DQW scheme]
 {Short range neutrino oscillation obtained for 450 time-steps of SS-DQW.}\label{simus}
 \end{figure}

In figs.~\ref{simul} and \ref{simus} we have shown the oscillation probabilities obtained by numerical simulation of
SS-DQW for an initial state that mimics electron neutrino. Our choices for coin parameters to reproduce the oscillations in 
figs.~\ref{theoryl} and \ref{theorys} are $ \theta^1_2(m_1) = 0.001 $ rad, $\theta^1_2(m_2) = 0.00615654 $ rad, $ \theta^1_2(m_3) = 0.0664688 $ rad, 
and $\frac{k a}{\hbar} = 0.01$ rad.
Here, we show oscillation probability of $\nu_e(0) \to \nu_e(t)$ (blue),
$\nu_e (0) \to \nu_\mu(t)$ (green), $\nu_e(0) \to \nu_\tau(t)$(red).

 Both, the long range and short range neutrino flavor oscillations shown in 
 figs.~\ref{theoryl} and \ref{theorys} obtained from the real neutrino experiment and those from our SS-DQW simulation in 
 figs.~\ref{simul} and \ref{simus} respectively, are matching perfectly.

\section{Position-coin entanglement during the oscillation}

Now we will discuss the coin-position entanglement as a function of time-steps during the neutrino oscillation.  
If the neutrino is in a particular momentum eigenstate as usually considered in neutrino oscillation theory, the entanglement
between coin and position is always zero. As in this case the position space and coin space never mix. 
But in general neutrino can be in a superposition of momenta eigenstates while be in a particular flavor state. 
In this case we can take the $\alpha$-flavor state as 
\begin{align}\label{inidis}
 \ket{\nu_\alpha} = \sum_{k} p(\alpha, k) \ket{\nu_\alpha(k)} \otimes \ket{k} = \sum_{k,j} p(\alpha, k) U^*_{\alpha j} \ket{\nu_j(k)} \otimes \ket{k}
\end{align}where the sum is over all possible momenta and $p(\alpha, k)$ is the probability amplitude 
to be in a momentum eigenstate, as well as in the $\alpha$-flavor state. For our analysis we consider a Gaussian like
(because of the discreteness, this is not exactly Gaussian) distribution in the momentum space:
\begin{align}
p(\alpha, k) = \frac{e^{-\frac{\Delta}{2}(k - k_0)^2} }{\sqrt{\sum_k  e^{-\Delta (k - k_0)^2} }}~~\text{same for all}~\alpha \in \{e, \mu, \tau\}.
\end{align} The momenta range is determined by the parameter $\epsilon$ such that 
$k \in \Big[ k_0 - \frac{\hbar}{a} \epsilon,  k_0 + \frac{\hbar}{a} \epsilon\Big]$. $k_0$ is the central momentum and $\Delta$ controls 
the width of the Gaussian like distribution. Starting from the state in eq.~(\ref{inidis}), after time $t$ the system state will be
\begin{align} \label{instflavor}
 \ket{\psi_\nu(t)} = \sum_{k,j} p(\alpha, k) U^*_{\alpha j} e^{-\frac{i E_j(k) t}{\hbar}} \ket{\nu_{m_j}(k)} \otimes \ket{k} 
 \Rightarrow \rho(t) = \ket{\psi_\nu(t)} \bra{\psi_\nu(t)}~ \nonumber\\
 = \sum_{k,j, k',l} p(\alpha, k) p^*(\alpha, k') U^*_{\alpha j} U_{\alpha l} e^{-\frac{i [E_j(k) - E_l(k')] t}{\hbar}}
 \ket{\nu_{m_j}(k)}\bra{\nu_{m_l}(k')} \otimes \ket{k}\bra{k'}~.
\end{align}
The partial traced state (traced out the position-space or momentum space) =
\begin{align}
 \rho_c(t) = \text{Tr}_k [\rho(t)] 
 = \sum_{k, j, l} |p(\alpha, k)|^2 U^*_{\alpha j} U_{\alpha l} e^{-\frac{i [ E_j(k) - E_l(k)]t}{\hbar}}\
 \ket{\nu_{m_j}(k)}\bra{\nu_{m_l}(k)} 
\end{align}
Thus the entanglement is the entropy of $\rho_c(t) = - \text{Tr}_c[ \rho_c(t) \ln \rho_c(t) ]$, as the whole system state always remains pure.
But as $\rho_c(t)$ is defined on a six-dimensional coin space, and hence, it is in general a $6 \times 6$ matrix.
So, if we use $\ln() = \log_e()$ in the entanglement entropy formula, the maximum value of the entropy may exceed $1$.  
\vspace{1cm}

\begin{figure}[h]\centering
\includegraphics[width = 0.7\textwidth]{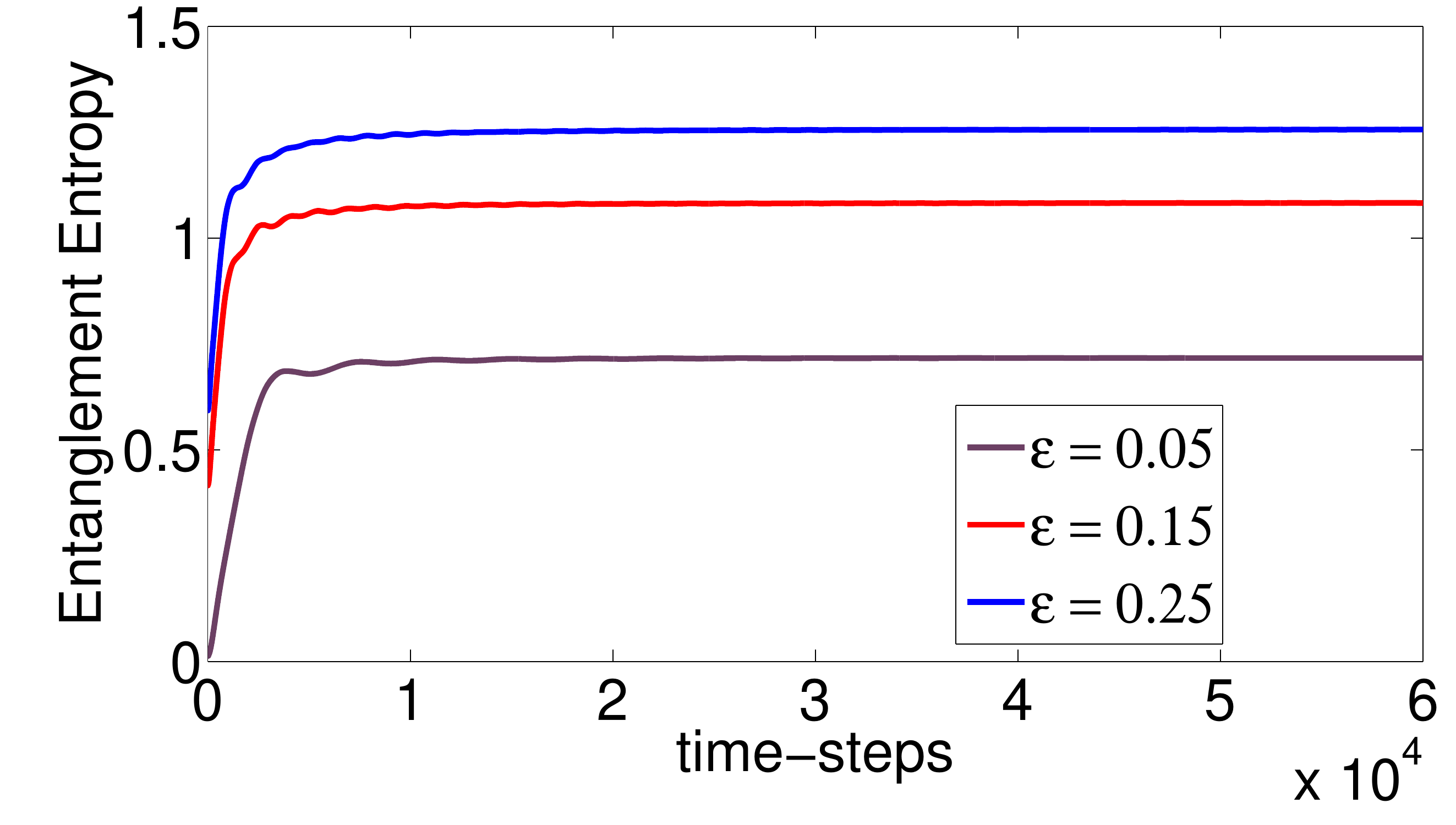}
\caption[Entanglement between spin and position degrees of freedom, during neutrino oscillation]
{Entropy as a measure of entanglement between spin and position degrees of freedom, during neutrino oscillation,
 simulated as a function of number of SS-DQW steps. With increase in number of steps, the entanglement entropy values reach saturation levels.
 As the spread in momentum space $\propto \epsilon$ increases the entanglement also increases.} \label{neuentangle}
\end{figure}

 In fig.~\ref{neuentangle} we have plotted the entanglement entropy as a function of SS-DQW steps and the parameter $\epsilon \in \{0.05, 0.15, 0.25\}$ unit. 
 The $\frac{k_0 a}{\hbar} = 0.01$ rad, minimum value between two $\frac{k a}{\hbar} = 0.001$ rad, $\Delta = 100$ unit. It is observed that with the 
 increase of the value of $\epsilon$ which is the spread or delocalization in momentum space, the entanglement entropy increases.

Our work is mainly for the neutrinos propagating through vacuum, and actually done in DQW framework. But for consistency with the other questions that we have addressed in this thesis, we have discussed this here
in the SS-DQW framework, without changing the primary concept of using higher dimensional coin Hilbert space.    
In a similar way, but in the framework of DQW, simulation of neutrino oscillation through vacuum and matters, has been investigated in ref.~\cite{molfetta5}.  

In some recent literatures like \cite{Banerjee1, Banerjee2, Banerjee3, Banerjee4, Banerjee5} very interesting quantum information perspectives of neutrino and its dynamics have been studied. 
Here our motive is to provide a simulation scheme which can be implemented easily in table-top set-ups and following these literatures we can study their theoretical predictions in our simulators.      

Our simulation scheme is applicable in other cases where the particle contains more than one degree of freedom. 
In these cases, the eigenstates of the evolution operator correspond to one degree of freedom, and the projective measurement operators formed by the 
eigenstates of the other degrees of freedom.


\chapter{Simulating Dirac particle dynamics in presence of general external potential and curved space-time}\label{3}

Dirac particle dynamics under general external potential and curved space-time,
is very important from the perspective of unification of fundamental theories. 
In the first chapter we have shown that the SS-DQW which matches with the DCA, 
is able to reproduce the free massive Dirac Hamiltonian in flat (1+1) space-time. 
Here our motivation is to see, how we can modify the conventional form of the 
SS-DQW operator such that it can capture all the potential effects on a massive Dirac particle dynamics. 
As the gauge potentials, curved space-time effects come through the space-time dependency, 
we can reproduce these effect using SS-DQW evolution by making the coin parameter space-time 
dependent and retaining the shift operators in the same form as used in previous chapters. 
Simulation of Dirac particle dynamics in the presence of the external 
abelian and nonabelian gauge field by DQW has been recently reported \cite{arnault,arnault2}. 
Some recent refs.~\cite{molfettamain, molfettacurve} in DQW framework, have shown that,
proper functional forms of coin parameters that depend on space-time coordinates can capture these external potential and curvature effects.
Two-step stroboscopic DQW with space-time dependent $U(2)$ coin operator was used to produce gravitational 
and gauge potential effects in single Dirac fermion \cite{molfettamain}, but their approach was unable to capture mass,
gravity and gauge potential in a single Hamiltonian.
A generalized single particle Dirac equation in curved space-time was later derived from a special DQW---grouped quantum
walk (GQW)---which needs prior unitary encoding and decoding at last\,\cite{qwcur, qwcur3, qwicur}.
DQW with $SU(2)$ coin parameters which are spatially independent but depend randomly on time-steps, 
has also been studied in the context of random artificial gauge fields~\cite{randomqw}. 
The randomized coin parameters which mimic random gravitational and gauge 
fields act as transition knobs from non-classical probability distributions to classical
probability distributions. A DQW with a single evolution step which contains four spatial shift
operations---mimics the Dirac evolution under the influence of gravitational waves in $(2+1)$ 
dimension---was also recently reported in ref.~\cite{grawave}.
 
 But our approach shows that a slight modification of inhomogeneous single-step SS-DQW can capture all background gauge potentials, gravitational effects
 in a single massive Dirac Hamiltonian, and we do no need to do encoding or decoding like the grouped QW.

Dirac equation (\ref{curvedirham}) in curved (1+1) or (2+1) dimension in presence of general background $U(1)$ gauge potentials: $A_\mu(x, t)$ is 
\begin{align}\label{curgauge1}
\bigg(i \hbar \sum_{\mu, a} e^\mu_{(a)} \gamma^{(a)}~ [\partial_\mu + \Gamma_\mu  - i A_\mu] - m c^2 \bigg)~\psi = 0~. 
\end{align}
Here the wavevector $\ket{\psi} \in \mathbb{C}^2 \otimes \mathcal{H}_x$, but in order to include a general nonabelian $U(N)$ background potential 
the wavevector should be defined on the Hilbert space: $\mathbb{C}^2 \otimes \mathbb{C}^N \otimes \mathcal{H}_x$. In the nonabelian case the 
Dirac equation in curved space-time takes the form:
\begin{align}\label{curgauge}
\Bigg(i \hbar \sum_{\mu, a} e^\mu_{(a)} \gamma^{(a)} \otimes \Lambda_0 \cdot [\partial_\mu + \Gamma_\mu\otimes \Lambda_0]  
+ \hbar \sum_{\mu, a, q} e^\mu_{(a)} \gamma^{(a)} \otimes  \Lambda_q ~ A^q_\mu - m c^2 \Bigg)~\psi = 0,  
\end{align} where $\Lambda_0 = \mathds{1}_N$. In practice we omit the summation signs and use Einstein's sum-convention.
One important point is that, the dynamical evolution of these gauge potentials are not discussed here, they simply act as background potentials. 
The functions $A^0_\mu(x,t)$ for $\mu = 1, 2$ correspond to the abelian gauge potentials and one common example of which is the electromagnetic potential. 
The functions $A^{q \neq 0}_\mu(x,t)$ for $\mu = 1, 2$ correspond to the non-abelian gauge potentials and common examples of this include potential effects due to 
weak and strong forces. In this eq.~(\ref{curgauge}) the Dirac particle charges are included in the potential functions. These potentials appear in eq.~(\ref{curgauge})
to preserve the invariance of this equation under the group transformation: 
\begin{align}\psi(x,t) \to \exp\Big(-i \alpha^q(x,t)~\sigma_0 \otimes \Lambda_q\Big)~\psi(x,t),\end{align} 
where the associated generators are $\{\Lambda_q\}_{q=0}^{N^2-1}$ and the group parameters are functions of the 
coordinates $(x, t)$.

The Schrodinger like equation: $H \ket{\psi} = i \hbar \frac{\partial}{\partial t} \ket{\psi}$ does not follow directly from the eq.~(\ref{curgauge}),
a particular transformation on the wavefunction can do this job. In ref.~\cite{oliveira}, under the assumption: $e^0_{(j)} = 0$ for all $j \in \{1, 2, 3\}$, 
if we use the replacement:
\begin{align}
\psi(x,t)~~\text{to}~~|\det(g)|^\frac{1}{4} \Big[e^0_{(0)}\Big]^\frac{1}{2} \psi(x,t)\end{align}
where $\det(g)$ is the determinant of the metric $g_{\mu \nu}$ of the background space-time,
the eq.~(\ref{curgauge}) takes the form:
\begin{align}\label{dircurgen}
 i \frac{\partial}{\partial t} \ket{\psi} = \Bigg( - \hbar \sigma_0 \otimes \Lambda_q \otimes A_0^q + c \gamma^{(0)} \gamma^{(a)} \otimes \Lambda_0 \otimes \frac{e^j_{(a)}}{e^0_{(0)}} p_j
 - \frac{i \hbar c}{2} \gamma^{(0)} \gamma^{(a)} \otimes \Lambda_0 \otimes \frac{\partial}{\partial x_j}\Bigg[\frac{e^j_{(a)}}{e^0_{(0)}}\Bigg] \nonumber\\
 - \hbar \gamma^{(0)} \gamma^{(a)} \otimes \Lambda_q \otimes \frac{e^j_{(a)}}{e^0_{(0)}} A_j^q + \gamma^{(0)} \otimes \Lambda_0 \otimes \frac{m c^2}{e^0_{(0)}}~\Bigg)\ket{\psi} = H \ket{\psi}~.~~~~~~~~~~~~~~~~~~~~
\end{align}Here we have considered $\Big[\gamma^{(0)}\Big]^2 = \sigma_0$. 
This is a Schrodinger equation and we will treat the generator of the time evolution: $H$ as our Dirac Hamiltonian. 
Throughout our analysis we will consider the sign-convention of the Minkowski metric: 
\begin{align}
 \eta^{(0)(0)} = 1, ~\eta^{(j)(j)} = -1~~\forall~~j \in \{1, 2\}.
\end{align}

\section{General SS-DQW and the problem in its continuum-limit}

Here we start from a general SS-DQW operator in $(1+1)$ dimensional space-time
where the coin operators are in general both position and time-step dependent, i.e., inhomogeneous in space-time. 
 The coin operators are now 
\begin{align}\label{coinssqw}
  C_j(t,\delta t) = \sum_{x} e^{-i \theta^0_j(x,t, \delta t)}\left(\begin{array}{ccc}
                                    F_j(x,t, \delta t) &  & G_j(x,t, \delta t) \\\\
                                   -G^*_j(x,t, \delta t) &   & F^*_j(x,t, \delta t) \\
                                  \end{array} \right) \otimes \ket{x}\bra{x}
 \end{align} for $j = 1, 2$ and subject to the condition $|F_j(x,t, \delta t)|^2 + |G_j(x,t, \delta t)|^2  = 1$, 
    $\theta^0_j(x,t, \delta t)$ are real for all $x, t, \delta t$. $F_j, G_j$ are similar explicit functions of $\theta^q_j(x, t, \delta t)$ for
    $q \in \{1, 2, 3\}$ as in th eq.~(\ref{dqwcoinp}). 
    The shift operators $S_+$, $S_-$ are the same as in the previous chapters.

Then the whole SS-DQW evolution operator at time-step $t$ takes the from 
 \begin{align}\label{unissqwform}
  U_{SQW}(t, \delta t) =  \ket{\uparrow}\bra{\uparrow}\otimes U_{00}(t, \delta t) 
 +  \ket{\uparrow}\bra{\downarrow} \otimes U_{01}(t, \delta t) \nonumber\\
 +  \ket{\downarrow}\bra{\uparrow} \otimes U_{10}(t, \delta t) 
 + \ket{\downarrow}\bra{\downarrow} \otimes U_{11}(t, \delta t),
 \end{align}
 where
 \begin{align}
 U_{00}(t, \delta t) = \sum_x e^{-i [\theta^0_1(x,t,\delta t) + \theta^0_2(x,t,\delta t)]}
 F_2(x,t,\delta t) F_1(x,t,\delta t) \ket{x+a}\bra{x} \nonumber\\
- e^{-i [\theta^0_1(x,t,\delta t) + \theta^0_2(x-a,t,\delta t)]} G_2(x-a,t,\delta t)G_1^*(x,t,\delta t) \ket{x}\bra{x},
\nonumber \end{align} \begin{align}
U_{01}(t, \delta t) =  \sum_x e^{-i [\theta^0_1(x,t,\delta t) + \theta^0_2(x,t,\delta t)]} F_2(x,t,\delta t)
G_1(x,t,\delta t) \ket{x+a}\bra{x} \nonumber\\
+ e^{-i [\theta^0_1(x,t,\delta t) + \theta^0_2(x-a,t,\delta t)]} G_2(x-a,t,\delta t)F_1^*(x,t,\delta t) \ket{x}\bra{x},
\nonumber\end{align} \begin{align}
U_{10}(t, \delta t) =  \sum_x - e^{-i [\theta^0_1(x,t,\delta t) + \theta^0_2(x,t,\delta t)]} G_2^*(x,t,\delta t) F_1(x,t,\delta t)
\ket{x}\bra{x}  \nonumber\\
- e^{-i [\theta^0_1(x,t,\delta t) + \theta^0_2(x-a,t,\delta t)]} F^*_2(x-a,t,\delta t) G_1^*(x,t,\delta t)\ket{x-a}\bra{x},
\nonumber\end{align} \begin{align}
U_{11}(t, \delta t) =  \sum_x - e^{-i [\theta^0_1(x,t,\delta t) + \theta^0_2(x,t,\delta t)]} G_2^*(x,t,\delta t)G_1(x,t,\delta t)
\ket{x}\bra{x} \nonumber\\
+ e^{-i [\theta^0_1(x,t,\delta t) + \theta^0_2(x-a,t,\delta t)]} F^*_2(x-a,t,\delta t)F^*_1(x,t,\delta t)\ket{x-a}\bra{x}.
\end{align}
At the continuum limit: $\delta t \to 0$, $a \to 0$,
the unitary operator given in eq.~(\ref{unissqwform}) should be equal to the identify operator in order to make the 
evolution of any system state consistent. In other words, the effective Hamiltonian $H$ defined by $U_{SQW} = e^{- \frac{i H \delta t}{\hbar}}$, 
will not be a bounded operator at the continuum unless $U_{SQW}$ is identity both in position 
and the internal degrees of freedom, at that limit. But at the limit $\delta t \to 0, a \to 0$ we get
\begin{align}\label{unissqwform2}
 U_{00}(t, 0) = \sum_x e^{-i [\theta^0_1(x,t,0) + \theta^0_2(x,t,0)]} \Big[
 F_2(x,t,0) F_1(x,t,0) - G_2(x,t,0)G_1^*(x,t,0) \Big] \ket{x}\bra{x},
\nonumber\\
U_{01}(t,0) =  \sum_x e^{-i [\theta^0_1(x,t,0) + \theta^0_2(x,t,0)]} \Big[ F_2(x,t,0)
G_1(x,t,0) +  G_2(x,t,0)F_1^*(x,t,0) \Big] \ket{x}\bra{x},
\nonumber\\
U_{10}(t, 0) =  \sum_x - e^{-i [\theta^0_1(x,t,0) + \theta^0_2(x,t,0)]} \Big[ G_2^*(x,t,0) F_1(x,t,0)
+ F^*_2(x,t,0) G_1^*(x,t,0) \Big]\ket{x}\bra{x},
\nonumber\\
U_{11}(t, 0) =  \sum_x - e^{-i [\theta^0_1(x,t,0) + \theta^0_2(x,t,0)]} \Big[ G_2^*(x,t,0)G_1(x,t,0)
- F^*_2(x,t,0)F^*_1(x,t,0) \Big]\ket{x}\bra{x},
\end{align} 
where we have assumed the limit exist.
The eq.~(\ref{unissqwform2}) implies $ U_{SQW}(t, 0)$ is not equal to identity unless we impose some extra condition on the functions 
$\theta^q_j(x,t,0)$ for all $q$, $j$, $x$, $t$. In refs.~\cite{molfettamain, molfettacurve}, some relations among the coin parameters $\theta^q_j(x,t,0)$ has been found which 
makes the whole evolution operator identify at the continuum limit, but this procedure reduces the total number of controllable parameters.
Their approach is based on DQW framework and all the abelian potential effects 
and curved space-time effect are not captured in a same massive Dirac Hamiltonian. In refs.~\cite{arnault} and \cite{arnault2} using DQW framework
abelian and nonabelian gauge potentials, respectively, has been included in massive Dirac Hamiltonian, but curved space-time effects has not been 
included there. In the next section we will choose a procedure to get rid of these problems. 

\section{Modified SS-DQW operator}\label{modisec}

The way we choose to get rid of this problem of mismatch between the SS-DQW evolution operator at the continuum limit and identify, 
is to modify the SS-DQW evolution operator in such a way that the modified version will automatically 
becomes identity operator at the continuum limit. We define our new (modified) evolution operator as 
\begin{align}\label{newmod}
 \mathscr{U}(t, \delta t) =  U^\dagger_{SQW}(t, 0) \cdot U_{SQW}(t, \delta t) \coloneqq \exp\bigg(- \frac{i \mathscr{H}(t) \delta t}{\hbar} \bigg)~.
\end{align}
Just like in previous chapters we consider $a = c \times \delta t$, so $\delta t \to 0$ $\Rightarrow$ $a \to 0$. $\mathscr{H}(t)$ is now the effective Hamiltonian at time-step $t$. 
It is evident that $U^\dagger_{SQW}(t, 0) = C^\dagger_1(t,0) \cdot C^\dagger_2(t, 0)$ is nothing but only a coin operation as the shift operators becomes identity 
= $\sigma_0 \otimes \sum_x \ket{x}\bra{x}$, in this case. So this modification does not change the form of the SS-DQW evolution operator in the homogeneous case, 
as for that case $U^\dagger_{SQW}(t, 0)$ in eq.~(\ref{ssqwope}) is an identity operator, where we considered $\lim\limits_{\delta t \to 0} \theta^1_2(\delta t) = 0$.
In inhomogeneous case it is difficult to diagonalize the evolution operator simply by going to the Fourier space and derive the effective Hamiltonian. Because here the
coin operators are inhomogeneous in position and time, and hence, nondiagonalizable in the same basis in which the shift operators are diagonalizable. 
So we derive the effective Hamiltonian by using Taylor expansion in $\delta t$ for every $x$, $t$ under the assumption that 
all coin parameters are smooth functions of $\delta t$ and also $x$, $t$.
One important point is that, we are working in a discrete space-time, so this smoothness of the functions actually 
mean that the envelop functions which approximate the functions, are smooth.  Using the Taylor expansion formula in eq.~(\ref{newmod}) we get
\begin{align}
 \mathscr{U}(t, \delta t) = \sigma_0 \otimes \sum_x \ket{x}\bra{x}  - \frac{i \mathscr{H}(t)}{\hbar} \delta t + \mathcal{O}(\delta t^2).
 \end{align}
Similarly, from the Taylor series expansion of the coin parameters we can write 
\begin{align}
 F_j(x,t, \delta t) =  F_j(x,t,0) + \delta t~f_j(x,t) + \mathcal{O}(\delta t^2),\nonumber\\
 G_j(x,t, \delta t) =   G_j(x,t,0) + \delta t~g_j(x,t) + \mathcal{O}(\delta t^2),\nonumber\\
 \theta^0_j(x,t,\delta t) =  \theta^0_j(x,t,0) + \delta t~\vartheta^0_j(x,t) + \mathcal{O}(\delta t^2)~.
 \end{align}
 Imposing the condition that $|F_j(x,t,\delta t)|^2 + |G_j(x,t, \delta t)|^2 = 1$ for all $x$, $t$, $\delta t$; 
 as the coefficient of $(\delta t)^n$ should be separately zero for each possible value of $n \in \mathbb{N}$, 
 we get \begin{align}\label{cond1}
 \Re\big[F_j(x,t,0)f^*_j(x,t) + G_j(x,t,0)g^*_j(x,t)\big] = 0.
 \end{align} From the conditions: $|F_j(x + a,t,0)|^2 + |G_j(x + a,t, 0)|^2$ =  $|F_j(x - a,t,0)|^2 + |G_j(x - a,t, 0)|^2$ 
  =  $|F_j(x,t,0)|^2 + |G_j(x,t, 0)|^2 = 1$ we can get a difference equation:    
   \begin{align}
  F_j(x + a,t,0) F^*_j(x + a,t,0) - F_j(x,t,0) F^*_j(x,t,0) \nonumber\\
  +~G_j(x + a,t,0) G^*_j(x + a,t,0) - G_j(x,t,0) G^*_j(x,t,0) = 0  
   \end{align}
 which, after expansion upto the first order in $a$ gives the condition: 
  \begin{align}\label{cond2}
  \Re\big[F_j(x,t,0) \partial_x F^*_j(x,t,0) + G_j(x,t,0) \partial_x G_j^*(x,t,0)  \big] = 0, 
\end{align}where we have defined 
\begin{align*}
\partial_x F^*_j(x,t,0) \coloneqq \lim_{a \to 0} \frac{1}{a}\big[ F^*_j(x + a,t,0) - F^*_j(x,t,0) \big]
= \lim_{a \to 0} \frac{1}{a}\big[ F^*_j(x,t,0) - F^*_j(x-a,t,0) \big].
\end{align*}
The similar definition will be used for the functions $F_j(x,t,0)$, $G_j(x,t,0)$, $G^*_j(x,t,0)$, $\theta^0_j(x,t,0)$ for all $j \in \{1, 2\}.$ 

By explicit calculation we obtain a effective Hamiltonian:
\begin{align}\label{derihammain}
\mathscr{H}(t) = \sum_{r=0}^3  \sigma_r \otimes \sum_x \Xi_r(x,t) \ket{x}\bra{x} + 
   c~\sum_{r=1}^3  \sigma_r \otimes \sum_x \Theta_r(x,t) \ket{x}\bra{x}~p
\end{align}  See Appendix \ref{modoper} and \ref{hamcalcu} for detailed calculations.
This can be matched with the Dirac Hamiltonian in (1+1) dimensional curved space-time under the 
influence of background abelian gauge potential only, given in eq.~(\ref{curgauge1}). 
In the following section we will discuss this in detail.

\subsection{Comparison of the Derived Hamiltonian with the Dirac Hamiltonian in ($\mathbf{1+1}$) dimensional curved space-time}
In strictly $(1+1)$ dimensional space-time and for abelian potentials, the Dirac Hamiltonian corresponding to the eq.~(\ref{curgauge1}) takes the form
 \begin{align}\label{11dimham}
 H =  - \hbar \sigma_0 \otimes  A_{0} 
 + c\gamma^{(0)} \gamma^{(a)} \otimes \Bigg[\frac{e^1_{(a)}}{e^0_{(0)}}\Bigg]  \hat{p}_1   
- \frac{i \hbar c}{2}\gamma^{(0)} \gamma^{(a)} \otimes \frac{\partial}{\partial x} \Bigg[ \frac{e^{1}_{(a)}}{e^{0}_{(0)}} \Bigg] \nonumber\\
- \hbar \gamma^{(0)} \gamma^{(a)} \otimes \Bigg[\frac{e^1_{(a)}}{e^0_{(0)}}\Bigg]  A_{1}
+ c^2 \gamma^{(0)} \otimes \frac{m}{e^{0}_{(0)}}
\end{align} where $(a) \in \{(0), (1)\}$.
So to compare this Hamiltonian with our derived Hamiltonian given in eq.~(\ref{derihammain}) one possible choice is
\begin{align}\label{onechoice}
 \theta^1_2(x,t,0) = - 2 \theta^1_1(x,t,0),~~\theta^q_j(x,t, \delta t) = 0~\forall~q \in \{2, 3\}, j \in \{1, 2\},~~ \nonumber\\
 e^1_{(0)} = 0~~\text{and}~~\gamma^{(0)} = \sigma_1,~\gamma^{(1)} = -i \sigma_2~. 
\end{align}
For detailed calculation please look at the Appendix \ref{appcoinchoice}.
For this choice the terms of the Hamiltonian given in  eq.~(\ref{derihammain}) become 
   \begin{align}\Theta_1  = 0,~~\Theta_2 = 0 ,~~
               \Theta_3 =  \cos[2 \theta^1_1(x,t,0)],~~
                \Xi_0 =  \hbar [\vartheta^0_1(x,t) + \vartheta^0_2(x,t)]  - \frac{\hbar c}{2} \partial_x \theta^0_2(x,t,0),~ \nonumber\\~ 
               \Xi_1 =  \hbar[\vartheta^1_1(x,t) + \vartheta^1_2(x,t)] +  \hbar c \partial_x \theta^1_1(x,t,0),~~ 
               \Xi_2 =  \frac{\hbar c}{2} \sin[2 \theta^1_1(x,t,0)] \partial_x \theta^0_2(x,t,0),\nonumber\\
 \Xi_3  =   i \hbar c  \sin[2 \theta^1_1(x,t,0)]\partial_x \theta^1_1(x,t,0) - \frac{\hbar c}{2} \cos[2\theta^1_1(x,t,0)]
 \big[2 \partial_x \theta^0_1(x,t,0) + \partial_x \theta^0_2(x,t,0)\big].
\end{align}
   
After omitting all the zero-valued terms, the Hamiltonian in eq.~(\ref{11dimham}) becomes
\begin{align}\label{11hamil}
 H =  - \hbar \sigma_0 \otimes  A_{0} 
 + c\sigma_3 \otimes \Bigg[\frac{e^1_{(1)}}{e^0_{(0)}}\Bigg]
  \hat{p}_1 - \frac{i \hbar c}{2}\sigma_3  \otimes \frac{\partial}{\partial x} \Bigg[ \frac{e^{1}_{(1)}}{e^{0}_{(0)}} \Bigg]
- \hbar \sigma_3 \otimes \Bigg[\frac{e^1_{(1)}}{e^0_{(0)}}\Bigg]  A_{1} + c^2 \sigma_1 \otimes \frac{m}{e^{0}_{(0)} }~.
\end{align} 
Now to properly compare the Hamiltonians given in  eq.~(\ref{derihammain}) and eq.~(\ref{11dimham}),
we have to identify
\begin{align}
\partial_x \theta^0_2(x,t,0) = 0,~~
   \Bigg[\frac{e^1_{(1)}}{e^0_{(0)}}\Bigg] = \cos[2 \theta^1_1(x,t,0)], \nonumber\\
   ~~\frac{m c^2}{e^0_{(0)}} = \hbar[\vartheta^1_1(x,t) + \vartheta^1_2(x,t)] + \hbar c \partial_x \theta^1_1(x,t,0),~ 
   A_{0} = - [ \vartheta^0_1(x,t) + \vartheta^0_2(x,t)],~~\nonumber\\ 
  A_{1} \Bigg[\frac{e^1_{(1)}}{e^0_{(0)}}\Bigg] = c ~\partial_x \theta^0_1(x,t,0)  \Rightarrow A_1 =  c \sec[2 \theta^1_1(x,t,0)]~\partial_x \theta^0_1(x,t,0),
  ~~~~~~~~~~~~\nonumber  \end{align} \begin{align}
    \text{Metric} ~ = \left( \begin{array}{cc}
                              g^{00} & g^{01} \\
                              g^{10} & g^{11} \\
                             \end{array} \right)
   = \left( \begin{array}{cc}
                             \Big[e^0_{(0)}\Big]^2 & 0 \\
                             0 & -  \Big[e^1_{(1)}\Big]^2 \\
                            \end{array}\right) = \Big[e^0_{(0)}\Big]^2\left( \begin{array}{cc}
                             1 & 0 \\
                             0 & -  \cos^2[2 \theta^1_1(x,t,0)] \\
                            \end{array}\right) .
 \end{align}
 In case we want to study the fundamental particle, the mass $m$ should be taken to be position-time independent. Thus, we can choose 
 \begin{align}
 e^0_{(0)} =  m c^2 \big( \hbar[\vartheta^1_1(x,t) + \vartheta^1_2(x,t)] + \hbar c \partial_x \theta^1_1(x,t,0) \big)^{-1}~.  
 \end{align}
 In condensed matter studies many kinds of emergent particles are possible whose masses may depend on both the time and position steps, so we can  
 set
 \begin{align}
  e^0_{(0)} = 1 ~\Rightarrow~ m c^2 = \hbar[\vartheta^1_1(x,t) + \vartheta^1_2(x,t)]+ \hbar c \partial_x \theta^1_1(x,t,0)~. 
 \end{align}
 As $\theta^1_1(x,t,0)$ can be an arbitrary function of $x$, $t$ but 
 $ - 1 \leq \cos[ 2\theta^1_1(x,t,0)] \leq 1 $, $g^{11}$ term of any metric can be captured 
by this through some constant value scaling.

\section{Numerical simulation} 
In this article our main purpose is to unify all the possible background potential effects in a single particle massive Dirac Hamiltonian.
For proper depiction one should do numerical analysis for all possible common mathematical forms of the metric and gauge potentials.
So that he/she can predict the mathematical forms of metric and gauge potentials corresponding to the experimentally 
observed phenomena where the metric and gauge potential functions are unknown.
Here, we have given examples of few common mathematical forms of metrics and external gauge potentials. Our numerical results are obtained by
 considering $\hbar = 1$ unit, $c = 1$ unit, $\delta t \coloneqq \frac{1}{L}$ unit and $a \coloneqq \frac{1}{L}$ unit.
 For the validity of the approximation used to derive the effective Hamiltonian, we should have~$L >>1$. 
We choose to work with the mass = $m = 0.04$ unit. 
Below we have shown probability profiles as functions of time-steps (SS-DQW evolution steps) and position-steps for different metrics and abelian gauge potentials. 
This probability is the existential probability of the quantum particle (walker), irrespective of its coin state, i.e., we have traced over the 
whole coin space while we calculated the probabilities. 

\subsection{A non-static metric case}
 
Here we will take $L = 150$.

\begin{enumerate}
 \item  {\bf Fig.~\ref{fig1} is for curved space-time with $U(1)$ potential:} 
    
  \begin{figure}[h]\centering
 \includegraphics[width = 13.5cm]{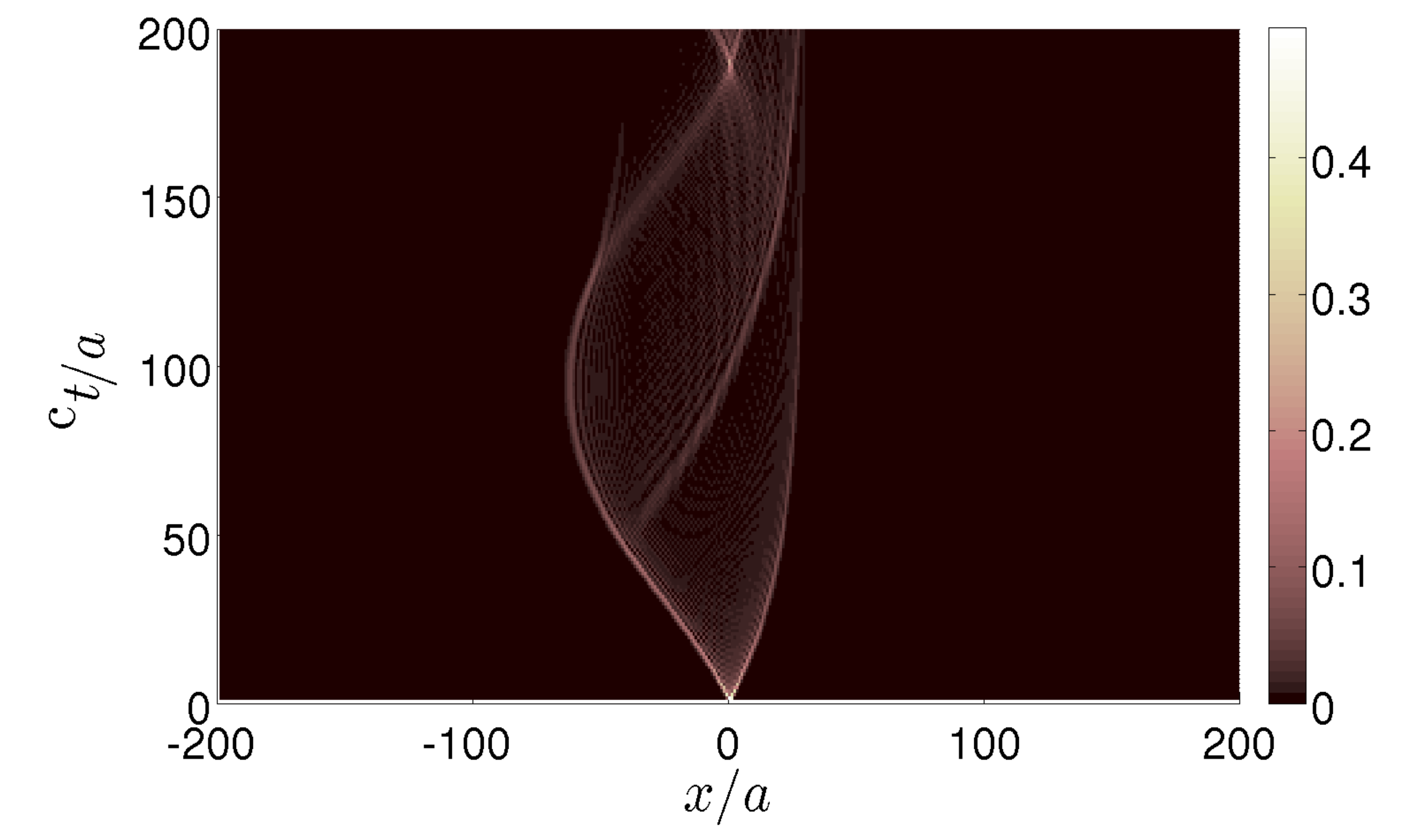} 
 \caption[Probability of finding a particle in a non-static metric system with gauge potentials]
 {(Color online) Probability  as function of 200 time steps of the modified SS-DQW on a flat-lattice with 400 lattice points. 
 The probability is for a non-static metric system:~$g^{00} = t^{-2}, g^{01} = g^{10} = 0, g^{11}
 = - \frac{t^{-2}}{2} \big[ \cos 4x + \sin 4x \big]^2$ in presence of $U(1)$ gauge potential with mass = 0.04 unit. 
 The initial state used for the evolution is $\frac{1}{\sqrt{2}}\big[ \ket{\uparrow} + i \ket{\downarrow} \big] \otimes \ket{x = 0}$.}
 \label{fig1}
\end{figure} 
 
    $e^0_{(0)} = \frac{1}{t}$, $e^1_{(1)} = \frac{1}{\sqrt{2} t} \big[ \cos 4x + \sin 4x \big]$, 
 the coin parameter functions are:\begin{align}
 ~ \theta^0_1(x,t,0) = -1000 x t, ~ \vartheta^0_1(x,t) = -0.03 x,~ \theta^0_2(x,t,0) = 0,~ \vartheta^0_2(x,t) = 0,~~ \nonumber\\
 \theta^1_1(x,t,0) = \frac{\pi}{8} + 2 x \Rightarrow \partial_x \theta^1_1(x,t,0) =  2,~ \vartheta^1_1(x,t) = - 2,
 ~ \vartheta^1_2(x,t) = 0.04 t,~ \nonumber
  \end{align}\begin{align}
 \Rightarrow~\text{our rotation angles are:} \nonumber\\
 \theta^1_1(x,t,\delta t) = \frac{\pi}{8} + 2 x - \frac{2}{L},~~
 \theta^1_2(x,t,\delta t) = - \frac{\pi}{4} - 4 x + \frac{0.04 t}{L}, \nonumber\\
 \text{our phases are:} \nonumber\\
 \theta^0_1(x,t,\delta t) = - 1000 x t - \frac{0.03 x}{L},~~
 \theta^0_2(x,t,\delta t) = 0.
\end{align}

 \item  {\bf Fig.~\ref{fig2} is for curved space-time without $U(1)$ potential:}
    
\begin{figure}[h]\centering
 \includegraphics[width = 13.5cm]{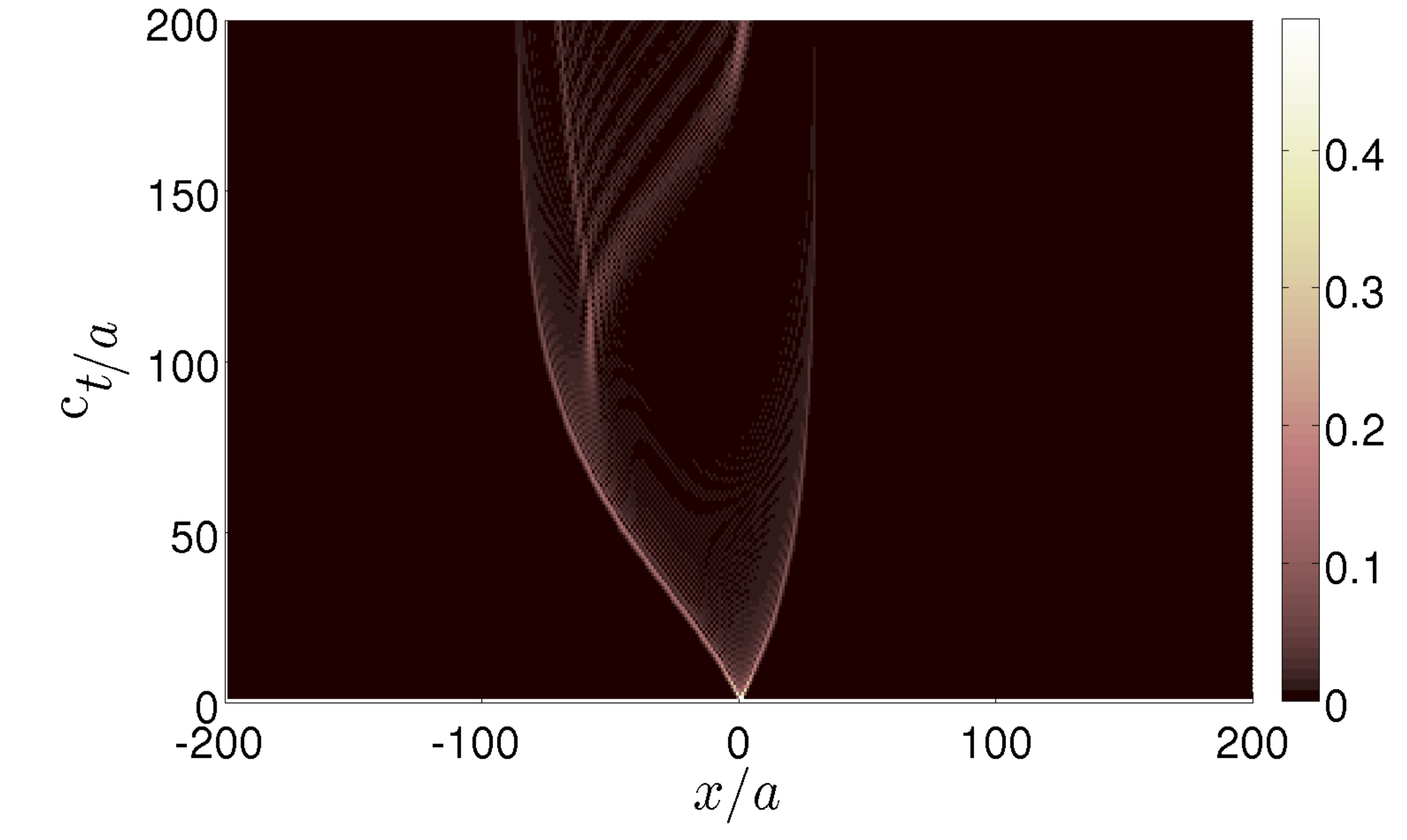} 
 \caption[Probability of finding a particle in a non-static metric system without gauge potentials]{(Color online) 
 Probability  as function of 200 time steps of the modified SS-DQW on a flat-lattice with 400 lattice points.
 The probability is for a non-static metric system:~$g^{00} = t^{-2}, g^{01} = g^{10} = 0,
 g^{11} = - \frac{t^{-2}}{2} \big[ \cos 4x + \sin 4x \big]^2$ in absence of gauge potential with mass = 0.04 unit.
 The initial state used for the evolution is $\frac{1}{\sqrt{2}}\big[ \ket{\uparrow} + i \ket{\downarrow} \big] \otimes \ket{x = 0}$.}
\label{fig2} 
\end{figure}

    $e^0_{(0)} = \frac{1}{t}$, $e^1_{(1)} = \frac{1}{\sqrt{2} t} \big[ \cos 4x + \sin 4x \big]$, 
 the coin parameter functions are:
\begin{align}
 ~ \theta^0_1(x,t,0) = 0, ~ \vartheta^0_1(x,t) = 0, ~\theta^0_2(x,t,0) = 0,~\vartheta^0_2(x,t) = 0,~~ \nonumber\\
 \theta^1_1(x,t,0) = \frac{\pi}{8} + 2 x \Rightarrow \partial_x \theta^1_1(x,t,0) =  2,~ \vartheta^1_1(x,t) = - 2, 
 ~ \vartheta^1_2(x,t) = 0.04 t ,~ \nonumber
 \end{align}\begin{align}
 \Rightarrow~\text{our rotation angles are:} \nonumber\\
 \theta^1_1(x,t,\delta t) = \frac{\pi}{8} + 2 x - \frac{2}{L},~~
 \theta^1_2(x,t,\delta t) = - \frac{\pi}{4} - 4 x + \frac{0.04 t}{L}, \nonumber\\
 \text{our phases are:} \nonumber\\
 \theta^0_1(x,t,\delta t) = 0,~~
 \theta^0_2(x,t,\delta t) = 0.
\end{align}
        
\item {\bf Fig.~\ref{fig3} is for flat space time without $U(1)$ potential:}
  
  \begin{figure}[h]\centering
 \includegraphics[width = 13.5cm]{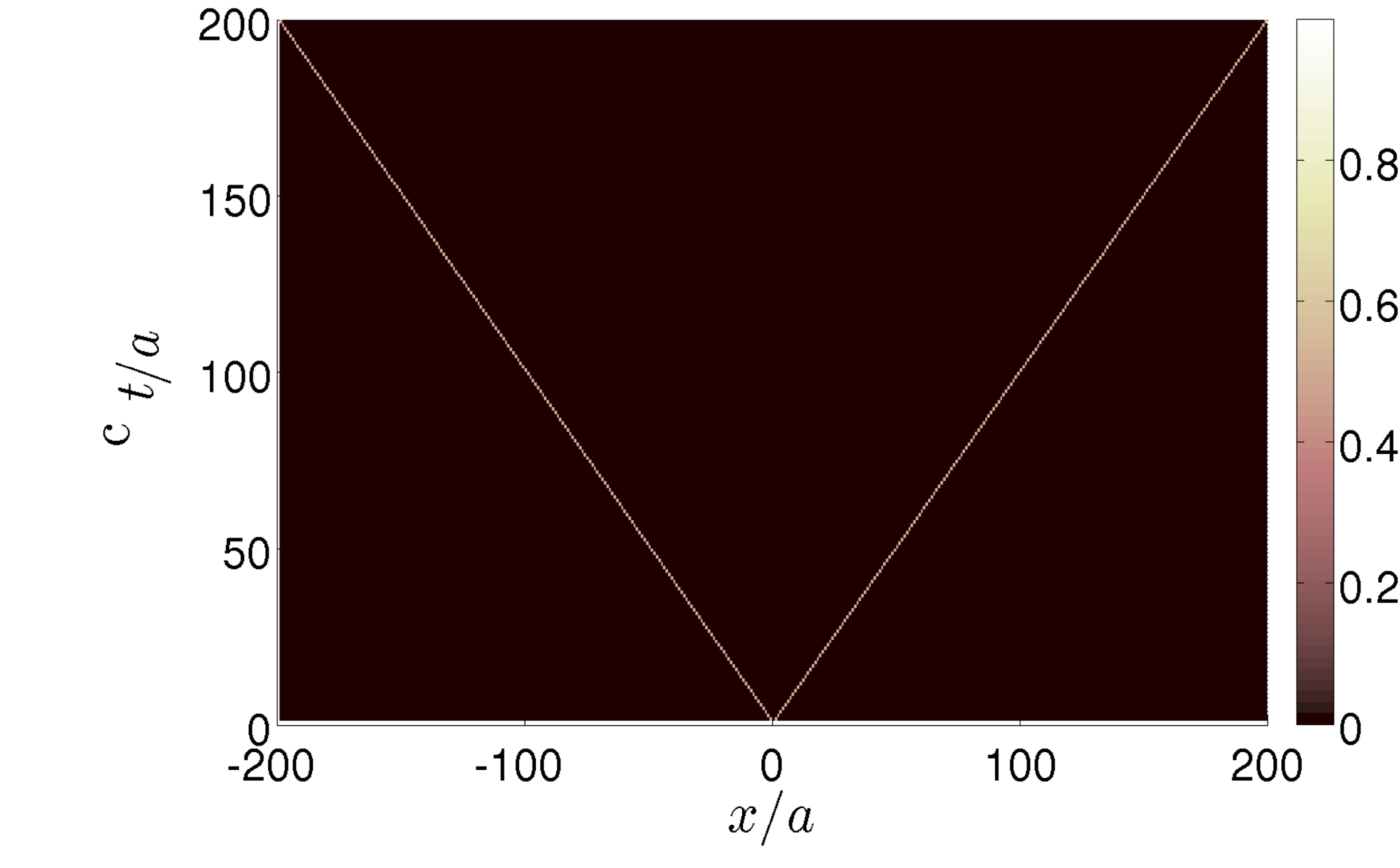} 
  \caption[Probability of finding a particle in a Minkowski metric system without gauge potentials]
  {(Color online) Probability as function of 200 time steps of the modified SS-DQW on a flat-lattice with 400 lattice points.
  The probability is for Minkowski metric system in absence of gauge potential with mass = 0.04 unit 
  and the initial state used for the evolution is $\frac{1}{\sqrt{2}}\big[ \ket{\uparrow} + i \ket{\downarrow} \big] \otimes \ket{x = 0}$.}
\label{fig3} 
\end{figure}
    
  $e^0_{(0)} = 1$, $e^1_{(1)} = 1$, the coin parameter functions are:
\begin{align}
 ~ \theta^0_1(x,t,0) = 0,~\vartheta^0_1(x,t) = 0,~\theta^0_2(x,t,0) = 0,~ \vartheta^0_2(x,t) = 0,~~ \nonumber\\
 \theta^1_1(x,t,0) = 0 \Rightarrow \partial_x \theta^1_1(x,t,0) =  0,~ \vartheta^1_1(x,t) = 0, ~ \vartheta^1_2(x,t) = 0.04,~ \nonumber
  \end{align}\begin{align}
 \Rightarrow~\text{our rotation angles are:} \nonumber\\
 \theta^1_1(x,t,\delta t) = 0,~~
 \theta^1_2(x,t,\delta t) = \frac{0.04}{L},\nonumber\\
 \text{our phases are:} \nonumber\\
 \theta^0_1(x,t,\delta t) = 0,~~
 \theta^0_2(x,t,\delta t) = 0.
\end{align}
     
 \end{enumerate}

\subsection{A static metric case} 
Here we will take $L = 250$.  

\begin{enumerate}
 \item 

 {\bf Fig.~\ref{fig4} is for curved space-time without $U(1)$ potential:}

\begin{figure}[h]\centering
 \includegraphics[width = 13.5cm]{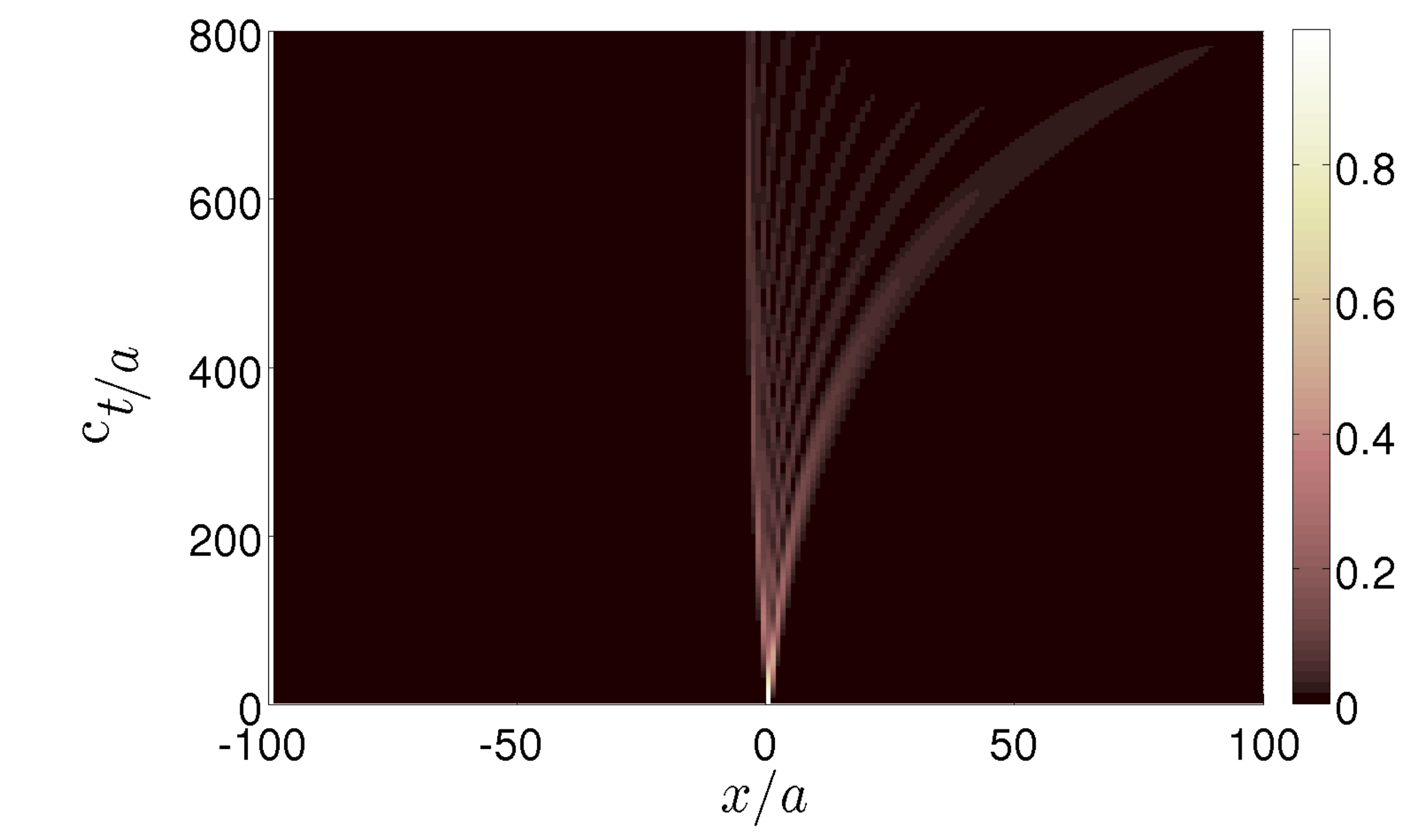}~~~~ 
 \caption[Probability of finding a particle in a static metric system without gauge potentials]
 {(Color online) Probability as a function of 800 time steps of the modified SS-DQW in a 
 flat-lattice with 200 lattice points. The probability is for the metric system: $g^{00} = 1$, $g^{01} = g^{10} = 0$, $g^{11} = - (x+5a)^2$ with mass = 0.04 unit
 and the initial state used for the evolution is $\frac{1}{\sqrt{2}}\big[ \ket{\uparrow} + i \ket{\downarrow} \big] \otimes \ket{x = 0}$.}
 \label{fig4}
\end{figure} 

We choose to work with $e^0_{(0)} = 1,$ $e^1_{(1)} = x + 5 a.$ 
 
 The coin parameter functions are:\begin{align}
 ~ \theta^0_1(x,t,0) = 0, ~ \vartheta^0_1(x,t) = 0,~\theta^0_2(x,t,0) = 0,~ \vartheta^0_2(x,t) = 0, \nonumber\\
 \theta^1_1(x,t,0) = \frac{1}{2} \cos^{-1}[x + 5 a]
 \Rightarrow \partial_x \theta^1_1(x,t,0) =  - \frac{1}{2}  \big(1 - [x + 5a]^2 \big)^{- \frac{1}{2}}, \nonumber\\
 ~ \vartheta^1_1(x,t) =  \frac{1}{2} \big(1 -[x + 5 a]^2 \big)^{- \frac{1}{2}},~\vartheta^1_2(x,t) = 0.04,~\nonumber
 \end{align} \begin{align}
 \Rightarrow~\text{our rotation angles are:}\nonumber\\
 \theta^1_1(x,t,\delta t) = \frac{1}{2} \cos^{-1}[x + 5 a] 
 + \frac{\delta t}{2} \big(1 -[x + 5 a]^2 \big)^{- \frac{1}{2}}, \nonumber\\
 \theta^1_2(x,t,\delta t) = - \cos^{-1}[x + 5 a] + 0.04 \delta t,\nonumber\\
 \text{our phases are:}~~
 \theta^0_1(x,t,\delta t) = 0,~~
 \theta^0_2(x,t,\delta t) = 0.
\end{align}

In  Fig.~\ref{fig4}, the probability distribution which spread only to the right side of the origin is seen.

\item  {\bf Fig.~\ref{fig5} is for curved space-time with $U(1)$ potential:}

\begin{figure}[h]\centering
 \includegraphics[width = 13.5cm]{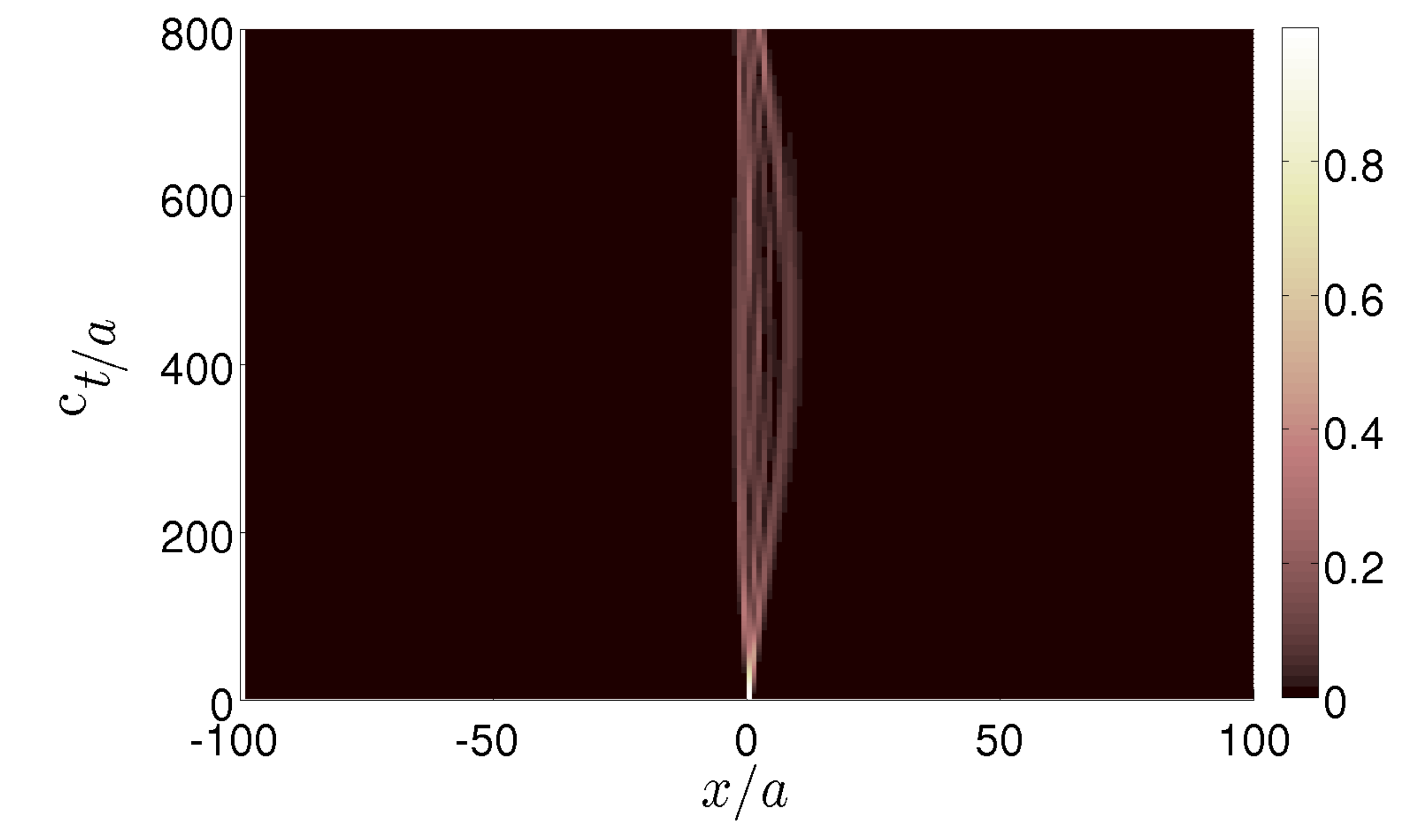}~~~~ 
 \caption[Probability of finding a particle in a static metric system with gauge potentials]
 {(Color online) Probability as a function of 800 time steps of the modified SS-DQW in a 
 flat-lattice with 200 lattice points. The probability is for the metric system: $g^{00} = 1$, $g^{01} = g^{10} = 0$, $g^{11} = - (x+5a)^2$ with mass = 0.04 unit
 and the initial state used for the evolution is 
 $\frac{1}{\sqrt{2}}\big[ \ket{\uparrow} + i \ket{\downarrow} \big] \otimes \ket{x = 0}$ in presence of gauge potential.}
 \label{fig5}
\end{figure}

In this case also, we choose to work with $e^0_{(0)} = 1,$ $e^1_{(1)} = x + 5 a.$ 

The gauge potential is captured by the parameters:
\begin{align}
 \theta^0_1(x,t,0) = - 1000 x t, ~ \vartheta^0_1(x,t) = - 0.03 x, ~\theta^0_2(x,t,0) = 0,~\vartheta^0_2(x,t) = 0. \nonumber
\end{align}
The other coin parameter functions are:\begin{align}
 \theta^1_1(x,t,0) = \frac{1}{2} \cos^{-1}[x + 5 a]
 \Rightarrow \partial_x \theta^1_1(x,t,0) =  - \frac{1}{2} \big(1 - [x + 5 a]^2 \big)^{- \frac{1}{2}}, \nonumber\\
 ~ \vartheta^1_1(x,t) =  \frac{1}{2}\big(1 - [x + 5 a]^2\big)^{-\frac{1}{2}},
 ~\vartheta^1_2(x,t) = 0.04,~\nonumber
  \end{align}\begin{align}
 \Rightarrow\text{our rotation angles are:} \nonumber\\
 \theta^1_1(x,t,\delta t) = \frac{1}{2}\cos^{-1}[x + 5 a]
 + \frac{\delta t}{2}\big(1 -[x + 5 a]^2\big)^{-\frac{1}{2}}, \nonumber\\ 
 \theta^1_2(x,t,\delta t) = -\cos^{-1}[x+ 5 a] + 0.04\delta t,\nonumber\\
 \text{our phases are:} \nonumber\\
 \theta^0_1(x,t,\delta t) = -1000 x t - \frac{0.03 x}{L},~~
 \theta^0_2(x,t,\delta t) = 0.
\end{align}

\vspace{0.7cm}

\item 
 {\bf Fig.~\ref{fig6} is for curved space-time without $U(1)$ potential:}

\begin{figure}[h]\centering
 \includegraphics[width = 13.5cm]{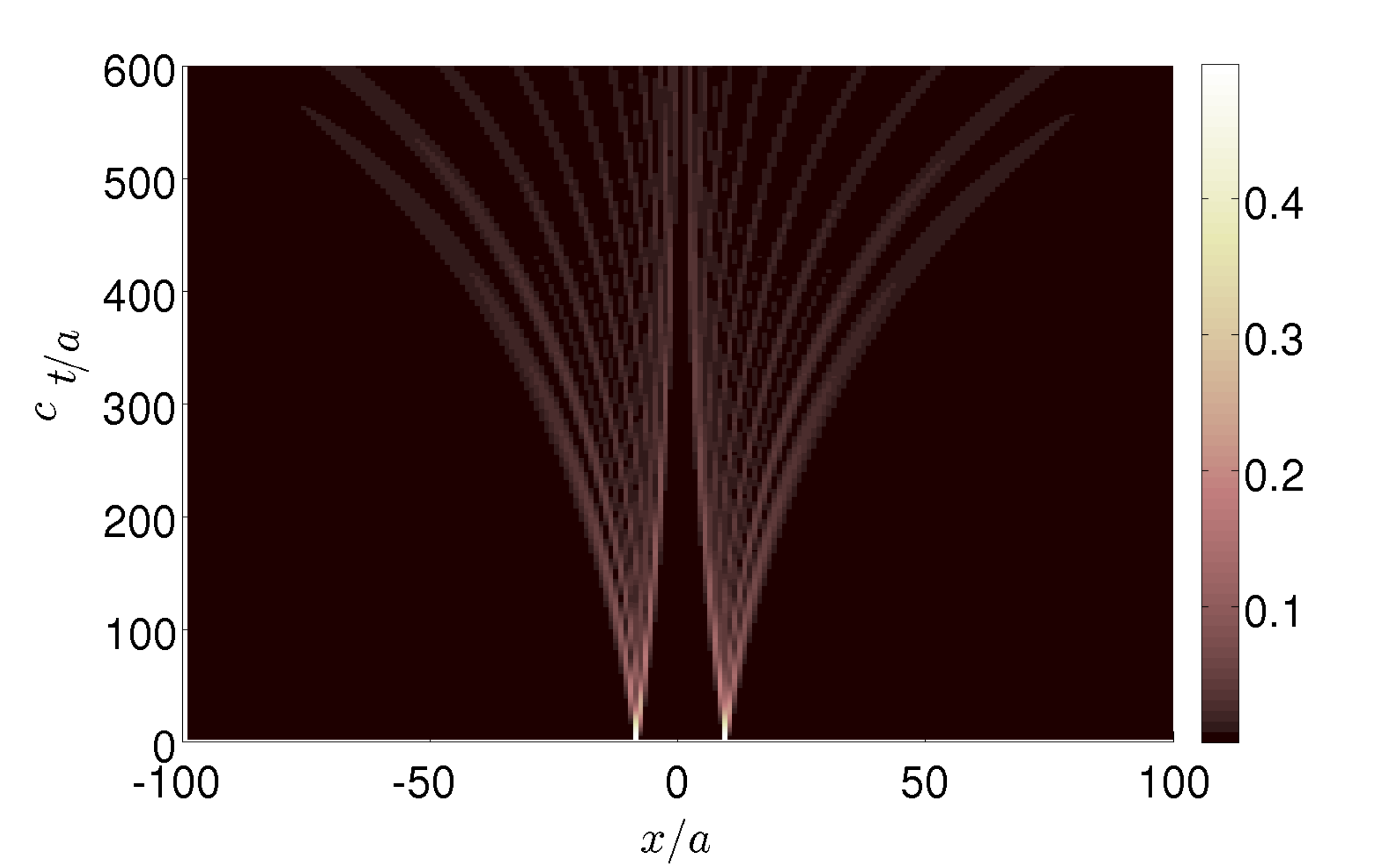}~~~~ 
 \caption[Probability of finding a particle in a static metric system without gauge potentials for delocalized initial state]
 {(Color online) Probability as a function of 600 time steps of the modified SS-DQW in a 
 flat-lattice with 200 lattice points. The probability is for the metric system: $g^{00} = 1$, $g^{01} = g^{10} = 0$, $g^{11} = - x^2$ with mass = 0.04 unit
 and the initial state used for the evolution is $\frac{1}{2}\big[ \ket{\uparrow} + i \ket{\downarrow} \big] \otimes \big(\ket{x = -9a} + \ket{x = 9a}\big)$.}
 \label{fig6}
\end{figure} 
Here we choose to work with $e^0_{(0)} = 1,$ $e^1_{(1)} = x.$ 

 The coin parameter functions are:\begin{align}
 ~ \theta^0_1(x,t,0) = 0, ~ \vartheta^0_1(x,t) = 0,~\theta^0_2(x,t,0) = 0,~ \vartheta^0_2(x,t) = 0, \nonumber\\
 \theta^1_1(x,t,0) = \frac{1}{2} \cos^{-1}[x]
 \Rightarrow \partial_x \theta^1_1(x,t,0) =  - \frac{1}{2}\big(1 - x^2 \big)^{- \frac{1}{2}}, \nonumber\\
 ~ \vartheta^1_1(x,t) =  \frac{1}{2} \big(1 -x^2\big)^{- \frac{1}{2}},~\vartheta^1_2(x,t) = 0.04,~\nonumber
 \end{align} \begin{align}
 \Rightarrow~\text{our rotation angles are:}\nonumber\\
 \theta^1_1(x,t,\delta t) = \frac{1}{2} \cos^{-1}[x] 
 + \frac{\delta t}{2} \big(1 - x^2 \big)^{- \frac{1}{2}}, \nonumber\\
 \theta^1_2(x,t,\delta t) = - \cos^{-1}[x] + 0.04 \delta t,\nonumber\\
 \text{our phases are:}~~
 \theta^0_1(x,t,\delta t) = 0,~~
 \theta^0_2(x,t,\delta t) = 0.
\end{align}

\end{enumerate}

{\bf Note:} For the static case the chosen vielbeins: $e^0_{(0)}$ is constant and $e^1_{(1)}$ is linear in position coordinate. 
In the non-static case we have chosen vielbeins: $e^0_{(0)}$ is inverse in time and $e^1_{(1)}$ is a combination of sinusoidal in position and inverse in time coordinate.
The choice of  $U(1)$ gauge potential is linear in both position and time coordinates. The presence of the gauge potential increases localization of probability profiles in positions.
The flat space-time metric case: $e^0_{(0)}$ = $e^1_{(1)}$ = 1, has been shown to get a comparable idea about the other plots.

\subsection{Simulating $(2 + 1)$ space-time dimension by $(1+1)$ space-time dimensional SS-DQW}\label{2dim}

In $(2 + 1)$ space-time dimension when one of the spatial momentum of the Dirac particle remains constant = $k_y$ unit
and all the operators in the Hamiltonian are simply function of the other spatial coordinate and time---the space-time
become {\it effectively} $(1+1)$ dimensional. Under this consideration the effective Dirac Hamiltonian 
in $(2 + 1)$ space-time dimension, corresponding to eq.~(\ref{curgauge1}) can be written as
\begin{align}\label{2dham}
 H =  - \hbar \sigma_0 \otimes  A_{0} 
 + \Big\{ c\sigma_0 \otimes q^1_{(0)} \hat{p}_1
 + c\sigma_0 \otimes q^2_{(0)} k_y
 + c\gamma^{(1)} \gamma^{(0)} \otimes q^1_{(1)} \hat{p}_1\nonumber\\
 + c\gamma^{(1)} \gamma^{(0)} \otimes q^2_{(1)} k_y 
 + c\gamma^{(2)} \gamma^{(0)} \otimes q^1_{(2)} \hat{p}_1
 + c\gamma^{(2)} \gamma^{(0)} \otimes q^2_{(2)} k_y \Big\}\nonumber\\
- \frac{i \hbar c}{2} \bigg\{\sigma_0 \otimes \frac{\partial}{\partial x} q^1_{(0)}
+ \gamma^{(1)} \gamma^{(0)} \otimes \frac{\partial}{\partial x} q^1_{(1)} 
+ \gamma^{(2)} \gamma^{(0)}  \otimes \frac{\partial}{\partial x}q^1_{(2)} \bigg\}\nonumber\\
- \hbar \Big\{\sigma_0 \otimes q^1_{(0)}  A_{1}
+ \sigma_0 \otimes q^2_{(0)}  A_{2} 
+ \gamma^{(1)} \gamma^{(0)} \otimes q^1_{(1)}  A_{1}\nonumber\\
+ \gamma^{(1)} \gamma^{(0)} \otimes q^2_{(1)}  A_{2} 
+ \gamma^{(2)} \gamma^{(0)} \otimes q^1_{(2)}  A_{1}
+ \gamma^{(2)} \gamma^{(0)} \otimes q^2_{(2)}  A_{2} \Big\}
+ c^2 \beta  \otimes \frac{m}{e^{0}_{(0)}},\nonumber\\
~~~~~~~~~~~~ \text{where}~~q^\mu_{(j)} \coloneqq \Bigg[\frac{e^\mu_{(j)}}{e^0_{(0)}}\Bigg]~~~~~~~~~~~
\end{align}

and we have taken all the operators in the Hamiltonian as the functions of $x$, $t$ only.
If we now consider
\begin{align}
\theta^q_j(x,t, \delta t) = 0~\forall~q \in \{2, 3\}, j \in \{1, 2\}, \nonumber\\
\gamma^{(0)} = \sigma_1,~ \gamma^{(1)} = - i \sigma_2,~ \gamma^{(2)} = i \sigma_3,~\Rightarrow~  
 \gamma^{(1)} \gamma^{(0)} = \sigma_3,~ \gamma^{(2)} \gamma^{(0)} =  \sigma_2~.
 \end{align} 
 In order to compare the Hamiltonian in eq.~(\ref{2dham}) with our Hamiltonian in 
 eq.~(\ref{derihammain}) derived from the modified SS-DQW, we have to make $e^1_{(0)} = 0$ which reduces the Hamiltonian in eq.~(\ref{2dham})
 to the form, \begin{align}\label{2dhamA}
 H =  - \hbar \sigma_0 \otimes  A_{0} 
 + \Big\{ c\sigma_0 \otimes q^2_{(0)} k_y
 + c\sigma_3 \otimes q^1_{(1)} \hat{p}_1
 + c\sigma_3 \otimes q^2_{(1)} k_y  \nonumber\\
 + c\sigma_2 \otimes q^1_{(2)} \hat{p}_1
 + c\sigma_2 \otimes q^2_{(2)} k_y \Big\} 
- \frac{i \hbar c}{2} \bigg\{ \sigma_3  \otimes \frac{\partial}{\partial x} q^1_{(1)}
+ \sigma_2  \otimes \frac{\partial}{\partial x} q^1_{(2)} \bigg\} \nonumber\\
- \hbar \Big\{ \sigma_0 \otimes q^2_{(0)}  A_{2} 
+ \sigma_3 \otimes q^1_{(1)}  A_{1}
+ \sigma_3 \otimes q^2_{(1)}  A_{2}\nonumber\\
+ \sigma_2 \otimes q^1_{(2)}  A_{1}
+ \sigma_2 \otimes q^2_{(2)}  A_{2}\Big\} 
+ c^2 \sigma_1  \otimes \frac{m}{e^{0}_{(0)} }.
\end{align}
 
In this case: \begin{align}\label{2dim1}
q^1_{(2)} = \Theta_2(x,t) = \frac{1}{2} \sin[2 \theta^1_1(x,t,0)] 
               + \frac{1}{2} \sin[2 \theta^1_1(x,t,0) + 2 \theta^1_2(x,t,0)],
\end{align}\begin{align}\label{2dim2}
 q^1_{(1)} = \Theta_3(x,t) = \frac{1}{2} \cos[2 \theta^1_1(x,t,0)] 
               + \frac{1}{2} \cos[2 \theta^1_1(x,t,0) + 2 \theta^1_2(x,t,0)],
               \end{align}
\begin{align}\label{2dim3}
 - \hbar A_0 +  q^2_{(0)} (k_y c - \hbar A_2) 
 = \Xi_0(x,t) =  \hbar [\vartheta^0_1(x,t) + \vartheta^0_2(x,t)]  - \frac{\hbar c}{2} \partial_x \theta^0_2(x,t,0),
\end{align}\begin{align}\label{2dim4}
  \frac{m  c^2 }{e^{0}_{(0)} }  =   \Xi_1(x,t) 
 =  \hbar[\vartheta^1_1(x,t) + \vartheta^1_2(x,t)] - \frac{\hbar c}{2} \partial_x \theta^1_2(x,t,0)
\end{align}\begin{align}\label{2dim5}
  q^2_{(1)} ( k_y c - \hbar  A_2) - \hbar q^1_{(1)} A_1  =  \nonumber\\
 - \hbar c \partial_x \theta^0_1(x,t,0) \Theta_3(x,t)  - \frac{\hbar c}{2}\partial_x \theta^0_2(x,t,0) \cos[2\theta^1_1(x,t,0) + 2\theta^1_2(x,t,0)], 
\end{align}\begin{align}\label{2dim6}
 q^2_{(2)} ( k_y c - \hbar  A_2) - \hbar q^1_{(2)} A_1  =   \nonumber\\
 - \hbar c  \partial_x \theta^0_1(x,t,0) \Theta_2(x,t) 
 - \frac{\hbar c}{2} \partial_x \theta^0_2(x,t,0) \sin[2\theta^1_2(x,t,0) + 2 \theta^1_1(x,t,0)]~.
\end{align}

The total number of variables in set $\Big\{A_0, A_1, A_2, m, e^0_{(0)}, e^1_{(1)}, e^1_{(2)}, e^2_{(0)}, e^2_{(1)}, e^2_{(2)} \Big\}$ 
of the set of the eqs.~(\ref{2dim1})-(\ref{2dim6}) are larger than the total 
number of the equations. So unique solution is not possible. One possible solution is 
\begin{multline}\label{quasical}
 A_0 = -[\vartheta^0_1(x,t) + \vartheta^0_2(x,t)],
 ~ A_1 = -  c \partial_x \theta^0_1(x,t,0),~ A_2 = -  c \partial_x \theta^0_2(x,t,0) + \frac{k_y c}{\hbar}, \\
 ~ \Bigg[\frac{e^2_{(0)}}{e^0_{(0)}}\Bigg] = \frac{1}{2},~
\Bigg[\frac{e^1_{(2)}}{e^0_{(0)}}\Bigg] =  \frac{1}{2} \sin[2 \theta^1_1(x,t,0)] 
               + \frac{1}{2} \sin[2 \theta^1_1(x,t,0) + 2 \theta^1_2(x,t,0)] ,\\
 \Bigg[\frac{e^1_{(1)}}{e^0_{(0)}}\Bigg] =  \frac{1}{2} \cos[2 \theta^1_1(x,t,0)] 
               + \frac{1}{2} \cos[2 \theta^1_1(x,t,0) + 2 \theta^1_2(x,t,0)], \\
       \Bigg[\frac{e^2_{(1)}}{e^0_{(0)}}\Bigg] = \frac{1}{2} \cos[2 \theta^1_1(x,t,0) + 2 \theta^1_2(x,t,0)],    \\
        \Bigg[\frac{e^2_{(2)}}{e^0_{(0)}}\Bigg] = \frac{1}{2} \sin[2 \theta^1_1(x,t,0) + 2 \theta^1_2(x,t,0)],~
  \frac{m c^2}{e^{0}_{(0)} }  =  \hbar[\vartheta^1_1(x,t) + \vartheta^1_2(x,t)] - \frac{\hbar c}{2} \partial_x \theta^1_2(x,t,0).              
\end{multline}
Therefore, the metric
 \begin{multline}
                         =  \left( \begin{array}{ccc}
                                   g^{00} & g^{01} & g^{02} \\
                                   g^{10} & g^{11} & g^{12} \\
                                   g^{20} & g^{21} & g^{22} \\
                                   \end{array} \right)
                                   = \left( \begin{array}{ccc}
                                   \Big[e^{0}_{(0)}\Big]^2 & 0 & e^{0}_{(0)} e^{2}_{(0)} \\ \\
                                   0 & - \Big[e^{1}_{(1)}\Big]^2 -  \Big[e^{1}_{(2)}\Big]^2 & - e^{1}_{(1)} e^{2}_{(1)} - e^{1}_{(2)} e^{2}_{(2)} \\ \\
                                  e^{0}_{(0)} e^{2}_{(0)} & - e^{1}_{(1)} e^{2}_{(1)} - e^{1}_{(2)} e^{2}_{(2)} 
                                  &  \Big[e^{2}_{(0)}\Big]^2 -  \Big[e^{2}_{(1)}\Big]^2 -  \Big[e^{2}_{(2)}\Big]^2 \\
                                   \end{array} \right)\\
                                   = \Big[e^{0}_{(0)}\Big]^2\left(\begin{array}{ccc}
                                            1  &  0   & \frac{1}{2}\\  \\
                                            0  & - \frac{1}{4} - \frac{1}{2} \cos^2 [\theta^1_2(x,t,0)] &  - \frac{1}{2} \cos^2 [\theta^1_2(x,t,0)] \\  \\
                                            \frac{1}{2} & - \frac{1}{2} \cos^2 [\theta^1_2(x,t,0)] & 0 \\
                                           \end{array}\right),
                        \end{multline} 
where we have used the definition:  $ g^{\mu \nu} = e^\mu_{(0)} e^\nu_{(0)} - e^\mu_{(1)} e^\nu_{(1)} - e^\mu_{(2)} e^\nu_{(2)}$.
We should note here that the choice described in eqs.~(\ref{quasical}) 
implies that the effect of the momentum $k_y$ of the hidden coordinate express itself as a part of the gauge potential $A_2(x,t)$. 
Other choices are possible which may give rise to different metrics.

\section{Implementation of our scheme in qubit-system}\label{qubitsim}

The shift operations $S_\pm$ and the coin operations $C_j(t, \delta t)$ are {\it controlled-unitary} 
operations. The shift operations $S_\pm$ change the position distribution while the coin state acts as the controller,
and the coin operations $C_j(t,\delta t)$ change the coin states while positions act as controllers. Coin state is 
already represented by a qubit, but the position space is $\mathcal{N}$ dimensional if the total number of lattice sites are $\mathcal{N}$,
so in general it can be of any dimension. Here, we will represent the position states by $n$-qubit system such that 
the total number of position will now be $2^n$ and each position 
is indexed by the decimal value of the corresponding binary bits expression. 
Although the number $\mathcal{N} = 2^n$ are only a particular kind of numbers, any general number of lattice sites can be 
constructed by neglecting some extra degrees of freedom. Below we demonstrate this scheme by a simple example.

\vspace{0.5cm}

Suppose our working system is a {\it periodic} lattice with 4 lattice sites, i.e., lattice system is 
$\{\ket{x}~\text{such that}~x~\in~\mathbb{Z}_4\}$. We can build it
by 2-qubit only---representing each qubit in the computational basis 
$\{\ket{0} \equiv (1 ~ 0)^T,~ \ket{1} \equiv (0 ~ 1)^T\}$, where $\ket{0}$, $\ket{1}$ are also the eigenbasis of the conventional Pauli matrix $\sigma_3$.
So we can write the basis of the two-qubit system as $\{\ket{00}, \ket{01}, \ket{10}, \ket{11}\}$.
We will use the definition: position state $\ket{0} \coloneqq \ket{00}$, position state $\ket{a} \coloneqq \ket{01}$, 
position state $\ket{2a} \coloneqq \ket{10}$, position state $\ket{3a} \coloneqq \ket{11}$. 
So, in this representation \begin{multline}
              \sum_x \ket{x+a}\bra{x} = \ket{01}\bra{00} + \ket{10}\bra{01} + \ket{11}\bra{10} + \ket{00}\bra{11} 
              = \left( \begin{array}{cccc}
                        0 & 0 & 0 & 1 \\
                        1 & 0 & 0 & 0 \\
                        0 & 1 & 0 & 0 \\
                        0 & 0 & 1 & 0 \\
                       \end{array}\right) \nonumber \\
    =    \frac{1}{4} \Big[ (\sigma_0 + \sigma_3) \otimes (\sigma_1 - i \sigma_2) 
    + (\sigma_1 - i \sigma_2) \otimes (\sigma_1 + i \sigma_2) + (\sigma_0 - \sigma_3) \otimes (\sigma_1 - i \sigma_2) 
    + (\sigma_1 + i \sigma_2) \otimes (\sigma_1 + i \sigma_2) \Big]   \nonumber\\
    = \frac{1}{2} \Big[ \sigma_0 \otimes (\sigma_1 - i \sigma_2) 
    + \sigma_1  \otimes (\sigma_1 + i \sigma_2) \Big]. 
                \end{multline}
 Similarly, \begin{multline}
              \sum_x \ket{x-a}\bra{x} = \ket{00}\bra{01} + \ket{01}\bra{10} + \ket{10}\bra{11} + \ket{11}\bra{00}  
              = \left( \begin{array}{cccc}
                        0 & 1 & 0 & 0 \\
                        0 & 0 & 1 & 0 \\
                        0 & 0 & 0 & 1 \\
                        1 & 0 & 0 & 0 \\
                       \end{array}\right) \\
    = \frac{1}{2} \Big[ \sigma_0 \otimes (\sigma_1 + i \sigma_2) 
    + \sigma_1  \otimes (\sigma_1 - i \sigma_2) \Big]. \nonumber
                \end{multline} 
  \begin{align}\label{shifmati}
              \sum_x \ket{x}\bra{x} = \ket{00}\bra{00} + \ket{01}\bra{01} + \ket{10}\bra{10} + \ket{11}\bra{11}  
              = \left( \begin{array}{cccc}
                        1 & 0 & 0 & 0 \\
                        0 & 1 & 0 & 0 \\
                        0 & 0 & 1 & 0 \\
                        0 & 0 & 0 & 1 \\
                       \end{array}\right) 
    =  \sigma_0 \otimes \sigma_0.
                \end{align}               

We should note that, in the matrix representations in the above equations, the limiting condition: 
$\lim\limits_{a \to 0} \sum_x \ket{x+a}\bra{x}$ $\neq$ $\sum_x \ket{x}\bra{x}$,
because this kind of matrix representation is not possible while $a$ varies with $\delta t$. 
For simulation purpose, we are considering $a$ as a constant quantity so that our results obtained in this way should not 
differ from the continuum theoretical results for the large number ($\mathcal{N}$) of lattice sites.
In convention, we first discretize the continuum theory so that it becomes implementable in discrete lattice space-time, as
every simulator or computer has finite amount of memory which forbids to carry information about continuous space-time. After 
discretization we simulate and then we take the continuum limit in space-time in order to match it with the actual physical results.
This is valid when the effective wavelengths of the system is very larger than the discrete cut-off of the lattice space-time. 
This is possible if the total number of lattice sites are very larger so that existence of large wavelengths of the quantum system is possible.
So in this current case, this matrix representation is justified. Here, one may question that we are showing the scheme only for $\mathcal{N} = 4$,
but our scheme can simply be extended for large $\mathcal{N}$ without any complication. 
Just for the reader friendly demonstration we have considered only $\mathcal{N} = 4$ here. 

Now to represent the coin space we will use another qubit with a basis states $\in$ $\text{span}\{\ket{\uparrow}_c, \ket{\downarrow}_c\}$. 
In this case the shift operators take the forms: 
\begin{multline}
     S_+ = \ket{\uparrow}_c \bra{\uparrow} \otimes \sum_x \ket{x+a}\bra{x}
                                  + \ket{\downarrow}_c \bra{\downarrow}\otimes \sum_x \ket{x}\bra{x} \\ 
      =  \left(\begin{array}{cc}
                1_c & 0_c\\
                0_c & 0_c \\
               \end{array}\right) \otimes 
 \left( \begin{array}{cccc}
                        0 & 0 & 0 & 1 \\
                        1 & 0 & 0 & 0 \\
                        0 & 1 & 0 & 0 \\
                        0 & 0 & 1 & 0 \\
                       \end{array}\right) + \left(\begin{array}{cc}
                0_c & 0_c\\
                0_c & 1_c \\
               \end{array}\right) \otimes \left( \begin{array}{cccc}
                        1 & 0 & 0 & 0 \\
                        0 & 1 & 0 & 0 \\
                        0 & 0 & 1 & 0 \\
                        0 & 0 & 0 & 1 \\
                       \end{array}\right)  \nonumber\\
                                  = \frac{1}{4}(\sigma_{0c} + \sigma_{3c} )
                                  \otimes  \big[ \sigma_0 \otimes (\sigma_1 - i \sigma_2) 
    + \sigma_1  \otimes (\sigma_1 + i \sigma_2) \big] + \frac{1}{2}(\sigma_{0c} - \sigma_{3c} )
                                  \otimes \sigma_0 \otimes \sigma_0,  
    \end{multline}
    
    \begin{multline}
        S_- =  \ket{\uparrow}_c \bra{\uparrow} \otimes \sum_x \ket{x}\bra{x}
                                  + \ket{\downarrow}_c \bra{\downarrow} \otimes \sum_x \ket{x-a}\bra{x} \\
  =  \left(\begin{array}{cc}
                1_c & 0_c\\
                0_c & 0_c \\
               \end{array}\right) \otimes \left( \begin{array}{cccc}
                        1 & 0 & 0 & 0 \\
                        0 & 1 & 0 & 0 \\
                        0 & 0 & 1 & 0 \\
                        0 & 0 & 0 & 1 \\
    \end{array}\right)  + \left(\begin{array}{cc}
                0_c & 0_c\\
                0_c & 1_c \\
               \end{array}\right) \otimes \left( \begin{array}{cccc}
                        0 & 1 & 0 & 0 \\
                        0 & 0 & 1 & 0 \\
                        0 & 0 & 0 & 1 \\
                        1 & 0 & 0 & 0 \\
                       \end{array}\right)  \nonumber\\
               =   \frac{1}{2}(\sigma_{0c} + \sigma_{3c})\otimes \sigma_0 \otimes \sigma_0 
          + \frac{1}{4}(\sigma_{0c} - \sigma_{3c}) \otimes  \big[ \sigma_0 \otimes (\sigma_1 + i \sigma_2) 
    + \sigma_1  \otimes (\sigma_1 - i \sigma_2) \big].
    \end{multline}
The two coin operations for $j = 1, 2 $  are defined as
                   \begin{multline}
                         C_j(t,\delta t) =  \Big[e^{- i \theta^0_j(x = 0,t, \delta t)} e^{- i \theta^1_j(x = 0, t, \delta t) \sigma_{1c}} \otimes \ket{00}\bra{00} + 
                        e^{- i \theta^0_j(x = a, t, \delta t)} e^{- i \theta^1_j(x = a, t, \delta t) \sigma_{1c}} \otimes \ket{01}\bra{01}~~~~ \nonumber\\
                         + e^{- i \theta^0_j(x = 2a, t, \delta t)} e^{- i \theta^1_j(x = 2 a, t, \delta t) \sigma_{1c}} \otimes \ket{10}\bra{10} +
                         e^{- i \theta^0_j(x = 3a, t, \delta t)} e^{- i \theta^1_j(x = 3a, t, \delta t) \sigma_{1c}} \otimes \ket{11}\bra{11} \Big]~~~~ \nonumber\\
         = \frac{1}{4}\Big[ e^{- i \theta^0_j(x = 0, t, \delta t)} e^{- i \theta^1_j(x = 0, t, \delta t) \sigma_{1c}}
         \otimes (\sigma_0 + \sigma_3) \otimes (\sigma_0 + \sigma_3) \\
         + e^{- i \theta^0_j(x = a,t,\delta t)} e^{- i \theta^1_j(x = a,t,\delta t) \sigma_{1c}}
         \otimes (\sigma_0 + \sigma_3) \otimes (\sigma_0 - \sigma_3) \\
         + e^{- i \theta^0_j(x = 2a,t,\delta t)} e^{- i \theta^1_j(x = 2a,t,\delta t) \sigma_{1c}}
         \otimes (\sigma_0 - \sigma_3) \otimes (\sigma_0 + \sigma_3)  \\
         +e^{- i \theta^0_j(x = 3a,t,\delta t)} e^{- i \theta^1_j(x = 3a,t,\delta t) \sigma_{1c}}
         \otimes (\sigma_0 - \sigma_3) \otimes (\sigma_0 - \sigma_3)\Big].
 \end{multline}

Therefore the whole evolution operator: 

$\mathscr{U}(t, \tau)$ = $C_1^\dagger(t,0) \cdot C_2^\dagger(t,0) \cdot \big[ S_+ \cdot C_2(t, \delta t) \cdot S_- \cdot C_1(t, \delta t) \big]$
is implementable in a simple qubit system.

\section{Introducing nonabelian gauge potentials}
 In the section \ref{modisec}, we have shown how the modified form of the inhomogeneous SS-DQW with dim$(\mathcal{H}_c) = 2$ can capture simultaneously the 
 effects of space-time curvature and the abelian potentials in the massive single particle Dirac Hamiltonian. Now in order to include the general nonabelian potential 
 effects we will use the concept of higher dimensional coin Hilbert space. Along with that we have to properly choose the coin operators.  
 We will take the same route as done in ref.~\cite{arnault2} for DQW case.  

In order to include the effect of a general nonabelian $U(N)$ gauge potential such as potentials due to the weak force, strong force;
we need $2N$ dimensional coin operator instead of the 2 dimensional one. The background space is still described by the one dimensional lattice system. 
We will define the shift operators as follows.
\begin{align}
S_+ =  \sum_{x}  \ket{\uparrow}\bra{\uparrow} \otimes \mathds{1}_N \otimes \ket{x + a}\bra{x}
+ \ket{\downarrow}\bra{\downarrow} \otimes \mathds{1}_N \otimes \ket{x}\bra{x},~ \nonumber\\
S_- =  \sum_{x}  \ket{\uparrow}\bra{\uparrow} \otimes \mathds{1}_N \otimes \ket{x}\bra{x} + \ket{\downarrow}\bra{\downarrow}
\otimes \mathds{1}_N \otimes \ket{x-a}\bra{x}, 
\end{align}where $\mathds{1}_N$ is the $N \times N$ identity operator on the coin Hilbert space. The form of these shift operators imply that the 
later considered $N$-dimensional coin space does not control the positional movements. 
The total number of the generators for $U(N)$ group is $N^2$, so a general unitary matrix operator can be expressed as a linear combinations of the 
these $N^2$ generators. Below we will use this property when we define the coin operators.
The coin operators are defined as
\begin{multline}
C_j(t, \delta t) = \sum_x \Big[ e^{- i \sum_{q=0}^3 \theta_j^q(x, t, \delta t)~\sigma_q} \otimes \mathds{1}_N \Big]
\cdot \Big[\mathcal{C}_{Nj}(x, t, \delta t) \otimes \ket{x}\bra{x}\Big],~~\text{for}~~j \in \{1,2\}~~\text{where}~~\\
\mathcal{C}_{Nj}(x,t,\delta t) = \Big[ \ket{\uparrow}\bra{\uparrow} \otimes e^{- i \delta t \sum_{q=0}^{N^2 - 1} \omega_j^q(x,t) \Lambda_q}
+  \ket{\downarrow}\bra{\downarrow} \otimes  e^{- i \delta t \sum_{q=0}^{N^2 - 1} \Omega_j^q(x,t) \Lambda_q} \Big]
\end{multline}and, $\Lambda_q$ are the generators of $U(N)$ group with  
$\omega^q_j(x,t)$, $\Omega_j^q(x,t)$ are the corresponding coefficients. 
Now we will follow the same procedure used for the SS-DQW case with two-dimensional coin Hilbert space.
The modified evolution operator will be defined as
$\mathscr{U}(t, \delta t) = C^\dagger_1(t, 0) \cdot C^\dagger_2(t,0) \cdot S_+ \cdot C_2(t, \delta t) \cdot S_- \cdot C_1(t, \delta t)$. 
Using the similar kind of Taylor expansion of this modified evolution operator in $\delta t$ 
assuming $\omega^q_j(x,t)$, $\Omega_j^q(x,t)$ are smooth functions of $x$, we an derive the effective Hamiltonian as the following.  
\begin{align}\label{nonabham}
 H = \sum_{r=0}^3 \sigma_r \otimes \Lambda_0 \otimes \sum_x \Xi_r(x,t) \ket{x}\bra{x} + 
   c~\sum_{r=1}^3  \sigma_r \otimes \Lambda_0 \otimes \sum_x \Theta_r(x,t) \ket{x}\bra{x}~ \hat{p} \nonumber\\
   + \sum_{r=0}^3 \sigma_r \otimes \sum_x \sum_{q = 0}^{N^2 - 1} \Lambda_q \chi^q_r(x,t) \otimes \ket{x}\bra{x}
\end{align} where the terms  $\sum\limits_{q = 1}^{N^2 - 1} \Lambda_q~ \chi^q_r(x,t) \otimes \ket{x}\bra{x}$
carry the knowledge about the nonabelian gauge potentials, and they can be expressed in terms of the coin parameters as
\begin{align}\label{coeffN}
\chi^q_0(x,t) = \frac{\hbar}{2}\Big[ \omega^q_1(x,t) + \Omega^q_1(x,t) + \omega^q_2(x,t) + \Omega^q_2(x,t)\Big] , \nonumber\\
\chi^q_3(x,t) = \frac{\hbar}{2}\Big[ \omega^q_1(x,t) - \Omega^q_1(x,t) + \{\omega^q_2(x,t) - \Omega^q_2(x,t)\}\{|F_1(x,t,0)|^2 - |G_1(x,t,0)|^2\} \Big] , \nonumber\\
\chi^q_1(x,t) = \hbar \Re[G_1(x,t,0) F_1^*(x,t,0)]\big[\omega^q_2(x,t) - \Omega_2^q(x,t)\big],~~ \nonumber\\
\chi^q_2(x,t) = - \hbar \Im[G_1(x,t,0) F_1^*(x,t,0)]\big[\omega^q_2(x,t) - \Omega_2^q(x,t)\big].
\end{align}
For the detailed derivation please look at the Appendix \ref{nonabel}. 

If we want to compare this Hamiltonian in (\ref{nonabham}) with the Hamiltonian given in eq.~(\ref{dircurgen})
and to make it consistent with the abelian case given in eq.~(\ref{11hamil}) we have 
to make $\chi^q_1(x,t) = \chi^q_2(x,t) = 0$ for all $q$, $x$, $t$. 
Therefore, $\omega^q_2(x,t) - \Omega_2^q(x,t) = 0$ which makes the non-zero terms in eq.~(\ref{coeffN})
as the following.
\begin{align}
\chi^q_0(x,t) = \frac{\hbar}{2}\Big[ \omega^q_1(x,t) + \Omega^q_1(x,t) + 2 \omega^q_2(x,t)\Big],~~
\chi^q_3(x,t) = \frac{\hbar}{2}\Big[ \omega^q_1(x,t) - \Omega^q_1(x,t)\Big]~.
\end{align}
Other coin parameters have to chosen according to the eq.~(\ref{onechoice}). 

\section{Two-particle SS-DQW}

In the previous sections of this chapter, we have discussed the single-particle case where entanglement between coin and position degrees of freedom is local, so 
this can not be used for distant quantum communication. But for two particle case coin-position, coin-coin, position-position entanglements
between two particles are possible. These entanglements can show nonlocal features.  
Moreover, in this case indistinguishable nature of particles plays an important role. Here we will not discussed the dynamics of these kinds of 
quantum correlations, but show a way to develop a two-particle simulation scheme in our modified SS-DQW framework. 
Extension of single-particle DQW with entangled coin operation has been previously studied in ref.~\cite{andraca,liu, liu2}.
Two-particle quantum walk under position dependent or independent coin operations which are separable in their coin degrees 
of freedom, have been investigated in refs.~\cite{busch, gabris, omar, berry, carson, wang}. But their frameworks are different than ours.

 Two-particle dynamics are interesting when the particles interact with each other.
For indistinguishable particles, the corresponding dynamics is interesting even without any interaction. 
 In this case we confine ourselves to the two dimensional coin Hilbert spaces for the individual particles.
Hence the total coin Hilbert space: 
$$\mathcal{H}^\text{two}_c = \text{span}\{\ket{\uparrow \uparrow}, \ket{\uparrow \downarrow}, \ket{\downarrow \uparrow}, 
\ket{\downarrow \downarrow}\} \equiv \mathcal{H}^\text{first}_c \otimes \mathcal{H}^\text{second}_c,$$
where the first entries in the kets correspond to the first particle and the last entries in the kets correspond to the 
second particle. We define the shift operators that are separable with respect to the first and second particles, 
\begin{multline}\label{shift1}
S_+  =  S^\text{first}_+ \otimes S^\text{second}_+ = \ket{\uparrow \uparrow}\bra{\uparrow \uparrow} \otimes \sum_{x_1, x_2} \ket{x_1 + a, x_2 + a}\bra{x_1, x_2} \\
+ \ket{\uparrow \downarrow}\bra{\uparrow \downarrow} \otimes \sum_{x_1, x_2} \ket{x_1 + a, x_2}\bra{x_1, x_2}
+ \ket{\downarrow \uparrow}\bra{\downarrow \uparrow} \otimes \sum_{x_1, x_2} \ket{x_1, x_2 + a}\bra{x_1, x_2} \\
+ \ket{\downarrow \downarrow}\bra{\downarrow \downarrow} \otimes \sum_{x_1, x_2} \ket{x_1, x_2}\bra{x_1, x_2},
\end{multline}
\begin{multline}\label{shift2}
S_-  =  S^\text{first}_- \otimes S^\text{second}_- = \ket{\uparrow \uparrow}\bra{\uparrow \uparrow} \otimes \sum_{x_1, x_2} \ket{x_1, x_2}\bra{x_1, x_2} \\
+ \ket{\uparrow \downarrow}\bra{\uparrow \downarrow} \otimes \sum_{x_1, x_2} \ket{x_1, x_2-a}\bra{x_1, x_2}
+ \ket{\downarrow \uparrow}\bra{\downarrow \uparrow} \otimes \sum_{x_1, x_2} \ket{x_1-a, x_2}\bra{x_1, x_2} \\
+ \ket{\downarrow \downarrow}\bra{\downarrow \downarrow} \otimes \sum_{x_1, x_2} \ket{x_1-a, x_2-a}\bra{x_1, x_2},
\end{multline}
where the subscripts 1, 2 in $x_1$, $x_2$ are for the first and second particles, respectively.  The position Hilbert space 
$\mathcal{H}^\text{two}_x = \text{span}\big\{ \ket{x_1, x_2} : x_1, x_2 \in \mathbb{Z}~~\text{or}~~\mathbb{Z}_\mathcal{N}\big\}$. 
The interaction among the particles are introduced via the global coin operators which are in general not separable with 
respect to the particles. We define the coin operators as
\begin{align}\label{globcoin}
 C^\text{two}_j(t, \delta t) 
 = \sum_{x_1, x_2}  \exp \Bigg( - i \sum_{q, r = 0}^3 \theta^{qr}_j(x_1, x_2, t, \delta t)~\sigma_q \otimes \sigma_r \Bigg) \otimes \ket{x_1, x_2}\bra{x_1, x_2}.
\end{align} 
In this case also I will consider the Taylor expansion of the functions $\theta^{qr}_j(x_1, x_2, t, \delta t)$ with respect to the variable $\delta t$ assuming the functions are
smooth in their arguments. \begin{align}
                            \theta^{qr}_j(x_1, x_2, t, \delta t) = \theta^{qr}_j(x_1, x_2, t, 0) + \delta t~\vartheta^{qr}_j(x_1, x_2, t) + \mathcal{O}(\delta t^2)~.
                           \end{align} We will consider similar kind of Taylor expansions in variable $x_1$, $x_2$ also.

In this thesis I will discuss only the case when the time-steps of the both the particles are same, i.e., $t_1$ = $t_2$ = $t$. If they are different we should 
change the forms of the shifts operators and coin operators such that, it appears like the two-particle operation acts --- for $t_1$ time-steps 
with respect to the first particle and $t_2$ time-steps for the second one.

In the indistinguishable particle case we have to impose symmetrization or antisymmetrization,
on the possible state space and if necessary, on the possible measurement operators. 
A primary requirement for describing two indistinguishable particles is that, the two-particle evolution operators should remain same under the exchange of 
particle indices. The shift operators given in (\ref{shift1}), (\ref{shift2}) are already in symmetric form under the joint exchange of coins and positions 
of the particles. The coin operator in (\ref{globcoin}) remains unchanged under this exchange if $\theta^{qr}_j(x_1, x_2, t, \delta t)$ = $\theta^{rq}_j(x_2, x_1, t, \delta t)$
for all $q$, $r$, $x_1$, $x_2$, $t$, $\delta t$.

\subsection{Separable Coin Operations}  

For the separable case the whole unitary evolution operator is factorisable with respect to the two particles. 
\begin{align}\label{septwoop}
 \mathscr{U}^\text{two}(t, \delta t) = \mathscr{U}^\text{first}(t, \delta t) \otimes \mathscr{U}^\text{second}(t, \delta t) \nonumber\\
 \Rightarrow \exp\bigg(- i \frac{\delta t}{\hbar} \mathscr{H}^\text{two}(t)\bigg) = \exp\bigg(- i \frac{\delta t}{\hbar} \mathscr{H}^\text{first}(t)\bigg)
 \otimes \exp\bigg(- i \frac{\delta t}{\hbar} \mathscr{H}^\text{second}(t)\bigg) \nonumber\\
 \Rightarrow \sigma_0 \otimes \sigma_0 \otimes \mathds{1}_1 \otimes \mathds{1}_2 - i \frac{\delta t}{\hbar} \mathscr{H}^\text{two}(t) + \mathcal{O}(\delta t^2) \nonumber\\
 = \bigg[\sigma_0 \otimes \mathds{1}_1  - i \frac{\delta t}{\hbar} \mathscr{H}^\text{first}(t) + \mathcal{O}(\delta t^2)  \bigg] 
 \otimes \bigg[\sigma_0 \otimes \mathds{1}_2  - i \frac{\delta t}{\hbar} \mathscr{H}^\text{second}(t) + \mathcal{O}(\delta t^2)  \bigg] \nonumber\\
 \Rightarrow \mathscr{H}^\text{two}(t)
 = \mathscr{H}^\text{first}(t) \otimes [\sigma_0 \otimes \mathds{1}_2] +  [\sigma_0 \otimes \mathds{1}_1] \otimes \mathscr{H}^\text{second}(t),
\end{align}
where we have used the Taylor expansion in $\delta t$ and coefficients of $(\delta t)^n$ for all $n \in \mathbb{N}$ on both side of the eq.~(\ref{septwoop}) should be equal. 
We have used the notations: $\mathds{1}_j = \sum\limits_{x_j} \ket{x_j}\bra{x_j}$ for $j \in \{1, 2\}$.
This two-particle Hamiltonian is a simple sum of two noninteracting local Hamiltonians. 
For distinguishable particles, the two particle dynamics can be studied by studying the dynamics of any one of the particles. 
The shift operators are already in separable forms, and for this separable case the coin operations 
$C^\text{two}_j(t, \delta t)  = C^\text{first}_j(t, \delta t) \otimes C^\text{second}_j(t, \delta t)$. 
In the global coin operation of the form given in (\ref{globcoin}), 
among the sixteen parameters: $\big\{\theta^{qr}_j(x_1, x_2, t, \delta t)\big\}_{q,r = 0}^4$ only 
seven terms will be nonzero, and they should take the forms like the following. 
\begin{align}\label{septwocoin}
\theta^{00}_j(x_1, x_2, t, \delta t) = \theta^{00}_{j,\text{first}}(x_1, t, \delta t)~~ \theta^{00}_{j,\text{second}}(x_2, t, \delta t), \nonumber\\
\theta^{0r}_{j}(x_1, x_2, t, \delta t) = \theta^{00}_{j,\text{first}}(x_1, t, \delta t)~~\theta^{0r}_{j}(x_2, t, \delta t)~~\forall~r \in \{1,2,3\}, \nonumber\\
\theta^{q0}_j(x_1, x_2, t, \delta t) = \theta^{q0}_{j}(x_1, t, \delta t)~~ \theta^{00}_{j,\text{second}}(x_2, t, \delta t)~~\forall~q \in \{1,2,3\}.
\end{align}

\subsection{Entangling Coin Operations} 
For the case of entangled coin operators, we choose
\begin{align}
 \theta^{qr}_j(x_1, x_2, t, \delta t) = 0~~ \forall~ q, r \notin \{0, 1\}  
\end{align}

Using the similar Taylor expansion of $\mathscr{U}^\text{two}(t, \delta t)$ in $\delta t$ as in the single particle case,
we have derived the two-particle effective Hamiltonian in Appendix \ref{twopar}.
\begin{align}\label{twoenham}
 \mathscr{H}^\text{two}(t) = \sum_{x_1, x_2} \sum_{q,r = 0}^3 \Theta^1_{qr}(x_1, x_2, t) \big[\sigma_q \otimes \sigma_r\big] \otimes \ket{x_1, x_2}\bra{x_1, x_2}\big[ p_1 c \otimes \mathds{1}_2\big] \nonumber\\
 +  \Theta^2_{qr}(x_1, x_2, t) \big[\sigma_q \otimes \sigma_r\big] \otimes \ket{x_1, x_2}\bra{x_1, x_2}\big[\mathds{1}_1 \otimes p_2 c\big] \nonumber\\
 + \Xi_{qr}(x_1, x_2, t) \big[\sigma_q \otimes \sigma_r\big] \otimes \ket{x_1, x_2}\bra{x_1, x_2}
\end{align}
where only nonvanishing terms are
$\Theta^1_{30},$ $ \Theta^1_{20}, $ $\Theta^1_{31}, $ $\Theta^1_{21}, $ $\Theta^2_{03}, $ $\Theta^2_{02}, $ $\Theta^2_{13}, $ $\Theta^2_{12},$
 $\Xi_{30}, $ $\Xi_{20}, $ $\Xi_{31}, $ $\Xi_{21}, $ $\Xi_{03}, $ $\Xi_{02}, $ $\Xi_{13}, $ $\Xi_{12}, $ $\Xi_{00}, $ $\Xi_{01}, $ $\Xi_{10}, $ $\Xi_{11}.$
For details, please look at the eq.~(\ref{entantwoham}) in Appendix \ref{twopar}. 
The terms:  $\Theta^1_{30}$, $\Theta^1_{20}$, $\Theta^1_{31}$, $\Theta^1_{21}$, $\Theta^2_{03}$, $\Theta^2_{02}$, $\Theta^2_{13}$, $\Theta^2_{12}$
carry the effect of space-time curvature. As these terms are functions of the coordinates of both the particles, one can study how the presence of one 
particle influences the gravitational effect on another.

In a very recent ref.~\cite{lobo} two particle DQW has been studied where the coin operation is global and considers only the coulomb like interaction.
The similar kind of thing can be discussed in our case if we choose:  
$\vartheta^{00}_1(x_1, x_2, t) + \vartheta^{00}_2(x_1, x_2, t) \propto  |x_1 - x_2|^{-1}$. Because of the smoothness condition imposed in our 
Taylor expansion this choice may not be valid for all of its domain, but the main unitary operation $\mathscr{U}^\text{two}(t, \delta t)$ can be done 
without being worried about this issue.

One can question about the local implementation of this entangling coin operations, when the particles are far apart.
Entanglement is an outcome of a majority class of interactions, so entangling operation is unavoidable if one wants to describe nature.
This entanglement has nonlocal nature in a sense even if they are far apart they can be entangled. But this does not mean that when the entanglement is created they were 
far from each other, it can be created via some interaction while the particles are nearby. 
In quantum simulation, the particles are kept usually very near to each other, so spatially local two-particle controlled operations can implement our global coin operators.
We can also consider the coefficients of the interactions: $\theta_j(x_1, x_2, t, \delta t)$ vanish outside the light-cone for all $\delta t$, $j \in \{1, 2\}$ 
with the assumption that the function (or the envelop of this function) approximates some smooth function, so that our Taylor series expansion with respect to $\delta t$ remains valid. 

 
 \chapter{Conclusion and Future direction}
 
 As the DCA is derived from some basic assumptions, the established connection between DCA and SS-DQW in this thesis implies the algorithms based on DQW have more 
 fundamental aspects than other algorithms which are developed for simulation of Dirac particle dynamics. 
 The importance of our work is twofold, in one direction it develops discrete quantum walk framework to describe all fundamental particles dynamics, especially Dirac particles and   
in other direction it shows simulation schemes for fundamental particle phenomena in low energy table-top set-ups, which are otherwise difficult to realize 
in real high energy experimental set-ups. The discovery of the rich structures of a simple single-step SS-DQW (modified) which are expected to be implementable
in the state-of-art quantum simulators, is the positive side our study.

The SS-DQW was initially developed for realization of various topological phases. Thus our works can be extended to find the connection of the general Dirac particle dynamics and 
topological properties of the system. 
Other important aspect is that, one can try to draw a possible connection of quantum search algorithms and the relativistic quantum phenomena as 
in both the cases DQW has shown its significance. Using the results of one field it may be possible to develop another field.   
 Note that, in the analysis of the chapter \ref{3}, either for the single particle case or the two-particle case, particles are embedded in a flat lattice, 
only the choice of the parameters effectively make emerge the particle dynamics in such a way that the effective Hamiltonians look similar to the case in curved space-time.

Our whole approach is based on first quantization where particle annihilation or creation are not incorporated
and the indistinguishability of identical particles is not mathematically straightforward.
For more advanced theory we need to extend our SS-DQW schemes so that it can capture the various aspects of second quantization approach. 
Now we have understood what coin parameters correspond to what physical object in Dirac Hamiltonian.
This has to be applied when we develop the DQW simulation scheme for quantum field theory or more general theory. 
One approach for this kind of simulation is to consider the dynamics in open quantum system frameworks
which is recently considered by some refs.~\cite{Banerjee6, Banerjee7, Banerjee8}.



\begin{appendices}

\chapter{Single particle case}

\section{Hamiltonian from a unitary operator in coin space}
 For any unitary operator in two dimensional coin-space, we can use this following form except some global phase factor, 
 \begin{align}\label{unitappen}
  U  = \left( \begin{array}{cc}
               \mathscr{F}  &  \mathscr{G}  \\
               - \mathscr{G}^* & \mathscr{F}^* \\
              \end{array} \right)~~
              \text{subject to the condition :}~  |\mathscr{F}|^2 + |\mathscr{G}|^2 = 1~.
              \end{align}
Eigenvalues of $U$ are, \\
$\Re(\mathscr{F}) \pm i \sqrt{1 - [\Re(\mathscr{F})]^2 }  = \Re (\mathscr{F}) \pm i \sqrt{ |\mathscr{G}|^2 + [\Im(\mathscr{F})]^2 } 
= e^{\pm i \cos^{-1}[\Re(\mathscr{F})]}$,  and \\
the corresponding eigenvectors are, respectively $\ket{\phi^{\mp}(k)}=$  \begin{align}
                                               \frac{1}{\sqrt{2 |\mathscr{G}|^2 + 2 \Im^2 (\mathscr{F}) - 2 \Im(\mathscr{F}) \sqrt{1 - [\Re(\mathscr{F})]^2 }}} 
                                               \left(\begin{array}{c}
                                                      - \mathscr{G} \\
                                                      i \Big[ \Im(\mathscr{F}) - \sqrt{1 - [\Re(\mathscr{F})]^2 } \Big] \\
                                                     \end{array} \right), \nonumber\\
 \frac{1}{\sqrt{2 |\mathscr{G}|^2 + 2 [\Im (\mathscr{F})]^2 + 2 \Im(\mathscr{F}) \sqrt{1 - [\Re(\mathscr{F})]^2 }}} 
                                               \left(\begin{array}{c}
                                                    -  \mathscr{G} \\
                                                      i \Big[ \Im(\mathscr{F}) + \sqrt{1 - [\Re(\mathscr{F})]^2 } \Big] \\
                                                     \end{array} \right)~.\end{align}
 Denoting these eigenvectors by $(x_+ ~~ y_+)^T  $ and $ (x_- ~~ y_-)^T $ , respectively, we get,    
  \begin{align}
  U  = \left( \begin{array}{cc}
               x_+  &  x_-  \\
               y_+  &  y_-  \\
              \end{array} \right)\left( \begin{array}{cc}
               e^{i \cos^{-1}[\Re(\mathscr{F})]} &  0 \\
               0   &   e^{- i\cos^{-1}[\Re (\mathscr{F})]} \\
              \end{array} \right)\left( \begin{array}{cc}
               x^*_+  &  x^*_-  \\
               y^*_+  &  y^*_-  \\
              \end{array} \right)^T \nonumber\\
              H  = \frac{i \hbar}{\delta t} \ln ( U ) = - \frac{\hbar}{\delta t} \cos^{-1} [\Re(\mathscr{F})]\left( \begin{array}{cc}
               x_+  &  x_-  \\
               y_+  &  y_-  \\
              \end{array} \right)\left( \begin{array}{cc}
                1 &  0 \\
               0   &   -1 \\
              \end{array} \right)\left( \begin{array}{cc}
               x^*_+  &  x^*_-  \\
               y^*_+  &  y^*_-  \\
              \end{array} \right)^T \\
 =    - \frac{\hbar}{\delta t}\cos^{-1}[\Re(\mathscr{F})]\left( \begin{array}{cc}
                                                        |x_+|^2 - |x_-|^2   &    x_+ y^*_+ - x_- y^*_-  \\
                                                         x^*_+ y_+ - x^*_- y_-  &    |y_+|^2 - |y_-|^2  \\ 
                                                        \end{array} \right) \nonumber\\
            =   - \frac{\hbar}{\delta t}\cos^{-1}[\Re(\mathscr{F})]\Big[(|x_+|^2 - |x_-|^2)\sigma_3 
            + \Re(x_+ y^*_+ - x_- y^*_-)\sigma_1  - \Im(x_+ y^*_+ - x_- y^*_-) \sigma_2 \Big] \nonumber\\
       \Rightarrow H     = - \frac{\hbar\cos^{-1} [ \Re(\mathscr{F}) ]}{\delta t \sqrt{1 - [\Re(\mathscr{F})]^2 } }  \Big[ \Im(\mathscr{F}) \sigma_3
       + \Re(\mathscr{G}) \sigma_2 + \Im(\mathscr{G}) \sigma_1 \Big]~.
            \end{align}
   In the DQW and the SS-DQW, the unitary evolution operator defined on $\mathcal{H}_c \otimes \mathcal{H}_x$.           
In the space-time independent coin operators cases the evolution operator can be written in the form given in eq.~(\ref{unitappen}).
Hence, the whole evolution operator is diagonalizable in this same procedure.

 \section{Derivation of Schrödinger like equation form curved space-time Dirac equation}\label{schroder}
 Flat space-time Dirac equation is given by
 \begin{align*}
  \left(i \hbar \gamma^{(a)}\partial_{(a)}- m c^2 \right )\psi=0,
 \end{align*} where $\partial_{(a)}~\text{or later used}~\partial_\mu \in \left\{ \partial_{t},  c~ \partial_{x^i} ~\text{such that}~i=1, 2, 3.\right\}$.
 Generalization to the curved space-time is given by
 \begin{align}
  \left(i \hbar e^{\mu}_{(a)}\gamma^{(a)}\nabla_{\mu} - m c^2 \right)\psi=0,
  \label{Dirac Curved Eq appendix}
 \end{align}
 where $\nabla_\mu = \partial_\mu + \Gamma_\mu - i A_\mu$, 
 $\Gamma_{\mu}=-\frac{i}{4}S_{(c)(d)}e^{(c)\nu}\left(\frac{\partial e^{(d)}_{\nu}}{\partial x^{\mu}}-\Gamma^{\lambda}_{\mu\nu}e^{(d)}_{\lambda}\right)$,\\
 $\Gamma^{\sigma}_{\lambda\mu}=\frac{1}{2}g^{\nu\sigma}\left(\partial_{\lambda}g_{\mu\nu}+\partial_{\mu}g_{\lambda\nu}-\partial_{\nu}g_{\mu\lambda}\right)$,
 and $S_{(c)(d)}$ are the flat spinor matrices: $S_{(c)(d)}=\frac{i}{2}[\gamma_{(c)},\gamma_{(d)}]$, $A_\mu$ is the $U(1)$ potential.
 Now in view of the following relations,
 \begin{align*}
 \gamma_{(a)}S_{(b)(c)}=\frac{1}{2}[\gamma_{(a)},S_{(b)(c)}]+\frac{1}{2}\{\gamma_{(a)},S_{(b)(c)}\},~~
 [\gamma_{(a)},S_{(b)(c)}]=2i\left(\eta_{(a)(b)}\gamma_{(c)}-\eta_{(a)(c)}\gamma_{(b)}\right),\\
 \{\gamma_{(a)},S_{(b)(c)}\}=-2i\epsilon_{(a)(b)(c)(d)}\gamma^{(d)}\gamma_{5};\,\,\gamma_{5}=\gamma_{(0)}\gamma_{(1)}\gamma_{(2)}\gamma_{(3)},
 ~~~~~~~~~~~~~~~~~~~~~~~~~~~~~~~~\end{align*}
 it is possible to write eq.~(\ref{Dirac Curved Eq appendix}) as,
 \begin{align}
  \frac{i \hbar}{2}\gamma^{(a)} \bigg[ \bigg\{e^{\mu}_{(a)},\bigg(\frac{\partial}{\partial x^{\mu}} - i A_{\mu} \bigg) \bigg\} 
  + e^{\rho}_{(a)}\Gamma^{\mu}_{\mu\rho} \bigg]\psi
  +\frac{i \hbar}{2}\gamma^{(a)}\gamma_{5}\mathcal{B}_{(a)}\psi = m c^2 \psi,
  \label{dirac equation appendix}
 \end{align}
 where $\mathcal{B}_{(a)}=\frac{1}{2}\epsilon_{(a)(b)(c)(d)}e^{(b)\mu}e^{(c)\nu}\frac{\partial e^{(d)}_{\nu}}{\partial x^{\mu}}$.
 For $(1+1)$ and $(2+1)$ dimensions $\epsilon_{(a)(b)(c)(d)}$ is always zero, so $\mathcal{B}_{(a)}=0$.
 To derive the current density we need to derive also the dual equation satisfied by $\bar{\psi}=\psi^{\dagger}\beta$, where $\beta=\gamma^{(0)}$
 and it is given by the following equation, with the assumption that all the vielbeins are real,
 \begin{align}
  \frac{i\hbar}{2} \bigg[ \bigg\{ e^{\mu}_{(a)},\bigg(\frac{\partial}{\partial x^{\mu}} + i A_{\mu} \bigg) \bigg\} 
  + e^{\rho}_{(a)}\Gamma^{\mu}_{\mu\rho} \bigg] \bar{\psi}\gamma^{(a)}
  -\frac{i \hbar }{2}\gamma^{(a)}\gamma_{5}\mathcal{B}_{(a)}\bar{\psi} = - m c^2 \bar{\psi},
  \label{dual dirac Curved Eq appendix}
 \end{align}
 From eq.~(\ref{dirac equation appendix}) and eq.~(\ref{dual dirac Curved Eq appendix}) it is possible to derive the four vector current $j^{\mu}$, and they are 
 given as 
 \begin{align}
  j^{\mu}=\sqrt{-g}e^{\mu}_{(a)}\bar{\psi}\gamma^{(a)}\psi ~\Rightarrow~ j^{0}=\sqrt{-g}e^{0}_{(0)}\psi^{\dagger}\psi+\sqrt{-g}e^{0}_{(i)}\bar{\psi}\gamma^{(i)}\psi,
 \label{current}
 \end{align}
 where $g = \det(g_{\mu\nu})$ and the current is conserved, i.e., $\frac{\partial j^{\mu}}{\partial x^{\mu}} = 0$.
 We want to write the curved space-time Dirac equation in the following Schrödinger equation like form
 \begin{align}
  i \hbar \frac{\partial\chi}{\partial t} = H \chi,
  \label{Schrodinger equation}
 \end{align}
 where $H$ is the Hermitian Hamiltonian operator. So the probability density is given by, $j^{0}=\chi^{\dagger}\chi.$
 After we multiply eq.~(\ref{dirac equation appendix}) by $\beta$, we get a similar equation like eq.~(\ref{Schrodinger equation}), as given by
 \begin{align}
  \frac{i \hbar}{2}\alpha^{(a)}\left[\left\{e^{\mu}_{(a)},\bigg( \frac{\partial}{\partial x^{\mu}}-i A_{\mu} \bigg)\right\}
  + e^{\rho}_{(a)}\Gamma^{\mu}_{\mu\rho}\right]\psi = m c^2 \beta\psi, \nonumber
   \end{align}
  \begin{align}
\Rightarrow  \frac{i \hbar}{2}\left\{e^{0}_{(0)},\left(\frac{\partial}{\partial t}-iA_{0}\right)\right\}\psi
  = - \frac{i \hbar }{2}\alpha^{(a)}\left\{e^{i}_{(a)}, c \frac{\partial}{\partial x^{i}}-iA_{i}\right\}
  \psi - \frac{i \hbar }{2}\alpha^{(a)}e^{\rho}_{(a)}\Gamma^{\mu}_{\mu\rho}\psi + m c^2 \beta\psi
  \label{dirac equation after beta multiplication}
 \end{align}
 where $\alpha^{(a)}=\beta\gamma^{(a)}$. However this Hamiltonian is not
 hermitian and the current is also not same as eq.~(\ref{current}). In this case current is given by,
 \begin{equation}
 j^{0}=\sqrt{-g}e^{0}_{(0)}\psi^{\dagger}\psi.
 \label{second current}
 \end{equation}
 Comparisons of eq.~(\ref{current}) and eq.~(\ref{second current}) suggests that
 we must make nonunitary transformation (with the assumption $e^{0}_{(i)}=0$),
 \begin{align}
  \chi=(-g)^{\frac{1}{4}}\Big[e^{0}_{(0)}\Big]^{\frac{1}{2}}\psi.
 \end{align}
 Now we will use this transformation in eq.~(\ref{dirac equation after beta multiplication}) to write $\psi$ in terms of $\chi$.
 \begin{multline}
 \bigg\{e^{0}_{(0)}, \bigg(\frac{\partial}{\partial t}-iA_{0} \bigg)\bigg\}\psi
 = 2e^{0}_{(0)}\frac{\partial\psi}{\partial t}-2ie^{0}_{(0)}A_{0}\psi+\frac{\partial e^{0}_{(0)}}{\partial t}\psi 
 =(-g)^{-\frac{1}{4}}\bigg(-\Big[e^{0}_{(0)}\Big]^{-\frac{1}{2}}\frac{\partial e^{0}_{(0)}}{\partial t}\chi \\
 + 2\Big[e^{0}_{(0)}\Big]^{\frac{1}{2}}\frac{\partial\chi}{\partial t}
 +\frac{\partial e^{0}_{(0)}}{\partial t}\Big[e^{0}_{(0)}\Big]^{-\frac{1}{2}}\chi \bigg)
 + 2\Big[e^{0}_{(0)}\Big]^{\frac{1}{2}}\frac{\partial(-g)^{-\frac{1}{4}}}{\partial t}\chi - 2i\Big[e^{0}_{(0)}\Big]^{\frac{1}{2}}A_{0}(-g)^{-\frac{1}{4}}\chi  \\
  = (-g)^{-\frac{1}{4}}2\Big[e^{0}_{(0)}\Big]^{\frac{1}{2}}\frac{\partial\chi}{\partial t}
 + 2\Big[e^{0}_{(0)}\Big]^{\frac{1}{2}}\frac{\partial(-g)^{-\frac{1}{4}}}{\partial t}\chi  -  2i\Big[e^{0}_{(0)}\Big]^{\frac{1}{2}}A_{0}(-g)^{-\frac{1}{4}}\chi.
 \end{multline}
 Similarly,
 \begin{align}
  &\left\{e^{i}_{(a)},\frac{\partial}{\partial x^{i}}-iA_{i}\right\}\psi=2e^{i}_{(a)}\frac{\partial\psi}{\partial x^{i}}+\frac{\partial e^{i}_{(a)}}{\partial x^{i}}\psi
  -2ie^{i}_{(a)}A_{i}\psi\nonumber\\
  &=2e^{i}_{(a)}\left[\Big[e^{0}_{(0)}\Big]^{-\frac{1}{2}}\frac{\partial(-g)^{-\frac{1}{4}}}{\partial x^{i}}
  \chi+(-g)^{-\frac{1}{4}}\Big[e^{0}_{(0)}\Big]^{-\frac{1}{2}}\frac{\partial\chi}
  {\partial x^{i}}+(-g)^{-\frac{1}{4}}\frac{\partial\Big[e^{0}_{(0)}\Big]^{-\frac{1}{2}}}{\partial x^{i}}\chi\right]\nonumber\\
  &+\frac{\partial e^{i}_{(a)}}{\partial x^{i}}(-g)^{-\frac{1}{4}}\Big[e^{0}_{(0)}\Big]^{-\frac{1}{2}}\chi-2ie^{i}_{(a)}
  A_{i}(-g)^{-\frac{1}{4}}\Big[e^{0}_{(0)}\Big]^{-\frac{1}{2}}\chi
 \end{align}
 and,
 \begin{align}\label{conn1}
  \Gamma^{\mu}_{\mu\rho}=\frac{1}{2}g^{\mu\lambda}\left\{\frac{\partial g_{\lambda\mu}}{\partial x^{\rho}}+\frac{\partial g_{\lambda\rho}}{\partial x^{\mu}}
  -\frac{\partial g_{\mu\rho}}{\partial x^{\lambda}}\right\}
 =\frac{1}{2}\left\{g^{\mu\lambda}\frac{\partial g_{\lambda\mu}}{\partial x^{\rho}}+\frac{\partial g_{\lambda\rho}}{\partial x^{\lambda}}
  -\frac{\partial g_{\mu\rho}}{\partial x^{\rho}}\right\}=\frac{1}{2}g^{\mu\lambda}\frac{\partial g_{\lambda\mu}}{\partial x^{\rho}} .
 \end{align}
 We can evaluate this easily by using the following relation for any arbitrary matrix M,
 \begin{align}\label{conn2}
  \text{Tr}\left\{M^{-1}(x)\frac{\partial}{\partial x^{\lambda}}M(x)\right\}=\frac{\partial}{\partial x^{\lambda}}\ln [\det M(x)]
 \end{align}
 So, $\Gamma^{\mu}_{\mu\rho}=\frac{1}{2}\frac{\partial}{\partial x^{\rho}}\text{ln}\,g=\frac{1}{\sqrt{g}}\frac{\partial}{\partial x^{\rho}}\sqrt{g}$.
 Finally using all the relations described above, we can write,
 \begin{align}
 i \hbar \frac{\partial\chi}{\partial t} = \Big[e^{0}_{(0)}\Big]^{-1} \Bigg(
 - \hbar\Big[e^{0}_{(0)}\Big] A_{0} + \frac{i \hbar}{4} \Big[e^{0}_{(0)}\Big] \frac{ \partial \ln (-g)}{\partial t}
 - i\hbar \alpha^{(a)}e^{i}_{(a)}\Bigg[ - \frac{c}{4}\frac{\partial \ln (-g) }{\partial x^{i}}\nonumber\\
 +  c \frac{\partial}{\partial x^{i}}
 - \frac{c}{2}  \frac{\partial \ln e^{0}_{(0)}}{\partial x^{i}}\Bigg]
 -\frac{i\hbar}{2}\alpha^{(a)} c \frac{\partial e^{i}_{(a)}}{\partial x^{i}}
 - \hbar \alpha^{(a)}e^{i}_{(a)}A_{i} - \frac{i \hbar}{2}\alpha^{(a)}e^{\rho}_{(a)}\Gamma^{\mu}_{\mu\rho} + m c^2 \beta \Bigg)\chi
 \label{Schrodinger equation 2}
 \end{align}
 Now using $e^{0}_{(i)}=0$ (which will not make any lose of generalization as the number of independent vielbeins in the metric is 
 less than the total number of vielbeins---see ref.~\cite{oliveira} for details) and the properties in  eqs.~(\ref{conn1}), (\ref{conn2})
 we can show that second, third, and eighth terms of the above equation
 will cancel with each other. Finally we can write,
 \begin{align}\label{schrohamcur}
   i \hbar \frac{\partial\chi}{\partial t} = & \Big[e^{0}_{(0)}\Big]^{-1} \Bigg(
    - \hbar\Big[e^{0}_{(0)}\Big] A_{0} - i\hbar c ~ \alpha^{(a)}e^{i}_{(a)}\Bigg[
  \frac{\partial}{\partial x^{i}}
 - \frac{1}{2}  \frac{\partial \ln e^{0}_{(0)}}{\partial x^{i}}\Bigg] \nonumber\\ 
& -\frac{i\hbar}{2}\alpha^{(a)} c \frac{\partial e^{i}_{(a)}}{\partial x^{i}}
 - \hbar \alpha^{(a)}e^{i}_{(a)}A_{i} + m c^2 \beta \Bigg)\chi \nonumber
 \end{align}
 \begin{align}
 \Rightarrow i \hbar \frac{\partial\chi}{\partial t} =  
 - \hbar A_{0}  \chi - i\hbar c ~ \alpha^{(a)}  \frac{e^{i}_{(a)}}{e^0_{(0)}} \frac{\partial  \chi}{\partial x^{i}}
 - \frac{i \hbar c}{2} \alpha^{(a)}  \frac{\partial}{\partial x^{i}} \Bigg[\frac{e^i_{(a)}}{e^0_{(0)}}\Bigg] \chi
 - \hbar \alpha^{(a)}\Bigg[\frac{e^i_{(a)}}{e^0_{(0)}}\Bigg] A_{i} \chi + \beta \frac{m c^2 }{e^0_{(0)}} \chi~.
 \end{align}
 So in operator form the above eq.~(\ref{schrohamcur}) can be expressed as:
 \begin{align}
 H =  - \hbar \sigma_0 \otimes A_{0} + c ~ \alpha^{(a)} \otimes \Bigg[\frac{e^i_{(a)}}{e^0_{(0)}}\Bigg] \hat{p}_i
 - \frac{i \hbar c}{2} \alpha^{(a)} \otimes  \frac{\partial}{\partial x^{i}} \Bigg[\frac{e^i_{(a)}}{e^0_{(0)}}\Bigg]
 - \hbar \alpha^{(a)} \otimes \Bigg[\frac{e^i_{(a)}}{e^0_{(0)}}\Bigg] A_{i} + \beta \otimes \frac{m c^2 }{e^0_{(0)}}.
 \end{align}
 For nonabelian potentials we can directly increase the dimension of the spin Hilbert space, and we have to 
 replace $A_\mu$ by $\sum\limits_q A_{\mu q} \Lambda^q$. The terms $A_{\mu 0}$ now correspond to the abelian potentials
 and other correspond to the nonabelian parts.   
 
 
\section{Calculating the explicit form of single particle evolution operator} \label{modoper}

The modified evolution operator for inhomogeneous SS-DQW, can be written in coin basis as
   \begin{align}\label{unisnext}
  \mathscr{U}(t, \delta t) =  \ket{\uparrow}\bra{\uparrow} \otimes \mathscr{U}_{00}(t, \delta t) 
 + \ket{\uparrow}\bra{\downarrow} \otimes \mathscr{U}_{01}(t, \delta t)  \nonumber\\
 +  \ket{\downarrow}\bra{\uparrow} \otimes \mathscr{U}_{10}(t, \delta t) 
 +  \ket{\downarrow}\bra{\downarrow} \otimes \mathscr{U}_{11}(t, \delta t),
\end{align} where forms of the elements of the evolution operators in terms of the unmodified 
SS-DQW evolution operator element can be written as
\begin{align}\label{unisnext1}
\mathscr{U}_{00}(t, \delta t) = U^\dagger_{00}(t,0) U_{00}(t,\delta t) + U^\dagger_{10}(t,0) U_{10}(t, \delta t), \nonumber\\
\mathscr{U}_{01}(t, \delta t) = U^\dagger_{00}(t,0) U_{01}(t,\delta t) + U^\dagger_{10}(t,0) U_{11}(t, \delta t) ,\nonumber\\
\mathscr{U}_{10}(t, \delta t) =  U^\dagger_{01}(t,0) U_{00}(t,\delta t) + U^\dagger_{11}(t,0) U_{10}(t, \delta t), \nonumber\\
\mathscr{U}_{11}(t, \delta t) =  U^\dagger_{01}(t,0) U_{01}(t, \delta t) + U^\dagger_{11}(t,0) U_{11}(t, \delta t).
\end{align}
Next we are going to use the property of positional transition operators: \\
$\sum\limits_x \ket{x \pm a}\bra{x} = \exp\bigg( - \frac{ \pm i p a }{\hbar} \bigg)$, where 
$p$ is the momentum operator which is regarded as the generator of positional translation. 
  \begin{itemize}
   \item The first-row first-column term of SS-DQW evolution operator in coin-basis
\begin{align}
 U_{00}(t,\delta t) =  \sum_x e^{-i [\theta^0_1(x,t,\delta t) + \theta^0_2(x,t,\delta t)]}  F_2(x,t,\delta t)F_1(x,t,\delta t) \ket{x+a}\bra{x}  \nonumber\\
- e^{-i [\theta^0_1(x,t,\delta t) + \theta^0_2(x-a,t,\delta t)]} G_2(x-a,t,\delta t)G_1^*(x,t,\delta t) \ket{x}\bra{x} \nonumber\\
= \sum_x e^{- i [\theta^0_1(x-a,t,\delta t) + \theta^0_2(x-a,t,\delta t)]} F_2(x-a,t,\delta t)
F_1(x-a,t,\delta t) \ket{x} \bra{x} e^{-\frac{i \hat{p} a}{\hbar}} \nonumber\\
-  e^{ - i [\theta^0_1(x,t,\delta t) + \theta^0_2(x-a,t,\delta t)]} G_2(x-a,t,\delta t)G_1^*(x,t,\delta t) \ket{x}\bra{x}.
\end{align}
\item The first-row second-column term of SS-DQW evolution operator in coin-basis
\begin{align}
U_{01}(t,\delta t) =  \sum_x  e^{- i [\theta^0_1(x,t,\delta t) + \theta^0_2(x,t,\delta t)]} F_2(x,t,\delta t) G_1(x,t,\delta t) \ket{x+a}\bra{x} \nonumber\\ 
+  e^{ - i [\theta^0_1(x,t,\delta t) + \theta^0_2(x-a,t,\delta t)]} G_2(x-a,t,\delta t)F_1^*(x,t,\delta t) \ket{x}\bra{x} \nonumber\\
= \sum_x  e^{- i [\theta^0_1(x-a,t,\delta t) + \theta^0_2(x-a,t,\delta t)]} F_2(x-a,t,\delta t) G_1(x-a,t,\delta t) 
\ket{x}\bra{x} e^{\frac{-i\hat{p}a}{\hbar}} \nonumber\\
+  e^{ - i [\theta^0_1(x,t,\delta t) + \theta^0_2(x-a,t,\delta t)]} G_2(x-a,t,\delta t)F_1^*(x,t,\delta t) \ket{x}\bra{x}.
\end{align}
\item The second-row first-column term of SS-DQW evolution operator in coin-basis
\begin{align}
U_{10}(t,\delta t) =  \sum_x -  e^{- i [\theta^0_1(x,t,\delta t) + \theta^0_2(x,t,\delta t)]} G_2^*(x,t,\delta t) F_1(x,t,\delta t) \ket{x}\bra{x}\nonumber\\
-  e^{- i [\theta^0_1(x,t,\delta t) + \theta^0_2(x-a,t,\delta t)]} F^*_2(x-a,t,\delta t) G_1^*(x,t,\delta t)\ket{x-a}\bra{x} \nonumber\\
= \sum_x -  e^{- i [\theta^0_1(x,t,\delta t) + \theta^0_2(x,t,\delta t)]} G_2^*(x,t,\delta t) F_1(x,t,\delta t) \ket{x}\bra{x}\nonumber\\
-  e^{- i [\theta^0_1(x+a,t,\delta t) + \theta^0_2(x,t,\delta t)]} 
F^*_2(x,t,\delta t) G_1^*(x+a,t,\delta t)\ket{x}\bra{x} e^{\frac{i \hat{p} a}{\hbar}}. 
\end{align} 
\item The second-row second-column term of SS-DQW evolution operator in coin-basis
\begin{align}
U_{11}(t,\delta t) =  \sum_x -  e^{- i [\theta^0_1(x,t,\delta t) + \theta^0_2(x,t,\delta t)]} G_2^*(x,t,\delta t)G_1(x,t,\delta t) \ket{x}\bra{x}\nonumber\\
+  e^{-i [\theta^0_1(x,t,\delta t) + \theta^0_2(x-a,t,\delta t)]} F^*_2(x-a,t,\delta t)F^*_1(x,t,\delta t)\ket{x-a}\bra{x}\nonumber\\
= \sum_x -  e^{-i [\theta^0_1(x,t,\delta t) + \theta^0_2(x,t,\delta t)]} G_2^*(x,t,\delta t)G_1(x,t,\delta t) \ket{x}\bra{x} \nonumber\\
+  e^{- i [\theta^0_1(x+a,t,\delta t) + \theta^0_2(x,t,\delta t)]} F^*_2(x,t,\delta t)F^*_1(x+a,t,\delta t)\ket{x}\bra{x}e^{\frac{i \hat{p} a}{\hbar}}.
\end{align}
\end{itemize}

\subsubsection*{The first-row first-column term of our modified evolution operator in coin-basis}
\begin{multline}
 \mathscr{U}_{00}(t,\delta t) = \sum_x e^{i [\theta^0_1(x,t,0) + \theta^0_2(x,t,0)]} \Big[F^*_2(x,t,0) F^*_1(x,t,0) -  G^*_2(x,t,0)G_1(x,t,0) \Big]\\
\times \Big[ e^{- i [\theta^0_1(x-a,t,\delta t) + \theta^0_2(x-a,t,\delta t)]} F_2(x-a,t,\delta t)F_1(x-a,t,\delta t) 
\ket{x} \bra{x} e^{-\frac{i p a}{\hbar}} \\
- e^{- i [\theta^0_1(x,t,\delta t) + \theta^0_2(x-a,t,\delta t)]} G_2(x-a,t,\delta t)G_1^*(x,t,\delta t) \ket{x}\bra{x} \Big]\\
+ \sum_x -  e^{ i [\theta^0_1(x,t,0) + \theta^0_2(x,t,0)]} \Big[ G_2(x,t,0) F^*_1(x,t,0) + F_2(x,t,0) G_1(x,t,0) \Big]\\
 \times \Big[- e^{- i [\theta^0_1(x,t,\delta t) + \theta^0_2(x,t,\delta t)]} G_2^*(x,t,\delta t) F_1(x,t,\delta t) \ket{x}\bra{x} \\
 - e^{-i [\theta^0_1(x+a,t,\delta t) + \theta^0_2(x,t,\delta t)]} F^*_2(x,t,\delta t) G_1^*(x+a,t,\delta t)\ket{x}\bra{x} e^{\frac{i p a}{\hbar}} \Big]
\end{multline}
$\Rightarrow$
\begin{multline}
 \mathscr{U}_{00}(t,\delta t) - \sum_x \ket{x}\bra{x} =  
  - \frac{ i a}{\hbar}\sum_x \Big[ |F_2(x,t,0)|^2 |F_1(x,t,0)|^2 - |F_2(x,t,0)|^2 |G_1(x,t,0)|^2 \\
   - 2 \Re\{G_2^*(x,t,0) F_1(x,t,0) F_2(x,t,0) G_1(x,t,0)\} \Big] \ket{x}\bra{x}~p \nonumber\end{multline}
\begin{multline}
  +  \sum_x  \bigg\{ -i \delta t [\vartheta^0_1(x,t,0) + \vartheta^0_2(x,t,0)] + \delta t \Big[ F_1^*(x,t,0) f_1(x,t,0)\nonumber\\ + g_1^*(x,t,0) G_1(x,t,0) 
  + G_2^*(x,t,0) g_2(x,t,0) + F_2(x,t,0) f_2^*(x,t,0) \Big] \nonumber\\
  + 2 i  \delta t \Im\Big[ f_2(x,t,0) F_2^*(x,t,0) |F_1(x,t,0)|^2 - f_2(x,t,0) G_1(x,t,0) F_1(x,t,0) G_2^*(x,t,0) \nonumber\\
  + g_2^*(x,t,0) G_2(x,t,0) |F_1(x,t,0)|^2 
  + g_2^*(x,t,0) F_2(x,t,0) F_1(x,t,0) G_1(x,t,0)\Big]  \nonumber\end{multline}
\begin{multline}
  + a \partial_x F_2(x,t,0) \Big[F_1(x,t,0) G_1(x,t,0) G_2^*(x,t,0) - |F_1(x,t,0)|^2 F_2^*(x,t,0) \Big]\\ 
  + a \partial_x F_1(x,t,0) \Big[F_2(x,t,0) G_1(x,t,0) G_2^*(x,t,0)  - |F_2(x,t,0)|^2 F_1^*(x,t,0) \Big]\\
 + a \partial_x G_2(x,t,0) \Big[ F_2^*(x,t,0) F_1^*(x,t,0) G_1^*(x,t,0) - |G_1(x,t,0)|^2 G_2^*(x,t,0) \Big] \\
  + a \partial_x G_1^*(x,t,0) \Big[ G_2(x,t,0) F_1^*(x,t,0) F_2^*(x,t,0) + G_1(x,t,0) |F_2(x,t,0)|^2 \Big] \\
  + i a \partial_x \theta^0_1(x,t,0) \Big( |F_2(x,t,0)|^2 |F_1(x,t,0)|^2  \\
  - |F_2(x,t,0)|^2|G_1(x,t,0)|^2 - 2 \Re[F_2(x,t,0) F_1(x,t,0) G_2^*(x,t,0) G_1(x,t,0)] \Big) \\
  + i a \partial_x \theta^0_2(x,t,0) \Big( |F_2(x,t,0)|^2 |F_1(x,t,0)|^2 + |G_2(x,t,0)|^2 |G_1(x,t,0)|^2 \\
  - 2 \Re[F_2(x,t,0) F_1(x,t,0) G_2^*(x,t,0) G_1(x,t,0)]\Big) \bigg\} \ket{x}\bra{x} + \mathcal{O}(\delta t^2)~.
 \end{multline}
\subsubsection*{The first-row second-column term of our modified evolution operator in coin-basis}
\vspace{-2cm}

\begin{multline}
 \mathscr{U}_{01}(t,\delta t) = \sum_x  e^{i [\theta^0_1(x,t,0) + \theta^0_2(x,t,0)]} \bigg[ F_2^*(x,t,0) F_1^*(x,t,0)  -  G^*_2(x,t,0)G_1(x,t,0)\bigg]\\
 \bigg[ e^{-i [\theta^0_1(x-a,t,\delta t) + \theta^0_2(x-a,t,\delta t)]} F_2(x-a,t,\delta t) G_1(x-a,t,\delta t) \ket{x}\bra{x} e^{\frac{-i p a}{\hbar}} \\
+  e^{-i [\theta^0_1(x,t,\delta t) + \theta^0_2(x-a,t,\delta t)]} G_2(x-a,t,\delta t)F_1^*(x,t,\delta t) \ket{x}\bra{x} \bigg]\\
+ \sum_x e^{i [\theta^0_1(x,t,0) + \theta^0_2(x,t,0)]} \bigg[- G_2(x,t,0) F^*_1(x,t,0) - F_2(x,t,0) G_1(x,t,0)\bigg]\\
\bigg[ -  e^{-i [\theta^0_1(x,t,\delta t) + \theta^0_2(x,t,\delta t)]} G_2^*(x,t,\delta t)G_1(x,t,\delta t) \ket{x}\bra{x} \\
+ e^{-i [\theta^0_1(x+a,t,\delta t) + \theta^0_2(x,t,\delta t)]} F^*_2(x,t,\delta t)F^*_1(x+a,t,\delta t)\ket{x}\bra{x}e^{\frac{i p a}{\hbar}} \bigg] 
\end{multline}
$\Rightarrow$
\begin{multline}
  \mathscr{U}_{01}(t, \delta t) = \sum_x - \frac{i a}{\hbar}
  \big[ 2 |F_2(x,t,0)|^2 G_1(x,t,0) F_1^*(x,t,0)  \nonumber\\
  - F_2(x,t,0) G_2^*(x,t,0) [G_1(x,t,0)]^2
  + F_2^*(x,t,0) [ F_1^*(x,t,0)]^2 G_2(x,t,0)  \big] \ket{x}\bra{x}~p \end{multline}
\begin{multline}
  + \sum_x \bigg\{ - a \partial_x G_1(x,t,0) \Big[|F_2(x,t,0)|^2 F_1^*(x,t,0)
      - G_2^*(x,t,0) G_1(x,t,0) F_2(x,t,0) \Big]\nonumber\\
  -  a \partial_x F_2(x,t,0) \Big[ F_2^*(x,t,0) F_1^*(x,t,0) G_1(x,t,0) - G_2^*(x,t,0) [G_1(x,t,0)]^2  \Big] \\
   - a \partial_x G_2(x,t,0) \Big[ F_2^*(x,t,0) [F_1^*(x,t,0)]^2 - F_1^*(x,t,0) G_1(x,t,0) G_2^*(x,t,0) \Big] \\
   - a \partial_x F_1^*(x,t,0) \Big[G_2(x,t,0) F_1^*(x,t,0) F_2^*(x,t,0) + G_1(x,t,0) |F_2(x,t,0)|^2 \Big] \\
   + i a \partial_x \theta^0_1(x,t,0) \Big[ 2 |F_2(x,t,0)|^2 G_1(x,t,0) F_1^*(x,t,0) + G_2(x,t,0) F_2^*(x,t,0) [F_1^*(x,t,0)]^2 \nonumber\\
   - G_2^*(x,t,0) F_2(x,t,0) [G_1(x,t,0)]^2 \Big]  + i a \partial_x \theta^0_2(x,t,0) \Big[ G_1(x,t,0) F_1^*(x,t,0) \nonumber\\ 
   \big[ |F_2(x,t,0)|^2 - |G_2(x,t,0)|^2 \big] - F_2(x,t,0) G_2^*(x,t,0) [G_1(x,t,0)]^2  \nonumber\\
  + G_2(x,t,0) F_2^*(x,t,0) [F_1^*(x,t,0)]^2 \Big]\end{multline}
\begin{multline} 
 + \delta t\Big[ g_1(x,t,0) F_1^*(x,t,0) - f_1^*(x,t,0) G_1(x,t,0)\Big] \\
  + \delta t g_2^*(x,t,0) \Big[ F_2(x,t,0) [G_1(x,t,0)]^2 + F_1^*(x,t,0) G_1(x,t,0) G_2(x,t,0) \Big]\\
  - \delta t f_2^*(x,t,0) [G_2(x,t,0) [F_1^*(x,t,0)]^2 + G_1(x,t,0) F_2(x,t,0) F_1^*(x,t,0)] \\
  +\delta t g_2(x,t,0) \Big[[F_1^*(x,t,0)]^2 F_2^*(x,t,0) - F_1^*(x,t,0) G_1(x,t,0) G_2^*(x,t,0) \Big] \\
  + \delta t f_2(x,t,0) \Big[ G_1(x,t,0) F_1^*(x,t,0) F_2^*(x,t,0) - [G_1(x,t,0)]^2 G_2^*(x,t,0) \Big] \bigg\} \ket{x}\bra{x} \\
  + \mathcal{O}(\delta t^2)~.
 \end{multline}
\subsubsection*{The second-row first-column term of our modified evolution operator in coin-basis}
\vspace{-2cm}

\begin{multline}
 \mathscr{U}_{10}(t, \delta t) 
 = \sum_x e^{i [\theta^0_1(x,t,0) + \theta^0_2(x,t,0)]} \bigg[ F^*_2(x,t,0) G^*_1(x,t,0) +  G^*_2(x,t,0)F_1(x,t,0) \bigg] \\
 \bigg[ e^{-i [\theta^0_1(x-a,t,\delta t) + \theta^0_2(x-a,t,\delta t)]} F_2(x-a,t,\delta t) F_1(x-a,t,\delta t) 
 \ket{x}\bra{x} e^{-\frac{i \hat{p} a}{\hbar}}\\
- e^{-i [\theta^0_1(x,t,\delta t) 
+ \theta^0_2(x-a,t,\delta t)]} G_2(x-a,t,\delta t)G_1^*(x,t,\delta t) \ket{x}\bra{x} \bigg] \\
+ \sum_x e^{i [\theta^0_1(x,t,0) + \theta^0_2(x,t,0)]} 
\bigg[- G_2(x,t,0)G^*_1(x,t,0) +  F_2(x,t,0) F_1(x,t,0) \bigg] \\
\bigg[ - e^{-i [\theta^0_1(x,t,\delta t) + \theta^0_2(x,t,\delta t)]} G_2^*(x,t,\delta t) F_1(x,t,\delta t) \ket{x}\bra{x}\\
- e^{-i [\theta^0_1(x+a,t,\delta t) + \theta^0_2(x,t,\delta t)]}
F^*_2(x,t,\delta t) G_1^*(x+a,t,\delta t)\ket{x}\bra{x} e^{\frac{i \hat{p} a}{\hbar}}\bigg] 
\end{multline}
\begin{multline}
\Rightarrow 
  \mathscr{U}_{10}(t, \delta t) = \sum_x - \frac{i a}{\hbar}
  \big[ 2 |F_2(x,t,0)|^2 G^*_1(x,t,0) F_1(x,t,0) \nonumber\\
  - F^*_2(x,t,0) G_2(x,t,0) [G_1^*(x,t,0)]^2  + F_2(x,t,0) [ F_1(x,t,0)]^2 G^*_2(x,t,0)  \big] \ket{x}\bra{x}~\hat{p}
\\
    + \delta t \Big[f_1(x,t,0) G_1^*(x,t,0) - g_1^*(x,t,0) F_1(x,t,0) \Big] \nonumber\\
    + \delta t f_2(x,t,0) \Big[F_1(x,t,0) F_2^*(x,t,0) G_1^*(x,t,0) + [F_1(x,t,0)]^2 G_2^*(x,t,0)  \Big] \nonumber\\
    - \delta t g_2(x,t,0) \Big[ F_2^*(x,t,0) [G_1^*(x,t,0)]^2 + F_1(x,t,0) G_1^*(x,t,0) G_2^*(x,t,0) \Big] \nonumber\\
   - \delta t g_2^*(x,t,0) \Big[ F_2(x,t,0) [F_1(x,t,0)]^2 - F_1(x,t,0) G_2(x,t,0) G_1^*(x,t,0) \Big] \nonumber\\
   - \delta t f_2^*(x,t,0) \Big[ G_1^*(x,t,0) F_1(x,t,0) F_2(x,t,0) - G_2(x,t,0) [G_1^*(x,t,0)]^2  \Big]\end{multline}
\begin{multline}
  + \sum_x \bigg\{- a \partial_x F_2(x,t,0) \Big[ F_1(x,t,0) F_2^*(x,t,0) G_1^*(x,t,0) + [F_1(x,t,0)]^2 G_2^*(x,t,0)  \Big]\\
   - a \partial_x F_1(x,t,0) \Big[ |F_2(x,t,0)|^2 G_1^*(x,t,0) + F_2(x,t,0) F_1(x,t,0) G_2^*(x,t,0)\Big] \\
   + a \partial_x G_2(x,t,0) \Big[ F_2^*(x,t,0) [G_1^*(x,t,0)]^2 + G_1^*(x,t,0) G_2^*(x,t,0) F_1(x,t,0) \Big]  \\
    - a \partial_x G_1^*(x,t,0) \Big[ |F_2(x,t,0)|^2 F_1(x,t,0) - F_2^*(x,t,0) G_2(x,t,0) G_1^*(x,t,0) \Big] \\
    + i a \partial_x \theta^0_1(x,t,0) \Big[ 2 |F_2(x,t,0)|^2 F_1(x,t,0) G_1^*(x,t,0) + [F_1(x,t,0)]^2 F_2(x,t,0) G_2^*(x,t,0)  \\
    - F_2^*(x,t,0) [G_1^*(x,t,0)]^2 G_2(x,t,0) \Big] 
   + i a \partial_x \theta^0_2(x,t,0) \Big [ |F_2(x,t,0)|^2 F_1(x,t,0) G_1^*(x,t,0) \\
    - |G_2(x,t,0)|^2 F_1(x,t,0) G_1^*(x,t,0) + [F_1(x,t,0)]^2 F_2(x,t,0) G_2^*(x,t,0) \\
   - G_2(x,t,0) [G_1^*(x,t,0)]^2 F_2^*(x,t,0) \Big] \bigg\} \ket{x}\bra{x} + \mathcal{O}(\delta t^2)~.
 \end{multline}
 \subsubsection*{The second-row second-column term of our modified evolution operator in coin-basis}
 \vspace{-3cm}
 
    \begin{multline}
 \mathscr{U}_{11}(t, \delta t)
 = \sum_x  e^{i [\theta^0_1(x,t,0) + \theta^0_2(x,t,0)]} \bigg[ F^*_2(x,t,0) G^*_1(x,t,0) +  G^*_2(x,t,0)F_1(x,t,0)\bigg] \\
\bigg[ e^{-i [\theta^0_1(x-a,t,\delta t) + \theta^0_2(x-a,t,\delta t)]} F_2(x-a,t,\delta t) G_1(x-a,t,\delta t)\\
\ket{x}\bra{x} e^{\frac{-i\hat{p}a}{\hbar}} 
+ e^{-i [\theta^0_1(x,t,\delta t) + \theta^0_2(x-a,t,\delta t)]} G_2(x-a,t,\delta t)F_1^*(x,t,\delta t) \ket{x}\bra{x}\bigg] \\
+ \sum_x  e^{i [\theta^0_1(x,t,0) + \theta^0_2(x,t,0)]} \bigg[ - G_2(x,t,0)G^*_1(x,t,0) +  F_2(x,t,0)F_1(x,t,0) \bigg] \\
\bigg[ - e^{-i [\theta^0_1(x,t,\delta t) + \theta^0_2(x,t,\delta t)]} G_2^*(x,t,\delta t)G_1(x,t,\delta t) \ket{x}\bra{x}\\
+ e^{-i [\theta^0_1(x+a,t,\delta t) + \theta^0_2(x,t,\delta t)]} F^*_2(x,t,\delta t)F^*_1(x+a,t,\delta t)\ket{x}\bra{x}e^{\frac{i \hat{p} a}{\hbar}} \bigg]
\end{multline}
\begin{multline} \Rightarrow \nonumber
  \mathscr{U}_{11}(t, \delta t) - \sum_x \ket{x}\bra{x}
  = \sum_x \frac{-i a}{\hbar} \big[ |F_2(x,t,0)|^2 |G_1(x,t,0)|^2 - |F_2(x,t,0)|^2 |F_1(x,t,0)|^2 \nonumber\\
  + 2 \Re\{F_1(x,t,0) F_2(x,t,0) G_1(x,t,0) G^*_2(x,t,0) \} \big] \ket{x}\bra{x}~\hat{p} \end{multline}
  \begin{multline}
  + \delta t \Big[ g_1(x,t,0) G_1^*(x,t,0) + F_1(x,t,0) f_1^*(x,t,0)  + g^*_2(x,t,0) G_2(x,t,0)  \nonumber\\
+ f_2(x,t,0) F_2^*(x,t,0) \Big]  + 2 i \delta t \Im \Big[ g_2(x,t,0) F_1^*(x,t,0) F_2^*(x,t,0) G_1^*(x,t,0) \nonumber\\
  + g_2(x,t,0) G_2^*(x,t,0) |F_1(x,t,0)|^2 + f_2(x,t,0) G_1(x,t,0) F_1(x,t,0) G_2^*(x,t,0)\nonumber\\
  - f_2(x,t,0) F_2^*(x,t,0) |F_1(x,t,0)|^2 \Big]
  - i \delta t [\vartheta^0_1(x,t,0) + \vartheta^0_2(x,t,0)]\end{multline}
\begin{multline}
 + \sum_x \bigg\{ - a \partial_x G_2(x,t,0) \Big[ F_1^*(x,t,0) F_2^*(x,t,0) G_1^*(x,t,0) 
 + |F_1(x,t,0)|^2 G_2^*(x,t,0) \Big] \\
  - a \partial_x F_2(x,t,0) \Big[ F_2^*(x,t,0) |G_1(x,t,0)|^2  + G_1(x,t,0) F_1(x,t,0) G_2^*(x,t,0) \Big] \\
  - a \partial_x G_1(x,t,0) \Big[ |F_2(x,t,0)|^2 G_1^*(x,t,0)  + F_1(x,t,0) F_2(x,t,0) G_2^*(x,t,0) \Big] \\
  + a \partial_x F_1^*(x,t,0) \Big[ F_1(x,t,0) |F_2(x,t,0)|^2 - F_2^*(x,t,0) G_2(x,t,0) G_1^*(x,t,0) \Big] \\
+ i a \partial_x \theta^0_1(x,t,0) \Big[ |F_2(x,t,0)|^2 |G_1(x,t,0)|^2 - |F_2(x,t,0)|^2 |F_1(x,t,0)|^2\\
  + 2 \Re[F_1(x,t,0) F_2(x,t,0) G_1(x,t,0) G_2^*(x,t,0)] \Big] \\
  + i a \partial_x \theta^0_2(x,t,0) \Big[|F_2(x,t,0)|^2 |G_1(x,t,0)|^2 + |G_2(x,t,0)|^2 |F_1(x,t,0)|^2 \\
  +   2 \Re[F_1(x,t,0) F_2(x,t,0) G_1(x,t,0) G_2^*(x,t,0)]  \Big] \bigg\} \ket{x}\bra{x} + \mathcal{O}(\delta t^2)~. 
 \end{multline}
  

  \section{Calculating the operator terms of the effective Hamiltonian for the single particle}\label{hamcalcu}
   
   Here we will use the definition of the effective Hamiltonian $\mathscr{H}$. From 
\begin{align}
\mathscr{U}(t, \delta t)  =  \exp \bigg(- i \frac{\mathscr{H}(t)\delta t}{\hbar} \bigg)
\end{align}
we can write using the Taylor series expansion in $\delta t$,
\begin{align}
 \mathscr{U}(t, \delta t) = \sigma_0 \otimes \sum_x \ket{x}\bra{x} - i \frac{\mathscr{H}(t)\delta t}{\hbar} + \mathcal{O}(\delta t^2). 
\end{align} Then the effective Hamiltonian can be calculated by the formula:
\begin{multline}\label{diraccuham}
 \mathscr{H}(t) = i \hbar \lim_{\delta t \to 0} \frac{1}{\delta t} \left(\begin{array}{cc}
                                      \mathscr{U}_{00}(t, \delta t) - \sum_x \ket{x}\bra{x} & \mathscr{U}_{01}(t, \delta t)  \\
                                       \mathscr{U}_{10}(t, \delta t)  & \mathscr{U}_{11}(t, \delta t) - \sum_x \ket{x}\bra{x} \\
                                                                  \end{array}\right)\\ \\
   = \sigma_0 \otimes i \hbar \lim_{\delta t \to 0} \frac{1}{2\delta t} \Big(  \mathscr{U}_{00}(t, \delta t) + \mathscr{U}_{11} (t, \delta t)
   - 2\sum_x \ket{x}\bra{x}\Big)\\
   + \sigma_3 \otimes i \hbar \lim_{\delta t \to 0} \frac{1}{2\delta t} \Big(\mathscr{U}_{00}(t, \delta t) - \mathscr{U}_{11}(t, \delta t) \Big) 
   + \sigma_1 \otimes i \hbar \lim_{\delta t \to 0} \frac{1}{2 \delta t} \Big(  \mathscr{U}_{01}(t, \delta t) + \mathscr{U}_{10}(t, \delta t) \Big) \\
   - \sigma_2 \otimes  \hbar \lim_{\delta t \to 0} \frac{1}{2 \delta t} \Big(  \mathscr{U}_{01}(t, \delta t) - \mathscr{U}_{10}(t, \delta t) \Big) \\  \\
   \coloneqq \sum_{r=0}^3  \sigma_r \otimes \sum_x \Xi_r(x,t) \ket{x}\bra{x} + 
   c~\sum_{r=0}^3  \sigma_r \otimes \sum_x \Theta_r(x,t) \ket{x}\bra{x}~p .
                                                               \end{multline}
   The operators $\sum\limits_x \Theta_r(x,t) \ket{x}\bra{x}, \sum\limits_x \Xi_r(x,t)\ket{x}\bra{x}$ are diagonal in the position basis, and they carry the information 
   of the space-time curvature and gauge potential effects.    
To calculate these terms we will use the properties given by the eqs.~(\ref{cond1}), (\ref{cond2}).
From the previous section \ref{modoper} we get the following.
\vspace{-1cm}

\subsubsection*{Coefficient of $\sigma_0$ is proportional to}
 \begin{multline}
  \mathscr{U}_{00}(t, \delta t) + \mathscr{U}_{11}(t, \delta t) - 2\sum_x \ket{x}\bra{x}  =
  \sum_x \bigg\{ 2 i \delta t [\vartheta^0_1(x,t) + \vartheta^0_2(x,t)]+ i a \partial_x \theta^0_2(x,t,0) \\ 
  - a i \Im\Big[F_2^*(x,t,0)\partial_x F_2(x,t,0) +  G_2^*(x,t,0)\partial_x G_2(x,t,0) \Big] \\
  + 2 i a |F_2(x,t,0)|^2 \Im \Big[F_1(x,t,0) \partial_x F_1^*(x,t,0) + G_1(x,t,0) \partial_x G_1^*(x,t,0) \Big] \\
  + 2 i a \Im \Big[ F_1^*(x,t,0) F_2^*(x,t,0) G_2(x,t,0) \partial_x G_1^*(x,t,0) 
  + F_2(x,t,0) \\ G_1(x,t,0) G_2^*(x,t,0) \partial_x F_1(x,t,0)  \Big]  \bigg\}
 \ket{x}\bra{x} + \mathcal{O}(\delta t^2). 
 \end{multline} 
\vspace{1cm}

 \subsubsection*{Coefficient of $\sigma_3$ is proportional to}
 \begin{multline}
  \mathscr{U}_{00}(t, \delta t) - \mathscr{U}_{11}(t, \delta t)  = \\
  - \frac{ 2 i a}{\hbar}\sum_x \Big[ |F_2(x,t,0)|^2 |F_1(x,t,0)|^2 - |F_2(x,t,0)|^2 |G_1(x,t,0)|^2 \nonumber\\ 
  - 2 \Re\{G_2^*(x,t,0) F_1(x,t,0) F_2(x,t,0) G_1(x,t,0)\} \Big] \ket{x}\bra{x}~p \end{multline}
\begin{multline}  
   + \sum_x  \bigg\{ 2 i \delta t \Im \Big[ F_1^*(x,t,0) f_1(x,t,0) \\
   + g_1^*(x,t,0) G_1(x,t,0) 
   + G_2^*(x,t,0) g_2(x,t,0) + F_2(x,t,0) f_2^*(x,t,0) \Big] \\
 + 4 i \delta t \Im\Big[ f_2(x,t,0) F_2^*(x,t,0) |F_1(x,t,0)|^2
- f_2(x,t,0) G_1(x,t,0) F_1(x,t,0) G_2^*(x,t,0) \nonumber\\
 + g_2^*(x,t,0) G_2(x,t,0) |F_1(x,t,0)|^2 + g_2^*(x,t,0) F_2(x,t,0) F_1(x,t,0) G_1(x,t,0)\Big] \end{multline}
\begin{multline}  
  + a \partial_x F_2(x,t,0) \Big[ 2 F_1(x,t,0) G_1(x,t,0) G_2^*(x,t,0) + F_2^*(x,t,0) |G_1(x,t,0)|^2 \\
  - |F_1(x,t,0)|^2 F_2^*(x,t,0) \Big]
  + 2 a |F_2(x,t,0)|^2 \Re \Big[ G_1(x,t,0) \partial_x G_1^*(x,t,0) - F_1(x,t,0) \partial_x F_1^*(x,t,0) \Big] \\
  + a \partial_x G_2(x,t,0) \Big[2 F_2^*(x,t,0) F_1^*(x,t,0) G_1^*(x,t,0)  + |F_1(x,t,0)|^2 G_2^*(x,t,0) \\
  - |G_1(x,t,0)|^2 G_2^*(x,t,0)\Big] 
  + 2 a \Re \Big[ F_2(x,t,0) G_1(x,t,0) G_2^*(x,t,0) \partial_x F_1(x,t,0) \\
  + F_1(x,t,0) F_2(x,t,0) G_2^*(x,t,0) \partial_x G_1(x,t,0) \Big] 
   +  i a \partial_x \theta^0_1(x,t,0)  \Big[ 2 |F_2(x,t,0)|^2 \\
   \big( |F_1(x,t,0)|^2 - |G_1(x,t,0)|^2 \big) 
   -  4 \Re[F_2(x,t,0) F_1(x,t,0) G_1(x,t,0) G_2^*(x,t,0)]\Big] \\
   + i a \partial_x \theta^0_2(x,t,0) \Big[ \big(|G_2(x,t,0)|^2 - |F_2(x,t,0)|^2 \big) \big(|G_1(x,t,0)|^2 
   - |F_1(x,t,0)|^2\big) \\
   - 4 \Re [F_2(x,t,0) F_1(x,t,0) G_1(x,t,0) G_2^*(x,t,0)] \Big] \bigg\} \ket{x}\bra{x} + \mathcal{O}(\delta t^2).
 \end{multline}
\vspace{2.5cm}
 
 \subsubsection*{Coefficient of $\sigma_1$ is proportional to}
   \begin{multline}
  \mathscr{U}_{01}(t, \delta t) + \mathscr{U}_{10}(t, \delta t) =  
  \sum_x - \frac{2 i a}{\hbar} \Re \Big[ 2 |F_2(x,t,0)|^2 G_1(x,t,0) F_1^*(x,t,0) \nonumber\\
 - F_2(x,t,0) G_2^*(x,t,0) [G_1(x,t,0)]^2 + F_2^*(x,t,0) [ F_1^*(x,t,0)]^2 G_2(x,t,0)  \Big] \ket{x}\bra{x}~p 
 \end{multline}\begin{multline}
   + \sum_x \bigg\{ 2 i \delta t \Im \Big[ g_1(x,t,0) F_1^*(x,t,0) - f_1^*(x,t,0) G_1(x,t,0)  \Big]  \nonumber\\
  +  2 i \delta t \Im  \Big[g_2^*(x,t,0) F_2(x,t,0) [G_1(x,t,0)]^2 + g_2^*(x,t,0) F_1^*(x,t,0) G_1(x,t,0) G_2(x,t,0)\nonumber\\
  -  f_2^*(x,t,0) G_2(x,t,0) [F_1^*(x,t,0)]^2 -  f_2^*(x,t,0) G_1(x,t,0) F_2(x,t,0) F_1^*(x,t,0) \nonumber\\
  +  g_2(x,t,0) [F_1^*(x,t,0)]^2 F_2^*(x,t,0) - g_2(x,t,0) F_1^*(x,t,0) G_1(x,t,0) G_2^*(x,t,0) \nonumber\\
  +  f_2(x,t,0) G_1(x,t,0) F_1^*(x,t,0) F_2^*(x,t,0)  - f_2(x,t,0) [G_1(x,t,0)]^2 G_2^*(x,t,0) \Big] \end{multline}
  \begin{multline}
 - 2 a \Re \Big[ F_1^*(x,t,0) |F_2(x,t,0)|^2 \partial_x G_1(x,t,0) - G_2^*(x,t,0) G_1(x,t,0) F_2(x,t,0) \partial_x G_1(x,t,0) \\ 
  + F_1^*(x,t,0) F_2^*(x,t,0) G_2(x,t,0) \partial_x F_1^*(x,t,0) 
  + G_1(x,t,0) |F_2(x,t,0)|^2 \partial_x F_1^*(x,t,0) \Big]\\
  - 2 a  \Re \Big[ F_1^*(x,t,0) G_1(x,t,0)\Big]
 \Big[ F_2^*(x,t,0) \partial_x F_2(x,t,0) - G_2^*(x,t,0) \partial_x G_2(x,t,0) \Big] \\
   - a G_2^*(x,t,0) \partial_x F_2(x,t,0) \big( [F_1(x,t,0)]^2 - [G_1(x,t,0)]^2 \big) \\
   -  a F_2^*(x,t,0) \partial_x G_2(x,t,0) \big( [F_1^*(x,t,0)]^2 - [G_1^*(x,t,0)]^2 \big) \\
   + i a \partial_x \theta^0_1(x,t,0) \Big[ 4 |F_2(x,t,0)|^2 \Re\big(G_1(x,t,0) F_1^*(x,t,0)\big) \\
   - 2 \Re\big(G_2^*(x,t,0) F_2(x,t,0) [G_1(x,t,0)]^2 \big)  + 2 \Re\big( G_2 F_2^* [F_1^*]^2 \big) \Big] \\
   + i a \partial_x \theta^0_2(x,t,0) \Big[ 2 |F_2(x,t,0)|^2 \Re\big( G_1(x,t,0) F_1^*(x,t,0)\big)\\
- 2 |G_2(x,t,0)|^2 \Re\big( G_1(x,t,0) F_1^*(x,t,0)\big) 
   + 2 \Re \big( [F_1(x,t,0)]^2 F_2(x,t,0) G_2^*(x,t,0) \big) \\
- 2 \Re\big( F_2(x,t,0) G_2^*(x,t,0) [G_1(x,t,0)]^2 \big) \Big] \bigg\} \ket{x}\bra{x} + \mathcal{O}(\delta t^2).
 \end{multline}
 \vspace{1.5cm}
 
 \subsubsection*{Coefficient of $\sigma_2$ is proportional to}
  \begin{multline}
  \mathscr{U}_{01}(t, \delta t) - \mathscr{U}_{10}(t, \delta t) =  
  \frac{2 a}{\hbar} \sum_x  \Im \Big[ 2 |F_2(x,t,0)|^2 G_1(x,t,0) F_1^*(x,t,0)\nonumber\\ 
  - F_2(x,t,0) G_2^*(x,t,0) [G_1(x,t,0)]^2 + F_2^*(x,t,0) [ F_1^*(x,t,0)]^2 G_2(x,t,0)  \Big] \ket{x}\bra{x}~p \end{multline}
  \begin{multline} + 2 \delta t \Re \Big[ g_1(x,t,0) F_1^*(x,t,0)\\ - f_1^*(x,t,0) G_1(x,t,0)
  + g_2^*(x,t,0) F_2(x,t,0) [G_1(x,t,0)]^2  \\
  + g_2^*(x,t,0) F_1^*(x,t,0) G_1(x,t,0) G_2(x,t,0)
  - f_2^*(x,t,0) G_2(x,t,0) [F_1^*(x,t,0)]^2 \\
  - f_2^*(x,t,0) G_1(x,t,0) F_2(x,t,0) F_1^*(x,t,0) \nonumber\\
  + g_2(x,t,0) F_2^*(x,t,0) [F_1^*(x,t,0)]^2 
  -  g_2(x,t,0) F_1^*(x,t,0) G_1(x,t,0) G_2^*(x,t,0) \nonumber\\
  + f_2(x,t,0) G_1(x,t,0) F_1^*(x,t,0) F_2^*(x,t,0) 
  - f_2(x,t,0) [G_1(x,t,0)]^2 G_2^*(x,t,0) \Big] \end{multline}
  \begin{multline}
  + \sum_x \bigg\{ - 2 i a \Im\Big[ \partial_x G_1(x,t,0) \big(|F_2(x,t,0)|^2 F_1^*(x,t,0)
  - G_2^*(x,t,0) G_1(x,t,0) F_2(x,t,0) \big)\Big] \\
 - 2 i a \Im \Big[ \partial_x F_1^*(x,t,0) \big( G_2(x,t,0) F_1^*(x,t,0) F_2^*(x,t,0) + G_1(x,t,0) |F_2(x,t,0)|^2 \big) \Big]\\
  - a \partial_x F_2(x,t,0) \Big[2 i F_2^*(x,t,0) \Im\big(F_1^*(x,t,0) G_1(x,t,0) \big) \\
  - G_2^*(x,t,0)  [F_1(x,t,0)]^2 -  G_2^*(x,t,0) [G_1(x,t,0)]^2  \Big] \\
   - a \partial_x G_2(x,t,0) \Big[ - 2 i G_2^*(x,t,0) \Im \big(F_1^*(x,t,0) G_1(x,t,0) \big) \\
   + F_2^*(x,t,0) \big( [F_1^*(x,t,0)]^2 + [G_1^*(x,t,0)]^2 \big) \Big] \\
  - 2 a  \partial_x \theta^0_1(x,t,0) \Big[2 |F_2(x,t,0)|^2 \Im \big(G_1(x,t,0) F_1^*(x,t,0)\big) \\
  + \Im \big( G_2(x,t,0) F_2^*(x,t,0) [F^*_1(x,t,0)]^2 - G_2^*(x,t,0) F_2(x,t,0) [G_1(x,t,0)]^2 \big) \Big] \\
  - 2 a \partial_x \theta^0_2(x,t,0) \Big[ \big( |F_2(x,t,0)|^2 - |G_2(x,t,0)|^2 \big) \Im \big(G_1(x,t,0) F_1^*(x,t,0) \big) \\
  + \Im \big( G_2(x,t,0) F_2^*(x,t,0) [F_1^*(x,t,0)]^2 \\
  - F_2(x,t,0) G_2^*(x,t,0) [G_1(x,t,0)]^2 \big) \Big] \bigg\} \ket{x}\bra{x} + \mathcal{O}(\delta t^2).
 \end{multline}

 \subsection{Explicit forms of the single Hamiltonian terms}

   The explicit form of the single-particle Hamiltonian terms defined in eq.~(\ref{diraccuham}) are as the following.
  \begin{multline}
  \Theta_3(x,t) = -   \Big[ |F_2(x,t,0)|^2 |G_1(x,t,0)|^2 - |F_2(x,t,0)|^2 |F_1(x,t,0)|^2 \\
  + 2 \Re\{F_1(x,t,0) F_2(x,t,0) G_1(x,t,0) G^*_2(x,t,0) \} \Big].
 \end{multline}
 \begin{align}
  \Theta_0(x,t) = 0.
 \end{align}
 \begin{multline}
  \Theta_1(x,t) =   \Re\Big[ 2 |F_2(x,t,0)|^2 G^*_1(x,t,0) F_1(x,t,0) - F^*_2(x,t,0) G_2(x,t,0) [G_1^*(x,t,0)]^2 \\
  + F_2(x,t,0) [ F_1(x,t,0)]^2 G^*_2(x,t,0)  \Big].
\end{multline}
 \begin{multline}
  \Theta_2(x,t) =   \Im \Big[ 2 |F_2(x,t,0)|^2 G^*_1(x,t,0) F_1(x,t,0) - F^*_2(x,t,0) G_2(x,t,0) [G_1^*(x,t,0)]^2 \\
  + F_2(x,t,0) [ F_1(x,t,0)]^2 G^*_2(x,t,0)  \Big] .
\end{multline}
 
\begin{align}
 \Xi_0(x,t) = \hbar [\vartheta^0_1(x,t) + \vartheta^0_2(x,t)] - \frac{\hbar c}{2} \partial_x \theta^0_2(x,t,0) \nonumber\\
 + \frac{\hbar c}{2} \Im\Big[F_2^*(x,t,0)\partial_x F_2(x,t,0) +  G_2^*(x,t,0)\partial_x G_2(x,t,0)\Big] \nonumber\\
 - \hbar c |F_2(x,t,0)|^2  \Im \Big[ F_1(x,t,0) \partial_x F_1^*(x,t,0) + G_1(x,t,0) \partial_x G_1^*(x,t,0) \Big]  \nonumber\\
 - \hbar c \Im \Big[ F_1^*(x,t,0) F_2^*(x,t,0) G_2(x,t,0) \partial_x G_1^*(x,t,0)\nonumber\\
 + F_2(x,t,0) G_1(x,t,0) G_2^*(x,t,0) \partial_x F_1(x,t,0)\Big].
\end{align} 
 \begin{multline}
  \Xi_3(x,t)  =  - \hbar \Im \Big[ F_1^*(x,t,0) f_1(x,t) + g_1^*(x,t) G_1(x,t,0) + G_2^*(x,t,0) g_2(x,t) + F_2(x,t,0) f_2^*(x,t) \Big] \\
  - 2 \hbar \Im\Big[ f_2(x,t) F_2^*(x,t,0) |F_1(x,t,0)|^2 - f_2(x,t) G_1(x,t,0) F_1(x,t,0) G_2^*(x,t,0) \\
  + g_2^*(x,t) G_2(x,t,0) |F_1(x,t,0)|^2 + g_2^*(x,t) F_2(x,t,0) F_1(x,t,0) G_1(x,t,0)\Big]\\
  + \frac{i \hbar c}{2} \partial_x F_2(x,t,0) \Big[ 2 F_1(x,t,0) G_1(x,t,0) G_2^*(x,t,0) + F_2^*(x,t,0) |G_1(x,t,0)|^2  \\
  - |F_1(x,t,0)|^2 F_2^*(x,t,0) \Big] + i \hbar c |F_2(x,t,0)|^2 \Re \Big[ G_1(x,t,0) \partial_x G_1^*(x,t,0)\\ - F_1(x,t,0) \partial_x F_1^*(x,t,0) \Big]
  + \frac{i \hbar c}{2} \partial_x G_2(x,t,0) \Big[2 F_2^*(x,t,0) F_1^*(x,t,0) G_1^*(x,t,0) \\ 
  + |F_1(x,t,0)|^2 G_2^*(x,t,0) - |G_1(x,t,0)|^2 G_2^*(x,t,0)\Big]
  + i \hbar c \Re \Big[ F_2(x,t,0)\\ G_1(x,t,0) G_2^*(x,t,0) \partial_x F_1(x,t,0) + F_1(x,t,0) F_2(x,t,0) G_2^*(x,t,0) \partial_x G_1(x,t,0) \Big]\\
   - \frac{\hbar c}{2} \partial_x \theta^0_1(x,t,0)  \Big[ 2 |F_2(x,t,0)|^2 \big(|F_1(x,t,0)|^2 - |G_1(x,t,0)|^2 \big) \\ 
   -  4 \Re[F_2(x,t,0) F_1(x,t,0) G_1(x,t,0) G_2^*(x,t,0)]\Big] - \frac{\hbar c}{2}\partial_x \theta^0_2(x,t,0) \\
   \Big[ \big(|G_2(x,t,0)|^2 - |F_2(x,t,0)|^2 \big)] \big(|G_1(x,t,0)|^2 - |F_1(x,t,0)|^2\big)  \\
   - 4 \Re [F_2(x,t,0) F_1(x,t,0) G_1(x,t,0) G_2^*(x,t,0)] \Big].
 \end{multline} 
 \begin{multline}
  \Xi_1(x,t) = 
  - \hbar \Im\Big[g_2^*(x,t) F_2(x,t,0) [G_1(x,t,0)]^2 + g_2^*(x,t) F_1^*(x,t,0) G_1(x,t,0) G_2(x,t,0) \\ 
  - f_2^*(x,t) G_2(x,t,0) [F_1^*(x,t,0)]^2 -  f_2^*(x,t) G_1(x,t,0) F_2(x,t,0) F_1^*(x,t,0) \\
  +  g_2(x,t)[F_1^*(x,t,0)]^2 F_2^*(x,t,0) - g_2(x,t) F_1^*(x,t,0) G_1(x,t,0) G_2^*(x,t,0) \\
  +  f_2(x,t) G_1(x,t,0) F_1^*(x,t,0) F_2^*(x,t,0) - f_2(x,t) [G_1(x,t,0)]^2 G_2^*(x,t,0) \Big]\\
  - i \hbar c \Re\Big[ F_1^*(x,t,0) |F_2(x,t,0)|^2 \partial_x G_1(x,t,0) - G_2^*(x,t,0) G_1(x,t,0)\\
   F_2(x,t,0) \partial_x G_1(x,t,0) + F_1^*(x,t,0) F_2^*(x,t,0) G_2(x,t,0) \partial_x F_1^*(x,t,0)\\
  + G_1(x,t,0) |F_2(x,t,0)|^2 \partial_x F_1^*(x,t,0) \Big] - i \hbar c  \Re \Big[ F_1^*(x,t,0) G_1(x,t,0)\Big] \\
  \Big[ F_2^*(x,t,0)\partial_x F_2(x,t,0) - G_2^*(x,t,0) \partial_x G_2(x,t,0) \Big] - \frac{i \hbar c}{2} G_2^*(x,t,0) \\ 
   \partial_x F_2(x,t,0) \big( [F_1(x,t,0)]^2 - [G_1(x,t,0)]^2 \big) -  \frac{i \hbar c}{2} F_2^*(x,t,0) \partial_x G_2(x,t,0) \\
   \big( [F_1^*(x,t,0)]^2 - [G_1^*(x,t,0)]^2 \big)  - \frac{\hbar c}{2} \partial_x \theta^0_1(x,t,0) \Big[4|F_2(x,t,0)|^2 \\
   \Re\big(G_1(x,t,0) F_1^*(x,t,0)\big) - 2 \Re\big(G_2^*(x,t,0) F_2(x,t,0) [G_1(x,t,0)]^2 \big) \\
  + 2 \Re\big( G_2(x,t,0) F_2^*(x,t,0) [F_1^*(x,t,0)]^2 \big) \Big]  - \frac{\hbar c}{2} \partial_x \theta^0_2(x,t,0) \Big[ 2 |F_2(x,t,0)|^2 \\
 \Re\big( G_1(x,t,0) F_1^*(x,t,0)\big) - 2 |G_2(x,t,0)|^2 \Re\big( G_1(x,t,0) F_1^*(x,t,0)\big) \\
   + 2 \Re \big( [F_1(x,t,0)]^2 F_2(x,t,0) G_2^*(x,t,0) \big) - 2 \Re\big( F_2(x,t,0) G_2^*(x,t,0) [G_1(x,t,0)]^2 \big) \Big]\\
- \hbar \Im \Big[ g_1(x,t) F_1^*(x,t,0) - f_1^*(x,t) G_1(x,t,0)\Big]. 
   \end{multline}
 \begin{multline}
  \Xi_2(x,t) = i \hbar c \Im\Big[ \partial_x G_1(x,t,0) \big(|F_2(x,t,0)|^2 F_1^*(x,t,0) - G_2^*(x,t,0) G_1(x,t,0) F_2(x,t,0) \big)\Big] \\
 + i \hbar c \Im\Big[\partial_x F_1^*(x,t,0) \big( G_2(x,t,0) F_1^*(x,t,0) F_2^*(x,t,0) + G_1(x,t,0) |F_2(x,t,0)|^2 \big) \Big]\\
  + \frac{\hbar c}{2} \partial_x F_2(x,t,0) \Big[2 i F_2^*(x,t,0) \Im\big(F_1^*(x,t,0) G_1(x,t,0) \big) - G_2^*(x,t,0)  [F_1(x,t,0)]^2 \\
 -  G_2^*(x,t,0) [G_1(x,t,0)]^2  \Big] + \frac{\hbar c}{2} \partial_x G_2(x,t,0) \Big[- 2 i G_2^*(x,t,0) \Im \big(F_1^*(x,t,0) G_1(x,t,0) \big) \\
   + F_2^*(x,t,0) \big( [F_1^*(x,t,0)]^2 + [G_1^*(x,t,0)]^2 \big) \Big]- \hbar  \Re \Big[ g_1(x,t) F_1^*(x,t,0) - f_1^*(x,t) G_1(x,t,0) \\
   + g_2^*(x,t) F_2(x,t,0) [G_1(x,t,0)]^2 + g_2^*(x,t) F_1^*(x,t,0) G_1(x,t,0) G_2(x,t,0) \\
   - f_2^*(x,t) G_2(x,t,0) [F_1^*(x,t,0)]^2 - f_2^*(x,t) G_1(x,t,0) F_2(x,t,0) F_1^*(x,t,0) \\
   + g_2(x,t) F_2^*(x,t,0) [F_1^*(x,t,0)]^2  -  g_2(x,t) F_1^*(x,t,0) G_1(x,t,0) G_2^*(x,t,0) \\
   + f_2(x,t) G_1(x,t,0) F_1^*(x,t,0) F_2^*(x,t,0) - f_2(x,t) [G_1(x,t,0)]^2 G_2^*(x,t,0) \Big] \\ 
  + \hbar c  \partial_x \theta^0_1(x,t,0) \Big[2 |F_2(x,t,0)|^2 \Im \big(G_1(x,t,0) F_1^*(x,t,0)\big)+ \Im \big( G_2(x,t,0) F_2^*(x,t,0) \\
 [F_1^*(x,t,0)]^2 - G_2^*(x,t,0) F_2(x,t,0) [G_1(x,t,0)]^2 \big) \Big] + \hbar c \partial_x \theta^0_2(x,t,0) \\
\Big[ \big( |F_2(x,t,0)|^2 - |G_2(x,t,0)|^2 \big) \Im \big(G_1(x,t,0) F_1^*(x,t,0) \big) + \Im \big( G_2(x,t,0) F_2^*(x,t,0) \\
[F_1^*(x,t,0)]^2 - F_2(x,t,0) G_2^*(x,t,0) [G_1(x,t,0)]^2 \big) \Big].
 \end{multline}

\section{Special coin operations}\label{appcoinchoice}
If we choose to work with 
$C_j(t, \delta t) = \exp\big(- i \theta^0_j(x,t,\delta t) \sigma_0 - i \theta^1_j(x,t, \delta t) \sigma_1\big)$, i.e., we are allowing the phase term and 
spin-rotation with respect to the $x$-axis, we have \begin{multline}
      F_j(x,t, \delta t) = \cos \theta_j^1(x,t,\delta t)\\ 
      \Rightarrow F_j(x,t,0) = \cos \theta_j^1(x,t,0),
      ~ f_j(x,t) = - \sin \theta^1_j(x,t,0)~\vartheta^1_j(x,t) , 
    \end{multline}\begin{multline}
    G_j(x,t, \delta t) = - i \sin \theta_j^1(x,t,\delta t) \\
      \Rightarrow G_j(x,t,0) = - i \sin \theta_j^1(x,t,0),
      ~ g_j(x,t) = - i \cos \theta^1_j(x,t,0)~\vartheta^1_j(x,t) ,
      \end{multline}
where we have considered the Taylor expansion: 
\begin{align}
\theta^q_j(x,t, \delta t) = \theta^q_j(x,t,0) + \delta t~\vartheta^q_j(x,t) + \mathcal{O}(\delta t^2)~~\text{for all}~q \in \{0, 1, 2, 3\}.
\end{align}
In this case the operator terms of the Hamiltonian in eq.~(\ref{diraccuham}) take the following forms.
\begin{align}\Theta_1(x,t)  = 0,~~\Theta_2(x,t) = \cos[\theta^1_2(x,t,0)] \sin[2 \theta^1_1(x,t,0) + \theta^1_2(x,t,0)], \nonumber\\
 \Theta_3(x,t) = \frac{1}{2} \cos[2 \theta^1_1(x,t,0)] 
               + \frac{1}{2} \cos[2 \theta^1_1(x,t,0) + 2 \theta^1_2(x,t,0)],\end{align}
\begin{align}
\Xi_0(x,t) =   \hbar [\vartheta^0_1(x,t) + \vartheta^0_2(x,t)]  - \frac{\hbar c}{2} \partial_x \theta^0_2(x,t,0),\nonumber\\
\Xi_1(x,t) =  \hbar[\vartheta^1_1(x,t) + \vartheta^1_2(x,t)] - \frac{\hbar c}{2} \partial_x \theta^1_2(x,t,0),
       \end{align}
 \begin{align}               
 \Xi_3(x,t)  = \frac{i \hbar c}{2}  \sin[2 \theta^1_1(x,t,0) + 2 \theta^1_2(x,t,0)]\partial_x \theta^1_2(x,t,0) \nonumber\\ 
 + i \hbar c \cos[\theta^1_2(x,t,0)] \sin[\theta^1_2(x,t,0) +  2 \theta^1_1(x,t,0)]\partial_x \theta^1_1(x,t,0)\nonumber\\
 - \frac{\hbar c}{2} \partial_x \theta^0_1(x,t,0)  \Big[  \cos[2\theta^1_1(x,t,0)] + \cos[2\theta^1_1(x,t,0) + 2\theta^1_2(x,t,0)] \Big] \nonumber\\
 - \frac{\hbar c}{2}\partial_x \theta^0_2(x,t,0) \cos[2\theta^1_1(x,t,0) + 2\theta^1_2(x,t,0)],
  \end{align}
 \begin{align}
 \Xi_2(x,t) = - i \hbar c \cos[\theta^1_2(x,t,0)]  \cos[2 \theta^1_1(x,t,0) + \theta^1_2(x,t,0)] \partial_x \theta^1_1(x,t,0) \nonumber\\
 - \frac{i \hbar c}{2} \cos[2 \theta^1_1(x,t,0) + 2 \theta^1_2(x,t,0) ]\partial_x \theta^1_2(x,t,0) \nonumber\\
 - \hbar c  \partial_x \theta^0_1(x,t,0)   \cos[\theta^1_2(x,t,0)] \sin[2 \theta^1_1(x,t,0) + \theta^1_2(x,t,0)] \nonumber\\ 
 - \frac{\hbar c}{2} \partial_x \theta^0_2(x,t,0)  \sin[2\theta^1_1(x,t,0) + 2\theta^1_2(x,t,0)].
\end{align}

\subsection{Further Choice}\label{furtherchoice}
For the choice: $\theta^1_2(x,t,0) = - 2 \theta^1_1(x,t,0)$ we get 
\begin{align}\Theta_1(x,t)  = 0,~~\Theta_2(x,t) = 0, 
 \Theta_3(x,t) = \cos[\theta^1_2(x,t,0)], \nonumber\\
\Xi_0(x,t) =   \hbar [\vartheta^0_1(x,t) + \vartheta^0_2(x,t)]  - \frac{\hbar c}{2} \partial_x \theta^0_2(x,t,0),\nonumber\\
\Xi_1(x,t) =  \hbar[\vartheta^1_1(x,t) + \vartheta^1_2(x,t)] - \frac{\hbar c}{2} \partial_x \theta^1_2(x,t,0), \nonumber\\
 \Xi_2(x,t) =  - \frac{\hbar c}{2} \partial_x \theta^0_2(x,t,0)  \sin[\theta^1_2(x,t,0)], \nonumber\\           
 \Xi_3(x,t)  = \frac{i \hbar c}{2}  \sin[\theta^1_2(x,t,0)]\partial_x \theta^1_2(x,t,0)
 - \hbar c \partial_x \theta^0_1(x,t,0) \cos[\theta^1_2(x,t,0)] \nonumber\\
 - \frac{\hbar c}{2}\partial_x \theta^0_2(x,t,0) \cos[\theta^1_2(x,t,0)].
  \end{align}
  
\section{Introducing nonabelian gauge potential in single particle SS-DQW}\label{nonabel}

In this case the modified evolution operator: \\
$\mathscr{U}(t, \delta t) = C^\dagger_1(t, 0) \cdot C^\dagger_2(t,0) \cdot S_+ \cdot C_2(t, \delta t) \cdot S_- \cdot C_1(t, \delta t)$, 
where
\begin{multline}
S_+ =  \sum_{x}  \ket{\uparrow}\bra{\uparrow} \otimes \mathds{1}_N \otimes \ket{x + a}\bra{x}
+ \ket{\downarrow}\bra{\downarrow} \otimes \mathds{1}_N \otimes \ket{x}\bra{x},\\
S_- =  \sum_{x}  \ket{\uparrow}\bra{\uparrow} \otimes \mathds{1}_N \otimes \ket{x}\bra{x} + \ket{\downarrow}\bra{\downarrow}
\otimes \mathds{1}_N \otimes \ket{x-a}\bra{x}, \\
C_j(t, \delta t) = \sum_x \bigg(\Big[e^{- i\sum_{q=0}^3 \theta_j^q(x, t, \delta t)~\sigma_q} \otimes \mathds{1}_N \Big]
\cdot \mathcal{C}_{Nj}(x, t, \delta t)\bigg) \otimes \ket{x}\bra{x},~\forall~j \in \{1,2\}~\text{with}~\\
\mathcal{C}_{Nj}(x,t,\delta t) = \Big[ \ket{\uparrow}\bra{\uparrow} \otimes e^{- i \delta t \sum_{q=0}^{N^2 - 1} \omega_j^q(x,t) \Lambda_q}
+  \ket{\downarrow}\bra{\downarrow} \otimes  e^{- i \delta t \sum_{q=0}^{N^2 - 1} \Omega_j^q(x,t) \Lambda_q} \Big].
\end{multline} 
 Note that the form of the $\mathcal{C}_{Nj}(x,t,\delta t)$ operators are chosen in such a way that $\mathcal{C}_{Nj}(x,t,0) = \sigma_0 \otimes \Lambda_0  =  \mathds{1}_{2N}$. 
As our main concern here is to derive the effective Hamiltonian which can be obtained by the Taylor expansion upto first order in $\delta t$ or $a$, here 
we will use the form: \begin{multline}
             \mathcal{C}_{Nj}(x,t,\delta t) =  \ket{\uparrow}\bra{\uparrow} \otimes e^{- i \delta t \sum_{q=0}^{N^2 - 1} \omega_j^q(x,t) \Lambda_q}
+  \ket{\downarrow}\bra{\downarrow} \otimes  e^{- i \delta t \sum_{q=0}^{N^2 - 1} \Omega_j^q(x,t) \Lambda_q}  \\
=  \ket{\uparrow}\bra{\uparrow} \otimes \Bigg[ \Lambda_0 - i \delta t \sum_{q=0}^{N^2 - 1} \omega_j^q(x,t) \Lambda_q \Bigg]
+  \ket{\downarrow}\bra{\downarrow} \otimes \Bigg[ \Lambda_0 - i \delta t \sum_{q=0}^{N^2 - 1} \Omega_j^q(x,t) \Lambda_q \Bigg] + \mathcal{O}(\delta t^2).  
            \end{multline}
            
Also we will not concern about effect of the positional translation on the functions $\omega_j^q(x,t)$, $\Omega_j^q(x,t)$ as they are 
already the coefficients of the first order term in $\delta t$. So, let us define \begin{align}
                                                                                C_j^\uparrow \coloneqq \sum_{q=0}^{N^2 - 1} \omega_j^q(x,t) \Lambda_q,~~
                                                                                C_j^\downarrow \coloneqq  \sum_{q=0}^{N^2 - 1} \Omega_j^q(x,t) \Lambda_q~.   
                                                                                  \end{align}
            In the following calculations we will always confine ourselves to the first order terms in $\delta t$ and $a$, while $C_j^\uparrow$, $C_j^\downarrow$ 
            terms are involved.           
Therefore in the basis $\{ \ket{\uparrow}, \ket{\downarrow}\}$ we can write 
\begin{multline}
 \mathcal{C}_{Nj}(x,t,\delta t) = \left( \begin{array}{cc}
                                          \Lambda_0 - i \delta t~ C_j^\uparrow & 0 \\
                                          0 & \Lambda_0 - i \delta t ~C_j^\downarrow \\
                                         \end{array} \right) 
 \Rightarrow   C_j(t, \delta t) = \\
 \sum_x e^{- i \theta^0_j(x,t,\delta t)} \left( \begin{array}{cc}
                                         F_j(x,t,\delta t) \big[\Lambda_0 - i \delta t~ C_j^\uparrow \big]
                                         & G_j(x,t, \delta t) \big[\Lambda_0 - i \delta t ~C_j^\downarrow \big]  \\ 
                                        - G_j^*(x,t,\delta t) \big[\Lambda_0 - i \delta t~ C_j^\uparrow \big]
                                         & F_j^*(x,t, \delta t) \big[\Lambda_0 - i \delta t ~C_j^\downarrow \big] \\
                                         \end{array} \right) \otimes \ket{x}\bra{x}                                    
\end{multline}  In this case
\begin{multline}
 S_- \cdot C_1(t, \delta t) = \sum_{x} e^{-i\theta_1^0(x,t,\delta t)} \bigg[ F_1(x,t, \delta t) \ket{\uparrow}\bra{\uparrow} \otimes \Lambda_0 \otimes \ket{x}\bra{x} 
 + G_1(x,t, \delta t) \ket{\uparrow}\bra{\downarrow} \otimes \Lambda_0 \otimes \ket{x}\bra{x}\\
 - G_1^*(x,t, \delta t) \ket{\downarrow}\bra{\uparrow}\otimes \Lambda_0 \otimes \ket{x-a}\bra{x} 
 + F_1^*(x, t, \delta t) \ket{\downarrow}\bra{\downarrow}\otimes \Lambda_0 \otimes \ket{x-a}\bra{x}\bigg] \\
 - i \delta t \sum_{x} e^{-i\theta_1^0(x,t,0)} \bigg[ F_1(x,t, 0) \ket{\uparrow}\bra{\uparrow} \otimes C_1^\uparrow
 + G_1(x,t, 0) \ket{\uparrow}\bra{\downarrow} \otimes C_1^\downarrow \\
 - G_1^*(x,t, 0) \ket{\downarrow}\bra{\uparrow}\otimes C_1^\uparrow  
 + F_1^*(x, t, 0) \ket{\downarrow}\bra{\downarrow}\otimes C_1^\downarrow\bigg]\otimes \ket{x}\bra{x},
\end{multline}
\begin{multline}
 \Rightarrow  C_2(t, \delta t) \cdot  S_- \cdot C_1(t, \delta t) = \\
  \sum_{x} \bigg[ e^{-i[ \theta_2^0(x,t,\delta t) + \theta_1^0(x,t,\delta t)]}
 F_2(x, t, \delta t) F_1(x,t, \delta t) \ket{\uparrow}\bra{\uparrow} \otimes \Lambda_0 \otimes \ket{x}\bra{x}\\
 -  e^{-i[\theta_2^0(x-a,t,\delta t) +\theta_1^0(x,t,\delta t) ]} G_2(x-a, t, \delta t) G^*_1(x,t, \delta t) \bigg]
 \ket{\uparrow}\bra{\uparrow} \otimes \Lambda_0 \otimes \ket{x-a}\bra{x} \\
 +\bigg[ e^{-i[ \theta_2^0(x,t,\delta t) + \theta_1^0(x,t,\delta t)]}
 F_2(x, t, \delta t) G_1(x,t, \delta t)\ket{\uparrow}\bra{\downarrow} \otimes \Lambda_0 \otimes \ket{x}\bra{x} \\
 +  e^{-i[\theta_2^0(x-a,t,\delta t) +\theta_1^0(x,t,\delta t) ]} G_2(x-a, t, \delta t) F^*_1(x,t, \delta t) \bigg]
 \ket{\uparrow}\bra{\downarrow} \otimes \Lambda_0 \otimes \ket{x-a}\bra{x} \\
 +\bigg[- e^{-i[ \theta_2^0(x,t,\delta t) + \theta_1^0(x,t,\delta t)]}
 G_2^*(x, t, \delta t) F_1(x,t, \delta t) \ket{\downarrow}\bra{\uparrow} \otimes \Lambda_0 \otimes \ket{x}\bra{x}\\
 -  e^{-i[\theta_2^0(x-a,t,\delta t) +\theta_1^0(x,t,\delta t) ]} F_2^*(x-a, t, \delta t) G^*_1(x,t, \delta t) \bigg]
 \ket{\downarrow}\bra{\uparrow} \otimes \Lambda_0 \otimes \ket{x-a}\bra{x}  \\
 + \bigg[- e^{-i[\theta_2^0(x,t,\delta t) + \theta_1^0(x,t,\delta t)]}
 G_2^*(x, t, \delta t) G_1(x,t, \delta t) \ket{\downarrow}\bra{\downarrow} \otimes \Lambda_0 \otimes \ket{x}\bra{x}\\
 +  e^{-i[\theta_2^0(x-a,t,\delta t) +\theta_1^0(x,t,\delta t) ]} F_2^*(x-a, t, \delta t) F^*_1(x,t, \delta t) \bigg]
 \ket{\downarrow}\bra{\downarrow} \otimes \Lambda_0 \otimes \ket{x-a}\bra{x}\\
 - i \delta t~ e^{-i[ \theta_2^0(x,t,0) + \theta_1^0(x,t,0)]} \Bigg\{ 
 F_2(x, t, 0) F_1(x,t, 0) \ket{\uparrow}\bra{\uparrow} \otimes \big[C_2^\uparrow + C_1^\uparrow \big] \\
 -   G_2(x, t, 0) G^*_1(x,t, 0)
 \ket{\uparrow}\bra{\uparrow} \otimes \big[C_2^\downarrow + C_1^\uparrow \big] 
 + F_2(x, t, 0) G_1(x,t, 0)\ket{\uparrow}\bra{\downarrow} \otimes \big[C_2^\uparrow + C_1^\downarrow \big] \\
 +  G_2(x, t, 0) F^*_1(x,t, 0) 
 \ket{\uparrow}\bra{\downarrow} \otimes \big[C_2^\downarrow + C_1^\downarrow \big] 
 -  G_2^*(x, t, 0) F_1(x,t, 0) \ket{\downarrow}\bra{\uparrow} \otimes \big[C_2^\uparrow + C_1^\uparrow \big]\\
 -   F_2^*(x, t, 0) G^*_1(x,t, 0) 
 \ket{\downarrow}\bra{\uparrow} \otimes \big[C_2^\downarrow + C_1^\uparrow \big] 
 - G_2^*(x, t, 0) G_1(x,t, 0) \ket{\downarrow}\bra{\downarrow} \otimes \big[C_2^\uparrow + C_1^\downarrow \big] \\
 + F_2^*(x, t, 0) F^*_1(x,t, 0) 
 \ket{\downarrow}\bra{\downarrow} \otimes \big[C_2^\downarrow + C_1^\downarrow \big] \Bigg\}\otimes \ket{x}\bra{x}.
\end{multline}
This expression implies that except the terms involving $C_j^\uparrow$, $C_j^\downarrow$, all terms are in a similar form of 
$\mathscr{U}(t, \delta t)$ for the abelian case: $\text{dim}(\mathcal{H}_c) = 2$, but here with the higher dimensional coin space. 
Thus following the same calculation as done previously for the abelian case, an extra term will add with the effective Hamiltonian, and that is the following.

\begin{multline}\label{additionnon}
 \hbar~C^\dagger_1(t, 0) \cdot  C^\dagger_2(t, 0) \cdot \sum_x e^{-i[ \theta_2^0(x,t,0) + \theta_1^0(x,t,0)]} \Bigg\{ 
 F_2(x, t, 0) F_1(x,t, 0) \ket{\uparrow}\bra{\uparrow} \otimes \big[C_2^\uparrow + C_1^\uparrow \big] \\
 -   G_2(x, t, 0) G^*_1(x,t, 0)
 \ket{\uparrow}\bra{\uparrow} \otimes \big[C_2^\downarrow + C_1^\uparrow \big] 
 + F_2(x, t, 0) G_1(x,t, 0)\ket{\uparrow}\bra{\downarrow} \otimes \big[C_2^\uparrow + C_1^\downarrow \big] \\
 +  G_2(x, t, 0) F^*_1(x,t, 0) 
 \ket{\uparrow}\bra{\downarrow} \otimes \big[C_2^\downarrow + C_1^\downarrow \big] 
 -  G_2^*(x, t, 0) F_1(x,t, 0) \ket{\downarrow}\bra{\uparrow} \otimes \big[C_2^\uparrow + C_1^\uparrow \big]\\
 -   F_2^*(x, t, 0) G^*_1(x,t, 0) 
 \ket{\downarrow}\bra{\uparrow} \otimes \big[C_2^\downarrow + C_1^\uparrow \big] 
 - G_2^*(x, t, 0) G_1(x,t, 0) \ket{\downarrow}\bra{\downarrow} \otimes \big[C_2^\uparrow + C_1^\downarrow \big] \\
 + F_2^*(x, t, 0) F^*_1(x,t, 0) 
 \ket{\downarrow}\bra{\downarrow} \otimes \big[C_2^\downarrow + C_1^\downarrow \big] \Bigg\}\otimes \ket{x}\bra{x} \\
= \hbar \sum_x \bigg( |F_1(x,t,0)|^2  \ket{\uparrow}\bra{\uparrow} \otimes\big[C_2^\uparrow + C_1^\uparrow \big]
+ |G_1(x,t,0)|^2  \ket{\uparrow}\bra{\uparrow} \otimes\big[C_2^\downarrow + C_1^\uparrow \big] \\
+ F_1^*(x,t,0)G_1(x,t,0) \ket{\uparrow}\bra{\downarrow} \otimes\big[C_2^\uparrow - C_2^\downarrow\big]
+ F_1(x,t,0)G_1^*(x,t,0) \ket{\downarrow}\bra{\uparrow} \otimes\big[C_2^\uparrow - C_2^\downarrow\big]\\
+ |G_1(x,t,0)|^2 \ket{\downarrow}\bra{\downarrow} \otimes\big[C_2^\uparrow + C_1^\downarrow \big]
+ |F_1(x,t,0)|^2 \ket{\downarrow}\bra{\downarrow} \otimes\big[C_2^\downarrow + C_1^\downarrow \big]
\bigg) \otimes \ket{x}\bra{x} \\
= \frac{\hbar}{2} \sum_x \bigg(  \sigma_0 \otimes \big\{ C_1^\uparrow + C_1^\downarrow + C_2^\uparrow + C_2^\downarrow \big\} \\
+ 2 \sigma_1 \otimes \Re\big[F_1^*(x,t,0) G_1(x,t,0)\big]\big\{C_2^\uparrow - C_2^\downarrow \big\}  
- 2 \sigma_2 \otimes \Im\big[F_1^*(x,t,0) G_1(x,t,0)\big]\big\{C_2^\uparrow - C_2^\downarrow \big\} \\
+ \sigma_3 \otimes \Big[C_1^\uparrow - C_1^\downarrow
+ \big\{|F_1(x,t,0)|^2 - |G_1(x,t,0)|^2\big\}\big\{C_2^\uparrow - C_2^\downarrow\big\} \Big]
 \bigg)\otimes \ket{x}\bra{x}~.
\end{multline}

For curved $(1+1)$ dimensional case gauge potentials are involved
only $\sigma_0$, $\sigma_3$ Pauli matrices as discussed for the special choice in the section \ref{furtherchoice} for the abelian case, and also evident from eq.~(\ref{dircurgen}).
So we have to choose 
\begin{align}
C_2^\uparrow = C_2^\downarrow~~\Rightarrow~~\omega_2^q(x,t)  = \Omega_2^q(x,t)~~\forall~ q, x, t. 
\end{align}
This consideration makes this additional term in eq.~(\ref{additionnon}) as the following. 
\begin{align}
 \frac{\hbar}{2} \sum_x \bigg( \sigma_0 \otimes \big\{ C_1^\uparrow + C_1^\downarrow + 2 C_2^\uparrow \big\}
+ \sigma_3 \otimes \big\{C_1^\uparrow - C_1^\downarrow\big\}
 \bigg)\otimes \ket{x}\bra{x} .
\end{align}

\chapter{Two-particle case}\label{twopar}
 

Here we will show the form of the two-particle SS-DQW evolution operator by explicit calculation. 
The two-particle shift operators are\begin{multline}\label{shifttwo}
S_+  =  S^\text{first}_+ \otimes S^\text{second}_+ =  \\ 
\bigg(\ket{\uparrow}\bra{\uparrow} \otimes \ket{x_1 + a}\bra{x_1} 
+ \ket{\downarrow}\bra{\downarrow} \otimes \ket{x_1}\bra{x_1} \bigg)
\otimes \bigg(\ket{\uparrow}\bra{\uparrow} \otimes \ket{x_2 + a}\bra{x_2} 
+ \ket{\downarrow}\bra{\downarrow} \otimes \ket{x_2}\bra{x_2} \bigg) \\
=  \ket{\uparrow \uparrow}\bra{\uparrow \uparrow} \otimes \sum_{x_1, x_2} \ket{x_1 + a, x_2 + a}\bra{x_1, x_2} 
+ \ket{\uparrow \downarrow}\bra{\uparrow \downarrow} \otimes \sum_{x_1, x_2} \ket{x_1 + a, x_2}\bra{x_1, x_2}\\
+ \ket{\downarrow \uparrow}\bra{\downarrow \uparrow} \otimes \sum_{x_1, x_2} \ket{x_1, x_2 + a}\bra{x_1, x_2} 
+ \ket{\downarrow \downarrow}\bra{\downarrow \downarrow} \otimes \sum_{x_1, x_2} \ket{x_1, x_2}\bra{x_1, x_2}, \\\\
S_-  =  S^\text{first}_- \otimes S^\text{second}_- = ~~~~~~~~~~~~~~~~~~~~ \\
\bigg(\ket{\uparrow}\bra{\uparrow} \otimes \ket{x_1}\bra{x_1} 
+ \ket{\downarrow}\bra{\downarrow} \otimes \ket{x_1-a}\bra{x_1} \bigg)
\otimes \bigg(\ket{\uparrow}\bra{\uparrow} \otimes \ket{x_2}\bra{x_2} 
+ \ket{\downarrow}\bra{\downarrow} \otimes \ket{x_2-a}\bra{x_2} \bigg) \\
=\ket{\uparrow \uparrow}\bra{\uparrow \uparrow} \otimes \sum_{x_1, x_2} \ket{x_1, x_2}\bra{x_1, x_2} 
+ \ket{\uparrow \downarrow}\bra{\uparrow \downarrow} \otimes \sum_{x_1, x_2} \ket{x_1, x_2-a}\bra{x_1, x_2}\\
+ \ket{\downarrow \uparrow}\bra{\downarrow \uparrow} \otimes \sum_{x_1, x_2} \ket{x_1-a, x_2}\bra{x_1, x_2} 
+ \ket{\downarrow \downarrow}\bra{\downarrow \downarrow} \otimes \sum_{x_1, x_2} \ket{x_1-a, x_2-a}\bra{x_1, x_2}.
\end{multline}
In the single particle SS-DQW analysis we have understood the importance of the spinor rotation with respect to the $x$-axis and the phase.
So in the two-particle case where $C_j(t, \delta t) = \sum_{x_1, x_2}  \exp \Bigg( - i \sum_{q, r = 0}^3 \theta^{qr}_j(x_1, x_2, t, \delta t)~
\sigma_q \otimes \sigma_r \Bigg) \otimes \ket{x_1, x_2}\bra{x_1, x_2}$
we choose $\theta^{qr}_j(x_1, x_2, t, \delta t) \neq 0$ only for $q, r \in \{0, 1\}$. Then 
\begin{align}\label{glocoin}
\sum_{q, r = 0}^3
 \theta^{qr}_j(x_1, x_2, t, \delta t)~ \sigma_q \otimes \sigma_r  = \left(\begin{array}{cccc}
                                                                          \theta^{00}_j &  \theta^{01}_j & \theta^{10}_j & \theta^{11}_j \\
                                                                          \theta^{01}_j &  \theta^{00}_j & \theta^{11}_j & \theta^{10}_j \\
                                                                           \theta^{10}_j &  \theta^{11}_j & \theta^{00}_j & \theta^{01}_j \\
                                                                            \theta^{11}_j &  \theta^{10}_j & \theta^{01}_j & \theta^{00}_j \\
                                                                          \end{array}\right)
\end{align}while for convenience we have omitted the arguments $x_1, x_2, t, \delta t$ in $\theta^{qr}(x_1, x_2, t, \delta t)$ in the above matrix. 
By diagonalize the matrix in eq.~(\ref{glocoin}) we get 
the eigenvalues: \begin{align}\label{lambdas}
                  \lambda^0_j = \theta^{00}_j + \theta^{01}_j + \theta^{10}_j + \theta^{11}_j,~~
\lambda^1_j = \theta^{00}_j + \theta^{01}_j - \theta^{10}_j - \theta^{11}_j, \nonumber\\
\lambda^2_j = \theta^{00}_j - \theta^{01}_j + \theta^{10}_j - \theta^{11}_j,~~
\lambda^3_j = \theta^{00}_j - \theta^{01}_j - \theta^{10}_j + \theta^{11}_j.
                 \end{align}

The corresponding eigenvectors are, respectively:\begin{align}\label{twoeigenvec}
          \ket{\psi_0} = \frac{1}{2}\Big(\ket{\uparrow \uparrow} + \ket{\uparrow \downarrow} + \ket{\downarrow \uparrow} + \ket{\downarrow \downarrow} \Big),~~
          \ket{\psi_1} = \frac{1}{2}\Big(-\ket{\uparrow \uparrow} - \ket{\uparrow \downarrow} + \ket{\downarrow \uparrow} + \ket{\downarrow \downarrow} \Big), \nonumber\\
          \ket{\psi_2} = \frac{1}{2}\Big(-\ket{\uparrow \uparrow} + \ket{\uparrow \downarrow} - \ket{\downarrow \uparrow} + \ket{\downarrow \downarrow} \Big),~~
          \ket{\psi_3} = \frac{1}{2}\Big(\ket{\uparrow \uparrow} - \ket{\uparrow \downarrow} - \ket{\downarrow \uparrow} + \ket{\downarrow \downarrow} \Big).
                                                 \end{align}

Therefore,
\begin{align}\label{cointwo} 
  \exp \Bigg( - i \sum_{q, r = 0}^3 \theta^{qr}_j(x_1, x_2, t, \delta t) \sigma_q \otimes \sigma_r \Bigg)
   = \sum_{q=0}^3 e^{- i\lambda^q_j(x_1, x_2, t, \delta t)} \ket{\psi_q}\bra{\psi_q} \nonumber\\
 \Rightarrow  C_j(t, \delta t) = \sum_{x_1, x_2} \sum_{q=0}^3 e^{- i\lambda^q_j(x_1, x_2, t, \delta t)} \ket{\psi_q}\bra{\psi_q} \otimes \ket{x_1, x_2}\bra{x_1, x_2}.
\end{align}

These eigenvectors leads to the following relations which will be used in the future analysis.  
\begin{multline}\label{propeigen}
 \sum_{q=0}^3 \ket{\psi_q}\bra{\psi_q} = \sigma_0 \otimes \sigma_0, ~  
 \sum_{q=0,3} \ket{\psi_q}\bra{\psi_q} -  \sum_{q=1,2} \ket{\psi_q}\bra{\psi_q} = \sigma_1 \otimes \sigma_1. \\ 
 \ket{\psi_0}\bra{\psi_1} + \ket{\psi_3}\bra{\psi_2} = - \frac{1}{2} \sigma_3 \otimes \sigma_0 + \frac{i}{2} \sigma_2 \otimes \sigma_1, 
 \ket{\psi_3}\bra{\psi_1} + \ket{\psi_0}\bra{\psi_2} = - \frac{1}{2} \sigma_0 \otimes \sigma_3 + \frac{i}{2} \sigma_1 \otimes \sigma_2, \\
 \ket{\psi_0}\bra{\psi_1} + \ket{\psi_2}\bra{\psi_3} = - \frac{1}{2} \sigma_3 \otimes \sigma_0 + \frac{i}{2} \sigma_2 \otimes \sigma_0,
 \ket{\psi_3}\bra{\psi_1} + \ket{\psi_2}\bra{\psi_0} = - \frac{1}{2} \sigma_0 \otimes \sigma_3 - \frac{i}{2} \sigma_0 \otimes \sigma_2, \\
 \ket{\psi_0}\bra{\psi_1} - \ket{\psi_2}\bra{\psi_3} = -\frac{1}{2} \sigma_3 \otimes \sigma_1 + \frac{i}{2} \sigma_2 \otimes \sigma_1, 
 \ket{\psi_2}\bra{\psi_0} - \ket{\psi_3}\bra{\psi_1} = -\frac{1}{2} \sigma_1 \otimes \sigma_3 - \frac{i}{2} \sigma_1 \otimes \sigma_2,\\
 \sum_{q=0,2} \ket{\psi_q}\bra{\psi_q} -  \sum_{q=1,3} \ket{\psi_q}\bra{\psi_q} =   \sigma_1 \otimes \sigma_0,
 \sum_{q=0,1} \ket{\psi_q}\bra{\psi_q} -  \sum_{q=2,3} \ket{\psi_q}\bra{\psi_q} = \sigma_0 \otimes \sigma_1,
\end{multline}

\section{Explicit calculation of the evolution operator}
\vspace{-1cm}

The whole evolution operator is $\mathscr{U}^\text{two}(t, \delta t) = [U^\text{two}(t, 0)]^\dagger \cdot U^\text{two}(t, \delta t)$ 
where $U^\text{two}(t, \delta t) =  S_+ \cdot C_2(t, \delta t) \cdot  S_- \cdot C_1(t, \delta t)$. 
 From the expressions in eqs.~(\ref{shifttwo}) and (\ref{cointwo}) we get: 
\begin{multline}
 S_- \cdot C_1(t, \delta t) = \frac{1}{2}\sum_{x_1, x_2} \Bigg[\ket{\uparrow \uparrow}\bra{\psi_0} \otimes \ket{x_1, x_2}\bra{x_1, x_2} 
+ \ket{\uparrow \downarrow}\bra{\psi_0} \otimes  \ket{x_1, x_2-a}\bra{x_1, x_2}\\
+ \ket{\downarrow \uparrow}\bra{\psi_0} \otimes  \ket{x_1-a, x_2}\bra{x_1, x_2} 
+ \ket{\downarrow \downarrow}\bra{\psi_0} \otimes  \ket{x_1-a, x_2-a}\bra{x_1, x_2} \Bigg]
 e^{- i\lambda^0_1(x_1, x_2, t, \delta t)} \\
+ \frac{1}{2} \sum_{x_1, x_2}\Bigg[-\ket{\uparrow \uparrow}\bra{\psi_1} \otimes  \ket{x_1, x_2}\bra{x_1, x_2} 
- \ket{\uparrow \downarrow}\bra{\psi_1} \otimes  \ket{x_1, x_2-a}\bra{x_1, x_2}\\
+ \ket{\downarrow \uparrow}\bra{\psi_1} \otimes  \ket{x_1-a, x_2}\bra{x_1, x_2} 
+ \ket{\downarrow \downarrow}\bra{\psi_1} \otimes \ket{x_1-a, x_2-a}\bra{x_1, x_2} \Bigg]
e^{- i\lambda^1_1(x_1, x_2, t, \delta t)} \\
+ \frac{1}{2} \sum_{x_1, x_2}\Bigg[-\ket{\uparrow \uparrow}\bra{\psi_2} \otimes  \ket{x_1, x_2}\bra{x_1, x_2} 
+ \ket{\uparrow \downarrow}\bra{\psi_2} \otimes  \ket{x_1, x_2-a}\bra{x_1, x_2}\\
- \ket{\downarrow \uparrow}\bra{\psi_2} \otimes \ket{x_1-a, x_2}\bra{x_1, x_2} 
+ \ket{\downarrow \downarrow}\bra{\psi_2} \otimes \ket{x_1-a, x_2-a}\bra{x_1, x_2} \Bigg]
 e^{- i\lambda^2_1(x_1, x_2, t, \delta t)} \\
+ \frac{1}{2}\sum_{x_1, x_2}\Bigg[\ket{\uparrow \uparrow}\bra{\psi_3} \otimes  \ket{x_1, x_2}\bra{x_1, x_2} 
- \ket{\uparrow \downarrow}\bra{\psi_3} \otimes  \ket{x_1, x_2-a}\bra{x_1, x_2}\\
- \ket{\downarrow \uparrow}\bra{\psi_3} \otimes  \ket{x_1-a, x_2}\bra{x_1, x_2} 
+ \ket{\downarrow \downarrow}\bra{\psi_3} \otimes  \ket{x_1-a, x_2-a}\bra{x_1, x_2} \Bigg]
e^{- i\lambda^3_1(x_1, x_2, t, \delta t)} \end{multline}\begin{multline}
\coloneqq \sum_{x_1, x_2}\ket{\uparrow \uparrow}\bra{\psi^{\uparrow \uparrow}} \otimes \ket{x_1, x_2}\bra{x_1, x_2}
+ \ket{\uparrow \downarrow}\bra{\psi^{\uparrow \downarrow}} \otimes \ket{x_1, x_2-a}\bra{x_1, x_2} \\
+ \ket{\downarrow \uparrow}\bra{\psi^{\downarrow \uparrow}} \otimes \ket{x_1-a, x_2}\bra{x_1, x_2}
+ \ket{\downarrow \downarrow}\bra{\psi^{\downarrow \downarrow}} \otimes \ket{x_1-a, x_2-a}\bra{x_1, x_2}
\end{multline}
where we have used the notations:
\begin{multline}\label{notation1}
 \bra{\psi^{\uparrow \uparrow}} = \frac{1}{2}\bigg[ e^{- i \lambda^0_1(x_1, x_2, t, \delta t) } \bra{\psi_0}\
  - e^{- i \lambda^1_1(x_1, x_2, t, \delta t) } \bra{\psi_1} - e^{- i \lambda^2_1(x_1, x_2, t, \delta t) } \bra{\psi_2} 
  + e^{- i \lambda^3_1(x_1, x_2, t, \delta t) } \bra{\psi_3} \bigg], \\
    \bra{\psi^{\uparrow \downarrow}} = \frac{1}{2}\bigg[ e^{- i \lambda^0_1(x_1, x_2, t, \delta t) } \bra{\psi_0}\
  - e^{- i \lambda^1_1(x_1, x_2, t, \delta t) } \bra{\psi_1} + e^{- i \lambda^2_1(x_1, x_2, t, \delta t) } \bra{\psi_2} 
 - e^{- i \lambda^3_1(x_1, x_2, t, \delta t) } \bra{\psi_3} \bigg], \\
\bra{\psi^{\downarrow \uparrow}} = \frac{1}{2}\bigg[ e^{- i \lambda^0_1(x_1, x_2, t, \delta t) } \bra{\psi_0}\
  + e^{- i \lambda^1_1(x_1, x_2, t, \delta t) } \bra{\psi_1} - e^{- i \lambda^2_1(x_1, x_2, t, \delta t) } \bra{\psi_2} 
  - e^{- i \lambda^3_1(x_1, x_2, t, \delta t) } \bra{\psi_3} \bigg], \\
\bra{\psi^{\downarrow \downarrow}} = \frac{1}{2}\bigg[ e^{- i \lambda^0_1(x_1, x_2, t, \delta t) } \bra{\psi_0}
  + e^{- i \lambda^1_1(x_1, x_2, t, \delta t) } \bra{\psi_1} + e^{- i \lambda^2_1(x_1, x_2, t, \delta t) } \bra{\psi_2} 
  + e^{- i \lambda^3_1(x_1, x_2, t, \delta t) } \bra{\psi_3} \bigg].
  \end{multline}

\begin{multline}
 \Rightarrow C_2(t, \delta t) \cdot  S_- \cdot C_1(t, \delta t)
 = \sum_{x_1, x_2}\frac{1}{2}\bigg[e^{- i\lambda^0_2(x_1, x_2, t, \delta t)}\ket{\psi_0} - e^{- i\lambda^1_2(x_1, x_2, t, \delta t)} \ket{\psi_1} 
 - e^{- i\lambda^2_2(x_1, x_2, t, \delta t)} \ket{\psi_2} \\
 + e^{- i\lambda^3_2(x_1, x_2, t, \delta t)} \ket{\psi_3} \bigg]
 \bra{\psi^{\uparrow \uparrow}} \otimes \ket{x_1, x_2}\bra{x_1, x_2}
 + \frac{1}{2}\bigg[e^{- i\lambda^0_2(x_1, x_2-a, t, \delta t)}\ket{\psi_0} - e^{- i\lambda^1_2(x_1, x_2-a, t, \delta t)} \ket{\psi_1} \\
 + e^{- i\lambda^2_2(x_1, x_2-a, t, \delta t)} \ket{\psi_2} 
 - e^{- i\lambda^3_2(x_1, x_2-a, t, \delta t)} \ket{\psi_3} \bigg]\bra{\psi^{\uparrow \downarrow}} \otimes \ket{x_1, x_2-a}\bra{x_1, x_2} \\
 + \frac{1}{2}\bigg[e^{- i\lambda^0_2(x_1-a, x_2, t, \delta t)}\ket{\psi_0} + e^{- i\lambda^1_2(x_1-a, x_2, t, \delta t)} \ket{\psi_1} 
 - e^{- i\lambda^2_2(x_1-a, x_2, t, \delta t)} \ket{\psi_2} 
 - e^{- i\lambda^3_2(x_1-a, x_2, t, \delta t)} \ket{\psi_3} \bigg]\\
 \bra{\psi^{\downarrow \uparrow}} \otimes \ket{x_1-a, x_2}\bra{x_1, x_2}
 + \frac{1}{2}\bigg[e^{- i\lambda^0_2(x_1-a, x_2-a, t, \delta t)}\ket{\psi_0} + e^{- i\lambda^1_2(x_1-a, x_2-a, t, \delta t)} \ket{\psi_1} \\
 + e^{- i\lambda^2_2(x_1-a, x_2-a, t, \delta t)} \ket{\psi_2} 
 + e^{- i\lambda^3_2(x_1-a, x_2-a, t, \delta t)} \ket{\psi_3} \bigg]
 \bra{\psi^{\downarrow \downarrow}} \otimes \ket{x_1-a, x_2-a}\bra{x_1, x_2}
\end{multline}
\begin{multline}
 \Rightarrow U^\text{two}(t, \delta t) =  S_+ \cdot C_2(t, \delta t) \cdot  S_- \cdot C_1(t, \delta t) = 
 \\ 
 \sum_{x_1, x_2}\frac{1}{4}
 \bigg[e^{- i\lambda^0_2(x_1, x_2, t, \delta t)} + e^{- i\lambda^1_2(x_1, x_2, t, \delta t)}
 + e^{- i\lambda^2_2(x_1, x_2, t, \delta t)}
 + e^{- i\lambda^3_2(x_1, x_2, t, \delta t)}\bigg] \\ \ket{\uparrow \uparrow} \bra{\psi^{\uparrow \uparrow}} \otimes \ket{x_1 + a, x_2 + a}\bra{x_1, x_2}
  + \frac{1}{4}\bigg[e^{- i\lambda^0_2(x_1, x_2-a, t, \delta t)} + e^{- i\lambda^1_2(x_1, x_2-a, t, \delta t)} \\
 - e^{- i\lambda^2_2(x_1, x_2-a, t, \delta t)} 
 - e^{- i\lambda^3_2(x_1, x_2-a, t, \delta t)} \bigg] \ket{\uparrow \uparrow}\bra{\psi^{\uparrow \downarrow}} \otimes \ket{x_1 + a , x_2}\bra{x_1, x_2} \\
  + \frac{1}{4}\bigg[e^{- i\lambda^0_2(x_1-a, x_2, t, \delta t)} - e^{- i\lambda^1_2(x_1-a, x_2, t, \delta t)} 
 + e^{- i\lambda^2_2(x_1-a, x_2, t, \delta t)}
 - e^{- i\lambda^3_2(x_1-a, x_2, t, \delta t)} \bigg]\\ \ket{\uparrow \uparrow}
 \bra{\psi^{\downarrow \uparrow}} \otimes \ket{x_1, x_2 + a}\bra{x_1, x_2}
 + \frac{1}{4}\bigg[e^{- i\lambda^0_2(x_1-a, x_2-a, t, \delta t)} - e^{- i\lambda^1_2(x_1-a, x_2-a, t, \delta t)} \\
 - e^{- i\lambda^2_2(x_1-a, x_2-a, t, \delta t)} 
 + e^{- i\lambda^3_2(x_1-a, x_2-a, t, \delta t)} \bigg]\ket{\uparrow \uparrow}
 \bra{\psi^{\downarrow \downarrow}} \otimes \ket{x_1, x_2}\bra{x_1, x_2} \nonumber
 \end{multline}\begin{multline}
+ \sum_{x_1, x_2}\frac{1}{4}\bigg[e^{- i\lambda^0_2(x_1, x_2, t, \delta t)} + e^{- i\lambda^1_2(x_1, x_2, t, \delta t)} 
 - e^{- i\lambda^2_2(x_1, x_2, t, \delta t)} 
 - e^{- i\lambda^3_2(x_1, x_2, t, \delta t)} \bigg] \\ \ket{\uparrow \downarrow}
 \bra{\psi^{\uparrow \uparrow}} \otimes \ket{x_1 + a, x_2}\bra{x_1, x_2}
 + \frac{1}{4}\bigg[e^{- i\lambda^0_2(x_1, x_2-a, t, \delta t)} + e^{- i\lambda^1_2(x_1, x_2-a, t, \delta t)}  \\
 + e^{- i\lambda^2_2(x_1, x_2-a, t, \delta t)} + e^{- i\lambda^3_2(x_1, x_2-a, t, \delta t)} \bigg] \ket{\uparrow \downarrow}
 \bra{\psi^{\uparrow \downarrow}} \otimes \ket{x_1 + a, x_2-a}\bra{x_1, x_2} \\
 + \frac{1}{4}\bigg[e^{- i\lambda^0_2(x_1-a, x_2, t, \delta t)} - e^{- i\lambda^1_2(x_1-a, x_2, t, \delta t)} 
 - e^{- i\lambda^2_2(x_1-a, x_2, t, \delta t)}
 + e^{- i\lambda^3_2(x_1-a, x_2, t, \delta t)} \bigg]\\ \ket{\uparrow \downarrow}
 \bra{\psi^{\downarrow \uparrow}} \otimes \ket{x_1, x_2}\bra{x_1, x_2}
 + \frac{1}{4}\bigg[e^{- i\lambda^0_2(x_1-a, x_2-a, t, \delta t)}- e^{- i\lambda^1_2(x_1-a, x_2-a, t, \delta t)} \\
 + e^{- i\lambda^2_2(x_1-a, x_2-a, t, \delta t)} 
 - e^{- i\lambda^3_2(x_1-a, x_2-a, t, \delta t)} \bigg] \ket{\uparrow \downarrow}
 \bra{\psi^{\downarrow \downarrow}} \otimes \ket{x_1, x_2-a}\bra{x_1, x_2} \nonumber 
\end{multline}
\begin{multline}
 + \sum_{x_1, x_2}\frac{1}{4}\bigg[e^{- i\lambda^0_2(x_1, x_2, t, \delta t)} - e^{- i\lambda^1_2(x_1, x_2, t, \delta t)} 
 + e^{- i\lambda^2_2(x_1, x_2, t, \delta t)}
 - e^{- i\lambda^3_2(x_1, x_2, t, \delta t)} \bigg]\ket{\downarrow \uparrow} \\
 \bra{\psi^{\uparrow \uparrow}} \otimes \ket{x_1, x_2 + a}\bra{x_1, x_2} 
 + \frac{1}{4}\bigg[e^{- i\lambda^0_2(x_1, x_2-a, t, \delta t)} - e^{- i\lambda^1_2(x_1, x_2-a, t, \delta t)} \\
 - e^{- i\lambda^2_2(x_1, x_2-a, t, \delta t)} 
 + e^{- i\lambda^3_2(x_1, x_2-a, t, \delta t)} \bigg]\ket{\downarrow \uparrow} \bra{\psi^{\uparrow \downarrow}} \otimes \ket{x_1, x_2}\bra{x_1, x_2} \\
 + \frac{1}{4}\bigg[e^{- i\lambda^0_2(x_1-a, x_2, t, \delta t)} + e^{- i\lambda^1_2(x_1-a, x_2, t, \delta t)} 
 + e^{- i\lambda^2_2(x_1-a, x_2, t, \delta t)} 
 + e^{- i\lambda^3_2(x_1-a, x_2, t, \delta t)} \bigg]\\ \ket{\downarrow \uparrow}
 \bra{\psi^{\downarrow \uparrow}} \otimes \ket{x_1-a, x_2 + a}\bra{x_1, x_2}
 + \frac{1}{4}\bigg[e^{- i\lambda^0_2(x_1-a, x_2-a, t, \delta t)} + e^{- i\lambda^1_2(x_1-a, x_2-a, t, \delta t)} \\
 - e^{- i\lambda^2_2(x_1-a, x_2-a, t, \delta t)} 
 - e^{- i\lambda^3_2(x_1-a, x_2-a, t, \delta t)} \bigg]\ket{\downarrow \uparrow}
 \bra{\psi^{\downarrow \downarrow}} \otimes \ket{x_1-a, x_2}\bra{x_1, x_2}\nonumber
\end{multline}
\begin{multline}
 + \sum_{x_1, x_2}\frac{1}{4}\bigg[e^{- i\lambda^0_2(x_1, x_2, t, \delta t)} - e^{- i\lambda^1_2(x_1, x_2, t, \delta t)} 
 - e^{- i\lambda^2_2(x_1, x_2, t, \delta t)}
 + e^{- i\lambda^3_2(x_1, x_2, t, \delta t)}\bigg] \ket{\downarrow \downarrow} \bra{\psi^{\uparrow \uparrow}} \\
\otimes \ket{x_1, x_2}\bra{x_1, x_2} 
 + \frac{1}{4}\bigg[e^{- i\lambda^0_2(x_1, x_2-a, t, \delta t)} - e^{- i\lambda^1_2(x_1, x_2-a, t, \delta t)}
 + e^{- i\lambda^2_2(x_1, x_2-a, t, \delta t)} \\
 - e^{- i\lambda^3_2(x_1, x_2-a, t, \delta t)} \bigg]\ket{\downarrow \downarrow}\bra{\psi^{\uparrow \downarrow}} \otimes \ket{x_1, x_2-a}\bra{x_1, x_2}
 + \frac{1}{4}\bigg[e^{- i\lambda^0_2(x_1-a, x_2, t, \delta t)}  \\
 + e^{- i\lambda^1_2(x_1-a, x_2, t, \delta t)} - e^{- i\lambda^2_2(x_1-a, x_2, t, \delta t)} 
 - e^{- i\lambda^3_2(x_1-a, x_2, t, \delta t)}\bigg] \ket{\downarrow \downarrow}
 \bra{\psi^{\downarrow \uparrow}} \otimes \ket{x_1-a, x_2}\bra{x_1, x_2} \\
 + \frac{1}{4}\bigg[e^{- i\lambda^0_2(x_1-a, x_2-a, t, \delta t)} + e^{- i\lambda^1_2(x_1-a, x_2-a, t, \delta t)} 
 + e^{- i\lambda^2_2(x_1-a, x_2-a, t, \delta t)} 
 + e^{- i\lambda^3_2(x_1-a, x_2-a, t, \delta t)} \bigg] \\ \ket{\downarrow \downarrow}
 \bra{\psi^{\downarrow \downarrow}} \otimes \ket{x_1-a, x_2-a}\bra{x_1, x_2}~.
\end{multline}
Using the property:
 \begin{align}
     \exp\bigg( \mp \frac{i p_j a}{\hbar} \bigg) = \sum_{x_j} \ket{x_j \pm a}\bra{x_j},~ \mathds{1}_j = \sum_{x_j} \ket{x_j}\bra{x_j},~
     ~~\text{for}~j \in \{1, 2\} \nonumber
    \end{align} and the relations given in eq.~(\ref{notation1}), we get   
\begin{multline}
 U^\text{two}(t, \delta t) =  \\ 
 \sum_{x_1, x_2} \Bigg\{ \frac{1}{8}
 \bigg[e^{- i\lambda^0_2(x_1-a, x_2-a, t, \delta t)} + e^{- i\lambda^1_2(x_1-a, x_2-a, t, \delta t)}
 + e^{- i\lambda^2_2(x_1-a, x_2-a, t, \delta t)}
 + e^{- i\lambda^3_2(x_1-a, x_2-a, t, \delta t)}\bigg] \ket{\uparrow \uparrow} \\ \bigg[ e^{- i \lambda^0_1(x_1-a, x_2-a, t, \delta t) } \bra{\psi_0}
  - e^{- i \lambda^1_1(x_1-a, x_2-a, t, \delta t) } \bra{\psi_1} - e^{- i \lambda^2_1(x_1-a, x_2-a, t, \delta t) } \bra{\psi_2} \\
  + e^{- i \lambda^3_1(x_1-a, x_2-a, t, \delta t) } \bra{\psi_3} \bigg] 
  \otimes \ket{x_1, x_2}\bra{x_1, x_2} e^{- \frac{ia}{\hbar}(p_1 \otimes \mathds{1}_2 + \mathds{1}_1 \otimes p_2)} \Bigg\} \nonumber\end{multline}
\begin{multline}
   + \Bigg\{ \frac{1}{8}\bigg[e^{- i\lambda^0_2(x_1-a, x_2-a, t, \delta t)} + e^{- i\lambda^1_2(x_1-a, x_2-a, t, \delta t)} 
  - e^{- i\lambda^2_2(x_1-a, x_2-a, t, \delta t)} 
  - e^{- i\lambda^3_2(x_1-a, x_2-a, t, \delta t)} \bigg] \ket{\uparrow \uparrow} \\
\bigg[ e^{- i \lambda^0_1(x_1-a, x_2, t, \delta t) } \bra{\psi_0}
  - e^{- i \lambda^1_1(x_1-a, x_2, t, \delta t) } \bra{\psi_1} + e^{- i \lambda^2_1(x_1-a, x_2, t, \delta t) } \bra{\psi_2} \\
 - e^{- i \lambda^3_1(x_1-a, x_2, t, \delta t) } \bra{\psi_3} \bigg] \otimes \ket{x_1, x_2}\bra{x_1, x_2}
 e^{- \frac{ia}{\hbar}(p_1 \otimes \mathds{1}_2)} \Bigg\} \nonumber\end{multline}
\begin{multline}
   + \Bigg\{ \frac{1}{8}\bigg[e^{- i\lambda^0_2(x_1-a, x_2-a, t, \delta t)} - e^{- i\lambda^1_2(x_1-a, x_2-a, t, \delta t)} 
  + e^{- i\lambda^2_2(x_1-a, x_2-a, t, \delta t)}
  - e^{- i\lambda^3_2(x_1-a, x_2-a, t, \delta t)} \bigg] \ket{\uparrow \uparrow} \\
  \bigg[ e^{- i \lambda^0_1(x_1, x_2-a, t, \delta t) } \bra{\psi_0}
  + e^{- i \lambda^1_1(x_1, x_2-a, t, \delta t) } \bra{\psi_1} - e^{- i \lambda^2_1(x_1, x_2-a, t, \delta t) } \bra{\psi_2} \\
  - e^{- i \lambda^3_1(x_1, x_2-a, t, \delta t) } \bra{\psi_3} \bigg] 
  \otimes \ket{x_1, x_2}\bra{x_1, x_2}e^{- \frac{ia}{\hbar}(\mathds{1}_1 \otimes p_2)} \Bigg\} \nonumber\end{multline}
\begin{multline}
  + \Bigg\{ \frac{1}{8}\bigg[e^{- i\lambda^0_2(x_1-a, x_2-a, t, \delta t)} - e^{- i\lambda^1_2(x_1-a, x_2-a, t, \delta t)} 
  - e^{- i\lambda^2_2(x_1-a, x_2-a, t, \delta t)} 
  + e^{- i\lambda^3_2(x_1-a, x_2-a, t, \delta t)} \bigg]\ket{\uparrow \uparrow} \\
  \bigg[ e^{- i \lambda^0_1(x_1, x_2, t, \delta t) } \bra{\psi_0}
  + e^{- i \lambda^1_1(x_1, x_2, t, \delta t) } \bra{\psi_1} + e^{- i \lambda^2_1(x_1, x_2, t, \delta t) } \bra{\psi_2} \\
  + e^{- i \lambda^3_1(x_1, x_2, t, \delta t) } \bra{\psi_3} \bigg] \otimes \ket{x_1, x_2}\bra{x_1, x_2} \Bigg\} \nonumber\end{multline}
\begin{multline}
 + \sum_{x_1, x_2} \Bigg\{\frac{1}{8}\bigg[e^{- i\lambda^0_2(x_1-a, x_2, t, \delta t)} + e^{- i\lambda^1_2(x_1-a, x_2, t, \delta t)} 
  - e^{- i\lambda^2_2(x_1-a, x_2, t, \delta t)} 
  - e^{- i\lambda^3_2(x_1-a, x_2, t, \delta t)} \bigg] \ket{\uparrow \downarrow} \\
  \bigg[ e^{- i \lambda^0_1(x_1-a, x_2, t, \delta t) } \bra{\psi_0}
  - e^{- i \lambda^1_1(x_1-a, x_2, t, \delta t) } \bra{\psi_1} - e^{- i \lambda^2_1(x_1-a, x_2, t, \delta t) } \bra{\psi_2} \\
  + e^{- i \lambda^3_1(x_1-a, x_2, t, \delta t) } \bra{\psi_3} \bigg]
  \otimes \ket{x_1, x_2}\bra{x_1, x_2}e^{- \frac{ia}{\hbar}(p_1 \otimes \mathds{1}_2)}  \Bigg\} \nonumber
  \end{multline}\begin{multline}
  +  \Bigg\{ \frac{1}{8}\bigg[e^{- i\lambda^0_2(x_1 -a, x_2, t, \delta t)} + e^{- i\lambda^1_2(x_1 - a, x_2, t, \delta t)} 
  + e^{- i\lambda^2_2(x_1 - a, x_2, t, \delta t)} + e^{- i\lambda^3_2(x_1 - a, x_2, t, \delta t)} \bigg] \ket{\uparrow \downarrow} \\
  \bigg[ e^{- i \lambda^0_1(x_1 - a, x_2 + a, t, \delta t) } \bra{\psi_0}\
  - e^{- i \lambda^1_1(x_1 - a, x_2 + a, t, \delta t) } \bra{\psi_1} + e^{- i \lambda^2_1(x_1-a, x_2 + a, t, \delta t) } \bra{\psi_2} \\
 - e^{- i \lambda^3_1(x_1-a, x_2 + a, t, \delta t) } \bra{\psi_3} \bigg] 
 \otimes \ket{x_1, x_2}\bra{x_1, x_2} e^{- \frac{ia}{\hbar}(p_1 \otimes \mathds{1}_2 - \mathds{1}_1 \otimes p_2)} \Bigg\} \nonumber
  \end{multline}\begin{multline}
  + \Bigg\{ \frac{1}{8}\bigg[e^{- i\lambda^0_2(x_1-a, x_2, t, \delta t)} - e^{- i\lambda^1_2(x_1-a, x_2, t, \delta t)} 
  - e^{- i\lambda^2_2(x_1-a, x_2, t, \delta t)}
  + e^{- i\lambda^3_2(x_1-a, x_2, t, \delta t)} \bigg] \ket{\uparrow \downarrow} \\
  \bigg[ e^{- i \lambda^0_1(x_1, x_2, t, \delta t) } \bra{\psi_0}\
  + e^{- i \lambda^1_1(x_1, x_2, t, \delta t) } \bra{\psi_1} - e^{- i \lambda^2_1(x_1, x_2, t, \delta t) } \bra{\psi_2} \\ 
  - e^{- i \lambda^3_1(x_1, x_2, t, \delta t) } \bra{\psi_3} \bigg] \otimes \ket{x_1, x_2}\bra{x_1, x_2} \Bigg\} \nonumber
  \end{multline}\begin{multline}
  + \Bigg\{ \frac{1}{8}\bigg[e^{- i\lambda^0_2(x_1-a, x_2, t, \delta t)}- e^{- i\lambda^1_2(x_1-a, x_2, t, \delta t)} 
  + e^{- i\lambda^2_2(x_1-a, x_2, t, \delta t)} 
  - e^{- i\lambda^3_2(x_1-a, x_2, t, \delta t)} \bigg] \ket{\uparrow \downarrow} \\
 \bigg[ e^{- i \lambda^0_1(x_1, x_2+a, t, \delta t) } \bra{\psi_0}
  + e^{- i \lambda^1_1(x_1, x_2+a, t, \delta t) } \bra{\psi_1} + e^{- i \lambda^2_1(x_1, x_2+a, t, \delta t) } \bra{\psi_2} \\
  + e^{- i \lambda^3_1(x_1, x_2+a, t, \delta t) } \bra{\psi_3} \bigg]
  \otimes \ket{x_1, x_2}\bra{x_1, x_2} e^{ i \frac{a}{\hbar}(\mathds{1}_1 \otimes p_2)}\Bigg\} \nonumber
  \end{multline}\begin{multline}
  + \sum_{x_1, x_2} \Bigg\{ \frac{1}{8}\bigg[e^{- i\lambda^0_2(x_1, x_2-a, t, \delta t)} - e^{- i\lambda^1_2(x_1, x_2-a, t, \delta t)} 
  + e^{- i\lambda^2_2(x_1, x_2-a, t, \delta t)}
  - e^{- i\lambda^3_2(x_1, x_2-a, t, \delta t)} \bigg]\ket{\downarrow \uparrow} \\
 \bigg[ e^{- i \lambda^0_1(x_1, x_2-a, t, \delta t) } \bra{\psi_0}
  - e^{- i \lambda^1_1(x_1, x_2-a, t, \delta t) } \bra{\psi_1} - e^{- i \lambda^2_1(x_1, x_2-a, t, \delta t) } \bra{\psi_2} \\
  + e^{- i \lambda^3_1(x_1, x_2-a, t, \delta t) } \bra{\psi_3} \bigg]
  \otimes \ket{x_1, x_2}\bra{x_1, x_2} e^{- i \frac{a}{\hbar}(\mathds{1}_1 \otimes p_2)}\Bigg\} \nonumber
  \end{multline}\begin{multline}
  + \Bigg\{ \frac{1}{8}\bigg[e^{- i\lambda^0_2(x_1, x_2-a, t, \delta t)} - e^{- i\lambda^1_2(x_1, x_2-a, t, \delta t)} 
  - e^{- i\lambda^2_2(x_1, x_2-a, t, \delta t)} 
  + e^{- i\lambda^3_2(x_1, x_2-a, t, \delta t)} \bigg]\ket{\downarrow \uparrow} \\ 
  \bigg[ e^{- i \lambda^0_1(x_1, x_2, t, \delta t) } \bra{\psi_0}\
  - e^{- i \lambda^1_1(x_1, x_2, t, \delta t) } \bra{\psi_1} + e^{- i \lambda^2_1(x_1, x_2, t, \delta t) } \bra{\psi_2} 
 - e^{- i \lambda^3_1(x_1, x_2, t, \delta t) } \bra{\psi_3} \bigg]
  \otimes \ket{x_1, x_2}\bra{x_1, x_2} \Bigg\} \nonumber
  \end{multline}\begin{multline}
  + \Bigg\{ \frac{1}{8}\bigg[e^{- i\lambda^0_2(x_1, x_2-a, t, \delta t)} + e^{- i\lambda^1_2(x_1, x_2-a, t, \delta t)} 
  + e^{- i\lambda^2_2(x_1, x_2-a, t, \delta t)} 
  + e^{- i\lambda^3_2(x_1, x_2-a, t, \delta t)} \bigg] \ket{\downarrow \uparrow} \\
  \bigg[ e^{- i \lambda^0_1(x_1+a, x_2-a, t, \delta t) } \bra{\psi_0}\
  + e^{- i \lambda^1_1(x_1+a, x_2-a, t, \delta t) } \bra{\psi_1} - e^{- i \lambda^2_1(x_1+a, x_2-a, t, \delta t) } \bra{\psi_2} \\
  - e^{- i \lambda^3_1(x_1+a, x_2-a, t, \delta t) } \bra{\psi_3} \bigg]
  \otimes \ket{x_1, x_2}\bra{x_1, x_2} e^{\frac{i a}{\hbar}(p_1 \otimes \mathds{1}_2 - \mathds{1}_1 \otimes p_2)}  \Bigg\}\nonumber
  \end{multline}\begin{multline}
  + \Bigg\{ \frac{1}{8}\bigg[e^{- i\lambda^0_2(x_1, x_2-a, t, \delta t)} + e^{- i\lambda^1_2(x_1, x_2-a, t, \delta t)}
  - e^{- i\lambda^2_2(x_1, x_2-a, t, \delta t)} 
  - e^{- i\lambda^3_2(x_1, x_2-a, t, \delta t)} \bigg]\ket{\downarrow \uparrow} \\
 \bigg[ e^{- i \lambda^0_1(x_1+a, x_2, t, \delta t) } \bra{\psi_0}
  + e^{- i \lambda^1_1(x_1+a, x_2, t, \delta t) } \bra{\psi_1} + e^{- i \lambda^2_1(x_1+a, x_2, t, \delta t) } \bra{\psi_2} \\
  + e^{- i \lambda^3_1(x_1+a, x_2, t, \delta t) } \bra{\psi_3} \bigg]
  \otimes \ket{x_1, x_2}\bra{x_1, x_2} e^{ \frac{i a}{\hbar}(p_1 \otimes \mathds{1}_2)} \Bigg\} \nonumber
 \end{multline}
 \begin{multline}
  + \sum_{x_1, x_2}\Bigg\{ \frac{1}{8}\bigg[e^{- i\lambda^0_2(x_1, x_2, t, \delta t)} - e^{- i\lambda^1_2(x_1, x_2, t, \delta t)} 
  - e^{- i\lambda^2_2(x_1, x_2, t, \delta t)}
  + e^{- i\lambda^3_2(x_1, x_2, t, \delta t)}\bigg] \ket{\downarrow \downarrow} \\
 \bigg[ e^{- i \lambda^0_1(x_1, x_2, t, \delta t) } \bra{\psi_0}\
  - e^{- i \lambda^1_1(x_1, x_2, t, \delta t) } \bra{\psi_1} - e^{- i \lambda^2_1(x_1, x_2, t, \delta t) } \bra{\psi_2} 
  + e^{- i \lambda^3_1(x_1, x_2, t, \delta t) } \bra{\psi_3} \bigg] \otimes \ket{x_1, x_2}\bra{x_1, x_2} \Bigg\}\nonumber
  \end{multline}\begin{multline}
 + \Bigg\{\frac{1}{8}\bigg[e^{- i\lambda^0_2(x_1, x_2, t, \delta t)} - e^{- i\lambda^1_2(x_1, x_2, t, \delta t)}
  + e^{- i\lambda^2_2(x_1, x_2, t, \delta t)} 
  - e^{- i\lambda^3_2(x_1, x_2, t, \delta t)} \bigg]\ket{\downarrow \downarrow} \\
  \bigg[ e^{- i \lambda^0_1(x_1, x_2 + a, t, \delta t) } \bra{\psi_0}\
  - e^{- i \lambda^1_1(x_1, x_2 + a, t, \delta t) } \bra{\psi_1} + e^{- i \lambda^2_1(x_1, x_2 + a, t, \delta t) } \bra{\psi_2} \\
 - e^{- i \lambda^3_1(x_1, x_2 + a, t, \delta t) } \bra{\psi_3} \bigg]
  \otimes \ket{x_1, x_2}\bra{x_1, x_2} e^{ \frac{i a}{\hbar}(\mathds{1}_1 \otimes p_2)}\Bigg\} \nonumber
  \end{multline}\begin{multline}
  + \Bigg\{ \frac{1}{8}\bigg[e^{- i\lambda^0_2(x_1, x_2, t, \delta t)} 
  + e^{- i\lambda^1_2(x_1, x_2, t, \delta t)} - e^{- i\lambda^2_2(x_1, x_2, t, \delta t)} 
  - e^{- i\lambda^3_2(x_1, x_2, t, \delta t)}\bigg] \ket{\downarrow \downarrow} \\
  \bigg[ e^{- i \lambda^0_1(x_1+a, x_2, t, \delta t) } \bra{\psi_0}\
  + e^{- i \lambda^1_1(x_1+a, x_2, t, \delta t) } \bra{\psi_1} - e^{- i \lambda^2_1(x_1+a, x_2, t, \delta t) } \bra{\psi_2} \\
  - e^{- i \lambda^3_1(x_1 +a, x_2, t, \delta t) } \bra{\psi_3} \bigg]\otimes \ket{x_1, x_2}\bra{x_1, x_2} e^{\frac{i a}{\hbar} (p_1 \otimes \mathds{1}_2)} \Bigg\} \nonumber
  \end{multline}\begin{multline}
  + \Bigg\{\frac{1}{8}\bigg[e^{- i\lambda^0_2(x_1, x_2, t, \delta t)} + e^{- i\lambda^1_2(x_1, x_2, t, \delta t)} 
  + e^{- i\lambda^2_2(x_1, x_2, t, \delta t)} 
  + e^{- i\lambda^3_2(x_1, x_2, t, \delta t)} \bigg] \ket{\downarrow \downarrow} \\
  \bigg[ e^{- i \lambda^0_1(x_1 + a, x_2 + a, t, \delta t) } \bra{\psi_0}
  + e^{- i \lambda^1_1(x_1 + a, x_2 + a, t, \delta t) } \bra{\psi_1} + e^{- i \lambda^2_1(x_1 + a, x_2 + a, t, \delta t) } \bra{\psi_2} \\
  + e^{- i \lambda^3_1(x_1 + a, x_2 + a, t, \delta t) } \bra{\psi_3} \bigg] 
  \otimes \ket{x_1, x_2}\bra{x_1, x_2} e^{\frac{i a}{\hbar}(p_1 \otimes \mathds{1}_2 + \mathds{1}_1 \otimes p_2)}  \Bigg\} ~.
 \end{multline} \begin{multline}
U^\text{two}(t, 0) = C_2(t, 0) \cdot C_1(t, 0) = \Bigg[ \sum_{x_1, x_2} \sum_{q=0}^3 e^{- i\lambda^q_2(x_1, x_2, t, 0)} \ket{\psi_q}\bra{\psi_q}
\otimes \ket{x_1, x_2}\bra{x_1, x_2} \Bigg] \cdot \\
\Bigg[ \sum_{x'_1, x'_2} \sum_{q'=0}^3 e^{- i\lambda^{q'}_1(x'_1, x'_2, t, 0)} \ket{\psi_{q'}}\bra{\psi_{q'}}
\otimes \ket{x'_1, x'_2}\bra{x'_1, x'_2} \Bigg] = \\
\sum_{x_1, x_2} \sum_{q=0}^3 e^{- i [ \lambda^q_2(x_1, x_2, t, 0) +  \lambda^q_1(x_1, x_2, t, 0) ]} \ket{\psi_q}\bra{\psi_q}
\otimes \ket{x_1, x_2}\bra{x_1, x_2} \\ 
\Rightarrow [U^\text{two}(t, 0)]^\dagger = \sum_{x_1, x_2} \sum_{q=0}^3 e^{ i [ \lambda^q_2(x_1, x_2, t, 0) +  \lambda^q_1(x_1, x_2, t, 0) ]} \ket{\psi_q}\bra{\psi_q}
\otimes \ket{x_1, x_2}\bra{x_1, x_2}
\end{multline} 

Here we will use the following Taylor expansion considering that the coin parameters are smooth functions of $x_1$, $x_2$, $\delta t$.
\begin{align}
\lambda^q_j(x_1, x_2, t, \delta t) = \lambda^q_j(x_1, x_2, t, 0) + \delta t~\tilde{\lambda^q_j}(x_1, x_2, t, 0) + \mathcal{O}(\delta t^2), \nonumber\\
\lambda^q_j(x_1 \pm a, x_2, t, 0) = \lambda^q_j(x_1, x_2, t, 0) \pm a~\partial_{x_1}\lambda^q_j(x_1, x_2, t, 0) + \mathcal{O}(a^2),\nonumber\\
\lambda^q_j(x_1, x_2 \pm a, t, 0) = \lambda^q_j(x_1, x_2, t, 0) \pm a~\partial_{x_2}\lambda^q_j(x_1, x_2, t, 0) + \mathcal{O}(a^2).
\end{align}
By Taylor expansion of the unmodified two-particle SS-DQW evolution operator upto first order both in $a$, $\delta t$ we get
 \begin{multline} U^\text{two}(t, \delta t) = 
 \\ 
 \sum_{x_1, x_2} \Bigg\{ \frac{1}{8}
 \bigg[ \sum_{q = 0}^3 e^{- i\lambda^q_2(x_1, x_2, t, 0)}\big[1 + ia \partial_{x_1}\lambda^q_2(x_1, x_2, t, 0)
 + ia \partial_{x_2}\lambda^q_2(x_1, x_2, t, 0) -i \delta t \tilde{\lambda^q_2}(x_1, x_2, t, 0) \big] \bigg]\ket{\uparrow \uparrow} \\
  \bigg[ e^{- i \lambda^0_1(x_1, x_2, t, 0) }\big[1 + ia \partial_{x_1}\lambda^0_1(x_1, x_2, t, 0) + ia \partial_{x_2}\lambda^0_1(x_1, x_2, t, 0) -i \delta t \tilde{\lambda^0_1}(x_1, x_2, t, 0) \big] \bra{\psi_0}\\
   - e^{- i \lambda^1_1(x_1, x_2, t, 0) }\big[1 + ia \partial_{x_1}\lambda^1_1(x_1, x_2, t, 0) + ia \partial_{x_2}\lambda^1_1(x_1, x_2, t, 0) -i \delta t \tilde{\lambda^1_1}(x_1, x_2, t, 0) \big]  \bra{\psi_1}\\
   - e^{- i \lambda^2_1(x_1, x_2, t, 0) }\big[1 + ia \partial_{x_1}\lambda^2_1(x_1, x_2, t, 0) + ia \partial_{x_2}\lambda^2_1(x_1, x_2, t, 0) -i \delta t \tilde{\lambda^2_1}(x_1, x_2, t, 0) \big]  \bra{\psi_2} \\
  + e^{- i \lambda^3_1(x_1, x_2, t, 0) }\big[1 + ia \partial_{x_1}\lambda^3_1(x_1, x_2, t, 0) + ia \partial_{x_2}\lambda^3_1(x_1, x_2, t, 0) -i \delta t \tilde{\lambda^3_1}(x_1, x_2, t, 0) \big]  \bra{\psi_3} \bigg] \\
   \otimes \ket{x_1, x_2}\bra{x_1, x_2} \bigg[ \mathds{1}_1 \otimes \mathds{1}_2  - \frac{ia}{\hbar}(p_1 \otimes \mathds{1}_2 + \mathds{1}_1 \otimes p_2) \bigg] \Bigg\} \nonumber
  \end{multline}\begin{multline}
    + \Bigg\{ \frac{1}{8}\bigg[e^{- i\lambda^0_2(x_1, x_2, t, 0)}\big[1 + ia \partial_{x_1}\lambda^0_2(x_1, x_2, t, 0) + ia \partial_{x_2}\lambda^0_2(x_1, x_2, t, 0) -i \delta t \tilde{\lambda^0_2}(x_1, x_2, t, 0) \big]\\
    + e^{- i\lambda^1_2(x_1, x_2, t, 0)}\big[1 + ia \partial_{x_1}\lambda^1_2(x_1, x_2, t, 0) + ia \partial_{x_2}\lambda^1_2(x_1, x_2, t, 0) -i \delta t \tilde{\lambda^1_2}(x_1, x_2, t, 0) \big] \\
   - e^{- i\lambda^2_2(x_1, x_2, t, 0)}\big[1 + ia \partial_{x_1}\lambda^2_2(x_1, x_2, t, 0) + ia \partial_{x_2}\lambda^2_2(x_1, x_2, t, 0) -i \delta t \tilde{\lambda^2_2}(x_1, x_2, t, 0) \big] \\
   - e^{- i\lambda^3_2(x_1, x_2, t, 0)}\big[1 + ia \partial_{x_1}\lambda^3_2(x_1, x_2, t, 0) + ia \partial_{x_2}\lambda^3_2(x_1, x_2, t, 0) -i \delta t \tilde{\lambda^3_2}(x_1, x_2, t, 0) \big] \bigg]  \ket{\uparrow \uparrow} \\
  \bigg[ e^{- i \lambda^0_1(x_1, x_2, t, 0) }\big[1 + ia \partial_{x_1}\lambda^0_1(x_1, x_2, t, 0) -i \delta t \tilde{\lambda^0_1}(x_1, x_2, t, 0) \big] \bra{\psi_0} \\
   - e^{- i \lambda^1_1(x_1, x_2, t, 0)}\big[1 + ia \partial_{x_1}\lambda^1_1(x_1, x_2, t, 0) -i \delta t \tilde{\lambda^1_1}(x_1, x_2, t, 0) \big] \bra{\psi_1} \\
   + e^{- i \lambda^2_1(x_1, x_2, t, 0) }\big[1 + ia \partial_{x_1}\lambda^2_1(x_1, x_2, t, 0) -i \delta t \tilde{\lambda^2_1}(x_1, x_2, t, 0) \big] \bra{\psi_2} \\
  - e^{- i \lambda^3_1(x_1, x_2, t, 0) }\big[1 + ia \partial_{x_1}\lambda^3_1(x_1, x_2, t, 0) -i \delta t \tilde{\lambda^3_1}(x_1, x_2, t, 0) \big] \bra{\psi_3} \bigg] \\
  \otimes \ket{x_1, x_2}\bra{x_1, x_2} \bigg[\mathds{1}_1 \otimes \mathds{1}_2 - \frac{ia}{\hbar}(p_1 \otimes \mathds{1}_2) \bigg] \Bigg\} \nonumber
  \end{multline}\begin{multline}
    + \Bigg\{ \frac{1}{8}\bigg[e^{- i\lambda^0_2(x_1, x_2, t, 0)}\big[1 + ia \partial_{x_1}\lambda^0_2(x_1, x_2, t, 0) + ia \partial_{x_2}\lambda^0_2(x_1, x_2, t, 0) -i \delta t \tilde{\lambda^0_2}(x_1, x_2, t, 0) \big] \\
    - e^{- i\lambda^1_2(x_1, x_2, t, 0)}\big[1 + ia \partial_{x_1}\lambda^1_2(x_1, x_2, t, 0) + ia \partial_{x_2}\lambda^1_2(x_1, x_2, t, 0) -i \delta t \tilde{\lambda^1_2}(x_1, x_2, t, 0) \big] \\ 
   + e^{- i\lambda^2_2(x_1, x_2, t, 0)}\big[1 + ia \partial_{x_1}\lambda^2_2(x_1, x_2, t, 0) + ia \partial_{x_2}\lambda^2_2(x_1, x_2, t, 0) -i \delta t \tilde{\lambda^2_2}(x_1, x_2, t, 0) \big] \\
   -e^{- i\lambda^3_2(x_1, x_2, t, 0)}\big[1 + ia \partial_{x_1}\lambda^3_2(x_1, x_2, t, 0) + ia \partial_{x_2}\lambda^3_2(x_1, x_2, t, 0) -i \delta t \tilde{\lambda^3_2}(x_1, x_2, t, 0) \big]  \bigg] \ket{\uparrow \uparrow} \\
   \bigg[ e^{- i \lambda^0_1(x_1, x_2, t, 0) }\big[1 + ia \partial_{x_2}\lambda^0_1(x_1, x_2, t, 0) -i \delta t \tilde{\lambda^0_1}(x_1, x_2, t, 0) \big] \bra{\psi_0}\\
   + e^{- i \lambda^1_1(x_1, x_2, t, 0) }\big[1 + ia \partial_{x_2}\lambda^1_1(x_1, x_2, t, 0) -i \delta t \tilde{\lambda^1_1}(x_1, x_2, t, 0) \big] \bra{\psi_1} \\
  - e^{- i \lambda^2_1(x_1, x_2, t, 0) }\big[1 + ia \partial_{x_2}\lambda^2_1(x_1, x_2, t, 0) - i \delta t \tilde{\lambda^2_1}(x_1, x_2, t, 0) \big] \bra{\psi_2} \\
   - e^{- i \lambda^3_1(x_1, x_2-a, t, \delta t) }\big[1 + ia \partial_{x_2}\lambda^3_1(x_1, x_2, t, 0) -i \delta t \tilde{\lambda^3_1}(x_1, x_2, t, 0) \big] \bra{\psi_3} \bigg] \\
   \otimes \ket{x_1, x_2}\bra{x_1, x_2}\bigg[ \mathds{1}_1 \otimes \mathds{1}_2 - \frac{ia}{\hbar}(\mathds{1}_1 \otimes p_2) \bigg] \Bigg\} \nonumber
  \end{multline}\begin{multline}
   + \Bigg\{ \frac{1}{8}\bigg[e^{- i\lambda^0_2(x_1, x_2, t, 0)}\big[1 + ia \partial_{x_1}\lambda^0_2(x_1, x_2, t, 0) + ia \partial_{x_2}\lambda^0_2(x_1, x_2, t, 0) -i \delta t \tilde{\lambda^0_2}(x_1, x_2, t, 0) \big] \\
   -e^{- i\lambda^1_2(x_1, x_2, t, 0)}\big[1 + ia \partial_{x_1}\lambda^1_2(x_1, x_2, t, 0) + ia \partial_{x_2}\lambda^1_2(x_1, x_2, t, 0) -i \delta t \tilde{\lambda^1_2}(x_1, x_2, t, 0) \big] \\
   -  e^{- i\lambda^2_2(x_1, x_2, t, 0)}\big[1 + ia \partial_{x_1}\lambda^2_2(x_1, x_2, t, 0) + ia \partial_{x_2}\lambda^2_2(x_1, x_2, t, 0) -i \delta t \tilde{\lambda^2_2}(x_1, x_2, t, 0) \big] \\ 
   +  e^{- i\lambda^3_2(x_1, x_2, t, 0)}\big[1 + ia \partial_{x_1}\lambda^3_2(x_1, x_2, t, 0) + ia \partial_{x_2}\lambda^3_2(x_1, x_2, t, 0) -i \delta t \tilde{\lambda^3_2}(x_1, x_2, t, 0) \big] \bigg]\ket{\uparrow \uparrow} \\
   \bigg[ e^{- i \lambda^0_1(x_1, x_2, t, 0) }\big[ 1 - i \delta t \tilde{\lambda^0_1}(x_1, x_2, t, 0) \big] \bra{\psi_0}
   + e^{- i \lambda^1_1(x_1, x_2, t, 0)}\big[ 1 - i \delta t \tilde{\lambda^1_1}(x_1, x_2, t, 0) \big]\bra{\psi_1} \\
   + e^{- i \lambda^2_1(x_1, x_2, t, 0)} \big[ 1 - i \delta t \tilde{\lambda^2_1}(x_1, x_2, t, 0) \big] \bra{\psi_2} 
   + e^{- i \lambda^3_1(x_1, x_2, t, 0) }\big[ 1 - i \delta t \tilde{\lambda^3_1}(x_1, x_2, t, 0) \big] \bra{\psi_3} \bigg] \\
   \otimes \ket{x_1, x_2}\bra{x_1, x_2} \Bigg\} \nonumber
  \end{multline}\begin{multline}
  + \sum_{x_1, x_2} \Bigg\{\frac{1}{8}\bigg[e^{- i\lambda^0_2(x_1, x_2, t, 0)}\big[1 + i a \partial_{x_1}\lambda^0_2(x_1, x_2, t, 0) - i \delta t \tilde{\lambda^0_2}(x_1, x_2, t, 0) \big]\\
  + e^{- i\lambda^1_2(x_1, x_2, t, 0)}\big[1 + i a \partial_{x_1}\lambda^1_2(x_1, x_2, t, 0) - i \delta t \tilde{\lambda^1_2}(x_1, x_2, t, 0) \big] \\ 
   - e^{- i\lambda^2_2(x_1, x_2, t, 0)}\big[1 + i a \partial_{x_1}\lambda^2_2(x_1, x_2, t, 0) - i \delta t \tilde{\lambda^2_2}(x_1, x_2, t, 0) \big]  \\
   - e^{- i\lambda^3_2(x_1, x_2, t, 0)}\big[1 + i a \partial_{x_1}\lambda^3_2(x_1, x_2, t, 0) - i \delta t \tilde{\lambda^3_2}(x_1, x_2, t, 0) \big] \bigg] \ket{\uparrow \downarrow} \\
   \bigg[ e^{- i \lambda^0_1(x_1, x_2, t, 0)}\big[1 + i a \partial_{x_1}\lambda^0_1(x_1, x_2, t, 0) - i \delta t \tilde{\lambda^0_1}(x_1, x_2, t, 0) \big] \bra{\psi_0} \\
   - e^{- i \lambda^1_1(x_1, x_2, t, 0)}\big[1 + i a \partial_{x_1}\lambda^1_1(x_1, x_2, t, 0) - i \delta t \tilde{\lambda^1_1}(x_1, x_2, t, 0) \big]  \bra{\psi_1} \\
    - e^{- i \lambda^2_1(x_1, x_2, t, 0) }\big[1 + i a \partial_{x_1}\lambda^2_1(x_1, x_2, t, 0) - i \delta t \tilde{\lambda^2_1}(x_1, x_2, t, 0) \big]  \bra{\psi_2} \\
   + e^{- i \lambda^3_1(x_1, x_2, t, 0) } \big[1 + i a \partial_{x_1}\lambda^3_1(x_1, x_2, t, 0) - i \delta t \tilde{\lambda^3_1}(x_1, x_2, t, 0) \big] \bra{\psi_3} \bigg] \\
   \otimes \ket{x_1, x_2}\bra{x_1, x_2}\bigg[ \mathds{1}_1 \otimes \mathds{1}_2 - \frac{ia}{\hbar}(p_1 \otimes \mathds{1}_2) \bigg]  \Bigg\} \nonumber
   \end{multline}\begin{multline}
   +  \Bigg\{ \frac{1}{8}\bigg[e^{- i\lambda^0_2(x_1, x_2, t, 0)}\big[1 + i a \partial_{x_1}\lambda^0_2(x_1, x_2, t, 0) - i \delta t \tilde{\lambda^0_2}(x_1, x_2, t, 0) \big] \\
   + e^{- i\lambda^1_2(x_1, x_2, t, 0)}  \big[1 + i a \partial_{x_1}\lambda^1_2(x_1, x_2, t, 0) - i \delta t \tilde{\lambda^1_2}(x_1, x_2, t, 0) \big] \\
   + e^{- i\lambda^2_2(x_1, x_2, t, 0)} \big[1 + i a \partial_{x_1}\lambda^2_2(x_1, x_2, t, 0) - i \delta t \tilde{\lambda^2_2}(x_1, x_2, t, 0) \big] \\
   + e^{- i\lambda^3_2(x_1, x_2, t, 0)}\big[1 + i a \partial_{x_1}\lambda^3_2(x_1, x_2, t, 0) - i \delta t \tilde{\lambda^3_2}(x_1, x_2, t, 0) \big] 
\bigg] \ket{\uparrow \downarrow} \\
   \bigg[ e^{- i \lambda^0_1(x_1, x_2, t, 0) } \big[1 + i a \partial_{x_1}\lambda^0_1(x_1, x_2, t, 0) -i a \partial_{x_2}\lambda^0_1(x_1, x_2, t, 0)  - i \delta t \tilde{\lambda^0_1}(x_1, x_2, t, 0) \big]\bra{\psi_0} \\
   - e^{- i \lambda^1_1(x_1, x_2, t, 0) } \big[1 + i a \partial_{x_1}\lambda^1_1(x_1, x_2, t, 0) -i a \partial_{x_2}\lambda^1_1(x_1, x_2, t, 0)  - i \delta t \tilde{\lambda^1_1}(x_1, x_2, t, 0) \big] \bra{\psi_1} \\
   + e^{- i \lambda^2_1(x_1, x_2, t, 0) }  \big[1 + i a \partial_{x_1}\lambda^2_1(x_1, x_2, t, 0) -i a \partial_{x_2}\lambda^2_1(x_1, x_2, t, 0)  - i \delta t \tilde{\lambda^2_1}(x_1, x_2, t, 0) \big]\bra{\psi_2} \\
  - e^{- i \lambda^3_1(x_1, x_2, t, 0) } \bra{\psi_3} \big[1 + i a \partial_{x_1}\lambda^3_1(x_1, x_2, t, 0) -i a \partial_{x_2}\lambda^3_1(x_1, x_2, t, 0)  - i \delta t \tilde{\lambda^3_1}(x_1, x_2, t, 0) \big] \bigg] \\
  \otimes \ket{x_1, x_2}\bra{x_1, x_2} \bigg[ \mathds{1}_1 \otimes \mathds{1}_2 - \frac{ia}{\hbar}(p_1 \otimes \mathds{1}_2 - \mathds{1}_1 \otimes p_2) \bigg] \Bigg\} \nonumber
  \end{multline}\begin{multline}
   + \Bigg\{ \frac{1}{8}\bigg[e^{- i\lambda^0_2(x_1, x_2, t, 0)}  \big[1 + i a \partial_{x_1}\lambda^0_2(x_1, x_2, t, 0) - i \delta t \tilde{\lambda^0_2}(x_1, x_2, t, 0) \big] \\
   - e^{- i\lambda^1_2(x_1, x_2, t, 0)} \big[1 + i a \partial_{x_1}\lambda^1_2(x_1, x_2, t, 0)  - i \delta t \tilde{\lambda^1_2}(x_1, x_2, t, 0) \big] \\ 
   - e^{- i\lambda^2_2(x_1, x_2, t, 0)} \big[1 + i a \partial_{x_1}\lambda^2_2(x_1, x_2, t, 0)  - i \delta t \tilde{\lambda^2_2}(x_1, x_2, t, 0) \big] \\
   + e^{- i\lambda^3_2(x_1, x_2, t, 0)}\big[1 + i a \partial_{x_1}\lambda^3_2(x_1, x_2, t, 0)  - i \delta t \tilde{\lambda^3_2}(x_1, x_2, t, 0) \big] \bigg] \ket{\uparrow \downarrow} \\
  \bigg[ e^{- i \lambda^0_1(x_1, x_2, t, 0) }\big[1 - i \delta t \tilde{\lambda^0_1}(x_1, x_2, t, 0) \big] \bra{\psi_0}
   + e^{- i \lambda^1_1(x_1, x_2, t, 0) }\big[1 - i \delta t \tilde{\lambda^1_1}(x_1, x_2, t, 0) \big] \bra{\psi_1} \\
   - e^{- i \lambda^2_1(x_1, x_2, t, 0)}\big[1 - i \delta t \tilde{\lambda^2_1}(x_1, x_2, t, 0) \big] \bra{\psi_2}  
  - e^{- i \lambda^3_1(x_1, x_2, t, 0) }[1 - i \delta t \tilde{\lambda^3_1}(x_1, x_2, t, 0) \big] \bra{\psi_3} \bigg] \\
  \otimes \ket{x_1, x_2}\bra{x_1, x_2} \Bigg\} \nonumber
  \end{multline}\begin{multline}
   + \Bigg\{ \frac{1}{8}\bigg[e^{- i\lambda^0_2(x_1, x_2, t, 0)} \big[1 + i a \partial_{x_1}\lambda^0_2(x_1, x_2, t, 0) - i \delta t \tilde{\lambda^0_2}(x_1, x_2, t, 0) \big] \\
  - e^{- i\lambda^1_2(x_1, x_2, t, 0)}  \big[1 + i a \partial_{x_1}\lambda^1_2(x_1, x_2, t, 0) - i \delta t \tilde{\lambda^1_2}(x_1, x_2, t, 0) \big] \\
   + e^{- i\lambda^2_2(x_1, x_2, t, 0)}  \big[1 + i a \partial_{x_1}\lambda^2_2(x_1, x_2, t, 0) - i \delta t \tilde{\lambda^2_2}(x_1, x_2, t, 0) \big] \\
   - e^{- i\lambda^3_2(x_1, x_2, t, 0)} \big[1 + i a \partial_{x_1}\lambda^3_2(x_1, x_2, t, 0) - i \delta t \tilde{\lambda^3_2}(x_1, x_2, t, 0) \big] \bigg] \ket{\uparrow \downarrow} \\
  \bigg[ e^{- i \lambda^0_1(x_1, x_2, t, 0) } \big[1 - i a \partial_{x_2}\lambda^0_1(x_1, x_2, t, 0) - i \delta t \tilde{\lambda^0_1}(x_1, x_2, t, 0) \big] \bra{\psi_0} \\
   + e^{- i \lambda^1_1(x_1, x_2, t, 0) }\big[1 - i a \partial_{x_2}\lambda^1_1(x_1, x_2, t, 0) - i \delta t \tilde{\lambda^1_1}(x_1, x_2, t, 0) \big] \bra{\psi_1} \\
   + e^{- i \lambda^2_1(x_1, x_2, t, 0) }\big[1 - i a \partial_{x_2}\lambda^2_1(x_1, x_2, t, 0) - i \delta t \tilde{\lambda^2_1}(x_1, x_2, t, 0) \big] \bra{\psi_2} \\
   + e^{- i \lambda^3_1(x_1, x_2, t, 0) } \big[1 - i a \partial_{x_2}\lambda^3_1(x_1, x_2, t, 0) - i \delta t \tilde{\lambda^3_1}(x_1, x_2, t, 0) \big]\bra{\psi_3} \bigg]\\
   \otimes \ket{x_1, x_2}\bra{x_1, x_2}\bigg[ \mathds{1}_1 \otimes \mathds{1}_2 + i \frac{a}{\hbar}(\mathds{1}_1 \otimes p_2) \bigg] \Bigg\} \nonumber
\end{multline}\begin{multline}
   + \sum_{x_1, x_2} \Bigg\{ \frac{1}{8}\bigg[e^{- i\lambda^0_2(x_1, x_2, t, 0)}\big[ 1  + i a \partial_{x_2} \lambda^0_2(x_1, x_2, t, 0) - i \delta t \tilde{\lambda_2^0}(x_1, x_2, t, 0) \big] \\
     - e^{- i\lambda^1_2(x_1, x_2, t, 0)}\big[ 1  + i a \partial_{x_2} \lambda^1_2(x_1, x_2, t, 0) - i \delta t \tilde{\lambda^1_2}(x_1, x_2, t, 0) \big] \\ 
   + e^{- i\lambda^2_2(x_1, x_2, t, 0)}\big[ 1  + i a \partial_{x_2} \lambda^2_2(x_1, x_2, t, 0) - i \delta t \tilde{\lambda^2_2}(x_1, x_2, t, 0) \big] \\ 
   - e^{- i\lambda^3_2(x_1, x_2, t, 0)} \bigg]\big[ 1  + i a \partial_{x_2} \lambda^3_2(x_1, x_2, t, 0) - i \delta t \tilde{\lambda^3_2}(x_1, x_2, t, 0) \big] \bigg]  \ket{\downarrow \uparrow} \\
  \bigg[ e^{- i \lambda^0_1(x_1, x_2, t, 0) }\big[ 1  + i a \partial_{x_2} \lambda^0_1(x_1, x_2, t, 0) - i \delta t \tilde{\lambda^0_1}(x_1, x_2, t, 0) \big] \bra{\psi_0} \\
   - e^{- i \lambda^1_1(x_1, x_2, t, 0) } \big[ 1  + i a \partial_{x_2} \lambda^1_1(x_1, x_2, t, 0) - i \delta t \tilde{\lambda^1_1}(x_1, x_2, t, 0) \big] \bra{\psi_1}\\ 
    - e^{- i \lambda^2_1(x_1, x_2, t, 0) }\big[ 1  + i a \partial_{x_2} \lambda^2_1(x_1, x_2, t, 0) - i \delta t \tilde{\lambda^2_1}(x_1, x_2, t, 0) \big]  \bra{\psi_2} \\
   + e^{- i \lambda^3_1(x_1, x_2, t, 0) } \big[ 1  + i a \partial_{x_2} \lambda^3_1(x_1, x_2, t, 0) - i \delta t \tilde{\lambda^3_1}(x_1, x_2, t, 0) \big] \bra{\psi_3} \bigg]\\
   \otimes \ket{x_1, x_2}\bra{x_1, x_2} \bigg[ \mathds{1}_1 \otimes \mathds{1}_2 - i \frac{a}{\hbar}(\mathds{1}_1 \otimes p_2) \bigg] \Bigg\} \nonumber
  \end{multline}\begin{multline}
   + \Bigg\{ \frac{1}{8}\bigg[e^{- i\lambda^0_2(x_1, x_2, t, 0)}\big[ 1  + i a \partial_{x_2} \lambda^0_2(x_1, x_2, t, 0) - i \delta t \tilde{\lambda^0_2}(x_1, x_2, t, 0) \big]\\ 
   - e^{- i\lambda^1_2(x_1, x_2, t, 0)}\big[ 1  + i a \partial_{x_2} \lambda^1_2(x_1, x_2, t, 0) - i \delta t \tilde{\lambda^1_2}(x_1, x_2, t, 0) \big] \\ 
   - e^{- i\lambda^2_2(x_1, x_2, t, 0)} \big[ 1  + i a \partial_{x_2} \lambda^2_2(x_1, x_2, t, 0) - i \delta t \tilde{\lambda^2_2}(x_1, x_2, t, 0) \big] \\
   + e^{- i\lambda^3_2(x_1, x_2, t, 0)}\big[ 1  + i a \partial_{x_2} \lambda^3_2(x_1, x_2, t, 0) - i \delta t \tilde{\lambda^3_2}(x_1, x_2, t, 0) \big] 
   \bigg]\ket{\downarrow \uparrow} \\ 
   \bigg[ e^{- i \lambda^0_1(x_1, x_2, t, 0)}\big[1 - i \delta t \tilde{\lambda^0_1}(x_1, x_2, t, 0) \big] \bra{\psi_0}
   - e^{- i \lambda^1_1(x_1, x_2, t, 0) }\big[1 - i \delta t \tilde{\lambda^1_1}(x_1, x_2, t, 0) \big] \bra{\psi_1} \\
   + e^{- i \lambda^2_1(x_1, x_2, t, 0) } \big[1 - i \delta t \tilde{\lambda^2_1}(x_1, x_2, t, 0) \big]\bra{\psi_2} 
  - e^{- i \lambda^3_1(x_1, x_2, t, 0) } \big[1 - i \delta t \tilde{\lambda^3_1}(x_1, x_2, t, 0) \big]\bra{\psi_3} \bigg] \\
   \otimes \ket{x_1, x_2}\bra{x_1, x_2} \Bigg\} \nonumber
  \end{multline}\begin{multline}
  + \Bigg\{ \frac{1}{8}\bigg[e^{- i\lambda^0_2(x_1, x_2, t, 0)}\big[ 1  + i a \partial_{x_2} \lambda^0_2(x_1, x_2, t, 0) - i \delta t \tilde{\lambda_2^0}(x_1, x_2, t, 0) \big] \\
   + e^{- i\lambda^1_2(x_1, x_2, t, 0)}\big[ 1  + i a \partial_{x_2} \lambda^1_2(x_1, x_2, t, 0) - i \delta t \tilde{\lambda_2^1}(x_1, x_2, t, 0) \big] \\ 
   + e^{- i\lambda^2_2(x_1, x_2, t, 0)}\big[ 1  + i a \partial_{x_2} \lambda^2_2(x_1, x_2, t, 0) - i \delta t \tilde{\lambda_2^2}(x_1, x_2, t, 0) \big] \\  
   + e^{- i\lambda^3_2(x_1, x_2, t, 0)} \big[ 1  + i a \partial_{x_2} \lambda^3_2(x_1, x_2, t, 0) - i \delta t \tilde{\lambda_2^3}(x_1, x_2, t, 0) \big] \bigg] \ket{\downarrow \uparrow} \\
  \bigg[e^{- i \lambda^0_1(x_1, x_2, t, 0)}\big[ 1 - i a \partial_{x_1} \lambda^0_1(x_1, x_2, t, 0)  + i a \partial_{x_2} \lambda^0_1(x_1, x_2, t, 0) - i \delta t \tilde{\lambda^0_1}(x_1, x_2, t, 0) \big] \bra{\psi_0} \\
   + e^{- i \lambda^1_1(x_1, x_2, t, 0) } \big[ 1 - i a \partial_{x_1} \lambda^1_1(x_1, x_2, t, 0)  + i a \partial_{x_2} \lambda^1_1(x_1, x_2, t, 0) - i \delta t \tilde{\lambda^1_1}(x_1, x_2, t, 0) \big]\bra{\psi_1} \\
  - e^{- i \lambda^2_1(x_1, x_2, t, 0) } \big[ 1 - i a \partial_{x_1} \lambda^2_1(x_1, x_2, t, 0)  + i a \partial_{x_2} \lambda^2_1(x_1, x_2, t, 0) - i \delta t \tilde{\lambda^2_1}(x_1, x_2, t, 0) \big] \bra{\psi_2} \\
   - e^{- i \lambda^3_1(x_1, x_2, t, 0) } \big[ 1 - i a \partial_{x_1} \lambda^3_1(x_1, x_2, t, 0)  + i a \partial_{x_2} \lambda^3_1(x_1, x_2, t, 0) - i \delta t \tilde{\lambda^3_1}(x_1, x_2, t, 0) \big] \bra{\psi_3} \bigg] \\
   \otimes \ket{x_1, x_2}\bra{x_1, x_2} \bigg[ \mathds{1}_1 \otimes \mathds{1}_2  +  \frac{i a}{\hbar}(p_1 \otimes \mathds{1}_2 - \mathds{1}_1 \otimes p_2) \bigg]  \Bigg\} \nonumber
\end{multline} \begin{multline}
  + \Bigg\{ \frac{1}{8}\bigg[e^{- i\lambda^0_2(x_1, x_2, t, 0)}\big[ 1 + i a \partial_{x_2} \lambda^0_2(x_1, x_2, t, 0) - i \delta t \tilde{\lambda^0_2}(x_1, x_2, t, 0) \big] \\
   + e^{- i\lambda^1_2(x_1, x_2, t, 0)}\big[ 1 + i a \partial_{x_2} \lambda^1_2(x_1, x_2, t, 0) - i \delta t \tilde{\lambda^1_2}(x_1, x_2, t, 0) \big] \\
  - e^{- i\lambda^2_2(x_1, x_2, t, 0)} \big[ 1 + i a \partial_{x_2} \lambda^2_2(x_1, x_2, t, 0) - i \delta t \tilde{\lambda^2_2}(x_1, x_2, t, 0) \big] \\
  - e^{- i\lambda^3_2(x_1, x_2, t, 0)}\big[ 1 + i a \partial_{x_2} \lambda^3_2(x_1, x_2, t, 0) - i \delta t \tilde{\lambda^3_2}(x_1, x_2, t, 0) \big] \bigg]\ket{\downarrow \uparrow} \\
  \bigg[ e^{- i \lambda^0_1(x_1, x_2, t, 0) }\big[ 1 - i a \partial_{x_1} \lambda^0_1(x_1, x_2, t, 0) - i \delta t \tilde{\lambda^0_1}(x_1, x_2, t, 0) \big] \bra{\psi_0} \\
   + e^{- i \lambda^1_1(x_1, x_2, t, 0) }\big[ 1 - i a \partial_{x_1} \lambda^1_1(x_1, x_2, t, 0) - i \delta t \tilde{\lambda^1_1}(x_1, x_2, t, 0) \big] \bra{\psi_1} \\
   + e^{- i \lambda^2_1(x_1, x_2, t, 0) }\big[ 1 - i a \partial_{x_1} \lambda^2_1(x_1, x_2, t, 0) - i \delta t \tilde{\lambda^2_1}(x_1, x_2, t, 0) \big] \bra{\psi_2} \\
  + e^{- i \lambda^3_1(x_1, x_2, t, 0) }\big[ 1 - i a \partial_{x_1} \lambda^3_1(x_1, x_2, t, 0) - i \delta t \tilde{\lambda^3_1}(x_1, x_2, t, 0) \big] \bra{\psi_3} \bigg] \\
   \otimes \ket{x_1, x_2}\bra{x_1, x_2} \bigg[ \mathds{1}_1 \otimes \mathds{1}_2 + \frac{i a}{\hbar}(p_1 \otimes \mathds{1}_2) \bigg] \Bigg\}\nonumber
  \end{multline}\begin{multline}
   + \sum_{x_1, x_2}\Bigg\{ \frac{1}{8}\bigg[e^{- i\lambda^0_2(x_1, x_2, t, 0)}\big[ 1 - i \delta t \tilde{\lambda^0_2}(x_1, x_2, t, 0)  \big]\\
   - e^{- i\lambda^1_2(x_1, x_2, t, 0)}\big[ 1 - i \delta t \tilde{\lambda^1_2}(x_1, x_2, t, 0)  \big]  
   - e^{- i\lambda^2_2(x_1, x_2, t, 0)}\big[ 1 - i \delta t \tilde{\lambda^2_2}(x_1, x_2, t, 0)  \big]\\
   + e^{- i\lambda^3_2(x_1, x_2, t, 0)} \big[ 1 - i \delta t \tilde{\lambda^3_2}(x_1, x_2, t, 0)  \big]\bigg] \ket{\downarrow \downarrow} 
  \bigg[ e^{- i \lambda^0_1(x_1, x_2, t, 0) }\big[ 1 - i \delta t \tilde{\lambda^0_1}(x_1, x_2, t, 0)  \big] \bra{\psi_0}\\
   - e^{- i \lambda^1_1(x_1, x_2, t, 0) }\big[ 1 - i \delta t \tilde{\lambda^1_1}(x_1, x_2, t, 0)  \big]  \bra{\psi_1} 
   - e^{- i \lambda^2_1(x_1, x_2, t, 0) } \big[ 1 - i \delta t \tilde{\lambda^2_1}(x_1, x_2, t, 0)  \big] \bra{\psi_2} \\
   + e^{- i \lambda^3_1(x_1, x_2, t, 0) }\big[ 1 - i \delta t \tilde{\lambda^3_1}(x_1, x_2, t, 0)  \big]  \bra{\psi_3} \bigg] \otimes \ket{x_1, x_2}\bra{x_1, x_2} \Bigg\}\nonumber
  \end{multline}\begin{multline}
  + \Bigg\{\frac{1}{8}\bigg[e^{- i\lambda^0_2(x_1, x_2, t, 0)} \big[ 1 - i \delta t \tilde{\lambda^0_2}(x_1, x_2, t, 0) \big] 
  - e^{- i\lambda^1_2(x_1, x_2, t, 0)}\big[ 1 - i \delta t \tilde{\lambda^1_2}(x_1, x_2, t, 0)  \big] \\
   + e^{- i\lambda^2_2(x_1, x_2, t, 0)}\big[ 1 - i \delta t \tilde{\lambda^2_2}(x_1, x_2, t, 0)  \big]  
   - e^{- i\lambda^3_2(x_1, x_2, t, 0)} \big[ 1 - i \delta t \tilde{\lambda^3_2}(x_1, x_2, t, 0)  \big] \bigg]\ket{\downarrow \downarrow} \\
  \bigg[ e^{- i \lambda^0_1(x_1, x_2, t, 0)} [1 - i a \partial_{x_2}\lambda^0_1(x_1, x_2, t, 0) - i \delta t \tilde{\lambda^0_1}(x_1, x_2, t, 0) ] \bra{\psi_0} \\
   - e^{- i \lambda^1_1(x_1, x_2, t, 0)} [1 - i a \partial_{x_2}\lambda^1_1(x_1, x_2, t, 0) - i \delta t \tilde{\lambda^1_1}(x_1, x_2, t, 0) ] \bra{\psi_1} \\
   + e^{- i \lambda^2_1(x_1, x_2, t, 0)} [1 - i a \partial_{x_2}\lambda^2_1(x_1, x_2, t, 0) - i \delta t \tilde{\lambda^2_1}(x_1, x_2, t, 0)] \bra{\psi_2} \\
 - e^{- i \lambda^3_1(x_1, x_2, t, 0) }  [1 - i a \partial_{x_2}\lambda^3_1(x_1, x_2, t, 0) - i \delta t \tilde{\lambda^3_1}(x_1, x_2, t, 0) ]\bra{\psi_3} \bigg]\\
   \otimes \ket{x_1, x_2}\bra{x_1, x_2}\bigg[\mathds{1}_1 \otimes \mathds{1}_2  + \frac{i a}{\hbar}(\mathds{1}_1 \otimes p_2) \bigg]\Bigg\} \nonumber
\end{multline}\begin{multline}
   + \Bigg\{ \frac{1}{8}\bigg[e^{- i\lambda^0_2(x_1, x_2, t, 0)}\big[ 1 - i \delta t \tilde{\lambda^0_2}(x_1, x_2, t, 0) \big] 
   + e^{- i\lambda^1_2(x_1, x_2, t, 0)} \big[ 1 - i \delta t \tilde{\lambda^1_2}(x_1, x_2, t, 0) \big] \\
   - e^{- i\lambda^2_2(x_1, x_2, t, 0)}\big[ 1 - i \delta t \tilde{\lambda^2_2}(x_1, x_2, t, 0) \big] 
   - e^{- i\lambda^3_2(x_1, x_2, t, 0)} \big[ 1 - i \delta t \tilde{\lambda^3_2}(x_1, x_2, t, 0) \big]\bigg] \ket{\downarrow \downarrow} \\
   \bigg[ e^{- i \lambda^0_1(x_1, x_2, t, 0)}\big[ 1 - i a \partial_{x_1}\lambda^0_1(x_1, x_2, t, 0)- i \delta t \tilde{\lambda^0_1}(x_1, x_2, t, 0) \big] \bra{\psi_0}\\
   + e^{- i \lambda^1_1(x_1, x_2, t, 0) }\big[ 1 - i a \partial_{x_1}\lambda^1_1(x_1, x_2, t, 0)- i \delta t \tilde{\lambda^1_1}(x_1, x_2, t, 0) \big] \bra{\psi_1} \\
   - e^{- i \lambda^2_1(x_1, x_2, t, 0) } \big[ 1 - i a \partial_{x_1}\lambda^2_1(x_1, x_2, t, 0)- i \delta t \tilde{\lambda^2_1}(x_1, x_2, t, 0) \big]\bra{\psi_2} \\ 
   - e^{- i \lambda^3_1(x_1, x_2, t, 0) }\big[ 1 - i a \partial_{x_1}\lambda^3_1(x_1, x_2, t, 0)- i \delta t \tilde{\lambda^3_1}(x_1, x_2, t, 0) \big] \bra{\psi_3} \bigg] \\
   \otimes \ket{x_1, x_2}\bra{x_1, x_2} \bigg[ \mathds{1}_1 \otimes \mathds{1}_2 + \frac{i a}{\hbar} (p_1 \otimes \mathds{1}_2)\bigg] \Bigg\} \nonumber
  \end{multline}\begin{multline}
   + \Bigg\{\frac{1}{8}\bigg[e^{- i\lambda^0_2(x_1, x_2, t, 0)} \big[ 1 - i \delta t \tilde{\lambda^0_2}(x_1, x_2, t, 0) \big]
   + e^{- i\lambda^1_2(x_1, x_2, t, 0)} \big[ 1 - i \delta t \tilde{\lambda^1_2}(x_1, x_2, t, 0) \big] \\
   + e^{- i\lambda^2_2(x_1, x_2, t, 0)}\big[ 1 - i \delta t \tilde{\lambda^2_2}(x_1, x_2, t, 0) \big] 
  + e^{- i\lambda^3_2(x_1, x_2, t, 0)} \big[ 1 - i \delta t \tilde{\lambda^3_2}(x_1, x_2, t, 0) \big] \bigg]\ket{\downarrow \downarrow} \\
   \bigg[ e^{- i \lambda^0_1(x_1, x_2, t, 0) }\big[ 1 - i a \partial_{x_1}\lambda^0_1(x_1, x_2, t, 0) -i a \partial_{x_2}\lambda^0_1(x_1, x_2, t, 0) - i \delta t \tilde{\lambda^0_1}(x_1, x_2, t, 0) \big] \bra{\psi_0} \\
   + e^{- i \lambda^1_1(x_1, x_2, t, 0) }\big[ 1 - i a \partial_{x_1}\lambda^1_1(x_1, x_2, t, 0) -i a \partial_{x_2}\lambda^1_1(x_1, x_2, t, 0) - i \delta t \tilde{\lambda^1_1}(x_1, x_2, t, 0) \big]\bra{\psi_1}  \\
  + e^{- i \lambda^2_1(x_1, x_2, t, 0) }\big[ 1 - i a \partial_{x_1}\lambda^2_1(x_1, x_2, t, 0) -i a \partial_{x_2}\lambda^2_1(x_1, x_2, t, 0) - i \delta t \tilde{\lambda^2_1}(x_1, x_2, t, 0) \big] \bra{\psi_2} \\
   + e^{- i \lambda^3_1(x_1, x_2, t, 0) }\big[ 1 - i a \partial_{x_1}\lambda^3_1(x_1, x_2, t, 0) -i a \partial_{x_2}\lambda^3_1(x_1, x_2, t, 0) - i \delta t \tilde{\lambda^3_1}(x_1, x_2, t, 0) \big]\bra{\psi_3} \bigg] \\
   \otimes \ket{x_1, x_2}\bra{x_1, x_2} \bigg[ \mathds{1}_1 \otimes \mathds{1}_2  + \frac{i a}{\hbar}(p_1 \otimes \mathds{1}_2 + \mathds{1}_1 \otimes p_2) \bigg] \Bigg\} ~, \nonumber
\end{multline}
Then considering the terms only upto the first order in $\delta t$ and $a$ we get 

\begin{multline}
 U^\text{two}(t, \delta t)  =  \\
 \sum_{x_1, x_2}\frac{1}{4} \bigg[ e^{- i \lambda^0_1(x_1, x_2, t, 0)}\big[e^{- i \lambda^0_2(x_1, x_2, t, 0)} +e^{- i \lambda^1_2(x_1, x_2, t, 0)} \big]
 [\ket{\uparrow \uparrow} + \ket{\uparrow \downarrow} - \ket{\downarrow \uparrow} - \ket{\downarrow \downarrow} ]\bra{\psi_0} \\
 -e^{- i \lambda^1_1(x_1, x_2, t, 0)}\big[e^{- i \lambda^0_2(x_1, x_2, t, 0)} +e^{- i \lambda^1_2(x_1, x_2, t, 0)} \big]
 [\ket{\uparrow \uparrow}  + \ket{\uparrow \downarrow} + \ket{\downarrow \uparrow} + \ket{\downarrow \downarrow} ]\bra{\psi_1} \\
 -e^{- i \lambda^2_1(x_1, x_2, t, 0)}\big[e^{- i \lambda^2_2(x_1, x_2, t, 0)} +e^{- i \lambda^3_2(x_1, x_2, t, 0)} \big]
 [\ket{\uparrow \uparrow} - \ket{\uparrow \downarrow} - \ket{\downarrow \uparrow} + \ket{\downarrow \downarrow} ]\bra{\psi_2} \\
 + e^{- i \lambda^3_1(x_1, x_2, t, 0)}\big[e^{- i \lambda^2_2(x_1, x_2, t, 0)} +e^{- i \lambda^3_2(x_1, x_2, t, 0)} \big]
 [\ket{\uparrow \uparrow} - \ket{\uparrow \downarrow} + \ket{\downarrow \uparrow} - \ket{\downarrow \downarrow} ]\bra{\psi_3} \bigg] \\
 \otimes \ket{x_1 x_2}\bra{x_1 x_2}\frac{- ia}{\hbar}(p_1 \otimes \mathds{1}_2) \nonumber
\end{multline}
\begin{multline}
 + \sum_{x_1, x_2}\frac{1}{4}\bigg[ e^{- i \lambda^0_1(x_1, x_2, t, 0)}\big[e^{- i \lambda^0_2(x_1, x_2, t, 0)} +e^{- i \lambda^2_2(x_1, x_2, t, 0)} \big]
 [\ket{\uparrow \uparrow} - \ket{\uparrow \downarrow} + \ket{\downarrow \uparrow} - \ket{\downarrow \downarrow} ]\bra{\psi_0} \\
 - e^{- i \lambda^1_1(x_1, x_2, t, 0)}\big[e^{- i \lambda^1_2(x_1, x_2, t, 0)} +e^{- i \lambda^3_2(x_1, x_2, t, 0)} \big]
 [\ket{\uparrow \uparrow} - \ket{\uparrow \downarrow} - \ket{\downarrow \uparrow} + \ket{\downarrow \downarrow} ]\bra{\psi_1} \\
 - e^{- i \lambda^2_1(x_1, x_2, t, 0)}\big[e^{- i \lambda^0_2(x_1, x_2, t, 0)} +e^{- i \lambda^2_2(x_1, x_2, t, 0)} \big]
 [\ket{\uparrow \uparrow} + \ket{\uparrow \downarrow} + \ket{\downarrow \uparrow} + \ket{\downarrow \downarrow} ]\bra{\psi_2} \\
 + e^{- i \lambda^3_1(x_1, x_2, t, 0)}\big[e^{- i \lambda^1_2(x_1, x_2, t, 0)} +e^{- i \lambda^3_2(x_1, x_2, t, 0)} \big]
 [\ket{\uparrow \uparrow} + \ket{\uparrow \downarrow} - \ket{\downarrow \uparrow} - \ket{\downarrow \downarrow} ]\bra{\psi_3} \\
 \otimes \ket{x_1 x_2}\bra{x_1 x_2}\frac{- ia}{\hbar}( \mathds{1}_1 \otimes p_2) \nonumber
  \end{multline}\begin{multline}
 + \sum_{x_1, x_2} \frac{1}{2}~ \bigg[ e^{- i [\lambda^0_1(x_1, x_2, t, 0) + \lambda^0_2(x_1, x_2, t, 0) ]} 
 [\ket{\uparrow \uparrow} + \ket{\uparrow \downarrow} + \ket{\downarrow \uparrow} + \ket{\downarrow \downarrow}]\bra{\psi_0}[ 1 - i \delta t~\tilde{\lambda^0_1}(x_1, x_2, t, 0)  ] \\
 -  e^{- i [\lambda^1_1(x_1, x_2, t, 0) + \lambda^1_2(x_1, x_2, t, 0) ]} 
 [\ket{\uparrow \uparrow}+ \ket{\uparrow \downarrow} - \ket{\downarrow \uparrow} - \ket{\downarrow \downarrow} ]\bra{\psi_1}[ 1 - i \delta t~\tilde{\lambda^1_1}(x_1, x_2, t, 0)  ] \\
-  e^{- i [\lambda^2_1(x_1, x_2, t, 0) + \lambda^2_2(x_1, x_2, t, 0) ]} 
 [\ket{\uparrow \uparrow} - \ket{\uparrow \downarrow} + \ket{\downarrow \uparrow}- \ket{\downarrow \downarrow}]\bra{\psi_2}[ 1 - i \delta t~\tilde{\lambda^2_1}(x_1, x_2, t, 0)  ]\\
 +  e^{- i [\lambda^3_1(x_1, x_2, t, 0) + \lambda^3_2(x_1, x_2, t, 0) ]} 
 [\ket{\uparrow \uparrow} - \ket{\uparrow \downarrow} - \ket{\downarrow \uparrow} + \ket{\downarrow \downarrow} ]\bra{\psi_3}[ 1 - i \delta t~\tilde{\lambda^3_1}(x_1, x_2, t, 0)  ]\bigg]
\otimes \ket{x_1 x_2}\bra{x_1 x_2} \nonumber
\end{multline}
\begin{multline}
 + \sum_{x_1, x_2} \frac{1}{4}\bigg[[i a \partial_{x_1} \lambda^0_1(x_1, x_2, t, 0)]~  e^{- i\lambda^0_1(x_1, x_2, t, 0) } [ e^{- i a\lambda^0_2(x_1, x_2, t, 0)} + e^{- i a\lambda^1_2(x_1, x_2, t, 0)} ] \\
 [\ket{\uparrow \uparrow} + \ket{\uparrow \downarrow} - \ket{\downarrow \uparrow} - \ket{\downarrow \downarrow} ]\bra{\psi_0} \\
 - [i a \partial_{x_1} \lambda^1_1(x_1, x_2, t, 0)]~  e^{- i\lambda^1_1(x_1, x_2, t, 0) } [ e^{- i a\lambda^0_2(x_1, x_2, t, 0)} + e^{- i a\lambda^1_2(x_1, x_2, t, 0)} ] \\
 [\ket{\uparrow \uparrow} + \ket{\uparrow \downarrow} + \ket{\downarrow \uparrow} + \ket{\downarrow \downarrow} ]\bra{\psi_1} \\
 - [i a \partial_{x_1} \lambda^2_1(x_1, x_2, t, 0)]~  e^{- i\lambda^2_1(x_1, x_2, t, 0) } [ e^{- i a\lambda^2_2(x_1, x_2, t, 0)} + e^{- i a\lambda^3_2(x_1, x_2, t, 0)} ] \\
 [\ket{\uparrow \uparrow} - \ket{\uparrow \downarrow} - \ket{\downarrow \uparrow} + \ket{\downarrow \downarrow} ]\bra{\psi_2} \\
 + [i a \partial_{x_1} \lambda^3_1(x_1, x_2, t, 0)]~  e^{- i\lambda^3_1(x_1, x_2, t, 0) } [ e^{- i a\lambda^2_2(x_1, x_2, t, 0)} + e^{- i a\lambda^3_2(x_1, x_2, t, 0)} ] \\
 [\ket{\uparrow \uparrow} - \ket{\uparrow \downarrow} + \ket{\downarrow \uparrow} - \ket{\downarrow \downarrow}]\bra{\psi_3} 
\bigg] \otimes \ket{x_1 x_2}\bra{x_1 x_2} \nonumber\end{multline} \begin{multline} 
 + \sum_{x_1, x_2} \frac{1}{4}\bigg[[i a \partial_{x_2} \lambda^0_1(x_1, x_2, t, 0)]~  e^{- i\lambda^0_1(x_1, x_2, t, 0) } [ e^{- i a\lambda^0_2(x_1, x_2, t, 0)} + e^{- i a\lambda^2_2(x_1, x_2, t, 0)} ] \\
 [\ket{\uparrow \uparrow} - \ket{\uparrow \downarrow} + \ket{\downarrow \uparrow} - \ket{\downarrow \downarrow}  ]\bra{\psi_0} \\ 
- [i a \partial_{x_2} \lambda^1_1(x_1, x_2, t, 0)]~  e^{- i\lambda^1_1(x_1, x_2, t, 0) } [ e^{- i a\lambda^1_2(x_1, x_2, t, 0)} + e^{- i a\lambda^3_2(x_1, x_2, t, 0)} ] \\
 [\ket{\uparrow \uparrow} - \ket{\uparrow\downarrow} - \ket{\downarrow \uparrow} + \ket{\downarrow \downarrow} ]\bra{\psi_1} \\
 - [i a \partial_{x_2} \lambda^2_1(x_1, x_2, t, 0)]~  e^{- i\lambda^2_1(x_1, x_2, t, 0) } [ e^{- i a\lambda^0_2(x_1, x_2, t, 0)} + e^{- i a\lambda^2_2(x_1, x_2, t, 0)} ] \\
 [\ket{\uparrow \uparrow} + \ket{\uparrow \downarrow} + \ket{\downarrow \uparrow} + \ket{\downarrow \downarrow} ]\bra{\psi_2} \\
+ [i a \partial_{x_2} \lambda^3_1(x_1, x_2, t, 0)]~  e^{- i\lambda^3_1(x_1, x_2, t, 0) } [ e^{- i a\lambda^1_2(x_1, x_2, t, 0)} + e^{- i a\lambda^3_2(x_1, x_2, t, 0)} ] \\
 [\ket{\uparrow \uparrow} + \ket{\uparrow \downarrow} - \ket{\downarrow \uparrow} - \ket{\downarrow \downarrow} ]\bra{\psi_3} 
\bigg] \otimes \ket{x_1 x_2}\bra{x_1 x_2} \nonumber
\end{multline}
\begin{multline}
 + \sum_{x_1, x_2} \frac{1}{2} \bigg[  e^{- i [\lambda^0_1(x_1, x_2, t, 0) + \lambda^0_2(x_1, x_2, t, 0)]} \\
 \Big([i a \partial_{x_1} \lambda^0_2(x_1, x_2, t, 0) + i a \partial_{x_2}\lambda^0_2(x_1, x_2, t, 0) - i \delta t~\tilde{\lambda^0_2}(x_1, x_2, t, 0)] \ket{\uparrow \uparrow} \bra{\psi_0} \\
 + [i a \partial_{x_1} \lambda^0_2(x_1, x_2, t, 0) - i \delta t~\tilde{\lambda^0_2}(x_1, x_2, t, 0)]\ket{\uparrow \downarrow} \bra{\psi_0} \\
 + [i a \partial_{x_2} \lambda^0_2(x_1, x_2, t, 0) - i \delta t~\tilde{\lambda^0_2}(x_1, x_2, t, 0)]\ket{\downarrow \uparrow} \bra{\psi_0} \\
 + [- i \delta t~\tilde{\lambda^0_2}(x_1, x_2, t, 0)]\ket{\downarrow \downarrow} \bra{\psi_0} \Big) 
- e^{- i [\lambda^1_1(x_1, x_2, t, 0) + \lambda^1_2(x_1, x_2, t, 0)]} \\
 \Big([i a \partial_{x_1} \lambda^1_2(x_1, x_2, t, 0) + i a \partial_{x_2}\lambda^1_2(x_1, x_2, t, 0) - i \delta t~\tilde{\lambda^1_2}(x_1, x_2, t, 0)] \ket{\uparrow \uparrow} \bra{\psi_1} \\
 + [i a \partial_{x_1} \lambda^1_2(x_1, x_2, t, 0) - i \delta t~\tilde{\lambda^1_2}(x_1, x_2, t, 0)]\ket{\uparrow \downarrow} \bra{\psi_1} \\
 - [i a \partial_{x_2} \lambda^1_2(x_1, x_2, t, 0) - i \delta t~\tilde{\lambda^1_2}(x_1, x_2, t, 0)]\ket{\downarrow \uparrow} \bra{\psi_1} \\
  - [- i \delta t~\tilde{\lambda^1_2}(x_1, x_2, t, 0)]\ket{\downarrow \downarrow} \bra{\psi_1}\Big) 
  - e^{- i [\lambda^2_1(x_1, x_2, t, 0) + \lambda^2_2(x_1, x_2, t, 0)]} \\
 \Big([i a \partial_{x_1} \lambda^2_2(x_1, x_2, t, 0) + i a \partial_{x_2}\lambda^2_2(x_1, x_2, t, 0) - i \delta t~\tilde{\lambda^2_2}(x_1, x_2, t, 0)] \ket{\uparrow \uparrow} \bra{\psi_2} \\
 -  [i a \partial_{x_1} \lambda^2_2(x_1, x_2, t, 0) - i \delta t~\tilde{\lambda^2_2}(x_1, x_2, t, 0)]\ket{\uparrow \downarrow} \bra{\psi_2} \\
 +  [i a \partial_{x_2} \lambda^2_2(x_1, x_2, t, 0) - i \delta t~\tilde{\lambda^2_2}(x_1, x_2, t, 0)]\ket{\downarrow \uparrow} \bra{\psi_2} \\
 - [- i \delta t~\tilde{\lambda^2_2}(x_1, x_2, t, 0)]\ket{\downarrow \downarrow} \bra{\psi_2} \Big)
 + e^{- i [\lambda^3_1(x_1, x_2, t, 0) + \lambda^3_2(x_1, x_2, t, 0)]} \\
 \Big([i a \partial_{x_1} \lambda^3_2(x_1, x_2, t, 0) + i a \partial_{x_2}\lambda^3_2(x_1, x_2, t, 0) - i \delta t~\tilde{\lambda^3_2}(x_1, x_2, t, 0)] \ket{\uparrow \uparrow} \bra{\psi_3} \\
 - [i a \partial_{x_1} \lambda^3_2(x_1, x_2, t, 0) - i \delta t~\tilde{\lambda^3_2}(x_1, x_2, t, 0)]\ket{\uparrow \downarrow} \bra{\psi_3} \\
 -  [i a \partial_{x_2} \lambda^3_2(x_1, x_2, t, 0) - i \delta t~\tilde{\lambda^3_2}(x_1, x_2, t, 0)]\ket{\downarrow \uparrow} \bra{\psi_3} \\
 + [- i \delta t~\tilde{\lambda^3_2}(x_1, x_2, t, 0)]\ket{\downarrow \downarrow} \bra{\psi_1}\Big)
  \bigg] \otimes \ket{x_1 x_2}\bra{x_1 x_2}~. 
 \end{multline}
Therefore from the from of the eigenvectors given in eq.~(\ref{twoeigenvec}) we get
\begin{multline}
 U^\text{two}(t, \delta t)  = 
  \frac{ i a}{2 \hbar}\sum_{x_1, x_2} \bigg[ e^{- i \lambda^0_1(x_1, x_2, t, 0)}\big[e^{- i \lambda^0_2(x_1, x_2, t, 0)} +e^{- i \lambda^1_2(x_1, x_2, t, 0)} \big]\ket{\psi_1}\bra{\psi_0} \\
 + e^{- i \lambda^1_1(x_1, x_2, t, 0)}\big[e^{- i \lambda^0_2(x_1, x_2, t, 0)} +e^{- i \lambda^1_2(x_1, x_2, t, 0)} \big]\ket{\psi_0}\bra{\psi_1} \\
 + e^{- i \lambda^2_1(x_1, x_2, t, 0)}\big[e^{- i \lambda^2_2(x_1, x_2, t, 0)} +e^{- i \lambda^3_2(x_1, x_2, t, 0)} \big]\ket{\psi_3}\bra{\psi_2} \\
 + e^{- i \lambda^3_1(x_1, x_2, t, 0)}\big[e^{- i \lambda^2_2(x_1, x_2, t, 0)} +e^{- i \lambda^3_2(x_1, x_2, t, 0)} \big]\ket{\psi_2} \bra{\psi_3} \bigg]
 \otimes \ket{x_1 x_2}\bra{x_1 x_2}(p_1 \otimes \mathds{1}_2) \nonumber
\end{multline}
\begin{multline}
 + \frac{ia}{2 \hbar} \sum_{x_1, x_2}\bigg[ e^{- i \lambda^0_1(x_1, x_2, t, 0)}\big[e^{- i \lambda^0_2(x_1, x_2, t, 0)} +e^{- i \lambda^2_2(x_1, x_2, t, 0)} \big]\ket{\psi_2} \bra{\psi_0} \\
 + e^{- i \lambda^1_1(x_1, x_2, t, 0)}\big[e^{- i \lambda^1_2(x_1, x_2, t, 0)} +e^{- i \lambda^3_2(x_1, x_2, t, 0)} \big]\ket{\psi_3}\bra{\psi_1} \\
 + e^{- i \lambda^2_1(x_1, x_2, t, 0)}\big[e^{- i \lambda^0_2(x_1, x_2, t, 0)} +e^{- i \lambda^2_2(x_1, x_2, t, 0)} \big]\ket{\psi_0}\bra{\psi_2} \\
 + e^{- i \lambda^3_1(x_1, x_2, t, 0)}\big[e^{- i \lambda^1_2(x_1, x_2, t, 0)} +e^{- i \lambda^3_2(x_1, x_2, t, 0)} \big]\ket{\psi_1}\bra{\psi_3} \bigg]
 \otimes \ket{x_1 x_2}\bra{x_1 x_2}(\mathds{1}_1 \otimes p_2) \nonumber
  \end{multline}\begin{multline}
 + \sum_{x_1, x_2}  \bigg[ e^{- i [\lambda^0_1(x_1, x_2, t, 0) + \lambda^0_2(x_1, x_2, t, 0) ]}\ket{\psi_0}\bra{\psi_0} 
 +  e^{- i [\lambda^1_1(x_1, x_2, t, 0) + \lambda^1_2(x_1, x_2, t, 0) ]} \ket{\psi_1}\bra{\psi_1}\\
 + e^{- i [\lambda^2_1(x_1, x_2, t, 0) + \lambda^2_2(x_1, x_2, t, 0) ]} \ket{\psi_2}\bra{\psi_2}
 +  e^{- i [\lambda^3_1(x_1, x_2, t, 0) + \lambda^3_2(x_1, x_2, t, 0) ]} \ket{\psi_3}\bra{\psi_3}\bigg]
\otimes \ket{x_1 x_2}\bra{x_1 x_2} \nonumber
  \end{multline}\begin{multline}
 - \frac{i a}{2} \sum_{x_1, x_2} \bigg[[\partial_{x_1} \lambda^0_1(x_1, x_2, t, 0)]~ 
 e^{- i\lambda^0_1(x_1, x_2, t, 0) } [ e^{- i a\lambda^0_2(x_1, x_2, t, 0)} + e^{- i a\lambda^1_2(x_1, x_2, t, 0)} ]\ket{\psi_1}\bra{\psi_0} \\
 + [ \partial_{x_1} \lambda^1_1(x_1, x_2, t, 0)]~  
 e^{- i\lambda^1_1(x_1, x_2, t, 0) } [ e^{- i a\lambda^0_2(x_1, x_2, t, 0)} + e^{- i a\lambda^1_2(x_1, x_2, t, 0)} ] \ket{\psi_0}\bra{\psi_1} \\
 + [\partial_{x_1} \lambda^2_1(x_1, x_2, t, 0)]~ 
 e^{- i\lambda^2_1(x_1, x_2, t, 0) } [ e^{- i a\lambda^2_2(x_1, x_2, t, 0)} + e^{- i a\lambda^3_2(x_1, x_2, t, 0)} ]\ket{\psi_3}\bra{\psi_2} \\
 + [ \partial_{x_1} \lambda^3_1(x_1, x_2, t, 0)]~ 
 e^{- i\lambda^3_1(x_1, x_2, t, 0) } [ e^{- i a\lambda^2_2(x_1, x_2, t, 0)} + e^{- i a\lambda^3_2(x_1, x_2, t, 0)} ]\ket{\psi_2}\bra{\psi_3} 
\bigg] \otimes \ket{x_1 x_2}\bra{x_1 x_2} \nonumber
  \end{multline}\begin{multline}
 - \frac{i a}{2} \sum_{x_1, x_2}\bigg[[\partial_{x_2} \lambda^0_1(x_1, x_2, t, 0)]~e^{- i\lambda^0_1(x_1, x_2, t, 0) } [ e^{- i a\lambda^0_2(x_1, x_2, t, 0)} + e^{- i a\lambda^2_2(x_1, x_2, t, 0)} ] \ket{\psi_2}\bra{\psi_0} \\ 
 +[\partial_{x_2} \lambda^1_1(x_1, x_2, t, 0)]~  e^{- i\lambda^1_1(x_1, x_2, t, 0) } [ e^{- i a\lambda^1_2(x_1, x_2, t, 0)} + e^{- i a\lambda^3_2(x_1, x_2, t, 0)}]\ket{\psi_3}\bra{\psi_1} \\
 + [\partial_{x_2} \lambda^2_1(x_1, x_2, t, 0)]~  e^{- i\lambda^2_1(x_1, x_2, t, 0) } [ e^{- i a\lambda^0_2(x_1, x_2, t, 0)} + e^{- i a\lambda^2_2(x_1, x_2, t, 0)}]\ket{\psi_0} \bra{\psi_2} \\
  + [\partial_{x_2} \lambda^3_1(x_1, x_2, t, 0)]~  e^{- i\lambda^3_1(x_1, x_2, t, 0) } [ e^{- i a\lambda^1_2(x_1, x_2, t, 0)} + e^{- i a\lambda^3_2(x_1, x_2, t, 0)}]\ket{\psi_1}\bra{\psi_3} 
\bigg] \otimes \ket{x_1 x_2}\bra{x_1 x_2}\nonumber
  \end{multline}\begin{multline}
  - i \delta t \sum_{x_1, x_2} \frac{1}{2} \bigg[ e^{- i [\lambda^0_1(x_1, x_2, t, 0) + \lambda^0_2(x_1, x_2, t, 0)]} 
[ \tilde{\lambda^0_1}(x_1, x_2, t, 0) + \tilde{\lambda^0_2}(x_1, x_2, t, 0)] \ket{\psi_0}\bra{\psi_0} \\
+ e^{- i [\lambda^1_1(x_1, x_2, t, 0) + \lambda^1_2(x_1, x_2, t, 0)]} 
[ \tilde{\lambda^1_1}(x_1, x_2, t, 0) + \tilde{\lambda^1_2}(x_1, x_2, t, 0)] \ket{\psi_1} \bra{\psi_1} \\
+ e^{- i [\lambda^2_1(x_1, x_2, t, 0) + \lambda^2_2(x_1, x_2, t, 0)]}
[ \tilde{\lambda^2_1}(x_1, x_2, t, 0) + \tilde{\lambda^2_2}(x_1, x_2, t, 0)] \ket{\psi_2} \bra{\psi_2} \\
 + e^{- i [\lambda^3_1(x_1, x_2, t, 0) + \lambda^3_2(x_1, x_2, t, 0)]} 
[ \tilde{\lambda^3_1}(x_1, x_2, t, 0) + \tilde{\lambda^3_2}(x_1, x_2, t, 0)] \ket{\psi_3} \bra{\psi_3} 
  \bigg] \otimes \ket{x_1 x_2}\bra{x_1 x_2} \nonumber
 \end{multline}
 \begin{multline}
 + \sum_{x_1, x_2} \frac{1}{2} \bigg[  e^{- i [\lambda^0_1(x_1, x_2, t, 0) + \lambda^0_2(x_1, x_2, t, 0)]} 
 \Big([i a \partial_{x_1} \lambda^0_2(x_1, x_2, t, 0)] [\ket{\psi_0} - \ket{\psi_1}] \bra{\psi_0} \\
 + [i a \partial_{x_2} \lambda^0_2(x_1, x_2, t, 0)][\ket{\psi_0} - \ket{\psi_2}] \bra{\psi_0} \Big) \\
- e^{- i [\lambda^1_1(x_1, x_2, t, 0) + \lambda^1_2(x_1, x_2, t, 0)]} 
 \Big([i a \partial_{x_1} \lambda^1_2(x_1, x_2, t, 0)] [\ket{\psi_0} - \ket{\psi_1}] \bra{\psi_1} \\
  + [i a \partial_{x_2} \lambda^1_2(x_1, x_2, t, 0)][\ket{\psi_3} - \ket{\psi_1}] \bra{\psi_1} \Big) \\
  - e^{- i [\lambda^2_1(x_1, x_2, t, 0) + \lambda^2_2(x_1, x_2, t, 0)]}
 \Big([i a \partial_{x_1} \lambda^2_2(x_1, x_2, t, 0)] [\ket{\psi_3} - \ket{\psi_2}] \bra{\psi_2} \\
 +  [i a \partial_{x_2} \lambda^2_2(x_1, x_2, t, 0)][\ket{\psi_0} - \ket{\psi_2}] \bra{\psi_2}  \Big)\\
 + e^{- i [\lambda^3_1(x_1, x_2, t, 0) + \lambda^3_2(x_1, x_2, t, 0)]} 
 \Big([i a \partial_{x_1} \lambda^3_2(x_1, x_2, t, 0)] [\ket{\psi_3} - \ket{\psi_2} ]\bra{\psi_3} \\
 +  [i a \partial_{x_2} \lambda^3_2(x_1, x_2, t, 0)][\ket{\psi_3} - \ket{\psi_1}  ]\bra{\psi_3} \Big) 
  \bigg] \otimes \ket{x_1 x_2}\bra{x_1 x_2} 
 \end{multline}
 
Therefore the modified two-particle SS-DQW evolution operator upto first order in $\delta t$, $a$  
\begin{multline}
=  \mathscr{U}^\text{two}(t, \delta t) = [U^\text{two}(t, 0)]^\dagger \cdot U^\text{two}(t, \delta t) 
= \sigma_0 \otimes \sigma_0 \otimes \mathds{1}_1 \otimes \mathds{1}_2 - i \frac{\delta t}{\hbar} \mathscr{H}^\text{two}(t) + \mathcal{O}(\delta t^2) \\ \\
 = \sigma_0 \otimes \sigma_0 \otimes \mathds{1}_1 \otimes \mathds{1}_2 
 - \frac{i \delta t}{2} \sum_{x_1, x_2} \sum_q [ \tilde{\lambda^q_1}(x_1, x_2, t, 0) + \tilde{\lambda^q_2}(x_1, x_2, t, 0)]
 \ket{\psi_q} \bra{\psi_q} \otimes \ket{x_1 x_2}\bra{x_1 x_2} \\
 + \frac{ i a}{2 \hbar}\sum_{x_1, x_2} \bigg[ e^{- i [\lambda^0_1(x_1, x_2, t, 0) - \lambda^1_1(x_1, x_2, t, 0) ] }
 \big[e^{- i [\lambda^0_2(x_1, x_2, t, 0) - \lambda^1_2(x_1, x_2, t, 0) ]}  + 1 \big]\ket{\psi_1}\bra{\psi_0} \\
 + e^{- i [\lambda^1_1(x_1, x_2, t, 0) -\lambda^0_1(x_1, x_2, t, 0) ]}
 \big[e^{- i [\lambda^1_2(x_1, x_2, t, 0) - \lambda^0_2(x_1, x_2, t, 0)]} + 1 \big]\ket{\psi_0}\bra{\psi_1} \\
 + e^{- i [\lambda^2_1(x_1, x_2, t, 0) - \lambda^3_1(x_1, x_2, t, 0)]}\big[e^{- i [\lambda^2_2(x_1, x_2, t, 0) - \lambda^3_2(x_1, x_2, t, 0) ]} 
 + 1 \big]\ket{\psi_3}\bra{\psi_2} \\
 + e^{- i [\lambda^3_1(x_1, x_2, t, 0) - \lambda^2_1(x_1, x_2, t, 0)]}\big[
 e^{- i [\lambda^3_2(x_1, x_2, t, 0) - \lambda^2_2(x_1, x_2, t, 0)]} + 1 \big]\ket{\psi_2} \bra{\psi_3} \bigg]
 \otimes \ket{x_1 x_2}\bra{x_1 x_2}(p_1 \otimes \mathds{1}_2)\nonumber
  \end{multline}\begin{multline}
 + \frac{ia}{2 \hbar} \sum_{x_1, x_2}\bigg[ e^{- i [\lambda^0_1(x_1, x_2, t, 0) - \lambda^2_1(x_1, x_2, t, 0) ]}
 \big[e^{- i [\lambda^0_2(x_1, x_2, t, 0) - \lambda^2_2(x_1, x_2, t, 0)  ]} + 1 \big]\ket{\psi_2} \bra{\psi_0} \\
 + e^{- i [\lambda^1_1(x_1, x_2, t, 0) -  \lambda^3_1(x_1, x_2, t, 0) ]}
 \big[e^{- i [\lambda^1_2(x_1, x_2, t, 0)-  \lambda^3_2(x_1, x_2, t, 0)]} + 1 \big]\ket{\psi_3}\bra{\psi_1} \\
 + e^{- i [\lambda^2_1(x_1, x_2, t, 0) -  \lambda^0_1(x_1, x_2, t, 0)]}
 \big[ e^{- i [\lambda^2_2(x_1, x_2, t, 0) -  \lambda^0_2(x_1, x_2, t, 0) ]}  + 1 \big]\ket{\psi_0}\bra{\psi_2} \\
 + e^{- i [\lambda^3_1(x_1, x_2, t, 0) -  \lambda^1_1(x_1, x_2, t, 0) ]}
 \big[ e^{- i [\lambda^3_2(x_1, x_2, t, 0) -  \lambda^1_2(x_1, x_2, t, 0) ]} + 1 \big]\ket{\psi_1}\bra{\psi_3} \bigg]
 \otimes \ket{x_1 x_2}\bra{x_1 x_2}(\mathds{1}_1 \otimes p_2) \nonumber
 \end{multline}
\begin{multline}
  - \frac{i a}{2} \sum_{x_1, x_2} \bigg[[\partial_{x_1} \lambda^0_1(x_1, x_2, t, 0)]~ 
 e^{- i[\lambda^0_1(x_1, x_2, t, 0)-\lambda^1_1(x_1, x_2, t, 0)]}
 [ e^{- i [\lambda^0_2(x_1, x_2, t, 0) -\lambda^1_2(x_1, x_2, t, 0) ]} + 1 ]\ket{\psi_1}\bra{\psi_0} \\
 + [ \partial_{x_1} \lambda^1_1(x_1, x_2, t, 0)]~  
 e^{- i[\lambda^1_1(x_1, x_2, t, 0) - \lambda^0_1(x_1, x_2, t, 0)]} 
 [ e^{- i [\lambda^1_2(x_1, x_2, t, 0) - \lambda^0_2(x_1, x_2, t, 0)]} + 1 ] \ket{\psi_0}\bra{\psi_1} \\
 + [\partial_{x_1} \lambda^2_1(x_1, x_2, t, 0)]~ 
 e^{- i[\lambda^2_1(x_1, x_2, t, 0) - \lambda^3_1(x_1, x_2, t, 0) ]}
 [ e^{- i [\lambda^2_2(x_1, x_2, t, 0) - \lambda^3_2(x_1, x_2, t, 0) ]} + 1 ]\ket{\psi_3}\bra{\psi_2} \\
 + [ \partial_{x_1} \lambda^3_1(x_1, x_2, t, 0)]~ 
 e^{- i[\lambda^3_1(x_1, x_2, t, 0) - \lambda^2_1(x_1, x_2, t, 0) ]}
 [ e^{- i [\lambda^3_2(x_1, x_2, t, 0) - \lambda^2_2(x_1, x_2, t, 0)]} + 1 ]\ket{\psi_2}\bra{\psi_3} 
\bigg] \\ \otimes \ket{x_1 x_2}\bra{x_1 x_2} \nonumber
  \end{multline}\begin{multline}
 - \frac{i a}{2} \sum_{x_1, x_2}\bigg[[\partial_{x_2} \lambda^0_1(x_1, x_2, t, 0)]~e^{- i[\lambda^0_1(x_1, x_2, t, 0) - \lambda^2_1(x_1, x_2, t, 0)] }
 [ e^{- i [\lambda^0_2(x_1, x_2, t, 0) - \lambda^2_2(x_1, x_2, t, 0)]} + 1 ] \ket{\psi_2}\bra{\psi_0} \\ 
 +[\partial_{x_2} \lambda^1_1(x_1, x_2, t, 0)]~  e^{- i[\lambda^1_1(x_1, x_2, t, 0) -\lambda^3_1(x_1, x_2, t, 0) ]}
 [ e^{- i [\lambda^1_2(x_1, x_2, t, 0) - \lambda^3_2(x_1, x_2, t, 0)]} + 1]\ket{\psi_3}\bra{\psi_1} \\
 + [\partial_{x_2} \lambda^2_1(x_1, x_2, t, 0)]~  e^{- i[\lambda^2_1(x_1, x_2, t, 0) -\lambda^0_1(x_1, x_2, t, 0) ]} 
 [e^{- i [\lambda^2_2(x_1, x_2, t, 0) - \lambda^0_2(x_1, x_2, t, 0) ]} + 1 ]\ket{\psi_0} \bra{\psi_2} \\
  + [\partial_{x_2} \lambda^3_1(x_1, x_2, t, 0)]~  e^{- i[\lambda^3_1(x_1, x_2, t, 0) - \lambda^1_1(x_1, x_2, t, 0)] } 
  [e^{- i [\lambda^3_2(x_1, x_2, t, 0) - \lambda^1_2(x_1, x_2, t, 0)]} + 1]\ket{\psi_1}\bra{\psi_3} \bigg] \\
  \otimes \ket{x_1 x_2}\bra{x_1 x_2}\nonumber
  \end{multline}\begin{multline}\label{entantwoham}
  + \frac{ia}{2}\sum_{x_1, x_2}  \sum_{q = 0}^3 \bigg[
 \Big(\partial_{x_1} \lambda^q_2(x_1, x_2, t, 0) + \partial_{x_2} \lambda^q_2(x_1, x_2, t, 0)\Big) \ket{\psi_q} \bra{\psi_q}\bigg] \otimes \ket{x_1 x_2}\bra{x_1 x_2} 
\\
 -  \bigg[  e^{- i [\lambda^0_1(x_1, x_2, t, 0) + \lambda^0_2(x_1, x_2, t, 0) - \lambda^1_1(x_1, x_2, t, 0) - \lambda^1_2(x_1, x_2, t, 0)]} 
 [\partial_{x_1} \lambda^0_2(x_1, x_2, t, 0)]\ket{\psi_1}\bra{\psi_0} \\
 + e^{- i [\lambda^0_1(x_1, x_2, t, 0) + \lambda^0_2(x_1, x_2, t, 0)- \lambda^2_1(x_1, x_2, t, 0) - \lambda^2_2(x_1, x_2, t, 0)]} 
 [\partial_{x_2} \lambda^0_2(x_1, x_2, t, 0)] \ket{\psi_2} \bra{\psi_0} \\
 + e^{- i [\lambda^1_1(x_1, x_2, t, 0) + \lambda^1_2(x_1, x_2, t, 0)- \lambda^0_1(x_1, x_2, t, 0) - \lambda^0_2(x_1, x_2, t, 0)]} 
 [\partial_{x_1} \lambda^1_2(x_1, x_2, t, 0)] \ket{\psi_0}\bra{\psi_1} \\
 + e^{- i [\lambda^1_1(x_1, x_2, t, 0) + \lambda^1_2(x_1, x_2, t, 0) - \lambda^3_1(x_1, x_2, t, 0) - \lambda^3_2(x_1, x_2, t, 0) ]} 
 [\partial_{x_2} \lambda^1_2(x_1, x_2, t, 0)]\ket{\psi_3}\bra{\psi_1} \\
 + e^{- i [\lambda^2_1(x_1, x_2, t, 0) + \lambda^2_2(x_1, x_2, t, 0)- \lambda^3_1(x_1, x_2, t, 0) - \lambda^3_2(x_1, x_2, t, 0)]}
 [\partial_{x_1} \lambda^2_2(x_1, x_2, t, 0)] \ket{\psi_3}\bra{\psi_2} \\
 + e^{- i [\lambda^2_1(x_1, x_2, t, 0) + \lambda^2_2(x_1, x_2, t, 0)- \lambda^0_1(x_1, x_2, t, 0) - \lambda^0_2(x_1, x_2, t, 0)]}
   [\partial_{x_2} \lambda^2_2(x_1, x_2, t, 0)]\ket{\psi_0} \bra{\psi_2}\\
 + e^{- i [\lambda^3_1(x_1, x_2, t, 0) + \lambda^3_2(x_1, x_2, t, 0)- \lambda^2_1(x_1, x_2, t, 0) - \lambda^2_2(x_1, x_2, t, 0)]} 
   [\partial_{x_1} \lambda^3_2(x_1, x_2, t, 0)]\ket{\psi_2}\bra{\psi_3} \\
 + e^{- i [\lambda^3_1(x_1, x_2, t, 0) + \lambda^3_2(x_1, x_2, t, 0)- \lambda^1_1(x_1, x_2, t, 0) - \lambda^1_2(x_1, x_2, t, 0)]} 
   [\partial_{x_2} \lambda^3_2(x_1, x_2, t, 0)]\ket{\psi_1}\bra{\psi_3} 
  \bigg]\\ \otimes \ket{x_1 x_2}\bra{x_1 x_2} ~.
 \end{multline}

Using the expressions given in (\ref{lambdas}), (\ref{propeigen})
and the Taylor expansion:

$\theta^{qr}_j(x_1, x_2, t, \delta t) =  \theta^{qr}_j(x_1, x_2, t, 0) + \delta t ~ \vartheta^{qr}_j(x_1, x_2, t) + \mathcal{O}(\delta t^2)$ we get following.

  $\sum_{q=0}^3 [\tilde{\lambda^q_1}(x_1, x_2, t, 0) + \tilde{\lambda^q_2}(x_1, x_2, t, 0)] \ket{\psi_q} \bra{\psi_q} = $ 
  \begin{multline}\label{twocalf}
  ~~~~~~~~~~~~~~~~~~ [\vartheta^{00}_1(x_1, x_2, t) + \vartheta^{00}_2(x_1, x_2, t)]~ \sum_{q=0}^3 \ket{\psi_q} \bra{\psi_q}\\
 + [\vartheta^{01}_1(x_1, x_2, t) + \vartheta^{01}_2(x_1, x_2, t)]~\Bigg(\sum_{q=0}^1 \ket{\psi_q} \bra{\psi_q} - \sum_{q=2}^3 \ket{\psi_q} \bra{\psi_q}\Bigg)\\
 + [\vartheta^{10}_1(x_1, x_2, t) + \vartheta^{10}_2(x_1, x_2, t)]~\Bigg(\sum_{q=0,2} \ket{\psi_q} \bra{\psi_q} - \sum_{q=1,3} \ket{\psi_q} \bra{\psi_q}\Bigg) \\
 + [\vartheta^{11}_1(x_1, x_2, t) + \vartheta^{11}_2(x_1, x_2, t)]~\Bigg(\sum_{q=0,3} \ket{\psi_q} \bra{\psi_q} - \sum_{q=1,2} \ket{\psi_q} \bra{\psi_q}\Bigg) \\
 = [\vartheta^{00}_1(x_1, x_2, t) + \vartheta^{00}_2(x_1, x_2, t)]~\sigma_0 \otimes \sigma_0 
 + [\vartheta^{01}_1(x_1, x_2, t) + \vartheta^{01}_2(x_1, x_2, t)]~\sigma_0 \otimes \sigma_1 \\
+  [\vartheta^{10}_1(x_1, x_2, t) + \vartheta^{10}_2(x_1, x_2, t)]~\sigma_1 \otimes \sigma_0
 + [\vartheta^{11}_1(x_1, x_2, t) + \vartheta^{11}_2(x_1, x_2, t)]~\sigma_1 \otimes \sigma_1~.
\end{multline}

    The coefficient of $(p_1 \otimes \mathds{1}_2)$ in (\ref{entantwoham})
 = $-\frac{ia}{\hbar}\sum_{q,r = 0}^3 \Theta^1_{qr}(x_1, x_2, t)~ \sigma_q \otimes \sigma_r \otimes \ket{x_1, x_2} \bra{x_1, x_2} \coloneqq$
\begin{multline}
 \frac{ia}{4 \hbar} \sum_{x_1, x_2}\Bigg( - \Re\big\{\mathscr{J}^+_1(x_1, x_2, t) + \mathscr{J}^-_1(x_1, x_2, t) \big\} \sigma_3 \otimes \sigma_0 \\
 + \Im\big\{\mathscr{J}^+_1(x_1, x_2, t) + \mathscr{J}^-_1(x_1, x_2, t) \big\} \sigma_2 \otimes \sigma_0
 - \Re\big\{\mathscr{J}^+_1(x_1, x_2, t) - \mathscr{J}^-_1(x_1, x_2, t) \big\} \sigma_3 \otimes \sigma_1 \\
 + \Im\big\{\mathscr{J}^+_1(x_1, x_2, t) - \mathscr{J}^-_1(x_1, x_2, t) \big\} \sigma_2 \otimes \sigma_1\Bigg) \otimes \ket{x_1, x_2} \bra{x_1, x_2}.
\end{multline}
 Hence, only nonvanishing terms are $\Theta^1_{30}(x_1, x_2, t)$, $\Theta^1_{20}(x_1, x_2, t)$, $\Theta^1_{31}(x_1, x_2, t)$, and $\Theta^1_{21}(x_1, x_2, t)$.

  The coefficient of $(\mathds{1}_1 \otimes p_2)$ in (\ref{entantwoham})
 = $-\frac{ia}{\hbar}\sum_{q,r = 0}^3 \Theta^2_{qr}(x_1, x_2, t)~ \sigma_q \otimes \sigma_r  \otimes \ket{x_1, x_2} \bra{x_1, x_2} \coloneqq$
 \begin{multline}
\frac{ia}{4 \hbar} \sum_{x_1, x_2}\Bigg( - \Re\big\{\mathscr{J}^+_2(x_1, x_2, t) + \mathscr{J}^-_2(x_1, x_2, t) \big\} \otimes \sigma_0 \otimes \sigma_3 \\
 + \Im\big\{\mathscr{J}^+_2(x_1, x_2, t) + \mathscr{J}^-_2(x_1, x_2, t) \big\} \otimes \sigma_0 \otimes \sigma_2
 - \Re\big\{\mathscr{J}^+_2(x_1, x_2, t) - \mathscr{J}^-_2(x_1, x_2, t) \big\} \otimes \sigma_1 \otimes \sigma_3 \\
 + \Im\big\{\mathscr{J}^+_2(x_1, x_2, t) - \mathscr{J}^-_2(x_1, x_2, t) \big\} \otimes \sigma_1 \otimes \sigma_2\Bigg) \otimes \ket{x_1, x_2} \bra{x_1, x_2}.
 \end{multline}

Hence, only nonvanishing terms are $\Theta^2_{03}(x_1, x_2, t)$, $\Theta^2_{02}(x_1, x_2, t)$, $\Theta^2_{13}(x_1, x_2, t)$, and $\Theta^2_{12}(x_1, x_2, t)$.

 Other terms in (\ref{entantwoham}) = 
\begin{multline}
 i a \sum_{x_1, x_2} \frac{1}{2}\big[\partial_{x_1} + \partial_{x_2}\big]\Big[ \theta^{00}_2(x_1, x_2, t, 0)~\sigma_0 \otimes \sigma_0 + \theta^{01}_2(x_1, x_2, t, 0)~\sigma_0 \otimes \sigma_1 \\
+ \theta^{10}_2(x_1, x_2, t, 0)~\sigma_1 \otimes \sigma_0
 + \theta^{11}_2(x_1, x_2, t, 0)~\sigma_1 \otimes \sigma_1 \Big]~\otimes \ket{x_1, x_2} \bra{x_1, x_2} \nonumber
 \end{multline}
 \begin{multline}
 - \frac{ia}{4} \sum_{x_1, x_2} \partial_{x_1}\theta^{00}_1(x_1, x_2, t, 0) \Bigg( - \Re\big\{\mathscr{J}^+_1(x_1, x_2, t) + \mathscr{J}^-_1(x_1, x_2, t) \big\} \otimes \sigma_3 \otimes \sigma_0 \\
 + \Im\big\{\mathscr{J}^+_1(x_1, x_2, t) + \mathscr{J}^-_1(x_1, x_2, t) \big\} \otimes \sigma_2 \otimes \sigma_0
 - \Re\big\{\mathscr{J}^+_1(x_1, x_2, t) - \mathscr{J}^-_1(x_1, x_2, t) \big\} \otimes \sigma_3 \otimes \sigma_1 \\
 + \Im\big\{\mathscr{J}^+_1(x_1, x_2, t) - \mathscr{J}^-_1(x_1, x_2, t) \big\} \otimes \sigma_2 \otimes \sigma_1\Bigg) \otimes \ket{x_1, x_2} \bra{x_1, x_2} \nonumber
\end{multline}
\begin{multline}
 - \frac{ia}{4} \sum_{x_1, x_2} \partial_{x_1}\theta^{01}_1(x_1, x_2, t, 0) \Bigg( - \Re\big\{\mathscr{J}^+_1(x_1, x_2, t) + \mathscr{J}^-_1(x_1, x_2, t) \big\} \otimes \sigma_3 \otimes \sigma_1 \\
 + \Im\big\{\mathscr{J}^+_1(x_1, x_2, t) + \mathscr{J}^-_1(x_1, x_2, t) \big\} \otimes \sigma_2 \otimes \sigma_1
 - \Re\big\{\mathscr{J}^+_1(x_1, x_2, t) - \mathscr{J}^-_1(x_1, x_2, t) \big\} \otimes \sigma_3 \otimes \sigma_0 \\
 + \Im\big\{\mathscr{J}^+_1(x_1, x_2, t) - \mathscr{J}^-_1(x_1, x_2, t) \big\} \otimes \sigma_2 \otimes \sigma_0\Bigg) \otimes \ket{x_1, x_2} \bra{x_1, x_2} \nonumber
\end{multline}
\begin{multline}
 - \frac{a}{4} \sum_{x_1, x_2} \partial_{x_1}\theta^{10}_1(x_1, x_2, t, 0) \Bigg(  \Im\big\{\mathscr{J}^+_1(x_1, x_2, t) + \mathscr{J}^-_1(x_1, x_2, t) \big\} \otimes \sigma_3 \otimes \sigma_0 \\
 + \Re\big\{\mathscr{J}^+_1(x_1, x_2, t) + \mathscr{J}^-_1(x_1, x_2, t) \big\} \otimes \sigma_2 \otimes \sigma_0
 + \Im\big\{\mathscr{J}^+_1(x_1, x_2, t) - \mathscr{J}^-_1(x_1, x_2, t) \big\} \otimes \sigma_3 \otimes \sigma_1 \\
 + \Re\big\{\mathscr{J}^+_1(x_1, x_2, t) - \mathscr{J}^-_1(x_1, x_2, t) \big\} \otimes \sigma_2 \otimes \sigma_1\Bigg) \otimes \ket{x_1, x_2} \bra{x_1, x_2} \nonumber
\end{multline}
\begin{multline}
 - \frac{a}{4} \sum_{x_1, x_2} \partial_{x_1}\theta^{11}_1(x_1, x_2, t, 0) \Bigg(  \Im\big\{\mathscr{J}^+_1(x_1, x_2, t) + \mathscr{J}^-_1(x_1, x_2, t) \big\} \otimes \sigma_3 \otimes \sigma_1 \\
 + \Re\big\{\mathscr{J}^+_1(x_1, x_2, t) + \mathscr{J}^-_1(x_1, x_2, t) \big\} \otimes \sigma_2 \otimes \sigma_1
 + \Im\big\{\mathscr{J}^+_1(x_1, x_2, t) - \mathscr{J}^-_1(x_1, x_2, t) \big\} \otimes \sigma_3 \otimes \sigma_0 \\
 + \Re\big\{\mathscr{J}^+_1(x_1, x_2, t) - \mathscr{J}^-_1(x_1, x_2, t) \big\} \otimes \sigma_2 \otimes \sigma_0\Bigg) \otimes \ket{x_1, x_2} \bra{x_1, x_2} \nonumber
\end{multline}
\begin{multline}
 - \frac{ia}{4} \sum_{x_1, x_2} \partial_{x_2}\theta^{00}_1(x_1, x_2, t, 0) \Bigg( - \Re\big\{\mathscr{J}^+_2(x_1, x_2, t) + \mathscr{J}^-_2(x_1, x_2, t) \big\} \otimes \sigma_0 \otimes \sigma_3 \\
 + \Im\big\{\mathscr{J}^+_2(x_1, x_2, t) + \mathscr{J}^-_2(x_1, x_2, t) \big\} \otimes \sigma_0 \otimes \sigma_2
 - \Re\big\{\mathscr{J}^+_2(x_1, x_2, t) - \mathscr{J}^-_2(x_1, x_2, t) \big\} \otimes \sigma_1 \otimes \sigma_3 \\
 + \Im\big\{\mathscr{J}^+_2(x_1, x_2, t) - \mathscr{J}^-_2(x_1, x_2, t) \big\} \otimes \sigma_1 \otimes \sigma_2\Bigg) \otimes \ket{x_1, x_2} \bra{x_1, x_2} \nonumber
\end{multline}
\begin{multline}
 - \frac{a}{4} \sum_{x_1, x_2} \partial_{x_2}\theta^{01}_1(x_1, x_2, t, 0) \Bigg(  \Im\big\{\mathscr{J}^+_2(x_1, x_2, t) + \mathscr{J}^-_2(x_1, x_2, t) \big\} \otimes \sigma_3 \otimes \sigma_0 \\
 + \Re\big\{\mathscr{J}^+_2(x_1, x_2, t) + \mathscr{J}^-_2(x_1, x_2, t) \big\} \otimes \sigma_0 \otimes \sigma_2
 + \Im\big\{\mathscr{J}^+_2(x_1, x_2, t) - \mathscr{J}^-_2(x_1, x_2, t) \big\} \otimes \sigma_1 \otimes \sigma_3 \\
 + \Re\big\{\mathscr{J}^+_2(x_1, x_2, t) - \mathscr{J}^-_2(x_1, x_2, t) \big\} \otimes \sigma_1 \otimes \sigma_2\Bigg) \otimes \ket{x_1, x_2} \bra{x_1, x_2} \nonumber
\end{multline}
\begin{multline}
 - \frac{ia}{4} \sum_{x_1, x_2} \partial_{x_2}\theta^{10}_1(x_1, x_2, t, 0) \Bigg( - \Re\big\{\mathscr{J}^+_2(x_1, x_2, t) + \mathscr{J}^-_2(x_1, x_2, t) \big\} \otimes \sigma_1 \otimes \sigma_3 \\
 + \Im\big\{\mathscr{J}^+_2(x_1, x_2, t) + \mathscr{J}^-_2(x_1, x_2, t) \big\} \otimes \sigma_1 \otimes \sigma_2
 - \Re\big\{\mathscr{J}^+_2(x_1, x_2, t) - \mathscr{J}^-_2(x_1, x_2, t) \big\} \otimes \sigma_0 \otimes \sigma_3 \\
 + \Im\big\{\mathscr{J}^+_2(x_1, x_2, t) - \mathscr{J}^-_2(x_1, x_2, t) \big\} \otimes \sigma_0 \otimes \sigma_2\Bigg) \otimes \ket{x_1, x_2} \bra{x_1, x_2} \nonumber
\end{multline}
\begin{multline}
 - \frac{a}{4} \sum_{x_1, x_2} \partial_{x_2}\theta^{11}_1(x_1, x_2, t, 0) \Bigg(\Im\big\{\mathscr{J}^+_2(x_1, x_2, t) + \mathscr{J}^-_2(x_1, x_2, t) \big\} \otimes \sigma_1 \otimes \sigma_3 \\
 + \Re\big\{\mathscr{J}^+_2(x_1, x_2, t) + \mathscr{J}^-_2(x_1, x_2, t) \big\} \otimes \sigma_1 \otimes \sigma_2
 + \Im\big\{\mathscr{J}^+_2(x_1, x_2, t) - \mathscr{J}^-_2(x_1, x_2, t) \big\} \otimes \sigma_0 \otimes \sigma_3 \\
 + \Re\big\{\mathscr{J}^+_2(x_1, x_2, t) - \mathscr{J}^-_2(x_1, x_2, t) \big\} \otimes \sigma_0 \otimes \sigma_2\Bigg) \otimes \ket{x_1, x_2} \bra{x_1, x_2} \nonumber
\end{multline}
\begin{multline}
 - \frac{i a}{4} \sum_{x_1, x_2} \partial_{x_1}\theta^{00}_2(x_1, x_2, t, 0) \Big[ - \mathscr{K}^{10}_1(x_1, x_2, t)~\sigma_3 \otimes \sigma_0 
 - \mathscr{K}^{10}_2(x_1, x_2, t)~\sigma_3 \otimes \sigma_1  \\
 + \mathscr{K}^{10}_3(x_1, x_2, t)~\sigma_2 \otimes \sigma_0 + \mathscr{K}^{10}_4(x_1, x_2, t)~\sigma_2 \otimes \sigma_1 \Big] \otimes \ket{x_1, x_2} \bra{x_1, x_2} \nonumber
\end{multline}
\begin{multline}
 - \frac{i a}{4} \sum_{x_1, x_2} \partial_{x_1}\theta^{01}_2(x_1, x_2, t, 0) \Big[ - \mathscr{K}^{10}_2(x_1, x_2, t)~\sigma_3 \otimes \sigma_0 
 - \mathscr{K}^{10}_1(x_1, x_2, t)~\sigma_3 \otimes \sigma_1  \\
 + \mathscr{K}^{10}_4(x_1, x_2, t)~\sigma_2 \otimes \sigma_0 + \mathscr{K}^{10}_3(x_1, x_2, t)~\sigma_2 \otimes \sigma_1 \Big] \otimes \ket{x_1, x_2} \bra{x_1, x_2} \nonumber
\end{multline}
\begin{multline}
 - \frac{a}{4} \sum_{x_1, x_2} \partial_{x_1}\theta^{10}_2(x_1, x_2, t, 0) \Big[ \mathscr{K}^{10}_3(x_1, x_2, t)~\sigma_3 \otimes \sigma_0 
 + \mathscr{K}^{10}_4(x_1, x_2, t)~\sigma_3 \otimes \sigma_1  \\
 + \mathscr{K}^{10}_1(x_1, x_2, t)~\sigma_2 \otimes \sigma_0 + \mathscr{K}^{10}_2(x_1, x_2, t)~\sigma_2 \otimes \sigma_1 \Big] \otimes \ket{x_1, x_2} \bra{x_1, x_2} \nonumber
\end{multline}
\begin{multline}
 - \frac{a}{4} \sum_{x_1, x_2} \partial_{x_1}\theta^{11}_2(x_1, x_2, t, 0) \Big[ \mathscr{K}^{10}_4(x_1, x_2, t)~\sigma_3 \otimes \sigma_0 
 + \mathscr{K}^{10}_3(x_1, x_2, t)~\sigma_3 \otimes \sigma_1  \\
 + \mathscr{K}^{10}_2(x_1, x_2, t)~\sigma_2 \otimes \sigma_0 + \mathscr{K}^{10}_1(x_1, x_2, t)~\sigma_2 \otimes \sigma_1 \Big] \otimes \ket{x_1, x_2} \bra{x_1, x_2} \nonumber
\end{multline}
\begin{multline}
 - \frac{i a}{4} \sum_{x_1, x_2} \partial_{x_2}\theta^{00}_2(x_1, x_2, t, 0) \Big[ - \mathscr{K}^{01}_1(x_1, x_2, t)~\sigma_0 \otimes \sigma_3 
 - \mathscr{K}^{01}_2(x_1, x_2, t)~\sigma_1 \otimes \sigma_3  \\
 + \mathscr{K}^{01}_3(x_1, x_2, t)~\sigma_0 \otimes \sigma_2 + \mathscr{K}^{01}_4(x_1, x_2, t)~\sigma_1 \otimes \sigma_2 \Big] \otimes \ket{x_1, x_2} \bra{x_1, x_2} \nonumber
\end{multline}
\begin{multline}
 - \frac{a}{4} \sum_{x_1, x_2} \partial_{x_2}\theta^{01}_2(x_1, x_2, t, 0) \Big[\mathscr{K}^{01}_3(x_1, x_2, t)~\sigma_0 \otimes \sigma_3 
 + \mathscr{K}^{01}_4(x_1, x_2, t)~\sigma_1 \otimes \sigma_3  \\
 + \mathscr{K}^{01}_1(x_1, x_2, t)~\sigma_0 \otimes \sigma_2 + \mathscr{K}^{01}_2(x_1, x_2, t)~\sigma_1 \otimes \sigma_2 \Big] \otimes \ket{x_1, x_2} \bra{x_1, x_2} \nonumber
\end{multline}
\begin{multline}
 - \frac{i a}{4} \sum_{x_1, x_2} \partial_{x_2}\theta^{10}_2(x_1, x_2, t, 0) \Big[ - \mathscr{K}^{01}_2(x_1, x_2, t)~\sigma_0 \otimes \sigma_3 
 - \mathscr{K}^{01}_1(x_1, x_2, t)~\sigma_1 \otimes \sigma_3  \\
 + \mathscr{K}^{01}_4(x_1, x_2, t)~\sigma_0 \otimes \sigma_2 + \mathscr{K}^{01}_3(x_1, x_2, t)~\sigma_1 \otimes \sigma_2 \Big] \otimes \ket{x_1, x_2} \bra{x_1, x_2} \nonumber
\end{multline}
\begin{multline}\label{twocall}
 - \frac{a}{4} \sum_{x_1, x_2} \partial_{x_2}\theta^{11}_2(x_1, x_2, t, 0) \Big[\mathscr{K}^{01}_4(x_1, x_2, t)~\sigma_0 \otimes \sigma_3 
 + \mathscr{K}^{01}_3(x_1, x_2, t)~\sigma_1 \otimes \sigma_3  \\
 + \mathscr{K}^{01}_2(x_1, x_2, t)~\sigma_0 \otimes \sigma_2 + \mathscr{K}^{01}_1(x_1, x_2, t)~\sigma_1 \otimes \sigma_2 \Big] \otimes \ket{x_1, x_2} \bra{x_1, x_2}~.
\end{multline}

Here we have used the notations:
 \begin{multline}
  \mathscr{J}^+_1(x_1, x_2, t) \coloneqq e^{-2i[\theta^{10}_1(x_1, x_2, t, 0) + \theta^{11}_1(x_1, x_2, t, 0)]} \Big( e^{-2i[\theta^{10}_2(x_1, x_2, t, 0) + \theta^{11}_2(x_1, x_2, t, 0)]} + 1\Big) \\
  =  \Big[ \cos[2\theta^{10}_1(x_1, x_2, t, 0) + 2\theta^{10}_2(x_1, x_2, t, 0)] \cos[2 \theta^{11}_1(x_1, x_2, t, 0) +  2 \theta^{11}_2(x_1, x_2, t, 0)] \\
  - \sin[2\theta^{10}_1(x_1, x_2, t, 0) + 2\theta^{10}_2(x_1, x_2, t, 0)] \sin[2 \theta^{11}_1(x_1, x_2, t, 0) +  2 \theta^{11}_2(x_1, x_2, t, 0)] \\
  - i \sin[2\theta^{10}_1(x_1, x_2, t, 0) + 2\theta^{10}_2(x_1, x_2, t, 0)] \cos[2 \theta^{11}_1(x_1, x_2, t, 0) +  2 \theta^{11}_2(x_1, x_2, t, 0)] \\
  - i \cos[2\theta^{10}_1(x_1, x_2, t, 0) + 2\theta^{10}_2(x_1, x_2, t, 0)] \sin[2 \theta^{11}_1(x_1, x_2, t, 0) +  2 \theta^{11}_2(x_1, x_2, t, 0)] \\
  + \cos[2\theta^{10}_1(x_1, x_2, t, 0)] \cos[2 \theta^{11}_1(x_1, x_2, t, 0)] - \sin[2\theta^{10}_1(x_1, x_2, t, 0)] \sin[2 \theta^{11}_1(x_1, x_2, t, 0)] \\
  - i\sin[2\theta^{10}_1(x_1, x_2, t, 0)] \cos[2 \theta^{11}_1(x_1, x_2, t, 0)] - i \cos[2\theta^{10}_1(x_1, x_2, t, 0)] \sin[2 \theta^{11}_1(x_1, x_2, t, 0)] \Big], 
 \end{multline}
 \begin{multline}
  \mathscr{J}^-_1(x_1, x_2, t) \coloneqq e^{-2i[\theta^{10}_1(x_1, x_2, t, 0) - \theta^{11}_1(x_1, x_2, t, 0)]} \Big( e^{-2i[\theta^{10}_2(x_1, x_2, t, 0) - \theta^{11}_2(x_1, x_2, t, 0)]} + 1\Big) \\
  =  \Big[ \cos[2\theta^{10}_1(x_1, x_2, t, 0) + 2\theta^{10}_2(x_1, x_2, t, 0)] \cos[2 \theta^{11}_1(x_1, x_2, t, 0) +  2 \theta^{11}_2(x_1, x_2, t, 0)] \\
  + \sin[2\theta^{10}_1(x_1, x_2, t, 0) + 2\theta^{10}_2(x_1, x_2, t, 0)] \sin[2 \theta^{11}_1(x_1, x_2, t, 0) +  2 \theta^{11}_2(x_1, x_2, t, 0)] \\
  - i \sin[2\theta^{10}_1(x_1, x_2, t, 0) + 2\theta^{10}_2(x_1, x_2, t, 0)] \cos[2 \theta^{11}_1(x_1, x_2, t, 0) +  2 \theta^{11}_2(x_1, x_2, t, 0)] \\
  + i \cos[2\theta^{10}_1(x_1, x_2, t, 0) + 2\theta^{10}_2(x_1, x_2, t, 0)] \sin[2 \theta^{11}_1(x_1, x_2, t, 0) +  2 \theta^{11}_2(x_1, x_2, t, 0)] \\
  + \cos[2\theta^{10}_1(x_1, x_2, t, 0)] \cos[2 \theta^{11}_1(x_1, x_2, t, 0)] + \sin[2\theta^{10}_1(x_1, x_2, t, 0)] \sin[2 \theta^{11}_1(x_1, x_2, t, 0)] \\
  - i\sin[2\theta^{10}_1(x_1, x_2, t, 0)] \cos[2 \theta^{11}_1(x_1, x_2, t, 0)] + i \cos[2\theta^{10}_1(x_1, x_2, t, 0)] \sin[2 \theta^{11}_1(x_1, x_2, t, 0)] \Big], 
 \end{multline}
 \begin{multline}
  \mathscr{J}^+_2(x_1, x_2, t) \coloneqq e^{-2i[\theta^{01}_1(x_1, x_2, t, 0) + \theta^{11}_1(x_1, x_2, t, 0)]} \Big( e^{-2i[\theta^{01}_2(x_1, x_2, t, 0) + \theta^{11}_2(x_1, x_2, t, 0)]} + 1\Big) \\
  =  \Big[ \cos[2\theta^{01}_1(x_1, x_2, t, 0) + 2\theta^{01}_2(x_1, x_2, t, 0)] \cos[2 \theta^{11}_1(x_1, x_2, t, 0) +  2 \theta^{11}_2(x_1, x_2, t, 0)] \\
  - \sin[2\theta^{01}_1(x_1, x_2, t, 0) + 2\theta^{01}_2(x_1, x_2, t, 0)] \sin[2 \theta^{11}_1(x_1, x_2, t, 0) +  2 \theta^{11}_2(x_1, x_2, t, 0)] \\
  - i \sin[2\theta^{01}_1(x_1, x_2, t, 0) + 2\theta^{01}_2(x_1, x_2, t, 0)] \cos[2 \theta^{11}_1(x_1, x_2, t, 0) +  2 \theta^{11}_2(x_1, x_2, t, 0)] \\
  - i \cos[2\theta^{01}_1(x_1, x_2, t, 0) + 2\theta^{01}_2(x_1, x_2, t, 0)] \sin[2 \theta^{11}_1(x_1, x_2, t, 0) +  2 \theta^{11}_2(x_1, x_2, t, 0)] \\
  + \cos[2\theta^{01}_1(x_1, x_2, t, 0)] \cos[2 \theta^{11}_1(x_1, x_2, t, 0)] - \sin[2\theta^{01}_1(x_1, x_2, t, 0)] \sin[2 \theta^{11}_1(x_1, x_2, t, 0)] \\
  - i\sin[2\theta^{01}_1(x_1, x_2, t, 0)] \cos[2 \theta^{11}_1(x_1, x_2, t, 0)] - i \cos[2\theta^{01}_1(x_1, x_2, t, 0)] \sin[2 \theta^{11}_1(x_1, x_2, t, 0)] \Big], 
 \end{multline}
 \begin{multline}
  \mathscr{J}^-_2(x_1, x_2, t) \coloneqq e^{-2i[\theta^{01}_1(x_1, x_2, t, 0) - \theta^{11}_1(x_1, x_2, t, 0)]} \Big( e^{-2i[\theta^{01}_2(x_1, x_2, t, 0) - \theta^{11}_2(x_1, x_2, t, 0)]} + 1\Big) \\
  =  \Big[ \cos[2\theta^{01}_1(x_1, x_2, t, 0) + 2\theta^{01}_2(x_1, x_2, t, 0)] \cos[2 \theta^{11}_1(x_1, x_2, t, 0) +  2 \theta^{11}_2(x_1, x_2, t, 0)] \\
  + \sin[2\theta^{01}_1(x_1, x_2, t, 0) + 2\theta^{01}_2(x_1, x_2, t, 0)] \sin[2 \theta^{11}_1(x_1, x_2, t, 0) +  2 \theta^{11}_2(x_1, x_2, t, 0)] \\
  - i \sin[2\theta^{01}_1(x_1, x_2, t, 0) + 2\theta^{01}_2(x_1, x_2, t, 0)] \cos[2 \theta^{11}_1(x_1, x_2, t, 0) +  2 \theta^{11}_2(x_1, x_2, t, 0)] \\
  + i \cos[2\theta^{01}_1(x_1, x_2, t, 0) + 2\theta^{01}_2(x_1, x_2, t, 0)] \sin[2 \theta^{11}_1(x_1, x_2, t, 0) +  2 \theta^{11}_2(x_1, x_2, t, 0)] \\
  + \cos[2\theta^{01}_1(x_1, x_2, t, 0)] \cos[2 \theta^{11}_1(x_1, x_2, t, 0)] + \sin[2\theta^{01}_1(x_1, x_2, t, 0)] \sin[2 \theta^{11}_1(x_1, x_2, t, 0)] \\
  - i\sin[2\theta^{01}_1(x_1, x_2, t, 0)] \cos[2 \theta^{11}_1(x_1, x_2, t, 0)] + i \cos[2\theta^{01}_1(x_1, x_2, t, 0)] \sin[2 \theta^{11}_1(x_1, x_2, t, 0)] \Big], 
 \end{multline}
 \begin{align}
  \mathscr{K}^{10}_1(x_1, x_2, t) \coloneqq \hspace{8cm}\nonumber\\
  2 \cos[2 \theta^{10}_1(x_1, x_2, t, 0) + 2 \theta^{10}_2(x_1, x_2, t, 0)] \cos[2 \theta^{11}_1(x_1, x_2, t, 0) + 2 \theta^{11}_2(x_1, x_2, t, 0)], \nonumber\\
  \mathscr{K}^{01}_1(x_1, x_2, t) \coloneqq  \hspace{8cm}\nonumber\\
  2 \cos[2 \theta^{01}_1(x_1, x_2, t, 0) + 2 \theta^{01}_2(x_1, x_2, t, 0)] \cos[2 \theta^{11}_1(x_1, x_2, t, 0) + 2 \theta^{11}_2(x_1, x_2, t, 0)], 
 \end{align}
 \begin{align}
  \mathscr{K}^{10}_2(x_1, x_2, t) \coloneqq  \hspace{8cm}\nonumber\\
 - 2 \sin[2 \theta^{10}_1(x_1, x_2, t, 0) + 2 \theta^{10}_2(x_1, x_2, t, 0)] \sin[2 \theta^{11}_1(x_1, x_2, t, 0) + 2 \theta^{11}_2(x_1, x_2, t, 0)], \nonumber\\
 \mathscr{K}^{01}_2(x_1, x_2, t) \coloneqq  \hspace{8cm}\nonumber\\
 - 2 \sin[2 \theta^{01}_1(x_1, x_2, t, 0) + 2 \theta^{01}_2(x_1, x_2, t, 0)] \sin[2 \theta^{11}_1(x_1, x_2, t, 0) + 2 \theta^{11}_2(x_1, x_2, t, 0)],
 \end{align}
  \begin{align}
  \mathscr{K}^{10}_3(x_1, x_2, t) \coloneqq  \hspace{8cm}\nonumber\\
 - 2 \sin[2 \theta^{10}_1(x_1, x_2, t, 0) + 2 \theta^{10}_2(x_1, x_2, t, 0)] \cos[2 \theta^{11}_1(x_1, x_2, t, 0) + 2 \theta^{11}_2(x_1, x_2, t, 0)],\nonumber\\
 \mathscr{K}^{01}_3(x_1, x_2, t) \coloneqq  \hspace{8cm}\nonumber\\
 - 2 \sin[2 \theta^{01}_1(x_1, x_2, t, 0) + 2 \theta^{01}_2(x_1, x_2, t, 0)] \cos[2 \theta^{11}_1(x_1, x_2, t, 0) + 2 \theta^{11}_2(x_1, x_2, t, 0)],
 \end{align}
   \begin{align}
  \mathscr{K}^{10}_4(x_1, x_2, t) \coloneqq  \hspace{8cm}\nonumber\\
 - 2 \cos[2 \theta^{10}_1(x_1, x_2, t, 0) + 2 \theta^{10}_2(x_1, x_2, t, 0)] \sin[2 \theta^{11}_1(x_1, x_2, t, 0) + 2 \theta^{11}_2(x_1, x_2, t, 0)],\nonumber\\
 \mathscr{K}^{01}_4(x_1, x_2, t) \coloneqq  \hspace{8cm}\nonumber\\
 - 2 \cos[2 \theta^{01}_1(x_1, x_2, t, 0) + 2 \theta^{01}_2(x_1, x_2, t, 0)] \sin[2 \theta^{11}_1(x_1, x_2, t, 0) + 2 \theta^{11}_2(x_1, x_2, t, 0)].
 \end{align}
 
 Collecting all the terms from eqs.~(\ref{twocalf})-(\ref{twocall}), the Hamiltonian can be written in the following form.
\begin{align}\label{twoenham}
 \mathscr{H}^\text{two}(t) = \sum_{x_1, x_2} \sum_{q,r = 0}^3 \Theta^1_{qr}(x_1, x_2, t) \big[\sigma_q \otimes \sigma_r\big] \otimes \ket{x_1, x_2}\bra{x_1, x_2}\big[ p_1 c \otimes \mathds{1}_2\big] \nonumber\\
 +  \Theta^2_{qr}(x_1, x_2, t) \big[\sigma_q \otimes \sigma_r\big] \otimes \ket{x_1, x_2}\bra{x_1, x_2}\big[\mathds{1}_1 \otimes p_2 c\big] \nonumber\\
 + \Xi_{qr}(x_1, x_2, t) \big[\sigma_q \otimes \sigma_r\big] \otimes \ket{x_1, x_2}\bra{x_1, x_2}
\end{align}
where only nonvanishing terms are
$\Theta^1_{30},$ $ \Theta^1_{20}, $ $\Theta^1_{31}, $ $\Theta^1_{21}, $ $\Theta^2_{03}, $ $\Theta^2_{02}, $ $\Theta^2_{13}, $ $\Theta^2_{12},$
 $\Xi_{30}, $ $\Xi_{20}, $ $\Xi_{31}, $ $\Xi_{21}, $ $\Xi_{03}, $ $\Xi_{02}, $ $\Xi_{13}, $ $\Xi_{12}, $ $\Xi_{00}, $ $\Xi_{01}, $ $\Xi_{10}, $ $\Xi_{11}.$

 \end{appendices}

\end{document}